\documentclass[twocolumn]{aastex631}

\usepackage{graphicx}

\usepackage{xcolor}
\usepackage{subfigure}
\usepackage{tikz}
\usepackage{csquotes}
\usepackage{scalefnt}
\usepackage{tabularx}
\usepackage{amssymb}
\usepackage{orcidlink}
\usepackage{array}
\usepackage{enumitem}
\usepackage{multirow}
\usepackage{pifont}
\usepackage{subfiles}
\usepackage{amsmath}

\usetikzlibrary{decorations.pathreplacing,positioning, arrows.meta}

\begin{document}

%\title{A search for Kilonovae Associated with Neutron Star-Black Hole Mergers: A case study of S230518h, GW230529, S230627c and the low-significance candidate S240422ed}
\title{Limits on the Ejecta Mass During the Search for Kilonovae Associated with Neutron Star-Black Hole Mergers: A case study of S230518h, GW230529, S230627c and the Low-Significance Candidate S240422ed}

\author[0000-0003-3224-2146]{M.~Pillas}
\affiliation{STAR Institute, Liege University, Allée du Six Août, 19C B-4000 Liege, Belgium}
\affiliation{Universit\'e de la Côte d'Azur, ARTEMIS, Nice, France}
 %ok

\author[0000-0002-7686-3334]{S.~Antier}
\affiliation{Universit\'e de la Côte d'Azur, ARTEMIS, Nice, France}
\affiliation{IJCLab, Univ Paris-Saclay, CNRS/IN2P3, Orsay, France} %ok

\author{K.~Ackley}
\affiliation{Department of Physics, University of Warwick, Gibbet Hill Road, Coventry CV4 7AL, UK}  % GOTO team

\author[0000-0002-2184-6430]{T.~Ahumada}
\affiliation{Division of Physics, Mathematics and Astronomy, California Institute of Technology, Pasadena, CA 91125, USA} %ok ZTF

%\author{V.~Aivazyan}
%\affiliation{E. Kharadze Georgian National Astrophysical Observatory, Mt.Kanobili, Abastumani, 0301, Adigeni, Georgia}

\author[0009-0006-4358-9929]{D.~Akl}
\affiliation{American University of Sharjah, Physics Department, PO Box 26666, Sharjah, UAE}%ok PWT

\author[0000-0001-8179-1147]{L.~de~Almeida}
\affiliation{Laboratório Nacional de Astrofísica, Rua Estado Unidos 154, 37504-364, Itajubá - MG, Brazil}%ok PWT

\author[0000-0003-3768-7515]{S.~Anand}
\affiliation{Division of Physics, Mathematics and Astronomy, California Institute of Technology, Pasadena, CA 91125, USA}%ok ZTF

\author[0009-0004-9687-3275]{C.~Andrade}
\affiliation{School of Physics and Astronomy, University of Minnesota, Minneapolis, MN 55455, USA}%ok HE

\author[0000-0002-8977-1498]{I.~Andreoni}
\affil{Department of Physics and Astronomy, University of North Carolina at Chapel Hill, Chapel Hill, NC 27599-3255, USA}%DECAM

\author[0000-0002-4924-444X]{K. A.~Bostroem}
\affiliation{Steward Observatory, University of Arizona, 933 North Cherry Avenue, Tucson, AZ 85721-0065, USA}
\altaffiliation{LSST-DA Catalyst Fellow}%SAGUARO

\author[0000-0002-8255-5127]{M.~Bulla}
\affiliation{Department of Physics and Earth Science, University of Ferrara, via Saragat 1, I-44122 Ferrara, Italy}
\affiliation{INFN, Sezione di Ferrara, via Saragat 1, I-44122 Ferrara, Italy}
\affiliation{INAF, Osservatorio Astronomico d’Abruzzo, via Mentore Maggini snc, 64100 Teramo, Italy}%HE

\author[0000-0002-2942-3379]{E.~Burns}
\affiliation{Department of Physics \& Astronomy, Louisiana State University, Baton Rouge, LA 70803, USA}%HE

\author[0000-0002-1270-7666]{T.~Cabrera}
\affil{McWilliams Center for Cosmology and Astrophysics, Department of Physics, Carnegie Mellon University, Pittsburgh, PA 15213, USA}%DECAM

\author[0000-0002-3118-8275]{S.~Chang}
\affiliation{SNU Astronomy Research Center, Astronomy Program, Department of Physics \& Astronomy, Seoul National University, 1 Gwanak-ro, Gwanak-gu, Seoul,  08826, Republic of Korea} %KMTNET

\author[0000-0003-4422-6426]{H.~Choi}
\affiliation{SNU Astronomy Research Center, Astronomy Program, Department of Physics \& Astronomy, Seoul National University, 1 Gwanak-ro, Gwanak-gu, Seoul,  08826, Republic of Korea} %KMTNET

%\author{O.~Burkhonov}
%\affiliation{Ulugh Beg Astronomical Institute, Uzbekistan Academy of Sciences, Astronomy str. 33, Tashkent 100052, Uzbekistan}

\author[0000-0002-9700-0036]{B.~O'Connor}
\affiliation{McWilliams Center for Cosmology and Astrophysics, Department of Physics, Carnegie Mellon University, Pittsburgh, PA 15213, USA}%DECAM

\author[0000-0002-8262-2924]{M. W. Coughlin}
\affiliation{School of Physics and Astronomy, University of Minnesota, Minneapolis, MN 55455, USA}%ok

\author[0000-0002-5150-5282]{W.~Corradi}
\affiliation{Laboratório Nacional de Astrofísica, Rua Estado Unidos 154, 37504-364, Itajubá - MG, Brazil}%ok

\author[0000-0002-2575-2618]{A.~R.~Gibbs}
\affiliation{Lunar and Planetary Laboratory, University of Arizona, 1629 East University Boulevard, Tucson, AZ 85721-0092, USA}%SAGUARO

\author[0000-0003-2374-307X]{T.~Dietrich}
\affiliation{Institut für Physik und Astronomie, Universität Potsdam, Haus 28, Karl-Liebknecht-Str. 24/25, 14476, Potsdam, Germany}
\affiliation{Max Planck Institute for Gravitational Physics (Albert Einstein Institute), Am Mühlenberg 1, Potsdam 14476, Germany}%ok HE

%\author{V.~Deloupy}
%\affiliation{DC20 Concordia Station}

\author[0000-0001-5729-1468]{D.~Dornic}
\affiliation{CPPM, Aix Marseille Univ, CNRS/IN2P3, CPPM, Marseille, France}%ok HE

\author{J.-G.~Ducoin}
\affiliation{CPPM, Aix Marseille Univ, CNRS/IN2P3, CPPM, Marseille, France}%ok HE

\author[0000-0002-3906-0997]{P.-A.~Duverne}
\affiliation{Université Paris Cité, CNRS, Astroparticule et Cosmologie, F-75013 Paris, France}%ok HE

\author{M.~Dyer}
\affiliation{Astrophysics Research Cluster, School of Mathematical and Physical Sciences, University of Sheffield, Sheffield, S3 7RH, UK}  % GOTO team

\author[0000-0001-5296-7035]{H.-B.~Eggenstein}
\affiliation{Volkssternwarte Paderborn, Im Schlosspark 13, 33104 Paderborn, Germany}

%\author{E.~G.~Elhosseiny}
%\affiliation{National Research Institute of Astronomy and Geophysics (NRIAG), 1 El-marsad St., 11421 Helwan, Cairo, Egypt}

\author{M.~Freeberg}
\affiliation{KNC, AAVSO, Hidden Valley Observatory(HVO), Colfax, WI.; iTelescope, NMS, Mayhill, NM.}

\author{M.~Fausnaugh}
\affiliation{Department of Physics and Kavli Institute for Astrophysics and Space Research, Massachusetts Institute of Technology, 77 Massachusetts
Ave, Cambridge, MA 02139, USA}%TESS

\author[0000-0002-7374-935X]{W.~Fong}
\affiliation{Center for Interdisciplinary Exploration and Research in Astrophysics (CIERA) and Department of Physics and Astronomy, Northwestern University, Evanston, IL 60208, USA}%SAGUARO

\author[0000-0003-4617-4738]{F.~Foucart}
\affiliation{Department of Physics \& Astronomy, University of New Hampshire, 105 Main St, Durham, NH 03824, USA}%ok

\author[0000-0002-7197-9004]{D.~Frostig}
\affil{Center for Astrophysics | Harvard \& Smithsonian, 60 Garden Street, Cambridge, MA 02138, USA}%winter

\author[0000-0003-1585-8205]{N.~Guessoum}
\affiliation{American University of Sharjah, Physics Department, PO Box 26666, Sharjah, UAE}%HE

%\author[0000-0002-7188-8428]{T.~Guillot}
%\affiliation{Laboratoire Lagrange, Université de Nice-Sophia Antipolis, Observatoire de la Côte d’Azur, 06304 NICE Cedex 04, France}

\author[0000-0002-7672-0480]{V.~Gupta}
\affiliation{School of Physics and Astronomy, University of Minnesota, Minneapolis, MN 55455, USA}%DECAM

%\author{E.~Gurbanov}
%\affiliation{N.Tusi Shamakhy Astrophysical Observatory Azerbaijan National Academy of Sciences, settl.Y. Mammadaliyev, AZ 5626, Shamakhy, Azerbaijan}

%\author[0009-0006-0547-1030]{P.~Gokuldass}
%\affiliation{Department of Physical Sciences, Embry-Riddle Aeronautical University, 1 Aerospace Boulevard, Daytona Beach, Fl 32114, USA}

%\author{R.~Hainich}
%\affiliation{University of Potsdam, Am Neuen Palais 10, 14469 Potsdam, Germany}

\author{P.~Hello}
\affiliation{IJCLab, Univ Paris-Saclay, CNRS/IN2P3, Orsay, France}%PWT

\author[0000-0002-0832-2974]{G.~Hosseinzadeh}
\affiliation{Steward Observatory, University of Arizona, 933 North Cherry Avenue, Tucson, AZ 85721-0065, USA}
\affiliation{Department of Astronomy Astrophysics, University of California, San Diego, 9500 Gilman Drive, MC 0424, La Jolla, CA 92093-0424, USA}%SAGUARO

%\author[0000-0003-4253-656X]{D. A. Howell}
%\affliliation{Las Cumbres Observatory, 6740 Cortona Drive, Suite 102, Goleta, CA 93117-5575, USA}
%\affiliation{Department of Physics, University of California, Santa Barbara, CA 93106-9530, USA}%SAGUARO

\author[0000-0001-7201-1938]{L.~Hu}
\affiliation{McWilliams Center for Cosmology and Astrophysics, Department of Physics, Carnegie Mellon University, Pittsburgh, PA 15213, USA}%DECAM

\author[0009-0009-2434-432X]{T.~Hussenot-Desenonges}
\affiliation{IJCLab, Univ Paris-Saclay, CNRS/IN2P3, Orsay, France}%okHE

\author[0000-0002-8537-6714]{M.~Im}
\affiliation{SNU Astronomy Research Center, Astronomy Program, Department of Physics \& Astronomy, Seoul National University, 1 Gwanak-ro, Gwanak-gu, Seoul,  08826, Republic of Korea} %KMTNET

\author[0000-0002-7778-3117]{R.~Jayaraman}
\affiliation{Department of Physics and Kavli Institute for Astrophysics and Space Research, Massachusetts Institute of Technology, 77 Massachusetts
Ave, Cambridge, MA 02139, USA}%TESS

\author[0009-0003-1280-0099]{M.~Jeong}
\affiliation{SNU Astronomy Research Center, Astronomy Program, Department of Physics \& Astronomy, Seoul National University, 1 Gwanak-ro, Gwanak-gu, Seoul,  08826, Republic of Korea} %KMTNET

\author[0000-0003-2758-159X]{V.~Karambelkar}
\affil{Cahill Center for Astrophysics, California Institute of Technology, Pasadena, CA 91125, USA}%winter

\author[0000-0003-0035-651X]{S.~Karpov}
\affiliation{FZU - Institute of Physics of the Czech Academy of Sciences, Na Slovance 1999/2, CZ-182 21, Praha, Czech Republic}%ok GRANDMA

\author[0000-0002-5619-4938]{M.~Kasliwal}
\affil{Cahill Center for Astrophysics, California Institute of Technology, Pasadena, CA 91125, USA}%winter

%\author[0000-0002-6653-0915]{R.~Inasaridze}
%\affiliation{E. Kharadze Georgian National Astrophysical Observatory, Mt.Kanobili, Abastumani, 0301, Adigeni, Georgia}

%\author{A.~Ishankar}
%\affiliation{Xinjiang Astronomical Observatory, Chinese Academy of Sciences, Urumqi, Xinjiang, 830011 PR China}

\author[0000-0002-5740-7747]{C.~D.~Kilpatrick}
\affiliation{Center for Interdisciplinary Exploration and Research in Astrophysics (CIERA), Northwestern University, Evanston, IL 60208, USA}%SAGUARO

\author[0000-0002-0070-1582]{S.~Kim}
\affiliation{SNU Astronomy Research Center, Astronomy Program, Department of Physics \& Astronomy, Seoul National University, 1 Gwanak-ro, Gwanak-gu, Seoul,  08826, Republic of Korea} %KMTNET

\author[0000-0001-5249-4354]{N.~Kochiashvili}
\affiliation{E. Kharadze Georgian National Astrophysical Observatory, Mt.Kanobili, Abastumani, 0301, Adigeni, Georgia}%okGRANDMA

%\author{P.~Jacquiery}
%\affiliation{Dunedin Astronomical Society (DAS), Royal Astronomical Society of New Zealand}

%\author[0009-0003-6181-4526]{T.~Jegou~du~Laz}
%\affiliation{Division of Physics, Mathematics, and Astronomy, California Institute of Technology, Pasadena, CA 91125, USA}

\author[0009-0000-4830-1484]{K.~Kunnumkai}
\affiliation{McWilliams Center for Cosmology and Astrophysics, Department of Physics, Carnegie Mellon University, Pittsburgh, PA 15213, USA}%DECAM

%\author[0000-0003-0106-4148]{A.~Klotz}
%\affiliation{IRAP, Université de Toulouse, CNRS, UPS, 14 Avenue Edouard Belin, F-31400 Toulouse, France}
%\affiliation{Université Paul Sabatier Toulouse III, Université de Toulouse, 118 route de Narbonne, 31400 Toulouse, France}

\author[0000-0002-8860-5826]{M.~Lamoureux}
\affiliation{Centre for Cosmology, Particle Physics and Phenomenology - CP3, Université Catholique de Louvain, B-1348 Louvain-la-Neuve, Belgium}%okHE

%\author[0000-0002-2321-1017]{N.~Leroy}
%\affiliation{IJCLab, Univ Paris-Saclay, CNRS/IN2P3, Orsay, France}

\author[0000-0003-0043-3925]{C.~U.~Lee}
\affiliation{Korea Astronomy and Space Science Institute, 776 Daedeokdae-ro, Yuseong-gu, Daejeon 34055, Korea}%KMTET

\author[0000-0002-4585-9981]{N.~Lourie}
\affiliation{Department of Physics and Kavli Institute for Astrophysics and Space Research, Massachusetts Institute of Technology, 77 Massachusetts
Ave, Cambridge, MA 02139, USA} 
%WINTER

\author{J.~Lyman}
\affiliation{Department of Physics, University of Warwick, Gibbet Hill Road, Coventry CV4 7AL, UK}  % GOTO team

\author{F.~Magnani}
\affiliation{CPPM, Aix Marseille Univ, CNRS/IN2P3, CPPM, Marseille, France}%ok PWT

\author[0000-0002-0967-0006]{M.~Mašek}
\affiliation{FZU - Institute of Physics of the Czech Academy of Sciences, Na Slovance 1999/2, CZ-182 21, Praha, Czech Republic}%ok GRANDMA

\author[0000-0001-6331-112X]{G.~Mo}
\affiliation{Department of Physics and Kavli Institute for Astrophysics and Space Research, Massachusetts Institute of Technology, 77 Massachusetts
Ave, Cambridge, MA 02139, USA} 
\affiliation{MIT LIGO Laboratory, Massachusetts Institute of Technology, Cambridge, MA 02139, USA}%TESS%winter

\author[0000-0002-3072-8671]{M.~Molham}
\affiliation{National Research Institute of Astronomy and Geophysics (NRIAG), 1 El-marsad St., 11421 Helwan, Cairo, Egypt}%ok GRANDMA

\author[0000-0002-0284-0578]{F.~Navarete}
\affiliation{SOAR Telescope/NSF’s NOIRLab, Avda Juan Cisternas 1500, 1700000, La Serena, Chile}%ok GRANDMA

\author{D.~O'Neill}
\affiliation{Department of Physics, University of Warwick, Gibbet Hill Road, Coventry CV4 7AL, UK}  % GOTO team

\author[0000-0002-2555-3192]{M.~Nicholl}
\affiliation{Astrophysics Research Centre, School of Mathematics and Physics, Queen's University Belfast, Belfast BT7 1NN, UK} %ATLAS

\author[0000-0002-1850-4587]{A.~H.~Nitz}
\affiliation{Syracuse University, 900 S Crouse Ave, Syracuse, NY 13244, USA}%ok PWT

\author[0000-0001-9109-8311]{K.~Noysena}
\affiliation{National Astronomical Research Institute of Thailand (Public Organization), 260, Moo 4, T. Donkaew, A. Mae Rim, Chiang Mai, 50180, Thailand}%ok GRANDMA

\author[0000-0002-6639-6533]{G.~S.H.~Paek}
\affiliation{SNU Astronomy Research Center, Astronomy Program, Department of Physics \& Astronomy, Seoul National University, 1 Gwanak-ro, Gwanak-gu, Seoul,  08826, Republic of Korea} 
\affiliation{Institute for Astronomy, University of Hawaii, 2680 Woodlawn Drive, Honolulu, HI 96822, USA}%KMTNET

\author[0000-0002-6011-0530]{A.~Palmese}
\affil{McWilliams Center for Cosmology and Astrophysics, Department of Physics, Carnegie Mellon University, Pittsburgh, PA 15213, USA}%DECAM

%\author[0000-0002-8560-4449]{J.~Peloton}
%\affiliation{IJCLab, Univ Paris-Saclay, CNRS/IN2P3, Orsay, France}

\author[0000-0002-9968-2464]{R.~Poggiani}
\affiliation{Dipartimento di Fisica, Universita di Pisa, Largo Bruno Pontecorvo, 3, 56127 Pisa, Italy}%HE

\author[0000-0001-5501-0060]{T.~Pradier}
\affiliation{Université de Strasbourg, CNRS, IPHC UMR 7178, F-67000 Strasbourg, France}%ok HE

%\author[0000-0002-3238-9597]{M.~Prouza}
%\affiliation{FZU - Institute of Physics of the Czech Academy of Sciences, Na Slovance 1999/2, CZ-182 21, Praha, Czech Republic}

\author{O.~Pyshna}
\affiliation{Astronomical Observatory Taras Shevshenko National University of Kyiv, Observatorna str. 3, Kyiv, 04053, Ukraine}%ok PWT

\author{Y.~Rajabov}
\affiliation{Ulugh Beg Astronomical Institute, Uzbekistan Academy of Sciences, Astronomy str. 33, Tashkent 100052, Uzbekistan}%ok PWT

\author[0000-0002-9267-6213]{J.~C.~Rastinejad}
\affiliation{Center for Interdisciplinary Exploration and Research in Astrophysics (CIERA) and Department of Physics and Astronomy, Northwestern University, Evanston, IL 60208, USA}%SAGUARO

%\author{F.~Rünger}
%\affiliation{University of Potsdam, Am Neuen Palais 10, 14469 Potsdam, Germany}

%\author[0000-0002-5162-4222]{T.~Sadibekova}
%\affiliation{Université Paris-Saclay, Université Paris Cité, CEA, CNRS, AIM, 91191, Gif-sur-Yvette, France}

%\author{N.~Sasaki}
%\affiliation{NEPA, Universidade do Estado do Amazonas (UEA), 69.152-510, Parintins, Brasil}

\author[0000-0003-4102-380X]{D.~J.~Sand}
\affiliation{Steward Observatory, University of Arizona, 933 North Cherry Avenue, Tucson, AZ 85721-0065, USA}%SAGUARO

\author[0000-0002-8249-8070]{P.~Shawhan}
\affiliation{University of Maryland, College Park, MD 20742, USA}%Review

\author[0000-0002-4022-1874]{M.~Shrestha}
\affiliation{Steward Observatory, University of Arizona, 933 North Cherry Avenue, Tucson, AZ 85721-0065, USA}%SAGUARO

\author{R.~Simcoe}
\affiliation{Department of Physics and Kavli Institute for Astrophysics and Space Research, Massachusetts Institute of Technology, 77 Massachusetts
Ave, Cambridge, MA 02139, USA} %WINTER

\author[0000-0002-8229-1731]{S.~J.~Smartt}
\affiliation{Astrophysics, Department of Physics,
 University of Oxford, Keble Road, Oxford, OX1 3RH, UK}
\affiliation{Astrophysics Research Centre, School of Mathematics and Physics, Queen's University Belfast, Belfast BT7 1NN, UK}%ATLAS

\author{D.~Steeghs}
\affiliation{Department of Physics, University of Warwick, Gibbet Hill Road, Coventry CV4 7AL, UK}  % GOTO team

\author[0000-0003-2434-0387]{R.~Stein}
\affil{Department of Astronomy, University of Maryland, College Park, MD 20742, USA}
\affil{Joint Space-Science Institute, University of Maryland, College Park, MD 20742, USA} 
\affil{Astrophysics Science Division, NASA Goddard Space Flight Center, Mail Code 661, Greenbelt, MD 20771, USA} %winter

\author[0000-0002-0504-4323]{H.~F.~Stevance}
\affiliation{Astrophysics, Department of Physics,
 University of Oxford, Keble Road, Oxford, OX1 3RH, UK}%ATLAS

\author{M.~Sun}
\affiliation{National Astronomical Research Institute of Thailand (Public Organization), 260, Moo 4, T. Donkaew, A. Mae Rim, Chiang Mai, 50180, Thailand}%ok GRANDMA

\author[0000-0003-1423-5516]{A.~Takey}
\affiliation{National Research Institute of Astronomy and Geophysics (NRIAG), 1 El-marsad St., 11421 Helwan, Cairo, Egypt}%ok GRANDMA

%\author{Y.~Tillayev}
%\affiliation{Ulugh Beg Astronomical Institute, Uzbekistan Academy of Sciences, Astronomy str. 33, Tashkent 100052, Uzbekistan}

%\author[0000-0001-5833-4052]{I.~Tosta~e~Melo}
%\affiliation{Department of Physics and Astronomy, University of Catania, 95125 Catania, Italy}

%\author{P.~Thierry}
%\affiliation{AGORA observatoire des Makes, AGORA, 18 Rue Georges Bizet, Observatoire des Makes, 97421 La Rivière, France}

\author[0009-0008-9546-2035]{A.~Toivonen}
\affiliation{School of Physics and Astronomy, University of Minnesota, Minneapolis, MN 55455, USA}%GW

\author[0000-0003-1835-1522]{D.~Turpin}
\affiliation{Université Paris-Saclay, Université Paris Cité, CEA, CNRS, AIM, 91191, Gif-sur-Yvette, France}%GRANDMA

\author{K.~Ulaczyk}
\affiliation{Department of Physics, University of Warwick, Gibbet Hill Road, Coventry CV4 7AL, UK}  % GOTO team

\author[0000-0002-9998-6732]{A.~Wold}
\affiliation{IPAC, California Institute of Technology, 1200 E. California Blvd., Pasadena, CA 91125, USA}%ZTF

\author[0009-0006-2797-3808]{T.~Wouters}
\affiliation{Institute for Gravitational and Subatomic Physics (GRASP),
Utrecht University, Princetonplein 1, 3584 CC Utrecht, The Netherlands}
\affiliation{Nikhef, Science Park 105, 1098 XG Amsterdam, The Netherlands}%GW
\begin{abstract}
Neutron star–black hole (NSBH) mergers, detectable via their gravitational-wave (GW) emission, are expected to produce kilonovae (KNe). Four NSBH candidates have been identified and followed-up by more than fifty instruments since the start of the fourth GW Observing Run (O4), in May 2023, up to July 2024; however, no confirmed associated KN has been detected. This study evaluates ejecta properties from multi-messenger observations to understand the absence of detectable KN: we use GW public information and joint observations taken from 05.2023 to 07.2024 (LVK, ATLAS, DECam, GECKO, GOTO, GRANDMA, SAGUARO, TESS, WINTER, ZTF). First, our analysis on follow-up observation strategies shows that, on average, more than 50\% of the simulated KNe associated with NSBH mergers reach their peak luminosity around one day after merger in the $g,r,i$- bands, which is not necessarily covered for each NSBH GW candidate. We also analyze the trade-off between observation efficiency and the intrinsic properties of the KN emission, to understand the impact on how these constraints affect our ability to detect the KN, and underlying ejecta properties for each GW candidate. In particular, we can only confirm the kilonova was not missed for 1\% of the GW230529 and S230627c sky localization region, given the large sky localization error of GW230529 and the large distance for S230627c and, their respective KN faint luminosities. More constraining, for S230518h, we infer the dynamical ejecta and post-merger disk wind ejecta $m_{dyn}, m_{wind}$ $<$ $0.03$ $M_\odot$ and the viewing angle $\theta$~$>$~$25^\circ$. Similarly, the non-astrophysical origin of S240422ed is likely further confirmed by the fact that we would have detected even a faint KN at the time and presumed distance of the S240422ed event candidate, within a minimum 45\% credible region of the sky area, that can be larger depending on the KN scenario.

\end{abstract}

\section{Introduction}
\label{Intro}

Gravitational and electromagnetic observations provide complementary information on compact object mergers, including neutron stars (NSs) and black holes (BHs), with implications for areas as disparate as dynamics in the strong-gravity regime, astrophysical environments, cosmology, dense matter, nucleosynthesis, ultrarelativistic particle acceleration and nuclear physics \citep{2019NatRP...1..585M,Burns_2020}. This was clearly demonstrated
on August 17th, 2017, when multi-messenger astronomy with gravitational waves (GWs) started, with the merger of a binary neutron star (BNS) system accompanied by a GW signal, GW170817, detected by the Advanced LIGO \citep{LIGO} and Virgo \citep{Virgo} detectors \citep{mma170817,Abbott_2017}, in coincidence with (1) a gamma-ray burst, GRB 170817A \citep{Goldstein_2017,grb_2017, Abbott_2017_2}, detected by the Fermi Gamma-ray Burst Monitor \citep{2009ApJ...702..791M} and INTEGRAL \citep{Savchenko_2017}, (2) a UVOIR transient, the so-called \textquote{kilonova} (KN) \citep{Abbott_2017_3, Abbott_2017_4, Pian_2017, 2017Natur.551...75S, 2017GCN.21529....1C, Metzger_2019}, and (3) other counterparts across the electromagnetic domain \citep{Alexander_2017, Troja_2017, 2017Sci...358.1579H, D_Avanzo_2018}. This first GW-based multi-messenger event opened new opportunities for studying various topics, from compact object physics to cosmology, among others \citep{Abbott_2017}. Over the last decade, much effort has been put into developing tools, infrastructure and observing collaborations to detect new multi-messenger events \citep{2024Mirro...6....6A} originating not only from BNS mergers, but also collisions between an NS and a BH (NSBH), as the latter are also expected to emit GWs and produce gamma-ray bursts (GRBs) and kilonovae (KNe) \citep{Metzger_2019, 2019MNRAS.486.5289B}.\\ %While the literature usually predicts these GRBs to be short ($<$2 seconds), we now also have some evidence for kilonovae associated with long GRBs \citep{Rastinejad_2022}. 

Produced by either NSBH or BNS mergers, a KN is an ultraviolet, optical, and infrared transient powered by the radioactive decay of heavy elements synthesized via $r$-process nucleosynthesis in the neutron-rich material released by the NS disruption(s) during the compact object merger \citep{Metzger_2019}. Bright electromagnetic emission is not always guaranteed, especially for NSBH mergers. For exemple, large mass ejection from NSBH mergers is possible as long as the NS is tidally disrupted by the BH companion outside of the innermost stable circular orbit of that BH ($R_{\rm ISCO}$) \citep{1974ApJ...192L.145L, Wiggins_2000, Pannarale_2011, PhysRevD.86.124007}. When tidal disruption occurs, most of the matter is rapidly accreted onto the BH, with a mass $m_{rem}$ remaining outside of the BH in the form of an accretion disk, a bound tidal tail, and unbound matter. The subsequent evolution of the BH-disk remnant leads to the ejection of slightly less neutron-rich material in the form of neutrino-driven and magnetically-driven winds but through secular ejecta and viscosity-driven winds \citep{Hayashi_2022, Hayashi_2023, gottlieb2023largescaleevolutionsecondslongrelativistic}. Any ejected neutron-rich matter %ejected into the surrounding interstellar medium 
then undergoes $r$-process nucleosynthesis, with the radioactive decay of the products of the $r$-process powering a KN \citep{Li_1998, Roberts_2011, Kasen_2013, Metzger_2019}.

%as it is required to have the NS disrupted before plunging into the BH, such that matter can be ejected and an accretion disk can be formed for a couple of seconds after the merger. 
%Binary configurations with rapidly spinning, light black holes, typically with M$\sim5$M$_{\odot}$, are favored to have kilonova emission as the neutrons star can be tidally disrupted.
%Binary configurations with rapidly spinning BHs are also favored to have KN emission, if the mass ratio is small enough, such that the neutrons star can be tidally disrupted. 

The quantity of the ejecta depends on intrinsic parameters of the binary such as the mass and spin of the NS and BH and the mass ratio of the binary system, as well as on the equation of state (EOS) of matter above nuclear saturation density \citep{Wiggins1999TidalIB, 1976ApJ...210..549L,PhysRevD.88.041503, Kyutoku_2015, Dietrich_2017}. Specifically, the ratio of the tidal disruption radius to innermost stable circular orbit ($R_{\rm ISCO}$) is known to increase as the component of the BH spin aligned with the angular momentum of the binary increases, as well as the mass ratio ($Q=M_{\rm BH}/M_{\rm NS}$) of the binary and/or the compactness of the NS ($C_{\rm NS}=GM_{\rm NS}/(R_{\rm NS}c^2)$) decreases \citep{Pannarale_2011, PhysRevD.86.124007, 2018PhRvD..98h1501F}. Here $M_{\rm BH}$ and $M_{\rm NS}$ are the BH and NS mass respectively and $R_{\rm NS}$ the NS radius. 
This increase facilitates tidal disruption, mass ejection, and disk formation. For BH spins anti-aligned with the orbital angular momentum, on the other hand, increasing BH spins decreases the likelihood of tidal disruption. If systems with large NS spins exist ($|\chi_{2z}| \gtrsim 0.3 $, \citealt{Dudi_2022}), they would also be more likely to lead to tidal disruption than systems with low or no NS spin.

KNe are widely studied, and numerous models associated with NSBH mergers are now avialable, with the aim of modeling different aspects of their production as accurately as possible. We focus here on %Anand 2021 - Bulla 2019 
\citet{2021NatAs...5...46A} - \citet{10.1093/mnras/stz2495} model, later called \textit{An21Bu19} that describes the KN from NSBH with two ejecta components during and after the merger: the \textit{dynamical ejecta}, $m_{dyn}$ that comes from the tidal disruption of the NS ($\mathcal{O}(ms)$) and unbound material or \textit{disk wind ejecta} from the post-merger accretion disk (longer timescales, referred as $m_{wind}$) \citep{2015PhRvD..92d4028K, 2019MNRAS.486.5289B}. 

%We modelize the KN by two types of ejecta components during and after an NSBH merger: the dynamical ejecta that comes primarily from the tidal disruption of the NS ($<$ 1s) and unbound material or \textit{disk wind ejecta} from the accretion disk post-merger (post 1~s, referred as $m_{wind}$) \citep{2015PhRvD..92d4028K, 2019MNRAS.486.5289B}. The \textit{dynamical ejecta} (referred to as $m_{dyn}$) component is mostly located in the equatorial plane, but can often cover only part of it, introducing a significant dependence on the viewing angle. This ejecta mass can reach $0.1$ M$_{\odot}$ and beyond for the most favorable configurations with sub-relativistic speeds at the 0.1-0.3 c level. This component peaks the day-week timescale at a luminosity of 10$^{40}$-10$^{42}$ erg.s$^{-1}$. On the other hand, the disk mass lies in the [0.01; 0.3] M$_{\odot}$ range with wind-driven outflows masses in the 10$^{-4}$-10$^{-3}$ M$_{\odot}$ range. The wind component peaks at the hour/day timescale with luminosities between 10$^{39}$-10$^{40}$ erg.s$^{-1}$ \citep{2019MNRAS.486.5289B}.

%GW200105 \citep{2021ApJ...915L...5A}
%GW190426\_152155 \citep{2024PhRvD.109b2001A}, GW190814 \citep{2020ApJ...896L..44A}, , and . 

%The first confirmed NSBH GW detection occurred during the O3 run of Advanced LIGO and Advanced Virgo observations named GW200115 

GWs from NSBH collisions have previously been detected by the LIGO-Virgo-KAGRA collaboration (LVK) during observing run O3, with the confident event %GW200105\textunderscore162426  \citep{2021ApJ...915L...5A}
GW200115\textunderscore042309 \citep{2021ApJ...915L...5A}, with 
possible NSBH mergers reported in real time and followed-up (GW190426\textunderscore152155 \citealt{li2020gw190426152155mergerneutronstarblack, 2024PhRvD.109b2001A} - 71 GCNs; GW190814 \citealt{Abbott_2020_GW190814, de_Wet_2021} - 126 GCNs), and later with further offline NSBH candidates \citep{2021ApJ...915L...5A}. %GW200105\textunderscore162426 is known to be the first confident NSBH detection while GW200105\textunderscore162426 is found to be a marginal detection, with a probability of astrophysical $<$ 0.5; still, it remains a candidate of interest.
%and offline or further low significant 
However, no electromagnetic (EM) counterparts were found. Under the assumption that these signals were of astrophysical origin, no GRB signals were observed likely because of orientation, sensitivity, lack of sky coverage or lack of tidal disruption. The lack of KN counterpart is either due to a binary configuration not allowing the launch of a KN, to a potential KN that would have been too faint to be detectable \citep{2021ApJ...921..156Z}, or incomplete coverage of the sky localization area \citep{2024arXiv240517558K, 2024arXiv241018274N}. \\

%, and under the assumption that these signals were of astrophysical origin, this lack of EM counterpart is either due to a binary configuration not allowing the launch of a KN, because a potential KN would have been too faint to be detectable \citep{2021ApJ...921..156Z} or due to non-efficient follow-up observations. \\

The fourth LIGO/Virgo/KAGRA observing run O4 is ongoing and up to 2024 July 24, one NSBH merger has been confirmed, GW230529  \citep{2024ApJ...970L..34A}, which is interpreted as a confident GW signal emitted by the merger of a NS with a compact object with mass between 2.5 and 4.5 M$_{\odot}$. %\citep{2024ApJ...970L..34A}
In addition, three NSBH merger candidates were announced (up to the end of July 2024): S230518h \citep{2023GCN.33813....1L}, observed during the engineering run prior to the start of O4, S230627c \citep{2024GCN.34086}, S240422ed, being now considered as marginal \citep{2024GCN.36236....1L,2024GCN.36240....1L}. However, none of these candidates led to the discovery of a confirmed EM counterpart despite follow-up observations, which were significant for some alerts, especially for S240422ed, when it was still categorized as a significant event before its false alarm rate was downgraded \citep{2024GCN.36812....1L}. Note that another NSBH candidate has been observed on 2025, February 6, S250206dm \citep{2025GCN.39175....1L}, but we do not discuss its follow-up here as its detection occurred too recently.

However, follow-up multi-messenger observations (optical with GW public information) can allow us to set constraints on the astrophysical scenario of compact binary mergers  \citep{2020MNRAS.492..863C,ahumada2024searchinggravitationalwaveoptical,Paek_2024}. In this work, \textbf{we focus on NSBH merger candidates} S230518h, GW230529, S230627c and
S240422ed \textbf{and their follow-up to infer ejecta mass of NSBH mergers} from non-detectable optical kilonovae
(KN) following gravitational wave candidate detections. A constraint on the viewing angle of the eject is also obtained for one of the candidates.

This multi-messenger approach is based on one hand from GW information for each public event from raw data for published events to information contained in the publicly distributed LVK alerts. On the other hand, it takes optical observations triggered by GW alerts thank to collaboration with the Asteroid Terrestrial-impact Last Alert System (ATLAS) \citep{2018PASP..130f4505T, 2024MNRAS.528.2299S}, the Dark Energy Camera (DECam) \citep{cabrera2024searchingelectromagneticemissionagn, 2024arXiv241113673K, 2024arXiv240910651K}, the Gravitational-wave Optical Transient Observer (GOTO) \citep{Steeghs_2022}, the Global Rapid Advanced Network Devoted to Multi-messenger Addicts (GRANDMA) \citep{Antier_2020}, the Gravitational-wave EM Counterpart Korean Observatory (GECKO) \citep{Im_Paek_Kim_Lim_2020, Paek_2024}, the Searches After Gravitational Waves Using ARizona Observatories (SAGUARO; \citealt{2019ApJ...881L..26L,2021ApJ...912..128P,2024ApJ...964...35H}), the Transiting Exoplanet Survey Satellite (TESS) \citep{2015JATIS...1a4003R, 2023ApJ...948L...3M}, the Wide-field infrared transient explorer (WINTER) \citep{WINTER2024, Lourie_2020}, and Zwicky Transient Facility (ZTF) \citet{ahumada2024searchinggravitationalwaveoptical}. 
%Secondly, public reports of additional observations were also included to this work as from BlackGEM \citep{Groot_2024}, Las Cumbres \citep{Brown_2013}, Magellan \citep{mardini2023metalpoorstarsobservedmagellan}, MASTER \citep{Kornilov_2011}, MeerLICHT \citep{2019IAUS..339..203P}, PRIME \citep{Yama_2023} and Swift/UVOT \citep{2005SSRv..120...95R}.\\ 

%For this work, we focus on NSBH merger populations and compare the optical data the community collected on S230518h, GW230529, S230627c and S240422ed. For each event, different levels of GW information are public, from raw data for published events to information contained in the publicly distributed LVK alerts. The method we develop throughout this article consists of testing consistencies between the EM and GW data.%, allowing us to highlight and quantify the tensions between GW theory and EM-follow-up data.

Deriving implications about merger properties from public GW alerts and optical KN follow-up involves several caveats. On the GW side, caveats arise from the limitations of the models to describe all types of NSBH mergers produced by the Universe. In particular, GW models used to analyze recent NSBH events \citep{Thompson_2020, Matas_2020} largely rely on a model for the disruption of the neutron star with $\sim$30\% relative error in the amplitude of the signal \citep{Lackey_2014}. These models also neglected orbital precession due to spin-orbit coupling as well as higher-order multipoles naturally present in unequal mass systems, though more recent work has improved on the first issue \citep{Thompson_2024}. In addition, several sources of uncertainties undermine our ability to strongly constrain the estimation of the parameters. Among those, the uncertainty about the astrophysical origin of the candidate, the accuracy of extrinsic parameters such as the sky localization or distance estimates, along with the measurements of the intrinsic properties of the mergers (chirp mass $\mathcal{M}$, total masses, and spins) are the most prominent. Similarly, merger properties inferred from these analyses can be biased by systematic uncertainties in KN models from semi-analytical codes or radiative transfer codes with approximate prescriptions for the ejecta and for key properties like heating rates, thermalization efficiencies and opacities \citep{Heinzel2021,Bulla2023,Tak2023,Tak2024,Brethauer2024,Fryer2024,Sarin2024,Jhawar2024}.
Finally, we can enumerate several challenges related to the optical follow-up, such as planning and covering the large sky localization search area down to a sufficient sensitivity to reach the expected brightness of the ejecta associated with the NSBH merger. Moreover, the possible diversity in KN and color evolution of the observations across the optical wavelengths also constitute sources of uncertainties. \\

In this work, we focus on NSBH merger populations and compare them with the optical data collected by the community on S230518h, GW230529, S230627c and S240422ed. We take into account the different caveats, comparing theoretical predictions and multi-messenger observations to provide an estimate for the different scenarios of NSBH mergers. In Sec.~\ref{sec:O4campaignmain}, we describe the current O4 observing campaign and highlight the NSBH GW candidates and GW230529. % in the context of the confirmed GW events and O4 alerts. 
We also summarize electromagnetic observations collected from S230518h, GW230529, S230627c, and S240422ed. In particular, we assess the impact and importance of observations occuring at the time of predicted brightness peak, and how it affects the detection. In Sec.~\ref{ejectaKN}, we describe various disruption models of the NS in NSBH binaries, to evaluate the range of ejecta masses that informs the r-process synthesis and KN emission. In Sec.~\ref{MM}, we describe the KN model used in our study and compare results with both observations of GW and optical data to constrain extrinsic and intrinsic parameters of the NSBH candidates and their astrophysical origin. In particular, we propose an indicator to evaluate the possibility of the presence of a KN in the optical observations triggered from the GW events. In Sec.~\ref{discussion}, we discuss our results from this study. 

\section{Gravitational-wave alerts during the O4 Observing Run}
\label{sec:O4campaignmain}
In the following
section, we describe the O4 observing run and the GW alerts that occurred during O4 up to July 24$^{th}$, 2024, and the origin of the dataset used through this paper.

\subsection{Sky localization and distance of the O4 alerts}

The O4 campaign started in May 2023 with only the two LIGO interferometers Hanford (H1) and Livingston (L1) being operational. KAGRA also collected data during the first 4 first weeks of O4 then exited the observation on June 20, 2023 at 23:00 UTC. The O4 campaign is split into three parts, O4a from the end of May 2023 to mid-January 2024 with H1 and L1, O4b starting mid-April 2024 with the Virgo detector joining the network, up to January 28, 2025 and finally O4c is planned for June 2025. The official end is currently planned for October 2025. Up to July 24th, 2024 LVK identified 134 significant CBC candidates, possibly from compact binary mergers (leading to a rate of $\sim$\,3 alerts per week). Among these, 18 were retracted within approximately 1 hour. Additionally, more than 2000 sub-threshold GW candidates were identified with a false alarm rate (FAR) less than $\sim$ 2 per day. None of these have been confirmed as significant or associated with any EM counterpart, despite the effort of a growing number of telescopes or networks of telescopes involved in GW follow-ups.

The O4 alert system is sending public alerts %to the astronomical community 
more rapidly than previous observational campaigns, with a latency of approximately $\sim$ 29 seconds post-merger discovery \citep{2023arXiv230804545S}. The duty cycle, with at least two GW detectors online, was 53.4\% during O4a. The median sky localization area for Hanford-Livingston (HL) detections was 1981 deg$^2$, which is comparable to previous runs. Virgo joined O4b on April 10th, 2024 with a sensitivity of 55\, Mpc. This helped to significantly reduce the sky localization area for some candidates, as low as 8 deg$^2$ (Bayesar, \citealt{Singer_2016}) updated to 5 deg$^2$ (Bilby, \citealt{Ashton_2019}) for the S240615dg BBH candidate \citep{2024GCN.36669,2024GCN.36704}. Unfortunately, no new BNS GW candidate has been confidently found up until now. Four NSBH candidates were detected with only one confirmed, GW230529 \citep{2024ApJ...970L..34A}. The latter was detected by the LIGO Livingston detector only, resulting in a poor sky localization which offered a challenging task for EM follow-up campaigns. %resulting in an almost all-sky localization and therefore a very challenging task for searching for EM counterparts. 
Among the candidates, S240422ed was the event with the most interest for follow-up with telescopes, as the nearest candidate supporting NSBH properties in O4, although its reduction in significance has dampened enthusiasm.

\begin{figure*}
    \centering
    \includegraphics[width=0.9\textwidth]{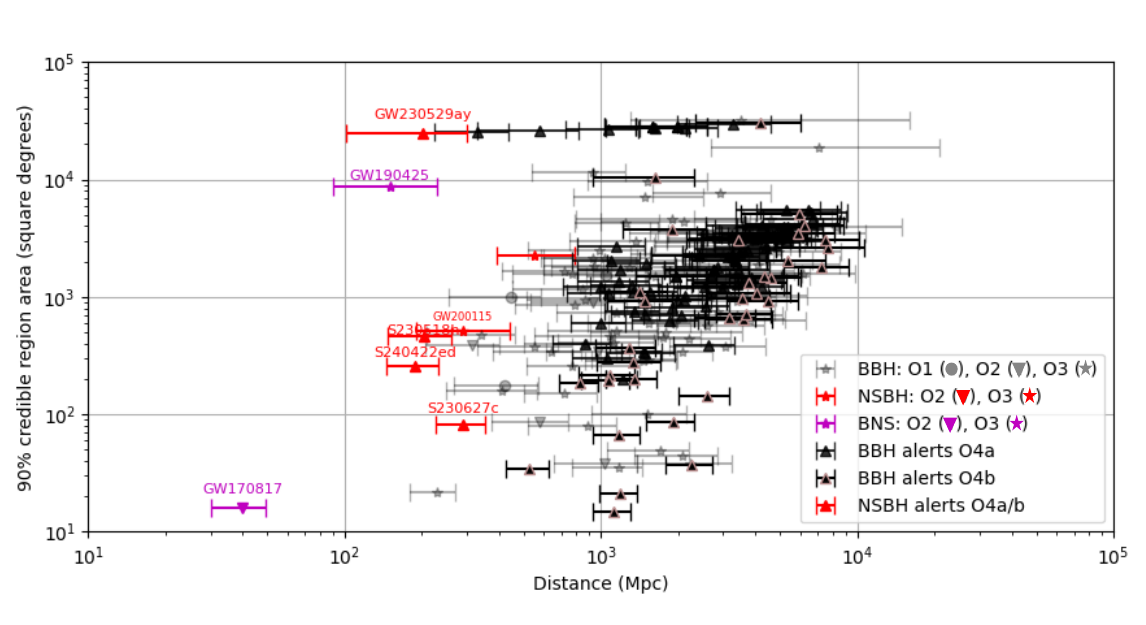}
    \caption{The most recently updated 90\% credible region area versus the most recently updated luminosity distance (posterior mean distance and posterior standard deviation of distance) for all LIGO/Virgo GW events/candidates of runs O1, O2, O3, and O4 a/b (up to 24/07/2024). O1 to O3 events belong to GWTC catalogs \citep{Abbott_2019,2024PhRvD.109b2001A,PhysRevX.13.041039}.}
    \label{fig:alertsO4plan}
\end{figure*}

In Fig.~\ref{fig:alertsO4plan}, we present the %most recently updated 
90\% credible region area versus the %most recently updated 
luminosity distance (posterior mean distance and posterior standard deviation of distance) for all GW events/candidates of runs O1, O2, O3 and O4 a/b up to 2024, July 24. We note a slight disagreement between the distance and the 90 \% credible region of BBH O4 alerts and the prediction from \citet{2023ApJ...958..158K}, with at least 50\% of the expected BBHs detected with a distance below 2 Gpc, while they represent in reality 35\% in O4a and 42\% in O4b. This is due to the actual sensitivity of the network as well as the difference in the duty cycle between prediction and reality (close events which are rare are likely to be missed if the duty cycle degrades). On the other hand, the fraction of localized events (below 500~deg$^2$ as 90~\% credible region) increased from 6\% in O4a to 36\% alerts in O4b. %represented only 6~\% in O4a, while they represent 36~\% of the total alerts in O4b.
In this way, we clearly see the valuable contribution of Virgo to the localization of GW alerts.

%\begin{figure}
%    \centering
%    \includegraphics[width=\columnwidth]{Figures/distanceO4.png}
%    \includegraphics[width=\columnwidth]{Figures/credibleregion.png}
%    \caption{Top: Cumulative density function of BBH Distance for BBH merger alerts during O4a (black line) and O4b (in blue). Bottom: O4 density profile of the 90 \% credible region of the alerts (all types) since the end of May 2024 up to 24-07-2024 (blue line for O4a) and in black for O4b. These distributions are not normalized by the duty cycle.}
%    \label{fig:alertsO4distancecrediblesection2}
%\end{figure}

\subsection{O4 NSBH GW candidate alerts and follow-up campaigns}
\label{followup}

In this article, we focus on several specific cases of GW alerts tagged as \textquote{NSBH} candidates, which prompted observations from the optical community. Details about a subset of optical follow-up in terms of time and spatial coverage can be found in Table~\ref{followupcoverage} (which corresponds to offline results of ATLAS, DECam, GECKO, GOTO, GRANDMA, SAGUARO, TESS, WINTER and ZTF) which is complemented from public results (see Appendix, Table~\ref{tab:coverage-public}). Coverage and upper limit computed thanks to refined analyses provided by collaborations listed Sec.~\ref{Intro} are used in the KN analysis of Secs.\ref{ejectaKN} and~\ref{ejectamass} (Table~\ref{followupcoverage}).
Those computed with public reports are not used in the KN analysis (Table~\ref{tab:coverage-public} in Appendix~\ref{Events}). Therefore, we present only the time and spatial coverage and upper limit with the refined analyses in this section (for inclusion of public report see the Appendix~\ref{Events}). Details of the GW alerts, including their properties, final classifications using low latency investigations from the LVK, and the subsequent multi-band follow-up campaigns conducted, can be also found in Appendix~\ref{Events}. Additionally, we emphasize the observational optical results and the contributions of various instruments from different electromagnetic collaborations involved in this joint effort.

\subsubsection{Summary}

\begin{table*}
\hspace*{-3cm}
\centering
\begin{tabular}{|l|ll|ll|ll|l|}
\hline
Filter & \multicolumn{2}{c|}{0 - 1 day} &  \multicolumn{2}{c|}{1 - 2 day} & \multicolumn{2}{c|}{2 - 6 day} & Instruments \\
& \% c.r & upper &  \% c.r & upper &  \% c.r & upper & \\
\hline
\multicolumn{8}{|c|}{\textbf{S230518h}}\\
\hline
%& clear & 5 \% & 17  & - & - & - & - & MASTER &yes \\
600 - 1000 nm  & 25\% & 16  & 25\% & 16  &  25\% & 16 & TESS  \\
%& $u$-band  & 2 \% & 19  & 33 \% & 18.5 &  11 \% & 18.5  & Swift/UVOT - MeerLICHT & yes \\
$R$-band  & 21\% & 21.5  & 18\% & 21.5 & - & - & GECKO \\ 
$o$-band & 44\% & 18  & - & - & 25\% & 19 & ATLAS  \\
$c$-band   & - & -  & 25\% & 20 & 47\% & 19.5 & ATLAS \\
%& $i$-band  & 1 \% & 21.5  &  33 \% & 19 &   11 \% & 19  & MeerLICHT - SWOPE & - \\
%& $q$-band   & - & -  & 37 \% & 20 &  15 \% & 20  & MeerLICHT & - \\
\hline
\multicolumn{8}{|c|}{\textbf{GW230529}}\\
\hline
%& Clear & 1\% & 21  & $<1$ \% & 20 &  1\% & 21 & MASTER - CSS & yes \\
$L$-band   &  10\% & 20 &  2\% & 19.5 & 2\% & 19 & GOTO  \\
$g$-band & 16\% & 20.5  & - & - & - & - & ZTF  \\
$r$-band & 12\% & 20.5  & - & - & - & - & ZTF  \\
$i$-band & 5\% & 20  & - & - & - & - & ZTF \\
$o$-band   & 2\% & 18  & 4\% & 19 & 23\% & 17.5 & ATLAS  \\
%ztf tess?
\hline
\multicolumn{8}{|c|}{\textbf{S230627c}}\\
\hline
%& clear & $<$ 1 \% & 16  & - & - & - & - & MASTER &yes \\
$L$-band & 45\% & 19 & 84\% & 19 & 23\% & 19 & GOTO \\
$g$-band & 88\% & 21 & - & - & - & - & ZTF  \\
$R$-band & 4\% & 18.5  & 2\% & 21 & 2\% & 21 & GRANDMA, GECKO \\
$r$-band & 88\% & 21 & - & - & - & - & ZTF \\
$o$-band & -  & - & 18\% & 18.5  & 17\% & 18  & ATLAS  \\
\hline
\multicolumn{8}{|c|}{\textbf{S240422ed}}\\
\hline
%& clear & 25\% & 19.3  & - & - & - & - & MASTER - CSS/SAGUARO - GRANDMA  & yes  \\
%& $B$-band & $<$1\%  & 19.5 & - & -  & - & -  & GRANDMA & yes \\
%& $g$-band & 54\% & 20.5 & 83\% & 20.5 & $<$1\% & 22.5 & Las Cumbres, GRANDMA, ZTF & yes \\
$g$-band & 53\% & 19.5 & 83\% & 20 & $<$1\% & 22.5 & GRANDMA, ZTF  \\
%& $q$-band & 97\%  & 19.3 & 94\% & 19.6 & 82\% & 19.9 & MeerLICHT, BlackGEM & - \\
$L$-band & 96\% & 19 &  96\% & 19.5 & 94\% & 20 & GOTO \\
$G$-band & 19\% \% & 19.5 & - & - & - & - & CSS/SAGUARO \\
$R$-band & 69 \% & 17  & 67\% & 21.5 & 22\% & 21.5 & GRANDMA, GECKO \\
$r$-band & 86 \% & 23 & 90\% & 23 & 71 \% & 23 & DECam, GRANDMA, ZTF \\
$i$-band & $<$ 1\% & 19.5 & 17\% & 20.3 & - & - & ZTF \\
$o$-band & 99\%  & 19 & 7\% & 18.5  & 99 \% & 18.5 & ATLAS \\
%& $i$-band & $<1\%$  & 19 & 16 \% & 20 & $<$1\% & 22  & ZTF \\
%& $I$-band & 2\%  & 16.8 & - & -  & - & -  & GRANDMA & yes \\
$z$-band & 75\%  & 22.5 & 81\% & 22.5  & 71\% & 23.0  & DECam \\
%& $J$-band & 22\%  & 17.8 & 14\% & 21.5 & - & -  & WINTER, PRIME & yes \\
$J$-band & 16\%  & 16.5 & - & - & - & -  & WINTER \\
\hline
\end{tabular}
\caption{Coverage fraction (\% c.r) of the most up-to-date sky localization area and order of magnitude of the upper limit for S230518h, GW230529, S230627c and S240422ed (only when totalling 2\% or more from 0 to 6 days post T$_0$ and per filter). We use results of refined analyses of \textquote{tiling and galaxy-targeting} observations provided by collaborations listed in Sec.~\ref{Intro}. We do not include serendipitous coverage from the follow-up of Gravitational wave candidates as they represent quasi-null coverage. Note that differences with coverage directly computed by ZTF and ATLAS collaboration likely arise from the fact that their coverage is computed within the 90\% credible region while we report the total probability coverage. In addition, ZTF also accounts for processing efficiency and ATLAS changed the skymap resolution to compute the coverage, which we did not do. Finally, PS1, DDOTI and LAST observations are not reported while observations have been made. Coverages with less than 2\% of sky localization area are taken into the kilonova upper limit computation if available.}
\label{followupcoverage}
\end{table*}

%As an example, Fig.~\ref{fig:observation-peak-time} summarizes the observations of the community in g, r, i, and J-band for the four candidates considered in this analysis. 

\paragraph{S230518h} The event, discovered prior to O4 (O4a engineering run), observed by the %full 
GW network on 2023-05-18 at 12:59:08 UTC, was classified as 86\% NSBH merger %(with a FAR of 3.2. 10$^{-10}$ Hz) 
(with a FAR of $\sim$1 per 98 yr). The most up-to-date sky localization was refined to 460 deg$^2$, and a distance of 204 $\pm$ 57 Mpc (see Fig.~\ref{fig:alertsO4plan}, \citealt{2023GCN.33884....1L}). In total, twenty teams reported follow-up observations of the event covering the neutrinos, gamma-rays, X-ray, UVOIR and radio domains. Among them, five teams reported follow-up on the event promptly in optical to 14 days (see Appendix~\ref{Events}). In the optical range, there was a total of 81\% coverage of the most recently updated sky localization for a magnitude limit ranging from 14.5 to 23.3 in various bands. This candidate has been particularly followed in $R$-band by GECKO, which covered 48\% of the Bilby skymap (37\% with KMTNet, and 11\% with another telescope, RASA36), and in $o$ and $c$-bands by ATLAS, which covered 69\% and 71\% of the skymap, respectively. About 100 counterpart candidates were reported but not definitively confirmed (see for example \citealt{paek2025geckofollowupobservationbinary}, submitted). % Paek et al. 2025 article about transient candidates in S230518h - we identified 128 transients from our GECKO data, but none were found to occur near galaxies at d=204 +- 57 Mpc.

\paragraph{GW230529} -- This event is highly significant (FAR of 1 per 160 yrs) and detected by 
LIGO L1 on 2023-05-29 at 18:15:00 UTC at 201$^{+102}_{-96}$ Mpc, yielding a broad initial sky localization of 24,100 deg$^2$ \citep{2024ApJ...970L..34A}. It is classified offline as an NSBH, with a support for the primary object being in the lower mass gap. It has been infrequently observed by the community due to the very poorly constrained sky localization. In total, 12 teams reported follow-up observations on the event in neutrino, gamma-rays, X-rays and optical. In particular, the optical follow-up started 0.18 days post trigger time (T$_0$) with GOTO (see Appendix~\ref{Events}). 
In the optical range, there was a total of 37\% coverage of the most recently updated sky localization for a magnitude limit ranging from 13.2 to 21.7 in various bands (see Figure~\ref{fig:skymap-mag-GW230529}a). This event has been particularly followed in $o$-band by ATLAS, which covered 24\% of the Bilby skymap. Several candidates were reported as ZTF \citep{ahumada2024searchinggravitationalwaveoptical}, but not definitively confirmed.
%

%\paragraph{S230518h} Tas described in Appendix~\ref{Events}. We can see that a few observations occured well before the shorted peak time but most of observations occurred between 1 and 2 days with tiles at T $\sim$ 1.2 days which is consistent with the peak time of 20$\%$ the scenerios. By accumulating, observations in i-band of S230518h covered the time peak of 43$\%$ of scenarios.

%Therefore, for each NSBH candidate, we extract the optical observations done during the followup thanks to Treasuremap tool \citep{2020ApJ...894..127W}.

% to be changed

\paragraph{S230627c} This event, detected by H1 and L1 on 2023-06-27 at 01:53:37.819 UTC, was classified 49\% NSBH and 48\% BBH \citep{2023GCN.34087....1L}. Despite the follow-up by multiple observatories due to its good localization (80 deg$^2$, 90\% confidence) and its 291~±~64 Mpc distance, no viable electromagnetic or neutrino counterparts were found. In total, ten teams reported follow-up observations on the event in gamma-rays, X-rays and optical range. Particularly, the optical follow-up started 2 h post T$_0$ with ZTF (see Appendix~\ref{Events}). In the optical range, there was a total of 96\% coverage of the most recently updated sky localization for a magnitude limit ranging from 16.3 to 21.3 in various bands (see Figure~\ref{fig:skymap-mag-S230627c} a). In particular, GRANDMA performed follow-up with a densely sampled galaxy-targeted strategy reaching a maximum depth of R$>$21.3 (at 3-$\sigma$ level) or GOTO observed 93\% of the total sky localization area from 0 to 6 days. %\citep{2023GCN.34130....1L}. %(see Appendix \ref{fig:GRANDMA-followup-S230627c}).  

\paragraph{S240422ed} This event triggered the most intensive follow-up campaign of O4 to date. This GW candidate was detected on April 22$^{nd}$, 2024, 21:35:13.417 UTC by H1, L1, and Virgo (V1) with an initial FAR of 1 in 10$^5$ years. This trigger has a 260 deg$^2$ localization and a distance of 188 ± 43 Mpc. Initially classified as $>$~99\% likely an NSBH merger, it was, two months later, reclassified as 93\% likely non-astrophysical \citep{2024GCN.36812....1L}.
%Initially $>$~99\% likely an NSBH merger, it now has a 265 deg$^2$ localization and a distance of 190 ± 40 Mpc. Two months later, it was reclassified as 93\% likely non-astrophysical. 
Follow-up across multiple wavelengths from gamma-ray to radio and neutrinos found 46 candidate counterparts but no confirmed association \citep{2024GCN.36263....1K}. Considering all observations in the optical, there was a total of
$\sim$99.7\%
coverage of the most recently updated sky localization for a magnitude limit ranging from 14.1 to 23.5 in various bands 
(see Fig.~\ref{fig:skymap-mag-S240422ed} a). 
S240422ed was well-followed by DECam \citep{cabrera2024searchingelectromagneticemissionagn, 2024arXiv240910651K, 2024arXiv241113673K} between 0 and 6 days post-T$_0$ with a total coverage of 84\% and 83\% at 23.1 mag and 22.6 mag (median), in $r$ and $z$-bands respectively (values without extinction correction). \\

\subsubsection{Details of optical observations used in this study}
For each alert, we collected optical follow-up observations triggered by GW events from refined analyses done by:

\begin{itemize}
    \item \textit{ATLAS} uses 4 identical telescopes as described in \citet{2018PASP..130f4505T}. The data are calibrated with respect to the ATLAS-REFCAT2 catalogue \citep{2018ApJ...867..105T}, and sources are processed with the ATLAS Transient Science Server \citep{2020PASP..132h5002S}. The GW follow-up is described in \citet{2024MNRAS.528.2299S}, which uses the \href{https://gocart.readthedocs.io/en/main/index.html}{gocart service} and \href{https://skytag.readthedocs.io/en/main/#how-to-cite-skytag}{skytag codes},
    \item \textit{DECam}, with the new survey program Gravitational Wave MultiMessenger Astronomy DECam Survey (GW-MMADS) (PIs: Andreoni \& Palmese) \citep{cabrera2024searchingelectromagneticemissionagn, 2024arXiv241113673K, 2024arXiv240910651K} and with raw data
    accessible from \href{https://astroarchive.noirlab.edu/portal/search/}{NOIRLab page},
    \item \textit{GECKO} using the KMTNet \href{https://github.com/jmk5040/KMTNet_ToO}{ToO program} (Jeong et al., in preparation) \citep{Im_Paek_Kim_Lim_2020, Paek_2024},
    \item \textit{GOTO} coverage using both targeted pointings as well as serendipitous coverage of the skymaps from the GOTO all-sky survey pointings. Depth information is derived from the GOTO kadmilos pipeline that processes images in real-time (Lyman et al., in preparation) using a photometric calibration against ATLAS-REFCAT2 sources. Difference imaging is also performed using deeper template observations. Images are then crossmatched to the GW events using a custom routine primarily built on healpix-alchemy \citep{Singer_2022} to provide total coverage for a given skymap,
    \item \textit{GRANDMA} \citep{Antier_2020}, which uses Skyportal to organize the GW observational campaign \citep{2023ApJS..267...31C} and \href{https://github.com/karpov-sv/stdpipe}{STDpipe} web service the detection, and photometric measurements, 
    \item \textit{SAGUARO} uses the CSS 1.5m telescope on Mt. Lemmon Observatory to search for GW counterparts and other (primarily) Arizona facilities for follow-up \citep{2019ApJ...881L..26L,2021ApJ...912..128P,2024ApJ...964...35H},
    \item \textit{TESS} \citep{2015JATIS...1a4003R, 2023ApJ...948L...3M} (only for S230518h) using standard photometric algorithms from \href{space mission data products}{https://tess.mit.edu/observations/sector-65/},
    \item WINTER GW follow up \citep{Frostig2022} using the WINTER data reduction pipeline implemented using \href{https://github.com/winter-telescope/}{mirar} [\href{https://doi.org/10.5281/zenodo.10888436}{ref}]. The sciences images were subtracted using reference images built from the UKIRT survey \citep{Dye2017} or VISTA survey \citep{10.1093/mnras/staa310}.
    \item \textit{ZTF} from measurements using techniques presented in \citet{ahumada2024searchinggravitationalwaveoptical}.
\end{itemize}
These observations are summarized in Table~\ref{followupcoverage} that shows the coverage of the GW sky localization in each observation filter, and in Appendix~\ref{Events}.
%We completed this dataset with observations from the optical community using the publicly reported information in GCNs and the Treasure Map \citep{Wyatt_2020}, assuming measurements were correct (see Table~\ref{followupcoverage} and Appendix~\ref{Events}) for BlackGEM, Las Cumbres, MASTER, MeerLICHT, PRIME, Swift/UVOT, SWOPE. \fixme{TO BE CHANGED}\\

The optimization of tiling the GW sky localization area across one or several telescopes was performed using several algorithms such as that of \citet{2023ApJS..267...31C}, resulting in the possibility to overlap observations of a GW sky localization with multiple telescopes/instruments as will be seen in Section~\ref{MM}; unfortunately, their usage does not enable the most efficient follow-up in terms of coverage due to parallel and independent observations. However, each team reported potential counterpart candidates, with a delay of hours to days, based on an algorithm that uses multiple detections, machine-learning, and cross-matching to a plethora of archival catalogs (see GCNs for instance). This non-controlled independent analysis and lack of joint infrastructure led to a diversity of reporting with a variety of specifications: time of the discovery, fading criteria, contamination of the host galaxy and subtraction, localization accuracy in arcsec, spectral analysis, etc. Therefore, this prevented the community from a more efficient observational campaign, despite having a more extensive dataset from the overlap of observations.
For instance, some candidates received a lot of redundant follow-up (multiple spectra confirming the same redshift, multiple photometry points conveying no evolution), although this redundancy may be good for confirming observations (e.g. discovery of counterparts). For example, we note a high popularity of X-ray transient candidate counterparts compared to optical counterparts. This preference may be attributed to several factors: the X-ray domain facilitates combining detection between GW and EM with significantly larger number of orphan GRB afterglows compared to on-axis jets, but significantly lower candidates than the optical transient sky, and also for timing association since the peak of detection can occur promptly to several days post GW and ease timing association. %The optimization of mapping of the GW sky localization area across one or several telescopes was perfrmed using several algorithms such as \citet{2023ApJS..267...31C}, resulting in the possibility to overlap observations in the international community as will be seen in Section~\ref{MM}, which prevented us from using the observation time efficiently even if it reached above 80 \% for 3 of the 4 NSBH candidates. Usually, each astronomical team reported with a delay of hours to days potential counterpart candidates based on an algorithm that uses multiple detections, machine-learning, and cross-matching to a plethora of archival catalogs (see GCNs for instance). The non-controlled independent analysis and lack of joint infrastructure led to a diversity of reporting with varied specifications: time of the discovery, fading criteria, contamination of the host galaxy and subtraction, localization accuracy in arcsec, spectral analysis, etc. This led to a highly complex ecosystem where the follow-up of source candidates was not optimal. For instance, some candidates received a lot of redundant follow-up (multiple spectra confirming the same redshift, multiple photometry points conveying no evolution), although this redundancy may be good for confirming observations (discovery of counterparts for instance). 
In contrast, some candidates did not receive attention to unambiguously classify them as non-viable transients.
Nevertheless, we consider for the following that \textbf{no kilonova} was observed %was detected and was present in images taken
during follow-up of the NSBH alert campaigns up to July 2024.

In the next section, we take a critical look at the overall follow-up strategy and investigate possible improvements for future events.% in order to maximize the probability of future KN counterpart detection.

\subsection{Optimization of Kilonova Detection}
\label{optimisationKN}
\begin{figure*}
    \centering
    \includegraphics[width=\textwidth]{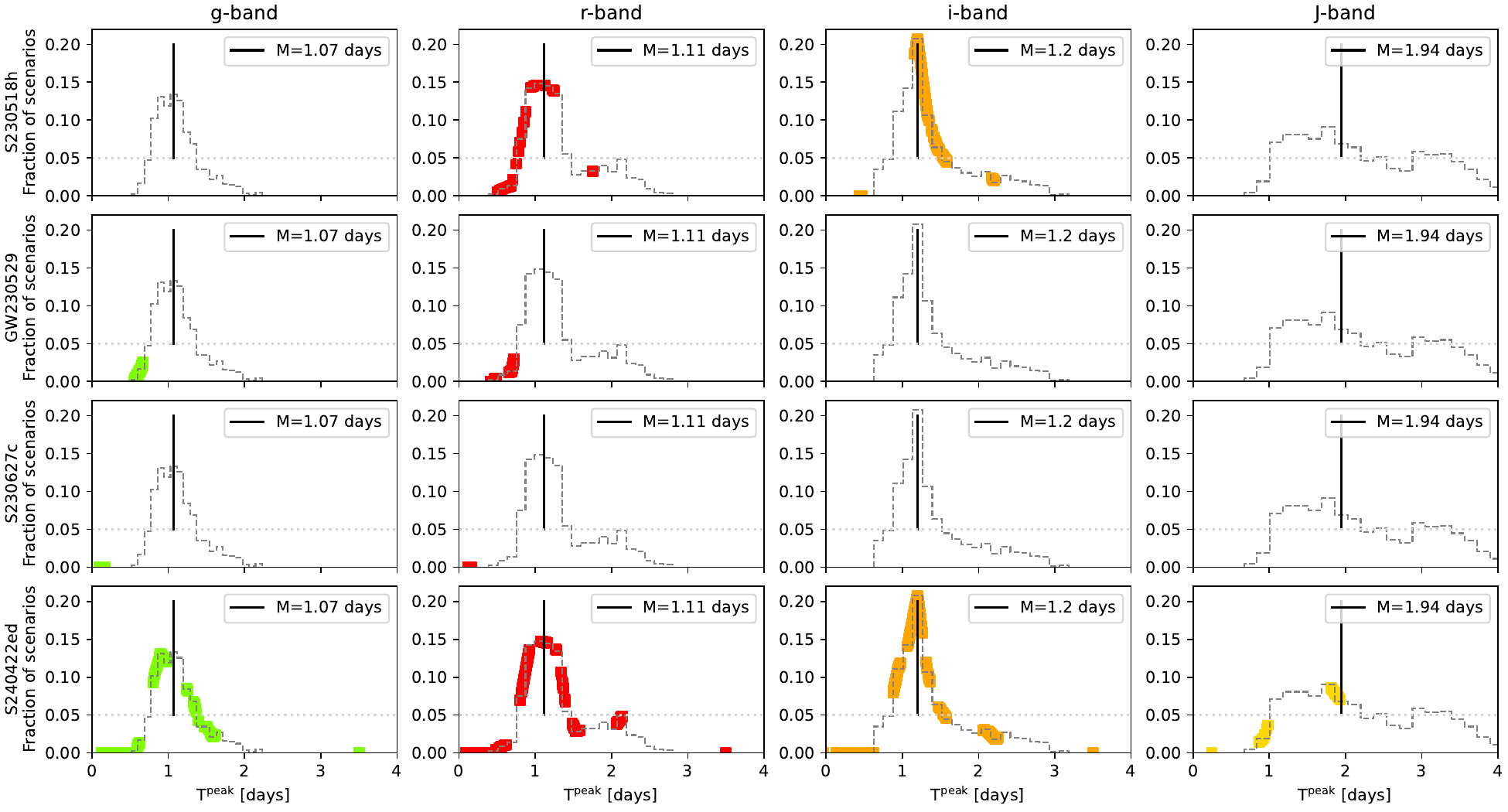}
    \caption{Comparison between the peak time luminosity of our simulated KN dataset in $g$ (first column), $r$ (second column), $i$ (third column), and $J$ (last column)-bands and the time of optical observations (using a mixed of private and public observations reported Table~\ref{followupcoverage} and Table~\ref{tab:coverage-public}) for S230518h (first row), GW230529 (second row), S230627c (third row), S240422ed (fourth row). The dashed gray line represents the distribution of peak time considering all m$_{dyn}$-m$_{wind}$-$\theta$ scenarios in the simulated dataset. The solid vertical black line represents the median of the peak time distribution considering only bins containing more than 5\% of the distribution (grey dashed line). Observations of the community are shown in color squares.}
    \label{fig:observation-peak-time}
\end{figure*}

\begin{figure*}
    \centering
    \includegraphics[width=\textwidth]{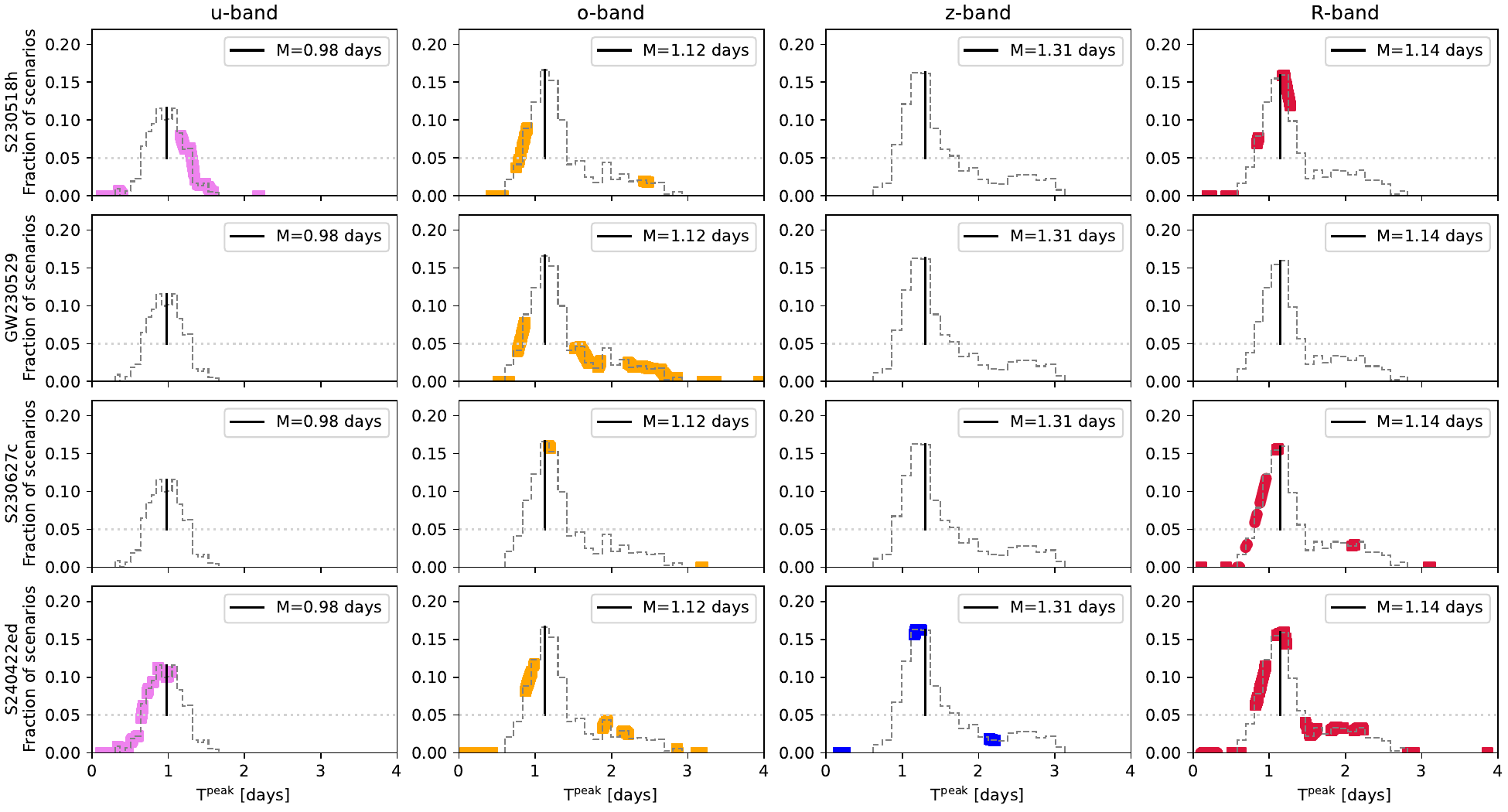}
    \caption{Comparison between the peak time luminosity of our kilonova population in $u$ (left column), $o$ (second column), $z$ (third column), and $R$ (fourth column)-bands and the time of optical observations (using a mixed of private and public observations reported Table~\ref{followupcoverage} and Table~\ref{tab:coverage-public}) for S230518h (first row), S230529ay (second row), S230627c (third row), S240422ed (fourth row). The dashed gray line represents the distribution of peak time considering all $m_{dyn}$-$m_{wind}$-$\theta$ scenarios. The solid black line represents the median of the peak time distribution considering only bins containing more than 5\% of the distribution. Observations of the community are shown in color squares. All observations in $o$-band are done by ATLAS.}
    \label{fig:opti-other-filters}
\end{figure*}

We explore the optimization of the detection of KNe. Here, our study focuses on the temporal efficiency of observations to discover KNe,
%efficiency of observations in time to discover KNe that was mainly studies, 
rather than assessing whether they covered sufficient sky area to ensure the source's location was observed \citep{2024arXiv240517558K}. We assume that the observations should seek to cover the \textquote{peak time} \textit{e.g.} the time that corresponds to the peak luminosity of a scenario, at least once in order to optimize the likelihood of counterpart detection. To do this, we compare the time of the observations taken by the community for each NSBH alert to the peak magnitude of 891 simulated KNe computed from \textit{An21Bu19}. %, projected at a distance of 200 Mpc. 
More precisely, we use synthetic light curves generated in 27 filters (corresponding to the filters of observations) from POSSIS spectra \citep{2021NatAs...5...46A, 10.1093/mnras/stz2495}. Here, we focus on observations in $u,g,G,r, R, o,i,z, c, L,tess,J$ and clear (also noted $C-$band) (and therefore compute the KNe light curves in these filters) during the O4 observational campaign (similar to the analysis of \citealt{10.1093/mnras/stab1090} for ZTF) as O4 NSBH events have been predominantly followed-up in these bands. Filters of observations and the corresponding filters used to compute the light curves from POSSIS can be found in Appendix~\ref{appendix-filter}, Table~\ref{tab:various-filter}. %Note that we collected observations in $R$-band together in $clear$ %as there is no corresponding light curve for the \textit{An21Bu19} model but they have comparable wavelengths.
%as the latter is not present in the 21 different bands in which the light curves are generated but $R$ and $clear$ have comparable wavelengths.
Details of the computation from \textit{An21Bu19} can be found in Section~\ref{MM}. Note that we do not take into account the duration of observations here but rather check if their midtime falls in the peak-time bins. %Comparing synthetic KNe to the light curve model is an important tool for follow-up of NS bearing events. 
Here we augmented the dataset from telescopes mentioned in Sec.~\ref{followup} with public information from GCNs and the Treasure Map \citep{Wyatt_2020} of additional observations from BlackGEM \citep{Groot_2024}, Las Cumbres \citep{Brown_2013}, Magellan and SWOPE \citep{mardini2023metalpoorstarsobservedmagellan}, MASTER \citep{Kornilov_2011}, MeerLICHT \citep{2019IAUS..339..203P}, PRIME \citep{Yama_2023} and Swift/UVOT \citep{2005SSRv..120...95R}, assuming the times of observations are correctly reported.

Each simulated KN defines a different possible \textquote{scenario} that corresponds to different properties of the dynamical and disk wind ejecta during and after the merger and viewing angle (see Table~\ref{ModelFramework}). The ejecta properties depend on the binary properties themselves. Fig.~\ref{fig:observation-peak-time} shows histograms of peak time for simulated KNe from NSBH mergers for $g$,$r$,$i$ and $J$ bands in dashed gray line. Fig.~\ref{fig:observation-peak-time} shows similar histograms for $u$,$o$,$z$ and $R$ bands. Distributions in other bands are provided in Fig.~\ref{fig:opti-CcLG-filters} and Fig.~\ref{fig:opti-tess-filters} of Appendix~\ref{optimisation-appendix}. The median of the histogram is shown with a solid black line. Here we use 20 bins to histrogram the distributions, meaning that for peak time distribution between $\sim$0.5 and 3 days, each bin corresponds to 3 hours. Most simulations show that the peak time for KNe occurs around 1 day post-merger: 12\% of synthetic KNe from the population in the $g$-band, 15\% in the $r$-band, and 14\% in the $i$-band. In contrast, the median of the distribution in the $J$-band peaks around 2 days post-merger with 6\% of the population reaching their maximum at that time. More precisely, observations taken from 0.9 to 1.4\,days post T$_0$ in $g,r,i$-bands can cover the peak time of $>$ 51\% of the synthetic KNe from our population, while fast observations prior to 0.5 day post T$_0$ represent only the peak luminosities for 0.2\% (and 0\%) in $r,i$-bands (and $g$-band). In $J$-band, observations taken from 1 day to 3 days after T$_0$ will be beneficial as the peak luminosity distribution of KNe is relatively flat \citep{Kawaguchi_2020}.%}
\\

We compare the observations of S230518h, GW230529, S230627c, and S230422ed (regardless of observation depth, which is discussed later in Sec.~\ref{MM}) with the distribution of KNe peak times, as shown in Fig.~\ref{fig:observation-peak-time} and in Fig.~\ref{fig:opti-other-filters} (and in Appendix~\ref{optimisation-appendix}, Fig.~\ref{fig:opti-CcLG-filters} and~\ref{fig:opti-tess-filters}). 

 \paragraph{S230518h} 
 Observations from 0 to 6 days of S230518h in the $i$-band, which occurred promptly or shortly after the peak (\textit{e.g.}, $>$1.2 days), covered 44\% of our simulated KNe (see the first row of Fig.~\ref{fig:observation-peak-time}). In $r$-band, the beginning of the histogram is well-covered (70\%), with observations occurring at the maximum of the peak time distribution. Observations using the most frequently employed filter, $tess$ filter, covered almost 100\% of the peak time of the KNe population (see Fig.~\ref{fig:opti-tess-filters} of Appendix~\ref{optimisation-appendix}). Finally, $R$- and $c$-bands observations of S230518h covered 34\% and 28\% of the peak time distribution respectively as seen in the first row of Fig.~\ref{fig:opti-other-filters} and of Fig.~\ref{fig:opti-CcLG-filters} in Appendix~\ref{optimisation-appendix}.

  \paragraph{GW230529} 
  Given the large sky localization error box, this event was followed by fewer telescopes and observations of GW230529 from 0 to 6 days in $g$-band and $r$-band cover only 2\% of the peak time of the synthetic KNe population (second row, Fig.~\ref{fig:observation-peak-time}). These observations primarily focus on times earlier than 0.5 days. Observations in the $o$-band span from 0 to 6 days post T$_0$, covering 37\% of the peak time of the KNe population as seen in the second row of Fig.~\ref{fig:opti-other-filters}. Finally, observations in $L$ and $G$-bands covered 44\% and 13\% of the peak time of synthetic KNe population, respectively (see Fig.~\ref{fig:opti-CcLG-filters}, Appendix~\ref{optimisation-appendix}).

 \paragraph{S230627c} Observations of S230627c, 100 Mpc further away than S230518h, GW230529, and S240422ed, did not cover the peak time of our KNe population in $r$-band and $g$-band, as they were performed prior to 0.5 days post T$_0$ (third row, Fig.~\ref{fig:observation-peak-time}). Observations in $o$-band and with the most frequently used filter, $R$-band, cover 17\% and 26\% of the peak time of the KNe population respectively (see the third row of Fig.~\ref{fig:opti-other-filters}. 
Finally, observations in $L$-band covered 16\% of the peak time distributions (see the third row of Fig.~\ref{fig:opti-CcLG-filters}, Appendix~\ref{optimisation-appendix}).

\paragraph{S240422ed} Finally, 
observations of S240422ed covered 62\%, 82\%, 73\%, and 18\% of the peak time of our KNe population in the $g,r,i, J$-bands, respectively (fourth row, Fig.~\ref{fig:observation-peak-time}). Overall, this candidate has been well observed in time and coverage by the community, though the coverage is not fully optimized. For example, $R$ and $u$-band observations covered respectively 78\% and 62\% of the peak time distribution (see the fourth row of Fig.~\ref{fig:opti-other-filters}); however, there are no observations in $R$-band between 1.26 and 1.37 days, while around 10\% of the KNe from our population peak during this interval. 
\\

In summary, observations at the predicted brightness peak of the KN are crucial for maximizing detection. Our study highlights not only the need for prompt imaging of NSBH mergers but also emphasizes the importance of 1-day post-merger observations in the UVOIR bands, particularly for telescopes taking shallower exposures. The KN brightness can vary by several magnitudes between early and peak-time observations (\textit{e.g.}, in the $g$-band: up to 8 mag, $r$-band: up to 9 mag, $i$-band: up to 9 mag, and $J$-band: up to 11 mag, in the \textit{An21Bu19} model). 
Although the community has effectively implemented prompt observation strategies, 1-day post-T$_0$ observations were performed in only 3 of the 4 cases (we include GW230529, although the 1-day post-T$_0$ observation strategy has only been done in $o$ and $L$-bands). We also advocate a more \textquote{flexible} approach for near-infrared and infrared observations given that the peak luminosity distribution is more uniform from 0 to 4 days \citep{Kawaguchi_2020}, although these bands are crucial to differentiate the ejecta properties (see \textit{e.g.} a similar effort in \citealt{Frostig2022}). In principle, the observational strategy can be further refined by taking into account additional information from the GW signal itself, e.g., classification probabilities, the chirp mass, or the viewing angle. This study could also be used in targeting follow-up, for instance, if an event is near the detection horizon of a given telescope, then it might be better to wait for the peak time to observe, but if the event is closer, then observations both at early time and at the peak may be well motivated, since constraining the rise and the fall of the light curve is important to best capture the KN properties. This EM follow-up coordination could be performed with tools such as \texttt{Teglon} software, which prioritizes fields and strategies for follow-up with different-sized fields of view \citep{2024arXiv240415441C}.

%\section{Ejecta mass}
\section{Ejecta mass estimates}
\label{ejectaKN}

In this section, we provide constrains on the ejecta mass produced during and after the merger of a NS and a BH for the four events.

\subsection{Definition of ejected matter and its dependance on the source properties}

%In this section, we will describe the ejected matter during and after the merger regarding the properties of the event. This will be then constrained by the information contained in each GW alert in the section.

For the predictions of the ejected matter, it is standard to differentiate between ejected matter produced during the merger over a timescale of a few milliseconds (\textit{dynamical} ejecta) and unbound material produced after the creation of an accretion disk around the %newly formed 
BH (\textit{disk wind} ejecta). The ejected total mass is then:

\begin{equation}
M_{\text{ej, total}} = m_{dyn} + m_{\text{wind}} = m_{dyn} + \xi \times m_{disk}
\end{equation} where  $m_{dyn}$ is the mass of dynamical ejecta,  $m_{wind}$ the mass of disk wind ejecta, $m_{disk}$ is the mass in the accretion torus after disruption, and $\xi$ the proportion of material from the disk that is eventually unbound. In our study,
we use fitting formulae calibrated to the result of merger and post-merger simulations to estimate $m_{dyn}$, $m_{disk}$, and $\xi$.
Specifically, we calculate $m_{dyn}$ using the fitting formula of (Eq.~9) of \cite{2020PhRvD.101j3002K}. We then use the fitting formula of (Eq.~4) of~\cite{2018PhRvD..98h1501F}  to calculate the total mass $m_{rem}$ of matter remaining outside of the BH after tidal disruption, which includes both the accretion torus and the dynamical ejecta (which is valid with a $\sigma=0.005M_\sun$ uncertainty of masses).  Finally, we get the torus mass using $m_{disk}=m_{rem}-m_{dyn}$. $\xi$ is computed using the average between the lower bound and upper bound from Eq.12 of \cite{2021ApJ...922..269R}. When using these fits, $M_{ej,total}$ is a function of the binary mass ratio, dimensionless BH spin, NS compactness, and NS baryon mass\footnote{Our code to produce the ejecta mass quantify is available at \url{https://github.com/MPillas/S240422ed/tree/main}}. We note that an alternative formula for the dynamical ejecta can be found in \cite{2024arXiv240415027K}, which is expected to perform similarly than \citet{2020PhRvD.101j3002K} when they apply within the range of parameters covered by existing numerical simulations. By comparison, the two formulae are fits using different functional forms, and a slightly different group of numerical simulations for calibration. 

In order to remain within the expected validity regime of the fitting formulae, we impose limits on the physical parameters of the binary shown in Table~\ref{ModelFramework}. The lower boundary of the NS is fixed at 1.2 solar mass, according to  \href{https://www3.mpifr-bonn.mpg.de/staff/pfreire/NS_masses.html}{a referential database of the pulsar measurements} \citep{Radiopulsars}. The upper bound, M$_{max,NS}$, depends on the EOS. For the reader's convenience, we highlight that the scenario yielding the largest ejecta mass (\( > 0.1 \, M_\odot \)) occurs when the mass ratio is $\sim$ 3, the black hole spin is large and aligned with the orbital angular momentum, and the neutron star has a low mass, e.g. \( m_1 = 3.6 \, M_\odot \), \( \chi_{1z} = 1.0 \), and \( m_2 = 1.2 \, M_\odot \). We might find larger ejected masses in scenario described in \citet{2013CQGra..30m5004L,2015ApJ...807L...3E}. 
%We maybe found more ejecta mass in scenario described in \citet{2013CQGra..30m5004L,2015ApJ...807L...3E}. 
By comparison to GW170817, the multi-messenger approach presented in Hussenot et al. (in preparation) provides an upper bound on the total ejecta mass of \( 0.08 \, M_\odot \)(95\% credible region).
%We also checked that the RISCO value (normalized by the mass of the BH) should be between 1 (maximally aligned spin) and 9 (maximally anti-aligned spin). 
Indeed some binary systems can reach $m_{wind}>0.1$ $M_\odot$. In addition, in cases when  $m_{dyn}$  or $m_{wind}$ is below $10^{-3}$ $M_\sun$, we ignore the corresponding component of the ejecta, as such values are consistent with zero within the errors of the numerical fits. 
We vary the spin component of the black hole aligned with the orbital angular momentum. As the anti-aligned spin scenario results in progressively null ejecta, we chose to explore \( \chi_{1z} = [-0.3, 0, 0.3, 0.8] \). We then produce maps for $m_{dyn}$  or $m_{wind}$, based on ($m_1$, $m_2$, $\chi_{1z}$, EOS) to evaluate the dependence on the ejecta of sources properties of the binary. To do this, we created a grid of $m_1$, $m_2$ and for each pair, we calculated %we created pixels bounded by a range of $m_1$, $m_2$, and we calculated for each pair 
the possible ejecta masses, assigning to each pair the median value of the predicted ejected masses. Results for the EOSs $SLy$ \citep{CHABANAT1998231}, involving rigid neutron stars and $H4$ \citep{PhysRevLett.67.2414}, that allows high tidal disruption of the NS, are shown in Fig.~\ref{fig:ejecta_masses}~\footnote{Note that, we also show results fixing $\xi=0.3$ in Appendix Sec.~\ref{appendix-ejecta-mass}, Fig.~\ref{fig:Map_ejecta_SLy_xi0-3}.}. The figure shows that the dynamical ejecta mass increases with high mass ratio, while the wind ejecta mass is increasing as lower mass is the neutron star and black hole. The contribution to the wind ejecta is greater than the dynamical ejecta in most of the cases, but can be the reverse with $\chi_{1z}$=0.8, $m_1$ $>$ $4.2$ $M_\sun$.\\

\begin{figure*}
    \centering
        \includegraphics[width=\textwidth]{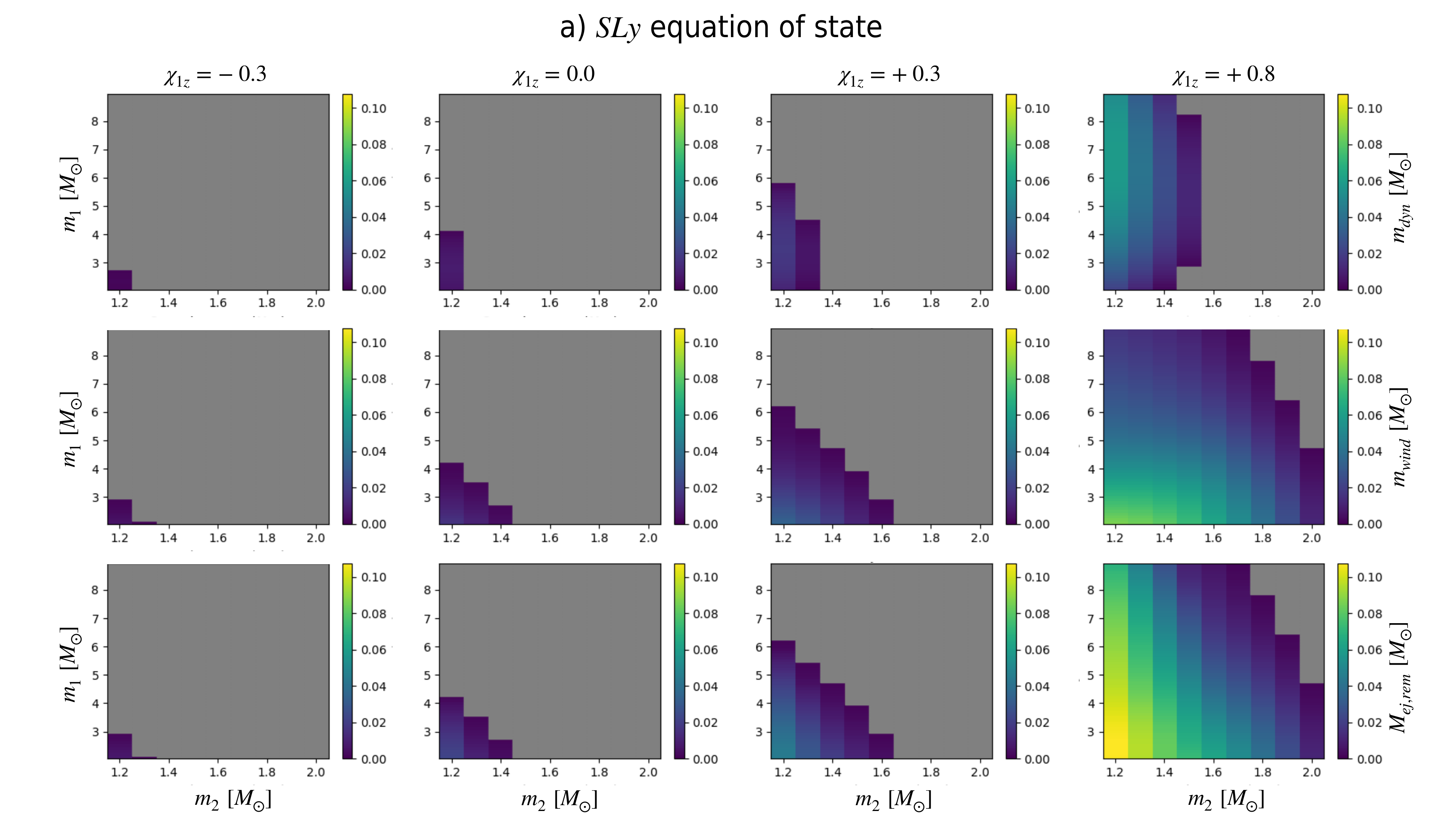}
        \hspace*{+0.2cm}
    \includegraphics[width=\textwidth]{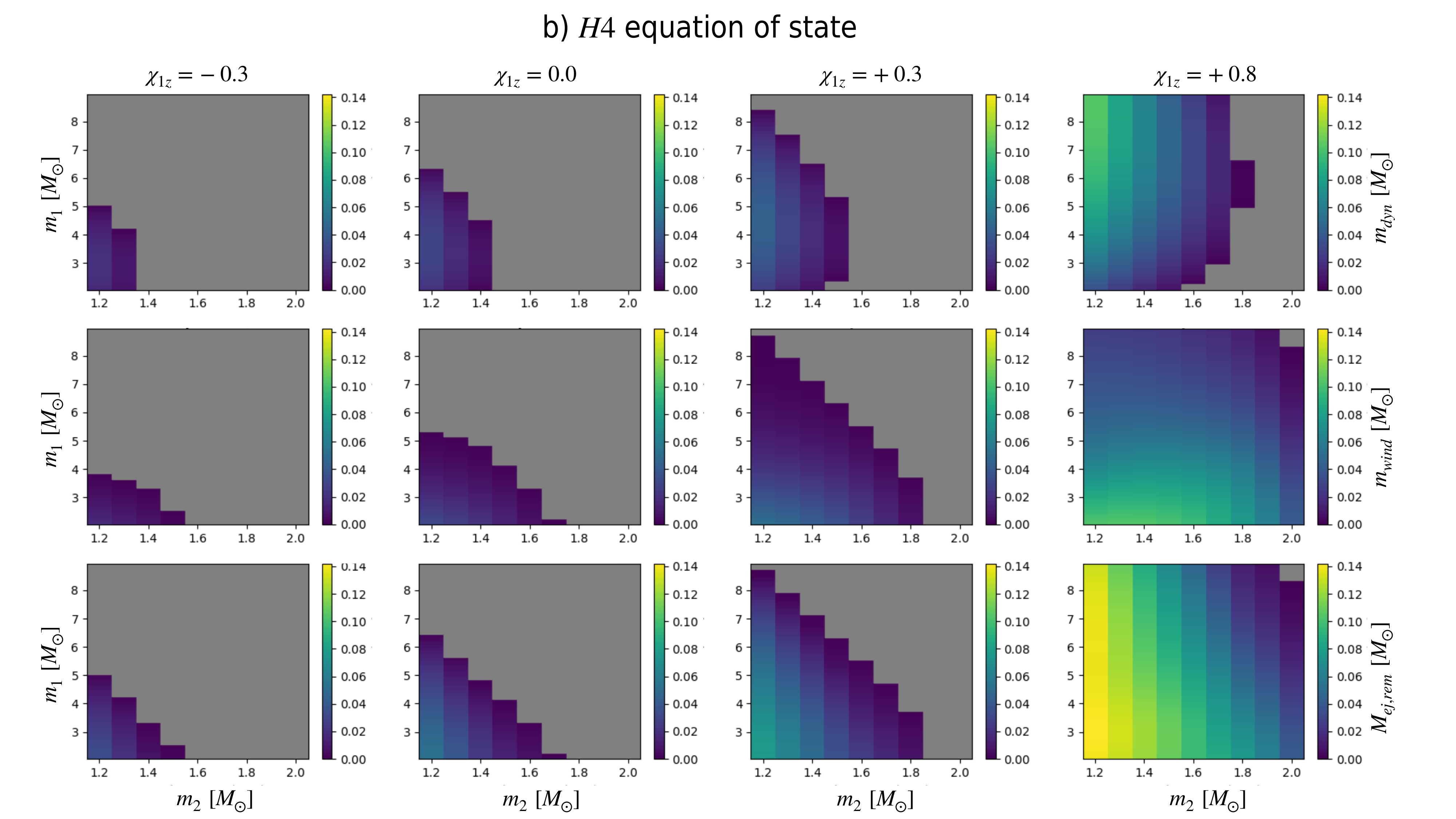}
    \caption{Dynamical (top), wind (middle), and total (bottom) ejecta masses given a certain spin component of the BH aligned with the orbital angular momentum. We consider no spin for the NS. The fraction $\xi$ of the disk that is eventually unbound is calculated as a function of the mass ratio, as described in the text. We show results a) using the $SLy$ EOS (for more compact NS) and b) using the $H4$ EOS (for less compact NS).}
    \label{fig:ejecta_masses}
\end{figure*}

In Table~\ref{ejectamass}, we show quantaties values of $m_{dyn}$, $m_{wind}$ and $M_{ej, total}$ depending on a) the $SLy$ and $H4$ EOSs b) dimensionless BH spins of $(0,0.8)$ c) an estimate of the masses given by the chirp mass $\Delta \mathcal{M}=0.2$~$M_\odot$.
We note that without significant BH spins, binary systems with a chirp mass $>$ 2.4 $M_\odot$ do not produce enough ejecta mass to power detectable KNe or GRBs. On the other hand, for rapidly rotating BHs, the total ejecta can go beyond $>$ 0.1 $M_\odot$ even for the more compact $SLy$ EOS. In addition, considering rapidly spinning BH, the SLy EOS, and the chirp mass is 2.8 $\pm$ 0.2 $M_\odot$ tells us that $M_{ej,total}<0.05M_\odot$ -- less than half the maximum value that would be inferred if the chirp mass was unconstrained. We also note that these quantities are fluctuating significantly with the spin of the BH and the EOS. For instance, taking again a chirp mass of 2.8 $\pm$ 0.2 $M_\odot$, the total ejecta mass can vary from 0.05 $M_\odot$  at maximum for $SLy$ to 0.11 $M_\odot$ for the $H4$ EOS.

\begin{table}
\begin{tabular}{| m{2cm} | m{4.2cm} |}
    \hline
   \textbf{ \scriptsize{Property}} &  \scriptsize{\textbf{Details}} \\
    \hline
        \multicolumn{2}{|c|}{\textbf{ \scriptsize{Source Properties of NSBH Event}}} \\
    \hline
     \scriptsize{BH Mass, $m_1$} & \([M_{max,NS} - 9.0] M_\odot\) \\
     \hline
    \scriptsize{NS Mass, $m_2$} & \([1.2 - M_{max,NS}] M_\odot\) \\    \hline
     \scriptsize{Spins} & 
    \begin{itemize}
        \item  BH Spin: \(\chi_{1z} \in \{-0.3, 0.0, 0.3, 0.8\}\)
        \item  NS Spin \(\chi_{2z}\): None
    \end{itemize} \\
    \hline
     \scriptsize{Equation of State of matter} & $SLy$, $H4$ \\% \hline \scriptsize{Compactness of the neutron star} & C \(> 0.14\) \ma{to be removed}\\
    \hline
\end{tabular}

\begin{tabular}{| m{2cm} | m{4.2cm} |}
    \hline
    %\textbf{Aspect} & \textbf{Details} \\
    %\hline
    \multicolumn{2}{|c|}{\textbf{ \scriptsize{Ejecta from the NS disruption ($m_{dyn}$}})} \\
    \hline
    Mass Range & \(0.0 - 0.1 M_\odot\) \\
    \hline
    \multicolumn{2}{|c|}{\textbf{ \scriptsize{Ejecta from the accretion disk ($m_{disk,wind}$}})} \\
    \hline
    Mass Range & \(0.0 - 0.15 M_\odot\) \\
    \hline
    Outflow $\xi$ & \(5\% - 40\%\) not accreted \\
    \hline
    \multicolumn{2}{|c|}{\textbf{ \scriptsize{Kilonova Light Curves}}} \\
    \hline
    Ejecta & 
    \begin{itemize}
        \item NSBH models computed with \texttt{POSSIS} \citet{2021NatAs...5...46A, 10.1093/mnras/stz2495} with m$_{dyn}$,m$_{wind} \in$ [0.01, 0.09] M$_\odot$ and $\theta \in$ [0, 90] degrees)
        \item 1D bolometric 
    \end{itemize} \\
    \hline
\end{tabular}
\caption{Diversity of the NSBH population, ejecta, and associated kilonovae studied in this work.}
\label{ModelFramework}
\end{table} 

\begin{table}
\label{tab:mej}
\centering
\begin{tabular}{|l|l|l|l|l|l|}
\hline
 & \footnotesize{$\mathcal{M}$} & \multicolumn{2}{c|}{\footnotesize{$\chi_{1z}$=0.0}}  & \multicolumn{2}{c|}{\footnotesize{$\chi_{1z}$=0.8}}    \\
 & \footnotesize{[$M_\odot$]} & \multicolumn{2}{c|}{} & \multicolumn{2}{c|}{} \\
\hline
& & \footnotesize{$SLy$} & \footnotesize{$H4$} & \footnotesize{$SLy$} & \footnotesize{$H4$} \\
\hline
\multirow{7}{*}{\small{$m_{dyn}$}}  & \footnotesize{any} & \footnotesize{$< 0.01$} & \footnotesize{$< 0.03 $} & \footnotesize{$< 0.06 $} & \footnotesize{$< 0.10 $ } \\
& 1.6 & \footnotesize{$< 0.01$} & \footnotesize{$< 0.03 $} & \footnotesize{$ < 0.04 $} & \footnotesize{$ < 0.07 $}  \\  
& 2.0 & \footnotesize{-} & \footnotesize{$ < 0.03 $} & \footnotesize{$< 0.06 $} & \footnotesize{$< 0.09 $}  \\  
& 2.4 & \footnotesize{-} & \footnotesize{-} & \footnotesize{$< 0.06 $}  & \footnotesize{$< 0.1$} \\  
& 2.8 & \footnotesize{-} & \footnotesize{-} & \footnotesize{$< 0.03 $} & \footnotesize{$< 0.08 $} \\  
& 3.2 & \footnotesize{-} & \footnotesize{-} & \footnotesize{-} & \footnotesize{$< 0.03 $} \\ 
\hline
\multirow{7}{*}{\small{$m_{wind}$}} & \footnotesize{any} & \footnotesize{$< 0.02$} & \footnotesize{$< 0.03 $} & \footnotesize{$< 0.09 $} & \footnotesize{$< 0.11 $ } \\
& 1.6 & \footnotesize{$< 0.01 $} & \footnotesize{$< 0.03$ }& \footnotesize{$< 0.09$} & \footnotesize{$< 0.11 $}  \\   
& 2.0 & \footnotesize{-} & \footnotesize{$< 0.01  $ }&\footnotesize{$< 0.06  $} & \footnotesize{$< 0.09$ } \\  
& 2.4 & \footnotesize{-} & \footnotesize{-} & \footnotesize{$< 0.04 $} & \footnotesize{$< 0.06 $}  \\  
& 2.8 & \footnotesize{-} & \footnotesize{-} &  \footnotesize{$< 0.02 $} & {\footnotesize$< 0.04 $} \\  
& 3.2 & \footnotesize{-} & \footnotesize{-} & \footnotesize{$< 0.01  $} & \footnotesize{$< 0.02$ } \\ 
\hline
\multirow{7}{*}{\small{$M_{ej,rem}$}} & \footnotesize{any} &  \footnotesize{$< 0.02$} &  \footnotesize{$< 0.05 $ }& \footnotesize{$< 0.11$} & \footnotesize{$< 0.16 $}  \\
& 1.6 & \footnotesize{$< 0.02  $} & \footnotesize{$< 0.05 $ } & \footnotesize{$< 0.11  $} & \footnotesize{$< 0.14 $}  \\   
& 2.0 & \footnotesize{$< 0.001  $} & \footnotesize{$< 0.04 $ } & \footnotesize{$ < 0.10 $} & \footnotesize{$< 0.14 $}  \\  
& 2.4 & \footnotesize{-} & \footnotesize{-} & \footnotesize{$ < 0.09  $} & \footnotesize{ $ < 0.14 $}  \\  
& 2.8 & \footnotesize{-} & \footnotesize{-} &  \footnotesize{$< 0.05  $} & \footnotesize{$ < 0.11 $}  \\  
& 3.2 & \footnotesize{-} & \footnotesize{-} &\footnotesize{$< 0.01  $} & \footnotesize{$ < 0.05 $}  \\ 
\hline
\end{tabular}

%\hspace*{-0.5cm}

\caption{Upper bound on the ejecta masses of the dynamical ejecta ($m_{dyn}$), the disk wind ejecta ($m_{wind}$) and the total ejecta $M_{ej,rem}$, in $M_\odot$, depending on the chirp mass of the system (with an uncertainty of 0.2 solar mass), the equation of state of matter and the spin of the primary component.}
\label{ejectamass}
\end{table}

\subsection{Upper limit on the ejecta mass}
\label{results_ejecta_mass}

In this section, we will provide broad upper limit (which corresponds to the optimistic scenario/less compact NS) boundaries considering a binary merger scenario with a high spinning BH and the EOS of matter $H4$ and narrower upper limit (or pessimistic), considering a binary merger scenario with no spinning BH and the EOS of matter $SLy$. We can provide the respective ejecta masses that will be used for the computation of synthetic light curves of KNe (see Sec.~\ref{kilonovaresults}). \textit{An21Bu19}  varies $m_{dyn}$ and $m_{wind}$ from 0.01 to 0.09 M$_{\odot}$. However, these masses can reach lower or higher values depending on the least and most favorable NSBH configurations. In these extreme scenarii, we use the boundaries of the grid at 0.01 and 0.09 M$_{\odot}$ respectively for the ejecta masses, which can over or under estimate the brightness of the associated KN. 

Despite the fact that no further information about the candidates S230518h, S230627c, or S240422ed are available, which hinders us to compute precise ejecta estimates from the ejecta mass grid in Table~\ref{tab:mej}, one can --under the assumption that the candidates have an astrophysical origin-- use EM follow-up observations to exclude regions of ejecta mass for the different alerts. In addition to EM follow-up observation, it is also possible to use information of GW230529 (chirp mass, individual masses, spin components aligned with the orbital momentum, viewing angle) beyond the initial circular to further constrain ejecta properties.

We constrain the ejecta mass ($m_{dyn}$, $m_{wind}$ and $M_{ej, total}$) of  S230518h, S230627c and S240422ed considering each event candidate's classification ("p-astro"; see \href{https://emfollow.docs.ligo.org/userguide/}{Public Alerts User Guide}) probabilities contained in the LVK alerts. Indeed, there are several low-latency compact-binary searches based on a matched filtering of a bank of waveform templates~\citep{Canton_2021, Aubin2021-ja, Chu2022-nw, 2024PhRvD.109d2008E}. In case of the detection of an event, these searches report publicly their source classifications in \href{https://gracedb.ligo.org/}{Gracedb}. 
The method~\citet{Villa-Ortega:2022qdo} used by the compact binary coalescence search pipelines PyCBC Live~\citep{Canton_2021} and SPIIR~\citep{Chu2022-nw} provides a way to use public candidate information to constrain source properties and provide an upper limit on the ejected matter during and after the merger. This approach does not rely on assumptions regarding the inclination angle, individual masses, or the black hole's spin. However, it assumes that the identifying template directly gives a point estimate of the source's detector-frame (redshifted) chirp mass that is already within $1\%$ of the true value~\citep{Villa-Ortega:2022qdo, Biscoveanu_2019} for BNS and NSBH sources. This approach also assumes an astrophysical origin of the event and that distance estimate from the alert is correct.

%For each candidate, the identifying template directly gives a point estimate of the source's detector-frame (redshifted) chirp mass that is already within $1\%$ of the true value~\citep{Villa-Ortega:2022qdo, Biscoveanu_2019} for BNS and NSBH sources. Each search uses a different method to translate this chirp mass into the publicly available BNS, NSBH, BBH, and astrophysical origin source classification probabilities~\citep{Kapadia_2020, Andres:2021vew, Villa-Ortega:2022qdo}. 

\paragraph{S230518h} 
We set the upper boundary to be $M_{ej,total,H4}<~0.11$~$M_\odot$, and the dynamical and wind ejected masses $m_{dyn,H4}<0.08$ $M_\odot$ and $m_{wind,H4}~<~0.04$~$M_\odot$, given the PyCBC Live relative source classifications of 95.9 \% NSBH and 4.1 \% BBH (see \href{https://gracedb.ligo.org/superevents/S230518h/}{Gracedb S230518h}, assuming the event is astrophysical) and using $\chi_{1z} =0.8$. For comparison, using the SLy EOS model, we obtain,  $M_{ej,total,SLy}<0.05$ $M_\odot$, $m_{dyn,SLy}<0.03$ $M_\odot$ and $m_{wind,SLy}<~0.02$ $M_\odot$. Secondly, no ejecta mass is produced in the case of a non-spinning BH given $SLy$ and $H4$ EOSs.
%Note that the viewing angle $\theta$ is unconstrained as well.

\paragraph{GW230529} The properties of the binary system of GW230529 have been published by the LVK collaboration \citep{2024ApJ...970L..34A}. Therefore, to estimate the ejecta masses of $m_{dyn}$, $m_{wind}$ and $M_{ej, total}$, we extract the posterior distributions of the masses and spin of the source computed with IMRPhenomXP waveform model \citep{Pratten_2021}\footnote{The posterior samples are released publicly in \citep{ligo_scientific_collaboration_2024_10845779}}. We artificially remove the lower edge of $m_{2,src}$ distribution ($<$~1.2 $M_\odot$) and then cut on the upper edge of  $m_{1,src}$~$>$~4.5 $M_\odot$ in order to keep a chirp mass of 1.9 $M_\odot$ to be consistent with our case study (see Table~\ref{ModelFramework}), since the chirp mass measurement is very precise. We restrict the spin components aligned with the angular momentum in [-0.8,0.8] as shown in Table~\ref{ModelFramework}.

As shown in Fig.~\ref{fig:GW230629ay_PE} of Appendix~\ref{appendix-ejecta-mass}, the posterior primary mass distribution of GW230529 is quite broad, and the spin's component aligned with the orbital momentum posterior follow their prior distributions. We expect a total mass ejecta $SLy$ $<$ 0.01 $M_\odot$ and $H4$ $<$ 0.01 $M_\odot$ taking the median value of the distribution. We select for our study (see Sec.~\ref{kilonovaresults}), $m_{dyn}$,~ $m_{wind}$~$<$~0.01~$M_\odot$, as it covers the standard deviation + median of the distribution (see Fig.~\ref{fig:GW230629ay_PE}).

Fixing, respectively, no spin component of the BH, and 0.8 of the spin component aligned with the orbital momentum, we obtain (for the median): a) for non-spinning BH,  $SLy$~$<$~0.01 $M_\odot$ and $H4$ $<$~0.01 $M_\odot$ b) for $\chi_{1z}=0.8$, $SLy$~$<$~0.06~$M_\odot$, $H4$~$<$~0.1~$M_\odot$. These results are consistent with Table~\ref{ejectamass}\footnote{Note that the case of $H4_{,\chi_{1z}=0.0}$ $<$~0.01 $M_\odot$ is less pessimistic than results in Table~\ref{ejectamass}, and may be explained by the fact the distribution spans from 0.0 to 0.4 $M_\odot$ and depends strongly on the distribution of the masses which is not uniform for GW230529.}, if we set a chirp mass $\mathcal{M}$ of 2.0$\pm$0.2, with $SLy$ $<$ 0.01 $M_\odot$ and $H4$ $<$ 0.04 $M_\odot$ for no spinning BH and $SLy$ $<$ 0.1 $M_\odot$ and $H4$ $<$ 0.14 $M_\odot$ for high spinning BH. 
%Finally, fixing the most probable scenario given the median of the masses, spin and $\theta$ posterior distribution of GW230529 ($m_1$=3.6, $m_2$=1.4, $spin_{1z}=-0.1$, $spin_{1z}=-0.0$, $\theta = 45^\circ$), \fixme{pb here ?}

Finally, fixing the most probable scenario given the median of the masses and spin posterior distribution of GW230529: $m_{1,2}$= (3.6, 1.4), $\chi_{1z,2z}$= (-0.1,0), we obtain m$_{dyn,H4}$, m$_{wind,H4}$ $<$ 0.01 $M_\odot$. By comparison, the total ejecta mass was computed in \citet{2024arXiv240503841C} and provided: $SLy < 0.01 M_\odot$, $H4 < 0.07 M_\odot$. Simulations of binaries similar to GW230529 close to the disruption limit indicate that in the regime the fitting formulae may be slightly pessimistic for the less compact NSs~\citep{Martineau:2024zur}, though that uncertainty is significantly lower than the variations discussed here between systems with different spins or EOSs.

\paragraph{S230627c} We expect a total ejecta mass (using $H4$ and $SLy$ EOSs), lower than 0.01 M$_\odot$ given the PyCBC Live p$_\mathrm{astro}$ of 49.3 \% NSBH and 50.7 \% BBH (see \href{https://gracedb.ligo.org/superevents/S230627c/}{Gracedb S230627c}, assuming the event is astrophysical). We assume for further computation that $M_{ej,total}< 0.01$ $M_\odot$, $m_{dyn}<0.01$ M$_\odot$ and $m_{wind} < 0.01$ M$_\odot$. %The inclination angle $\theta$ is unconstrained as well.

\paragraph{S240422ed} We set the upper boundary $M_{ej,total,H4}<0.14$~$M_\odot$, $m_{dyn,H4}<0.05$~$M_\odot$ and $m_{wind,H4}~<0.1$~$M_\odot$ given the PyCBC Live p$_\mathrm{astro}$ of 70 \% BNS and 30 \% NSBH (see \href{https://gracedb.ligo.org/superevents/S240422ed/}{Gracedb S240422ed}, assuming the event is astrophysical) and using $\chi_{1z} =0.8$. For comparison,  $M_{ej,total,SLy}~<~0.11$ $M_\odot$, $m_{ej,dyn,SLy}~<~0.03$~$M_\odot$ and $m_{ej,wind,SLy}~<~0.08$~$M_\odot$. If we consider the case of non-spinning BH, we obtain $M_{ej,total,H4}~<0.05$~$M_\odot$, $m_{dyn,H4}~<0.03$~$M_\odot$ and $m_{wind,H4}~<~0.02$~$M_\odot$. By comparison for non spinning and $SLy$ EOS,  $M_{ej,total}<~0.02$~$M_\odot$, $m_{dyn}~<~0.01~M_\odot$ and $m_{wind}~<~0.01$~$M_\odot$.

%\footnote{We note that these results are slightly different than Table~\ref{ejectamass} because the uncertainty of the chirp mass is 0.06 $M_\odot$, compared to 0.2 for the table.}. \\
%The inclination angle $\theta$ is unconstrained as well.

These upper limits on $m_{dyn}$ and $m_{wind}$ allows us to reduce the number of synthetic light curves to consider for each NSBH candidate in the section below. Note that we are not able to constrain the viewing angle $\theta$, upon which the light curves strongly depend, from LVK public alert information. 

\section{Observability of a kilonova associated with NSBH events}
\label{MM}

In this section, we assume the O4a NSBH candidates discussed in Section~\ref{sec:O4campaignmain} are of astrophysical origin and further adopt the hypothesis that we did not detect any KN associated with the GW events (see Sec.~\ref{sec:O4campaignmain}). %We describe in this section how we can place constraints on the ejecta mass during and after the merger and on the viewing angle, based on the multi-messenger observations from GWs and in the optical.
In this section we describe how constraints can be placed on the ejecta mass during and after the merger, as well as on the viewing angle, based on multi-messenger GW and EM (optical) observations.

\subsection{Method}
\label{methodkilonova}
%On one side, we select  The light curves have three free parameters: m$_{dyn}$, m$_{wind}$ and a viewing angle $\theta$. On another side, we extract optical observations from the entire EM community for each NSBH candidate.
%For each event we also extract the associated optical observations from the entire EM community thanks to Treasuremap tool \fixme{cite}. By comparison between these observations and the peak magnitude of the predicted light curves we first check that the community adopt the best strategy for the followup of these NSBHs. Indeed, depending of the ejecta scenario and the viewing angle of the event, the time of KN brightness peak may vary by several days. In order to optimize search strategies and ensure a detection, we should at least observe one field at the predicted peak time.Indeed, the kilonova lightcurves are characterized by their ejecta properties. Among them, the mass of the ejecta ($m_{dyn}$, $m_{wind}$ and $M_{ej, total}$) are informed by and related to the binary parameters using up-to-date NR simulations, as descrived in Sec~\ref{ejectaKN}. 

We aim to cross-match all information related to GW and optical observations of NSBH events to rule out the ejecta configurations %that do not correspond to 
not compatible with the observations and keep the remaining scenarios of NSBH mergers (possible masses of the compact object, viewing angle, ejecta masses). For each NSBH candidate, we use as inputs: a) optical observations of each NSBH candidate shown in Sec.~\ref{followup}, b) our sample of KN light curves computed with the \texttt{POSSIS} %\citep{2021NatAs...5...46A, 10.1093/mnras/stz2495} 
(\textit{An21Bu19}) model (also used in Sec.~\ref{optimisationKN}), and c) PyCBC Live p$_\mathrm{astro}$oi and the derived upper bound of the total ejecta mass, \textit{dynamical} ejecta, and \textit{wind} ejecta provided in Sec.~\ref{results_ejecta_mass}. The parameters of \textit{An21Bu19} model grid used here are summarized in Table~\ref{ModelFramework}.  Furthermore, the $m_{dyn}$ component is not spherical, introducing a significant dependence on the viewing angle $\theta$; which is left as a free parameter in the model, going from 0$^\circ$ to 90$^\circ$. Time- and wavelength-dependent opacities are computed as in \cite{10.1093/mnras/stz2495} and for the wind component estimated assuming a composition that is intermediate between lanthanide-poor and lanthanide-rich. We note, however, that the wind composition is a rather uncertain parameter that can strongly affect the KN appearance. Finally, \textit{An21Bu19} does not consider the dependence on ejecta velocity, which can vary in other models; and sets the half-opening angle of the lanthanide-rich component $\phi$ to 30$^\circ$.

To test the models against the observations, we proceed as follows: for each field observed by an optical telescope (defined by a specific field of view, filter, limiting magnitude, and epoch), we extract the corresponding pixels of the GW HEALPix skymap \citep{Gorski_2005}. We extract the GW distance associated with each of these pixels and compute the apparent magnitude of the selected synthetic KN light curves at this distance. We can hence compare the brightness of the simulated KNe with the limiting magnitude of the fields at the time of the observation. The \textit{An21Bu19} model based on POSSIS spectra computation generates 891 light curves in 27 different bands. We provide details about filters we used in Appendix~\ref{appendix-filter}, Table~\ref{tab:various-filter}. Then, if a simulated KN luminosity in the detector frame is brighter than the limiting magnitude of the observation, we conclude that the simulated KN emission and properties are not compatible with the observations (i.e., we assume that no such KN signal was in this location, and at this epoch). On the other hand, if the KN luminosity is as or less bright than the limiting magnitude of the observation, we cannot rule out the simulated KN properties and consider them as compatible with the observations. 
We also conduct investigations to confirm or rule out the presence of KNe in observed regions, taking into account time and location in the sky. We should mention that we ignore observations performed as follow-up of transient candidates in the field of GW localization as such follow-ups are typically performed with small field-of-view telescopes and contribute very little to the GW sky localization coverage.

We extract observations that occurred at time $t \in \Delta t = [0, 1]$, $[1, 2]$ and $[2, 6]$ days. For each time bin, we scan the GW skymap, collecting observations in that time+localization bin, and for each pixel we compute a scale reflecting the possibility of a KN being present at this location. This scale is shown in Eq.~\ref{eqn:eq1} where $M$ is a synthetic KN from \textit{An21Bu19} in the detector frame (for instance 22 apparent mag in $r$), $\rm m_{obs}$ is the observation apparent magnitude (for instance 20 apparent mag in $r$), $\mathrm{filt}$ the filter of the observations, $\Delta t$ the time bin considered ($\Delta t = [0, 1]$, $[1, 2]$ or $[2, 6]$ days) and $n_\mathrm{tot, KN}$ is the total number of synthetic KNe from the grid considered for each event.  

Figure~\ref{fig:cumulative-hist-prob-coverage} presents our final result: the fraction of incompatible KN scenarios as a function of the percentage of skymap coverage that rules them out, and to illustrate the method, Figs~\ref{fig:S230518h-skymap-mag-fraction},~\ref{fig:skymap-mag-GW230529},~\ref{fig:skymap-mag-S230627c} and~\ref{fig:skymap-mag-S240422ed} show some examples for each NSBH candidate. The top figure describes the limiting depth reached by observations over the skymap\footnote{If several fields cover this location, we show the one with the deepest sensitivity and ignore the shallower ones.}. In the middle figure, we only select the field of an epoch associated with the most probable sky localization among the localizations covered by the observations (colored diamonds) and compare its limiting magnitude to our synthetic KN light curves population computed at the distance corresponding to this localization. Note that, due to the small number of simulated light curves to consider in the case of GW230529 and S230627c, we show all light curves, contrary to S230518h and S240422ed for which we represented only light curves corresponding to $\theta=0^\circ, 45.57^\circ,90^\circ$. The bottom figure illustrates Eq.~\ref{eqn:eq1} over the GW skymap. 

Finally, for each synthetic KN type ($m_{dyn}-m_{wind}-\theta$), we compute the deterministic probability $\bar{P}_{\theta,m_{dyn},m_{wind}, \Delta t}$ that the source is located in a region delimited by the absence of emission from a KN of this type. As previously, we separate cases into time ranges $\Delta t$. 
Thus, $\bar{P}_{\theta,m_{dyn},m_{wind}, \Delta t}$ can be summarized by Eq.~\ref{eqn:eq2}.

We iterate this process for all the synthetic KNe in the population and we repeat this work for all NSBH cases and discuss the results in the section below.

\begin{widetext}
\begin{equation}
    \mathrm{S}_{\mathrm{KN}, \Delta t, \mathrm{ipix}} = \frac{1}{n_{\mathrm{tot, KN}}} \times \sum_{k=1}^{n_{\mathrm{tot, KN}}} \begin{cases} 1 & \mathrm{ if } \ M(\mathrm{filt},\theta,m_{dyn},m_{wind},t) > \mathrm{m_{obs}}(\mathrm{filt}, t, \mathrm{ipix}) \\
0 & \mathrm{ otherwise } \end{cases}
\label{eqn:eq1}
\end{equation}

\begin{equation}
\bar{P}_{\theta,m_{dyn},m_{wind}, \Delta t} = 
\sum_{\mathrm{ipix}} P(\mathrm{GW} \mid \mathrm{ipix}) \times \begin{cases}
1 & \mathrm{ if } \ M( \mathrm{filt}, \theta, m_{dyn}, m_{wind}, t) 
< \mathrm{m_{obs}}(\mathrm{filt}, t, \mathrm{ipix}) \\
0 & \mathrm{ otherwise }
\end{cases}
\label{eqn:eq2}
\end{equation}
\end{widetext}

\subsection{Detectability of kilonovae for NSBH events}
\label{kilonovaresults}

We present the results of our analysis to constrain properties of potential KNe and their viewing angles using optical observations taken during S230518h, GW230529, S230627c, and S240422ed GW candidates follow-up campaigns. We apply the method described in Sec.~\ref{methodkilonova}.

\begin{figure*}
    \centering
    \includegraphics[width=\textwidth]{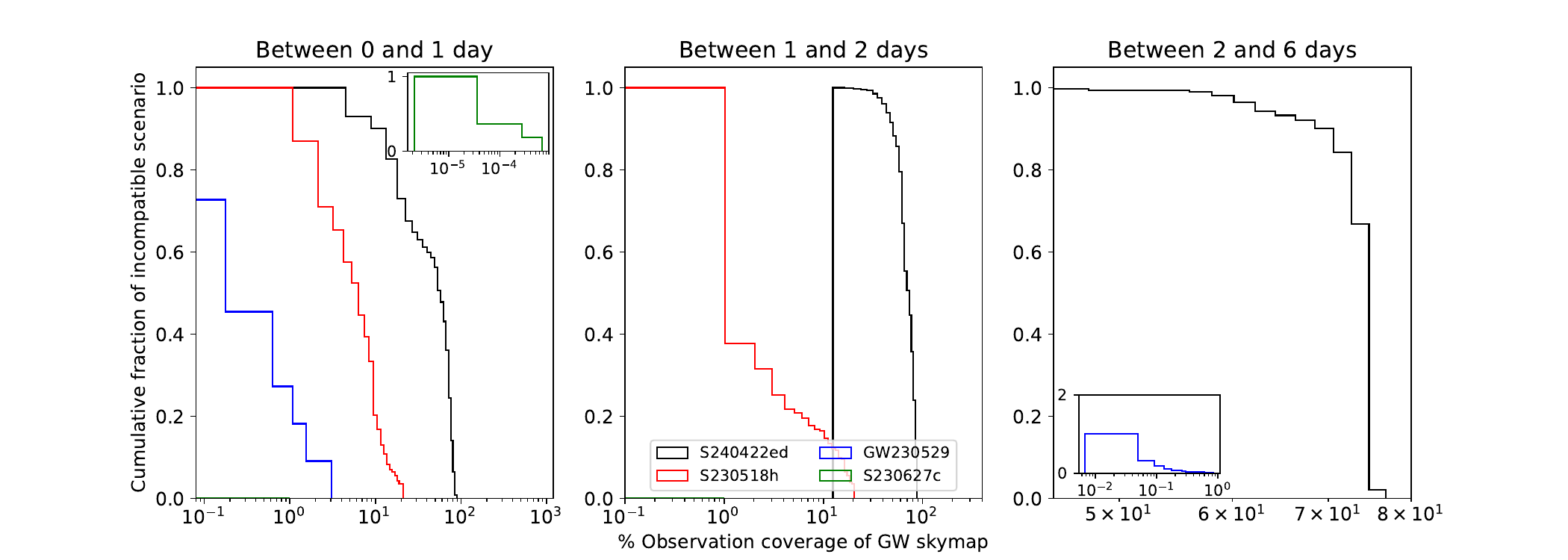}
    \caption{Cumulative histograms showing the fraction of KN scenario that are incompatible with optical observations between 0 and 1 day (left), between 1 and 2 days (middle), between 2 and 6 days (right), as a function of the GW skymap coverage that rules them out, for each NSBH candidates (each colored line). For instance, for S240422ed between 2 and 6 days, $\sim$100\% of KN scenarios are ruled out by observations covering more than 45\% of S240422ed skymap. The insets show the results for S230627c between 0 and 1 day and GW230529 between 2 and 6 days as the fraction of coverage is too small to be visible in the main figures.}
    \label{fig:cumulative-hist-prob-coverage}
\end{figure*} 

\paragraph{S230518h} We draw our conclusion using information on ejecta constraints derived from GW public information (Sec.~\ref{results_ejecta_mass}). In the [0, 1] day time epoch, the sensitivity of the observations over the sky can be seen in Fig.~\ref{fig:S230518h-skymap-mag-fraction} (and later epochs in Appendix~\ref{coverage-later-time}, Fig.~\ref{fig:appendix-S230518h-skymaps}). For example, most of the fields that overlap with the 90\% credible region reach a limiting depth of 22 mag in $R$-band, between 0 and 1 day post T$_0$. We proceed to compare these observations with the synthetic population of KNe: 
Fig.~\ref{fig:S230518h-skymap-mag-fraction} shows one example of the observations covering the most probable sky localization among all the field's coverage on top of the light curves compatible with the upper boundary on ejecta mass for S230518h, for different viewing angle $\theta$ = 0$^\circ$, 45$^\circ$, 90$^\circ$ and taken by time bin. %Fig.~\ref{fig:cumulative-hist-prob-coverage}
We then show the corresponding scale defined in Eq.~\ref{eqn:eq1} (1 - $\mathrm{S}_\mathrm{KN,\Delta t,\mathrm{ipix}}$) over the mapped sky based on observations (lower panel). We set an arbitrary threshold of 0.7 to indicate the absence of a KN in the observations (i.e., 70\% of our synthetic KN population is inconsistent with the observed data) : we rule out the presence of a KN over an area of 0.63 deg² within the 90\% S230518h credible region within the first day post T$_0$ and 0 thereafter. These regions correspond to areas where the GW event is localized with 0.04\%  confidence from 0 to 1 day and 0 beyond 1 day. 

We present final conclusions in Fig.~\ref{fig:cumulative-hist-prob-coverage}, which represents a cumulative histogram showing $\bar{P}_{\theta,m_{dyn},m_{wind}, \Delta t}$, i.e., the fraction of KNe over our synthetic population that is incompatible with optical observations as a function of the coverage that rules them out, taken by time bin, for each NSBH event. This histogram is derived from Eq.~\ref{eqn:eq2}. A fraction of 34\%  
of KNe are incompatible with observations covering more than 8\% 
of S230518h sky localization region between 0 and 1 day. While between 1 and 2 days, this number falls to 17\% 
and to 0\% between 2 and 6 days (i.e., we cannot exclude the presence of a KN after 2 days based on the observations).
In detail, between 0 to 4 days, it has not been possible to observe KNe emitted from an on-axis collision up to a viewing angle of $\theta = 25^\circ$, assuming a minimum confidence of 8\% for the presence of the source in this region.Finally low-mass ejecta ($m_{dyn},m_{wind} \leq 0.03 M_\odot$) seem to be favored against large ejected masses since the latter is overall incompatible with a larger region of the skymap (on average $>$7\%).

\begin{figure}
    \centering
    \includegraphics[width=1.0\linewidth]{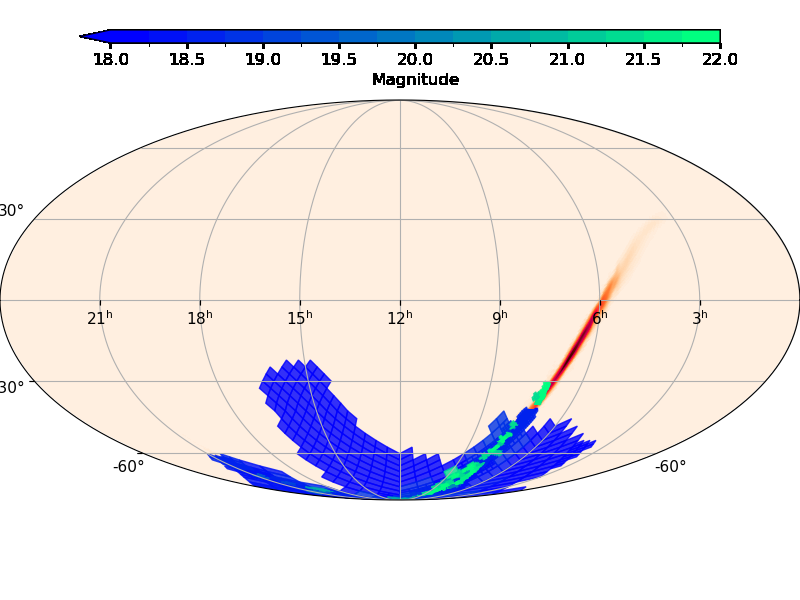}
    
    \vspace{-1.05cm}
    
    \hspace*{-1.2cm}
    \includegraphics[width=1.3\linewidth]{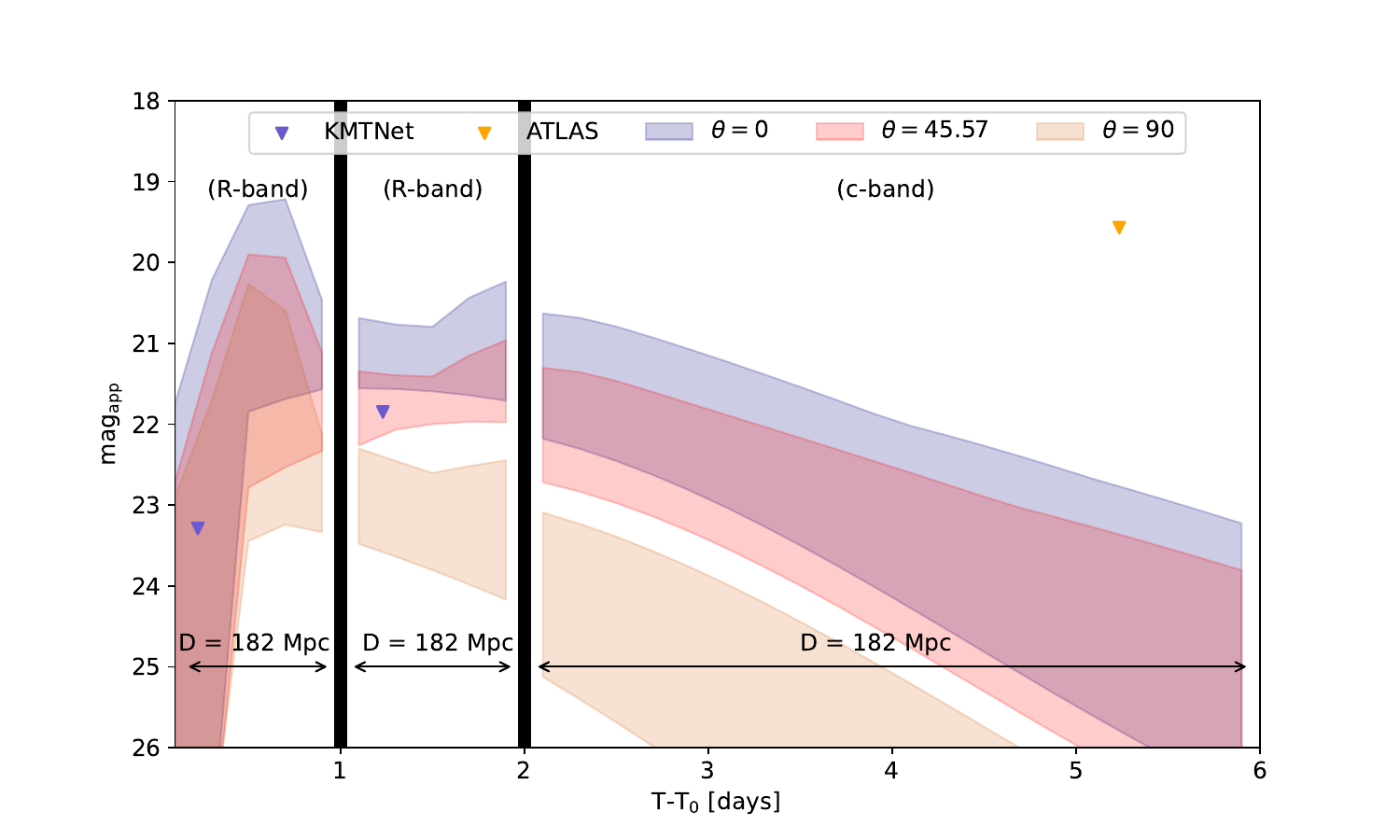}
    \vspace{-1.05cm}
    \includegraphics[width=1.0\linewidth]{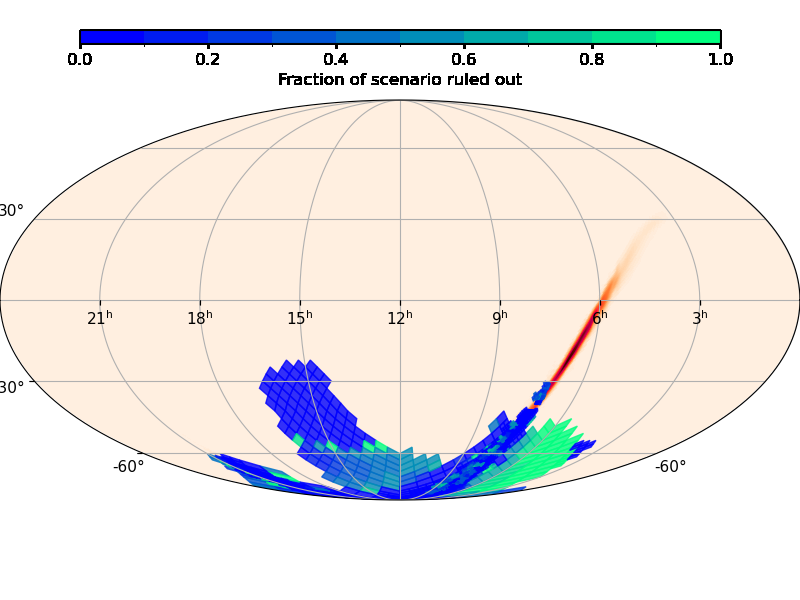}
    \caption{S230518h upper limit map over the sky. The upper limit corresponds to the deepest magnitude taken during the follow-up campaign in any filter (top).  Light curves compatible with the upper boundary on ejecta mass for S230518h (shaded regions include all selected light curves for different viewing angle $\theta$ = 0$^\circ$, 45$^\circ$, 90$^\circ$) with the observation covering the most probable sky localization  among all the field's coverage (and therefore light curves are computed at the corresponding distance), taken by time range (middle). S230518h 2-D histogram over the sky of the fraction of scenarios (in any filter) that rule out the presence of the KN in the observations. The most up-to-date GW skymap is shown in reddish color (bottom). Here both skymaps include observations from Table~\ref{followupcoverage} taken between 0 and 1 day.}
    \label{fig:S230518h-skymap-mag-fraction}
\end{figure}

\paragraph{GW230529} 

\begin{figure}
    \centering
    \includegraphics[width=1.0\linewidth]{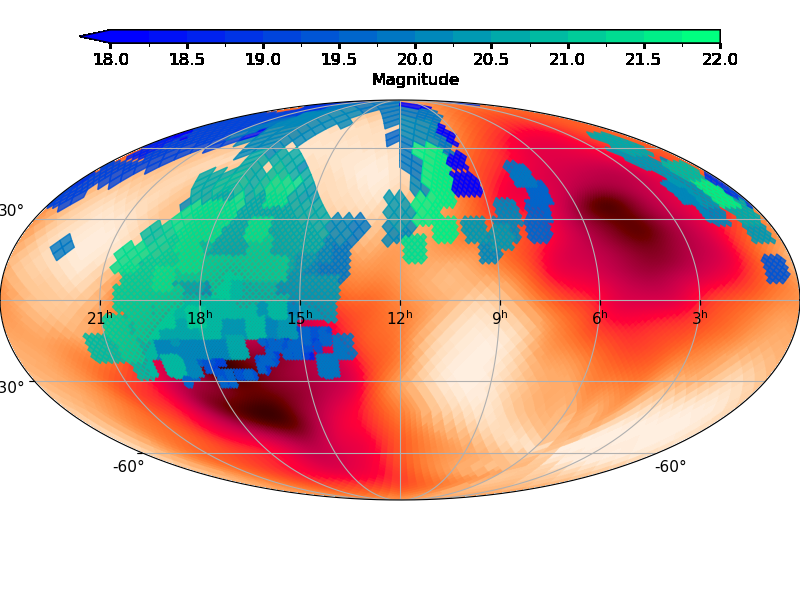}

    \vspace{-1.1cm}
    
    \includegraphics[width=\linewidth]{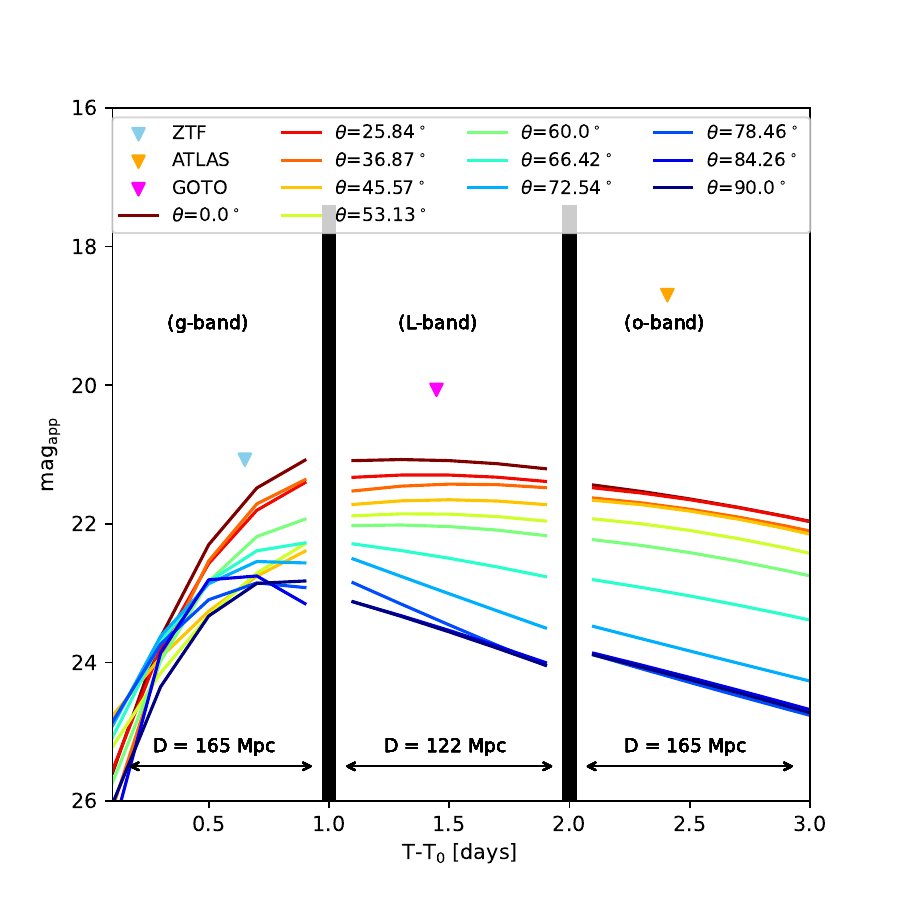}
    \includegraphics[width=1.0\linewidth]{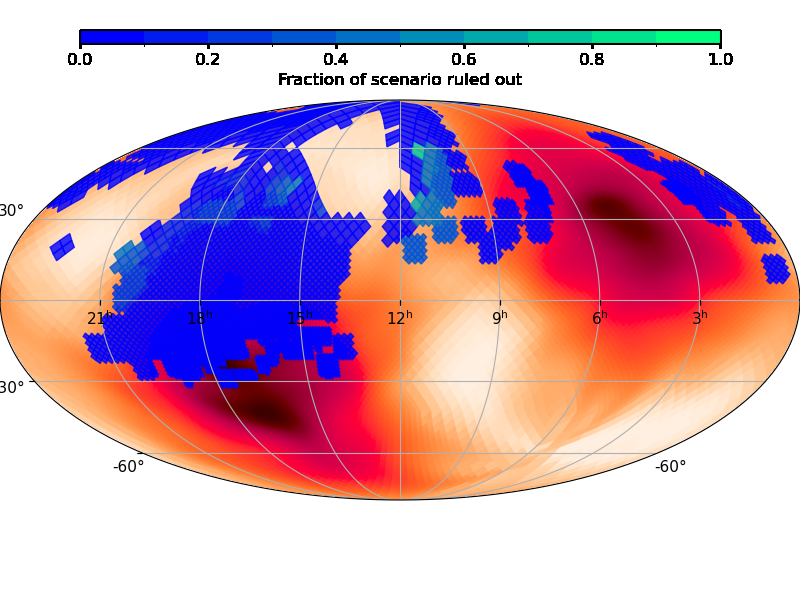}
    
    \vspace{-1.05cm}

    \caption{GW230529 upper limit map over the sky. The upper limit corresponds to the deepest magnitude taken during the follow-up campaign in any filter (top). Light curves compatible with the upper boundary on ejecta mass and PE of $\theta$ for GW230529 (we display each single selected light curves due their small number considered here) with the observation covering the most probable sky localization among all the field's coverage (and therefore light curves are computed at the corresponding distance), taken by time range (middle). GW230529 2-D histogram over the sky of the fraction of scenarios (in any filter) that rule out the presence of the KN in the observations (bottom). Here both skymaps include observations from Table~\ref{followupcoverage} taken between 0 and 1 day.}
    \label{fig:skymap-mag-GW230529}
\end{figure}

In addition to the constraints placed on GW230529 ejecta mass in Sec.~\ref{ejectaKN}, we extract the posterior distribution of the viewing angle $\theta$ from \citet{2024ApJ...970L..34A}, computed with IMRPhenomXP waveform model \citep{Pratten_2021}. However, the posterior distribution on $\theta$ is broadly unconstrained, and therefore we select the viewing angle to be in [0$^\circ$, 90$^\circ$].In the [0-1 day] time epoch, the sensitivity of the observations over the sky taken during the GW230529 follow-up campaign can be seen in Fig.~\ref{fig:skymap-mag-GW230529} (and later epochs in Appendix~\ref{coverage-later-time}, Fig.~\ref{fig:appendix-GW230529ay-skymaps}). Given the large skymap, observations partially covered GW230529 sky localization. For example, in $g$-band, observations reach a brightness limit of 20.5 mag. In particular, this occurred at the most probable sky localization among all the field's coverage, as shown in the middle panel of Fig.~\ref{fig:skymap-mag-GW230529}. We then show the corresponding scale defined in Eq.~\ref{eqn:eq1} (1 - $\mathrm{S}_\mathrm{KN,\Delta t,\mathrm{ipix}}$) over the mapped sky based on observations (lower panel). Over the skymap, there is no location within the GW230529 90\% credible region that rules out the presence of a KN, using the same threshold as for S230518h. The rare cases of scenario being incompatible are summarized as follows: between 0 and 1 day, viewing angles $\theta=0^\circ$, $\theta=25^\circ$ and $\theta=37^\circ$ are incompatible with observations that covered $\sim$3\%, $\sim$2\% and $\sim$1\% of GW230529 skymap, respectively. Other observations that constrain the viewing angle of the source cover $<$1\% of the skymap. As a conclusion, given the large size of the sky localization and the small regions covered by these observations, we cannot place constraints with certainty. However, under the assumption that the source's location was covered by observations ruling out viewing angles up to $\theta=37^\circ$ at that time, our result is consistent with the ZTF analysis that ruled out $\theta$ between 0 and 26$^\circ$ \citep{ahumada2024searchinggravitationalwaveoptical}.

\paragraph{S230627c} 
\begin{figure}
    \centering
    \includegraphics[width=\linewidth]{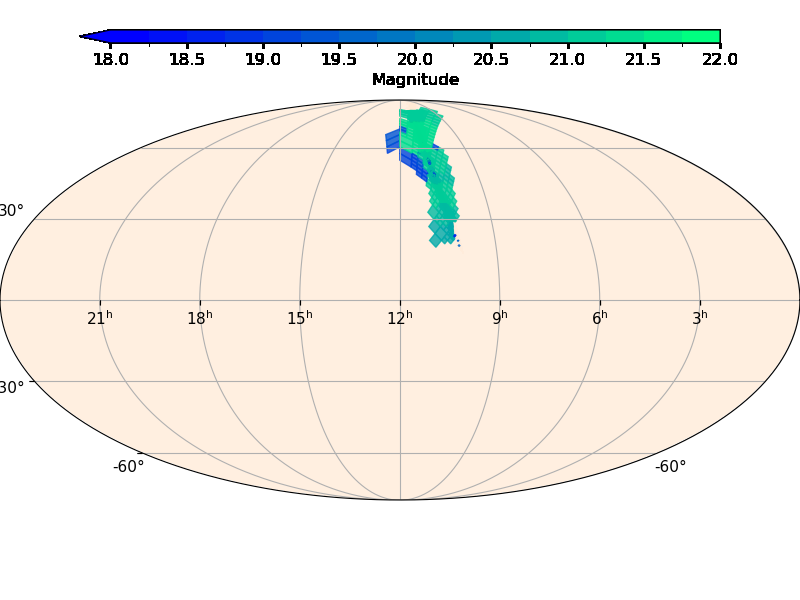}
    
    \vspace{-1.1cm}
    
    %\hspace*{-1.2cm}
\includegraphics[width=\linewidth]{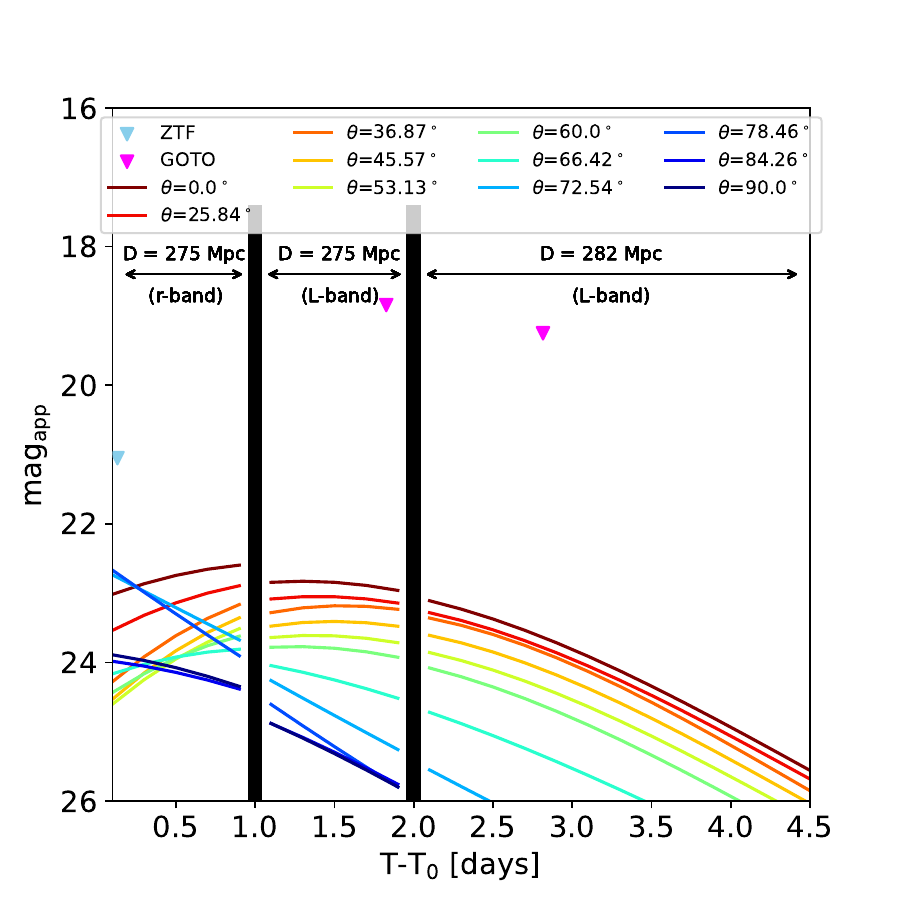}

    %\vspace{-0.2cm}
    
    \includegraphics[width=\linewidth]{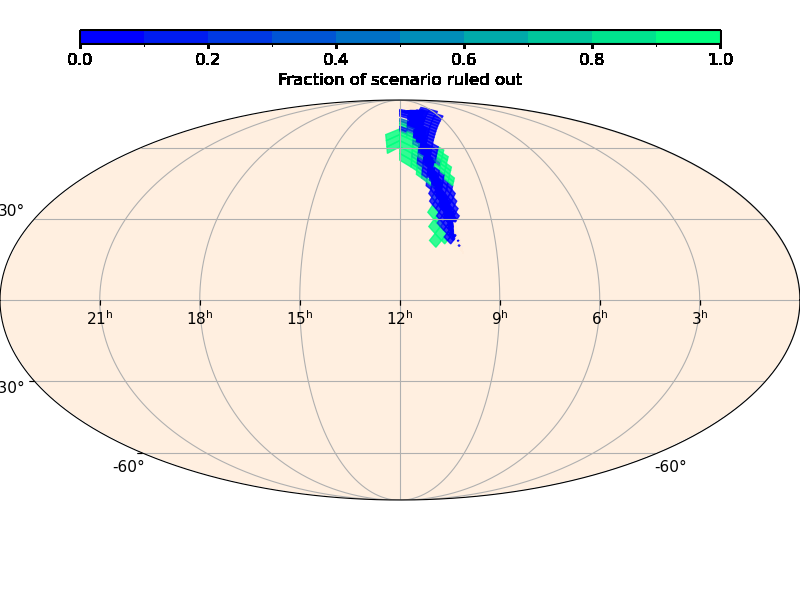}
    
    \vspace{-1.05cm}
    
    \caption{S230627c upper limit map over the sky. The upper limit corresponds to the deepest magnitude taken during the follow-up campaign in any filter (top). Light curves compatible with the upper boundary on ejecta mass for S230627c (we display each single selected light curves due their small number considered here)  with the observation covering the most probable sky localization among all the field's coverage (and therefore light curves are computed at the corresponding distance), taken by time range (middle). S230627c 2-D histogram over the sky of the fraction of scenarios (in any filter) that rule out the presence of the KN in the observations (bottom). Here both skymaps include observations from Table~\ref{followupcoverage} taken between 0 and 1 day.}
    \label{fig:skymap-mag-S230627c}
\end{figure}

We draw our conclusion using information on ejecta constraints derived from GW public information (Sec.~\ref{ejectaKN}). In the [0-1 day] time epoch, the sensitivity of
the observations over the sky taken during the S230627c follow-up campaign can be seen in Fig.~\ref{fig:skymap-mag-S230627c} (and later epochs in Appendix~\ref{coverage-later-time}, Fig.~\ref{fig:appendix-S230627c-skymaps}). Observations reached an upper limit of 21.3 mag in $r$-band during the first day. In particular, this occurred at the most probable sky localization among all the field's coverage, as shown in the middle figure of Fig.~\ref{fig:skymap-mag-S230627c}. We then show the corresponding scale defined in Eq.~\ref{eqn:eq1} (1 - $\mathrm{Scale}_\mathrm{KN,\Delta t,\mathrm{ipix}}$) over the mapped sky based on observations (lower panel). Overall, observations that assess no KNe compatible with GW cover less than 1\% of S230627c skymap as shown in Fig.~\ref{fig:cumulative-hist-prob-coverage}, therefore, we cannot place constraints on the viewing angle and ejecta properties for S230627c.

\paragraph{S240422ed} 
\begin{figure}
    \centering
    \includegraphics[width=\linewidth]{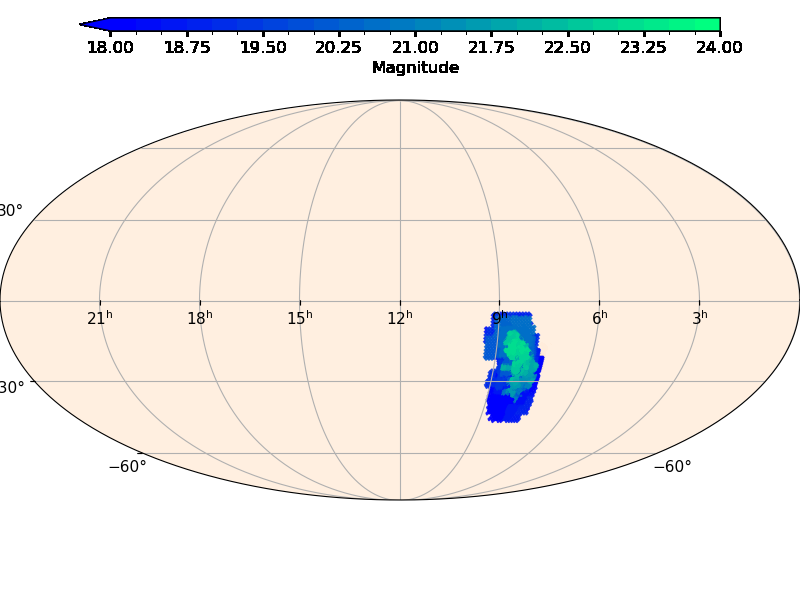}
    
    \vspace{-1.05cm}
    
    \hspace*{-1.cm}
    \includegraphics[width=1.3\linewidth]{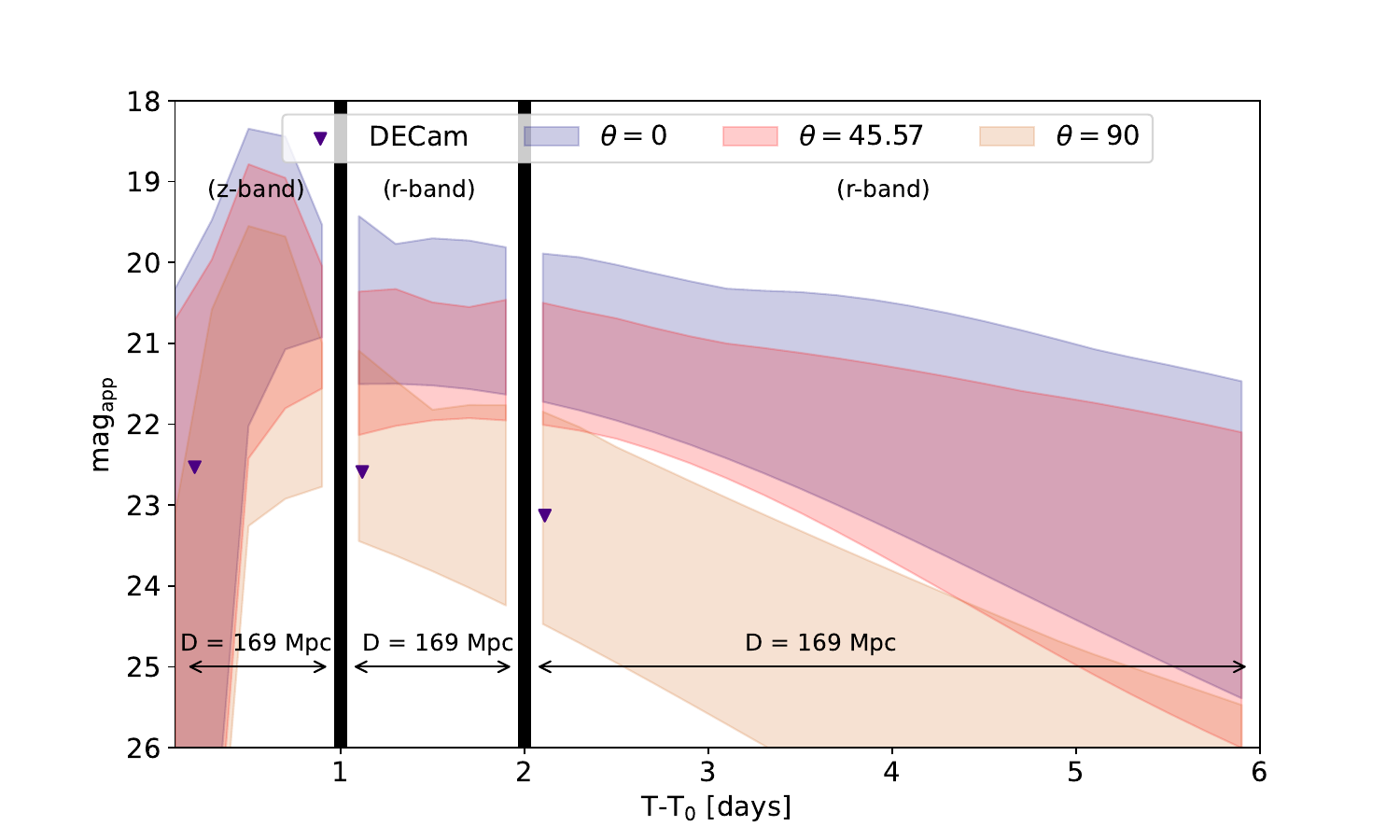}
    \includegraphics[width=\linewidth]{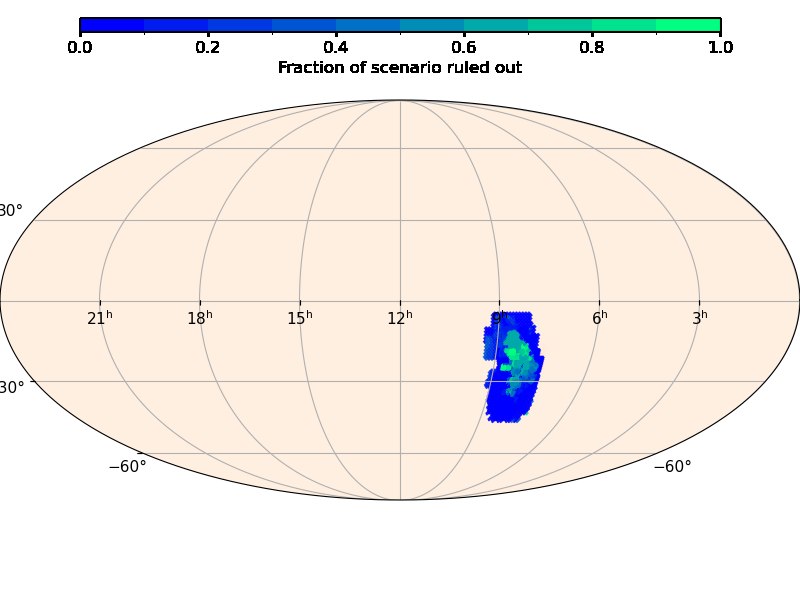}
    
    \vspace{-1.05cm}
    
    \caption{S240422ed upper limit map over the sky. The upper limit corresponds to the deepest magnitude taken during the follow-up campaign in any filter (top). All light curves from the model grid (shaded regions include all light curves for different viewing angle $\theta$ = 0$^\circ$, 45$^\circ$, 90$^\circ$) with the observation covering the most probable sky localization  among all the field's coverage (and therefore light curves are computed at the corresponding distance), taken by time range (middle). S240422ed 2-D histogram over the sky of the fraction of scenarios (in any filter) that rule out the presence of the KN in the observations. The most up-to-date GW skymap is shown in reddish color (bottom). Here both skymaps include observations from Table~\ref{followupcoverage} taken between 0 and 1 day.}
    \label{fig:skymap-mag-S240422ed}
\end{figure}

As mentioned in Sec.~\ref{followup}, the significance and classification of S240422ed was updated in June 2024, making it a low-significance candidate. Therefore, we decided to be as agnostic as possible and consider the entire grid of KN synthetic light curves regardless its PyCBC Live p$_\mathrm{astro}$ from Sec.~\ref{ejectaKN}. We looked at the astrophysical scenarios compatible with observations and finally checked if these scenarios are consistent with the S240422ed PyCBC Live p$_\mathrm{astro}$ publicly available. 
Moreover, S240422ed GW candidate was located near the Galactic plane therefore we corrected all limiting magnitudes for Milky Way extinction with the calibration of the Milky Way dust maps from \citet{Schlafly_2011} (which was not the case for the other events). We use the \texttt{piscola} package to compute the extinction correction and \texttt{sncosmo} to extract the filter transmission functions.

In the [0-1 day] time epoch, the sensitivity of the observations over the sky taken during the S240422ed follow-up campaign can be seen in Fig.~\ref{fig:skymap-mag-S240422ed} (and later epochs in Appendix~\ref{coverage-later-time} Fig.~\ref{fig:appendix-S240422ed-skymaps}). For example, most of the fields that overlap the 90\% credible region of the GW skymap reach a limiting depth max of 23.4 and 23.1 mag in $r$ and $z$-band respectively (values computed after the extinction correction), between 0 and 6 days post T$_0$ (DECam upper limit). We proceed to compare these observations with the synthetic population of KNe: Fig.~\ref{fig:skymap-mag-S240422ed} (middle) shows one example of the observations covering the most probable sky localization among all the fields’ coverage on top of all the light curves from the model grid, for
different viewing angle $\theta$ = 0$^\circ$, 45.57$^\circ$, 90$^\circ$ and taken by time bin. We then show the corresponding scale defined in Eq.~\ref{eqn:eq1} (1 - $\mathrm{S}_\mathrm{KN,\Delta t,\mathrm{ipix}}$) over the mapped sky based on observations (lower panel) and use a pre-defined threshold of 0.7. We
rule out the presence of a KN (i.e., 70\% of our synthetic KN population is inconsistent
with the observed data) over an area of 43
deg² within the 90\% credible region within the
first day post T$_0$, 153 deg² within the 90\% S240422ed
credible region between 1 and 2 days post T$_0$, and
178 deg² thereafter. These regions correspond to areas
where the GW event is localized with 13\% confidence
from 0 to 1 day, 60\% from 1 to 2 days, and 72\% beyond 2 days, respectively. 

We present final conclusions in Fig.~\ref{fig:cumulative-hist-prob-coverage}. 59\% of the simulated KNe are incompatible with observations covering more than 45\% of S240422ed skymap between 0 and 1 day. This value increases to 93\% between 1 and 2 days and reaches $\sim$100\% between 2 and 6 days (with 85\% of simulated KNe ruled out by more than 70\% of the skymap). In conclusion, while we do not entirely exclude the possibility of a KN located in a region of the skymap that did not allow us to place constraints, we can most likely rule out the presence of a KN over S240422ed sky localization. This absence of KN is not consistent with the PyCBC Live p$_\mathrm{astro}$ and the upper bound on the ejecta mass from Sec.~\ref{ejectaKN} as a null ejecta would correspond to a probability of BNS classification at least lower than $\sim$ 0.3.
%\sa{would correspond to a probability classification of XX\% BNS, XX\% NSBH and XX\% BBH}. \fixme{(for Alex: because according to the previous version we know that it means that the chirp mass should be above 2 M$_\odot$, could be place a constraint like therefore the pBNS should be lower than XXX?)}

%to a chirp mass $\mathcal{M} \geq 2.0~M_\odot$. %Although we do not exclude the fact that the KN models are not accurate enough, 
Although we do not exclude uncertainties being present with KN models, this result seems to point out that S240422ed was not of astrophysical origin, which would be consistent with the reduced significance of this candidate. We discuss this hypothesis in greater detail in the next section.

\section{Conclusion and discussion}
\label{discussion}
In conclusion, we studied three NSBH candidates S230518h, S230627c, and S230422ed, and one confirmed event GW230529, from the LVK O4 observing run. 

\paragraph{On Follow-up campaign during O4} Among these, two events have a probability exceeding 80\% to be classified as NSBH systems. The four events were located at distances ranging from 100 Mpc to 355 Mpc, considering uncertainties in localization distributions, slightly closer than the reference event GW200105 \citep{2021ApJ...915L...5A}, which was measured at 280 $\pm$ 110 Mpc. The 90\% sky localization areas for three of these events ranged between 80 deg$^2$ and 500 deg$^2$, enabling the optical follow-up community to achieve more than 80\% total coverage. These observations spanned magnitudes 17 to 23 mag across multiple filters, with prompt efforts initiated within a latency of 0 to 2 hours, extending up to 6 days post-detection. In contrast, GW230529, with its significantly larger localization area above 20,000 deg$^2$, due to being a single-detector observation, received only 37\% observational coverage with the follow-up. Across all four events, we cataloged participation from more than 40 collaborations/instruments across the full EM range and neutrinos. Despite global efforts, which collectively reported more than 50 candidates, none of them were finally confirmed as EM counterparts to the GW events.

These observations must be contextualized within the broader scope of GW alerts during O4 and previous detections from O1 to O3. Up to July 24, 2024, the observing campaign has seen a detection rate of about $\sim$3-4 BBH candidates per week surpassing a FAR threshold of 1 per 14 day, but with no BNS alerts reported so far. The arrival of Virgo in April 2024 significantly enhanced localization accuracy, with an estimated improvement of 35 \%.

\paragraph{On Follow-up strategy} In our study, we discuss the observational strategy that the optical community has followed regarding the O4 NSBH candidates and their efficiency. First, simulating diverse KNe, we find that the brightness peak time can be different with respect to color and the quantity of ejected masses from the binary system merger. More than 10\% of the peak luminosity of synthetic kilonovae in the $g,r,i$ bands is happening around one day, whereas the primary peak in the $J$-band occurs around two days post-merger representing  6 \% of the simulations. More precisely, observations conducted from 0 to 1.4 days cover the peak time of more than half of synthetic KNe in $g,r,i$ while the first 0.5 days account for less than 1\%. This highlights not only the necessity to image promptly after the merger time but also the importance of imaging 1 day post-merger, especially for shallower images, as the magnitude of the KN may significantly vary between prompt observations to the peak time. For instance, the apparent magnitude varies by more than 9 magnitudes in $g,r,i,J$ bands from prompt to peak time. The community has usually followed a prompt strategy at a cost of a \textquote{later time} one, although we could consider a more flexible approach for near and infrared bands for which the maximum of the peak time of the KN is more variable. 

Furthermore, these results show that providing additional measurements from the GW signal itself, such as the chirp mass or inclination, will be useful for the optical community in estimating a narrower range of time in which the maximum brightness is expected. This approaches the aim of maximizing the chance of a KN detection since we only select a subset of KNe consistent with the ejecta mass.

\paragraph{On the constraints placed for KNe emission during NSBH alerts} Finally, we conducted an end-to-end study to constrain the properties of KNe emitted by NSBH merger candidates detected during the O4 observing run, assuming their astrophysical origin as NSBHs. To achieve this, we extracted publicly available information on GW events, including event time, 3D localization, and p$_{astro}$. The latter was then used to estimate ejecta masses, both \textit{dynamical} and \textit{wind ejecta} contributions based on formulas outlined in \cite{2018PhRvD..98h1501F, 2020PhRvD.101j3002K,2021ApJ...922..269R}. The results were highly dependent on the BH spin projected onto the angular momentum and the EOS of densed matter employed -- both of which remain highly uncertain -- potentially leading to discrepancies of up to a factor of two in the total ejected mass. In light of these uncertainties, we adopted two approaches: a conservative model assuming compact NSs with no spin effects, and an opportunistic model allowing for more deformable NSs and significant BH spin. From these ejecta mass estimates, we selected compatible multi-band KN light curves from the synthetic population \textit{An21Bu19}. Additionally, we generated sensitivity skymaps over time, incorporating effective observational coverage while assuming the absence of detectable kilonova emission in the images. Through this, we evaluated the completeness of the observational coverage in both time and sky localization of ATLAS, DECam, GECKO, GOTO, GRANDMA, SAGUARO, TESS, WINTER and ZTF collaborations/instruments. Then, by projecting the subset of synthetic KNe onto 3D distances provided by LVK alerts, we measured compatibility with non-detections and observational upper limits. This allowed us to infer astrophysical properties of the ejecta, such as mass and viewing angle, for NSBH merger candidates observed during O4.

As a result, regarding the confirmed event GW230529 and S230627c candidate, we cannot place constraints on the ejecta or viewing angle of the source, due to telescope observational efficiency: in some cases, telescopes observed the most probable region of the GW event’s location but lacked sufficient depth in sensitivity. In other cases, telescopes achieved deep sensitivity but targeted areas outside the most probable region of the GW localization. However, there was likely not enough ejecta mass for telescopes to detect either of these events.

On S230518h and S240422ed, constraints can be placed. For S230518h, it has not been possible to observe KNe emitted from an on-axis collision up to a viewing angle of $25^\circ$, assuming a minimum confidence of 7\% for the presence of the source in this region and a low-mass ejecta ($m_{dyn,wind} \leq 0.03$~$M_\odot$) is favored against high ejected masses. Eventually, for the low-significance candidate S240422ed, observations ruled out the presence of a KN (with or without GWs), as all KN scenarios are ruled out by observations between 2 and 6 days covering more than 45\% of the GW skymap. 
The significance and the source classification of S240422ed have since been changed and seem to favor a non-astrophysical or at least a BNS rather than an NSBH origin. Therefore, using a grid of light curves modeling the KN production from BNS could be appropriate here. However, our result seems to be in favor of the non-astrophysical origin of the candidate, which would be consistent with the downgrade of S240422ed significance by LVK collaboration. In conclusion, based on these four events experiences, using a comparable available time on the telescopes and sensitivity, we state that first constraints on ejecta mass and viewing angle are possible for events below 200 Mpc, that produced a minimal ejecta above 0.1 solar mass, and which sky localization coverage is less 150 deg$^2$.

\paragraph{Debating on assessing KNe emission} 
In conclusion, while it may seem straightforward at first to determine whether a KN associated with a GW event should have been observed during follow-up campaigns, the reality is much more complex. The answer depends on multiple factors: when the observations were conducted, where they were located, which filters and instruments were used, and at which sensitivity. For every point in the sky, there is quadripartite information to consider: the probability of the GW signal being present, the associated distance (which can vary by more than 20 Mpc depending on sky location), the EM information in the observed filter, and the latency relative to the merger event. Balancing these aspects is the challenge of predicting KN emission, given the absence of critical properties of the merger (that may be obtained later with offline analysis): these include the viewing angle, the spin of the BH, the EOS of ultra-dense matter, and the fraction of the disk mass converted into ejecta. These uncertainties result in scenarios where a KN may or may not be produced and where its brightness might or might not be compatible with observations. Our results are also KN model-dependent (although state of the art): we are using these models to guide us in taking the best observations, however we do not expect the current models to fully represent reality (which also motivates making these observations). Therefore, we cannot exclude the possibility that \textit{An21Bu19} does not fully describe the diversity of KNe. 

To address these complexities, we established thresholds and developed a custom \textquote{scale} method to intercompare events and provide a global analysis of whether or not we missed the opportunity to detect KN from NSBH candidates. However, our approach is designed to be generic and can be refined as new detections are made. Improved localization will allow follow-up observations to be more concentrated in both time and sky area, while reduced measurement uncertainties in GW parameters will enable better selection of KNe consistent with the events.

\section*{Contribution}

%PWT
M. Pillas and S. Antier are the main contributors to the work including the analysis and writing across all sections, coordination and organization of contributors related to observational and GW and high energy results.

%PWT
M. Pillas is the main author of Sec.~II - 2.2, Sec.IV and Appendix of IV, V., C, D and E. M. Pillas is the main author of section IV with help of S. Antier, J-G Ducoin, T. Hussenot and M. Coughlin. M. Pillas is the main author of Sec.~V with contributions of S. Antier. P-A Duverne and Y. Rajabov are co-contributors to Sec.~I, with the coordination of S. Antier and M. Pillas. S. Antier and P. Hello are co-contributors of Sec.~II - 2.1 related to GW with the help of F. Magnani. S. Antier and F. Foucart are the main contributors of Sec.~III-3.1, and 3.2 with crucial contribution of A. H. Nitz. S. Antier is the main contributor of Sec.~A.1, A.3, A.4, and B with contributions from D. Akl, M. Pillas, O. Pynshna, R. Poggiani, J.-G. Ducoin and L. Almeida.

%GRANDMA astronomers
W. Corradi, H-B. Eiggenstein, M. Freeberg, S. Karpov, N. Kochiashvili, M. Masek, M. Molham, A. Takey, K. Noysena contributed to this work via observations and measurements taken by GRANDMA observatory

%GWandHighEnergy
C. Andrade, E. Burns, W. Corradi, T. Dietrich, T. Hussenot, D. Dornic, N. Guessoum, P. Hello, M. Lamoureux, F. Navarete, Ttable. Pradier, P. Shawhan, M. Sun, A. Toivenen, D. Turpin, T. Wouters  contributed to the review and edition of the full article to improve the quality of the article.

% ZTF
T. Ahumada, A. Wold, M W. Coughlin, M. Kasliwal and S. Anand contibruted to this work via observations and measurements taken by the Zwicky Transient Facility.

%DECAM
 I.~Andreoni, A.~Palmese, T.~Cabrera, L.~Hu, K.~Kunnumkai, B.~O'Connor to this work via observations and measurements taken by the DECAM. In addition, V. Gupta contributed to Sec.~II - 2.3 and IV - 4.1 by providing pointings of S240422ed followup from DECam telescope. 

%GOTO
K.Ackley, M.Dyer, D.O'Neill provided coverage and analysis information related to GOTO follow-up. K.Ulaczyk and J.Lyman were key to enable the data analysis. D.Steeghs contributed to manuscript review and provided additional text. 

%KMTNet/GECKO
M. Im, G. Paek, M. Jeong, S-W Chang, S. Kim, H. Choi, C.~U Lee contributed to this work via observations and measurements taken by GECKO.

%WINTER
D. Frostig, V. Karambelkar, M. Kasliwal, N. Lourie, G. Mo, R. Simcoe, and R. Stein contributed to this work via observations and measurements taken by WINTER. 

%TESS
M. Fausnaugh, R. Jayaraman, and G. Mo contributed to this work via observations and measurements taken by TESS on S230518h.

%ATLAS
M. Nicholl, S. J. Smartt, H. F. Stevance contributed to this work via observations and measurements taken from ATLAS.

%CSS/SAGUARO
K.~A.~Bostroem, W.~Fong, A. R. Gibbs, G.~Hosseinzadeh, C.~D.~Kilpatrick, J.~Rastinejad, D.~J.~Sand, M.~Shrestha contributed to this work via observations and measurements taken by SAGUARO and CSS.

\section*{Acknowledgements}
A.H.N. acknowledges support from NSF grant PHY-2309240. M.W.C, C.A. and V.G. acknowledge support from the National Science Foundation with grant numbers PHY-2308862 and PHY-2117997. The authors of MM and SK thanks to the following grants of the Ministry of Education of the Czech Republic LM2023032 and LM2023047, as well as EU/MEYS grants  CZ.02.1.01/0.0/0.0/16\_013/0001403, CZ.02.1.01/0.0/0.0/18\_046/0016007, CZ.02.1.01/0.0/0.0/ 16\_019/0000754, and CZ.02.01.01/00/22\_008/0004632. S.A. thanks the CNRS MITI and TAROT collaboration for providing support for this work. M.L. is a Postdoctoral Researcher of the Fonds de la Recherche Scientifique - M.P. acknowledges support from FNRS and IISN 4.4503. T.W. acknowledges funding through NWO under grant number OCENW.XL21.XL21.038. The work of F.\,N. is supported by NOIRLab, which is managed by the Association of Universities for Research in Astronomy (AURA) under a cooperative agreement with the National Science Foundation. AP is supported by NSF Grant No. 2308193. Based on observations at Cerro Tololo Inter-American Observatory, NSF’s NOIRLab (NOIRLab Prop. ID 2023B-851374 , PI: Andreoni \& Palmese), which is managed by the Association of Universities for Research in Astronomy (AURA) under a cooperative agreement with the National Science Foundation. W.C. acknowledges the support from CNPq - BRICS 440142/2022-9, FAPEMIG APQ 02493-22 and FNDCT/FINEP/REF 0180/22. KAO-NRIAG team acknowledges financial support from the Egyptian Science, Technology \& Innovation Funding Authority (STDF) under grant number 45779. F.F. gratefully acknowledges support from the Department of Energy, Office of Science, Office of Nuclear Physics, under contract number DE-AC02-05CH11231 and from NASA through grant 80NSSC22K0719. H.F. S. is supported by Schmidt Sciences. SJS acknowledges funding from STFC Grants ST/Y001605/1, ST/X006506/1, ST/T000198/1, a Royal Society Research Professorship and the Hintze Charitable Foundation. We thank I.M.G. and K.C. for the discussion regarding the ejecta mass computation. S.A and M.P thank Mansi Kasliwal for contributions on expertise and data related to ZTF for this work. This publication was made possible through the support of Grant 62192 from the John Templeton Foundation to LSST-DA: the opinions expressed in this publication are those of the author(s) and do not necessarily reflect the views of LSST-DA or the John Templeton Foundation. We acknowledge funding from the Daimler and Benz Foundation for the project ``NUMANJI" and from the European Union (ERC, SMArt, 101076369). Views and opinions expressed are those of the authors only and do not necessarily reflect those of the European Union or the European Research Council. Neither the European Union nor the granting authority can be held responsible for them. The Programme National des Hautes Énergies (PNHE) of CNRS/INSU co-funded by CNRS/IN2P3, CNRS/INP, CEA and CNES supported this work. N. Kochiashvili thanks V. Ayvazyan and R. Inasaridze for contribution of AbAO observations during O4: results are published in GCNs. S. A thanks les Makes Observatory, NOWT, ASTEP, UBAI-T60, OST-CDK, OPD-60, NOWT, TRT, Kilonova-catcher, FRAM, KAO, AbAO-T70, F. Rünger, P. Thierry, A. Klotz, A.~Ishankar, R.~Hainich, H.B.~Eggesntein, P.~Vignes, P.~Jacquiery, T.~Guillot, V.~Deloupy, O.~Burkhonov., E. Gurbanov, S. Ehgamberdiev, T. Sadibekova, I. Tosta e Melo, X. F. Wang, J. Zhu, X. Zeng, A. Iskandar, S. Pormente, P. Gokuldass, L. Abe, K. Agabi, J. Dibasso, for useful coordination and observation of the O4 GW candidates during O4, results are published in GCNs (GCN 34130, GCN 36284, 36299, 36326). M. Masek and S. Karpov thank M. Prouza for providing useful contributions to the observation of FRAM of electromagnetic counterpart follow-up during the O4 campaign. S.A thanks IJCLAB and the Skyportal team for set up great system of ICARE, especially M. Jouvin, J. Peloton, N. Leroy, T. Du Laz. A. Takey thank E.~G. Elhosseiny and M.~Abdelkareem for providing useful contributions to the observation of KAO of electromagnetic counterpart of candidates during the O4 campaign. W. Corradi, F. Navarete and T. de Almeida thank L. Fraga,  N. Sasaki for their contribution to the observations of OPD-60 of electromagnetic counterpart of candidates during the O4 campaign. S.A and M.P thank Y.~Tillayev, O.~Burkhonov for their contribution to the observations of UBAI electromagnetic counterpart of candidates during the O4 campaign. S.A and M.P thank O. Sokoliuk for discussion to improve the quality of the paper. M.I., G.S.H.P., M.J., S.W.C., H.C., C.U.L. and S.K. acknowledge the support from the National Research Foundation of Korea (NRF) grants, No. 2020R1A2C3011091, and No. 2021M3F7A1084525, funded by the Korea government (MSIT). This work has made use of data from the Asteroid Terrestrial-impact Last Alert System (ATLAS) project. The Asteroid Terrestrial-impact Last Alert System (ATLAS) project is primarily funded to search for near earth asteroids through NASA grants NN12AR55G, 80NSSC18K0284, and 80NSSC18K1575; byproducts of the NEO search include images and catalogs from the survey area. This work was partially funded by Kepler/K2 grant J1944/80NSSC19K0112 and HST GO-15889, and STFC grants ST/T000198/1 and ST/S006109/1. The ATLAS science products have been made possible through the contributions of the University of Hawaii Institute for Astronomy, the Queen’s University Belfast, the University of Oxford, the Space Telescope Science Institute, the South African Astronomical Observatory, and The Millennium Institute of Astrophysics (MAS), Chile. We thank the observers, especially Joh-Na Yoon, Hongjae Moon, Sumin Lee, Changgon Kim, Soojong Pak, and Jin-Guk Seo, of the Kyung Hee Astronomical Observatory, CBNUO, the SNU Astronomical Observatory, and KMTNet for performing the GECKO observations of several targets. This research has made use of the KMTNet system operated by the Korea Astronomy and Space Science Institute (KASI) at three host sites CTIO in Chile, SAAO in South Africa, and SSO in Australia. Data transfer from the host site to KASI was supported by the Korea Research Environment Open NETwork (KREONET).
D. Frostig's contribution to this material is based upon work supported by the National Science Foundation under Award No. AST-2401779. Time-domain research by the University of Arizona team and D.J.S.\ is supported by National Science Foundation (NSF) grants 2108032, 2308181, 2407566, and 2432036 and the Heising-Simons Foundation under grant \#2020-1864. This work was supported by Shota Rustaveli National Science Foundation of Georgia (SRNSFG) grant FR-24-7713.
MP and SA thanks the MOCpy's team \citep{matthieu_baumann_2024_14205461} and especially Manon Marchand and Francois-Xavier PINEAU for close communication to produce the skymap observations of the article.
MN is supported by the European Research Council (ERC) under the European Union’s Horizon 2020 research and innovation programme (grant agreement No.~948381) and by UK Space Agency Grant No.~ST/Y000692/1. IA is supported by NASA award 24-ADAP24-0159 and NSF award 2407924. The authors gratefully acknowledge the support of the NSF, STFC, INFN and CNRS for provision of computational resources. This material is based upon work supported by NSF's LIGO Laboratory which is a major facility fully funded by the National Science Foundation.

\bibliography{bibliographie}{}

\begin{thebibliography}{}
\expandafter\ifx\csname natexlab\endcsname\relax\def\natexlab#1{#1}\fi
\providecommand{\url}[1]{\href{#1}{#1}}
\providecommand{\dodoi}[1]{doi:~\href{http://doi.org/#1}{\nolinkurl{#1}}}
\providecommand{\doeprint}[1]{\href{http://ascl.net/#1}{\nolinkurl{http://ascl.net/#1}}}
\providecommand{\doarXiv}[1]{\href{https://arxiv.org/abs/#1}{\nolinkurl{https://arxiv.org/abs/#1}}}

\bibitem[{Aartsen {et~al.}(2020)Aartsen, Ackermann, Adams, Aguilar, Ahlers, Ahrens, Alispach, Andeen, Anderson, Ansseau, Anton, Argüelles, Auffenberg, Axani, Bagherpour, Bai, V., Barbano, Bartos, Barwick, Bastian, Baum, Baur, Bay, Beatty, Becker, Tjus, BenZvi, Berley, Bernardini, Besson, Binder, Bindig, Blaufuss, Blot, Bohm, Böser, Botner, Böttcher, Bourbeau, Bourbeau, Bradascio, Braun, Bron, Brostean-Kaiser, Burgman, Buscher, Busse, Carver, Chen, Cheung, Chirkin, Choi, Clark, Clark, Classen, Coleman, Collin, Conrad, Coppin, Corley, Correa, Countryman, Cowen, Cross, Dave, Clercq, DeLaunay, Dembinski, Deoskar, Ridder, Desiati, de~Vries, de~Wasseige, de~With, DeYoung, Diaz, Díaz-Vélez, Dujmovic, Dunkman, Dvorak, Eberhardt, Ehrhardt, Eller, Engel, Evenson, Fahey, Fazely, Felde, Filimonov, Finley, Fox, Franckowiak, Friedman, Fritz, Gaisser, Gallagher, Ganster, Garrappa, Gerhardt, Ghorbani, Glauch, Glüsenkamp, Goldschmidt, Gonzalez, Grant, Grégoire, Griffith, Griswold, Günder, Gündüz, Haack, Hallgren,
  Halliday, Halve, Halzen, Hanson, Haungs, Hebecker, Heereman, Heix, Helbing, Hellauer, Henningsen, Hickford, Hignight, Hill, Hoffman, Hoffmann, Hoinka, Hokanson-Fasig, Hoshina, Huang, Huber, Huber, Hultqvist, Hünnefeld, Hussain, In, Iovine, Ishihara, Jansson, Japaridze, Jeong, Jero, Jones, Jonske, Joppe, Kang, Kang, Kappes, Kappesser, Karg, Karl, Karle, Katz, Kauer, Keivani, Kellermann, Kelley, Kheirandish, Kim, Kintscher, Kiryluk, Kittler, Klein, Koirala, Kolanoski, Köpke, Kopper, Kopper, Koskinen, Kowalski, Krings, Krückl, Kulacz, Kurahashi, Kyriacou, Lanfranchi, Larson, Lauber, Lazar, Leonard, Leszczyńska, Liu, Lohfink, Mariscal, Lu, Lucarelli, Ludwig, Lünemann, Luszczak, Lyu, Ma, Madsen, Maggi, Mahn, Makino, Mallik, Mallot, Mancina, Mariş, Marka, Marka, Maruyama, Mase, Maunu, McNally, Meagher, Medici, Medina, Meier, Meighen-Berger, Merino, Meures, Micallef, Mockler, Momenté, Montaruli, Moore, Morse, Moulai, Muth, Nagai, Naumann, Neer, Nguyen, Niederhausen, Nisa, Nowicki, Nygren, Pollmann, Oehler,
  Olivas, O’Murchadha, O’Sullivan, Palczewski, Pandya, Pankova, Park, Peiffer, de~los Heros, Philippen, Pieloth, Pieper, Pinat, Pizzuto, Plum, Porcelli, Price, Przybylski, Raab, Raissi, Rameez, Rauch, Rawlins, Rea, Rehman, Reimann, Relethford, Renschler, Renzi, Resconi, Rhode, Richman, Robertson, Rongen, Rott, Ruhe, Ryckbosch, Cantu, Safa, Herrera, Sandrock, Sandroos, Santander, Sarkar, Sarkar, Satalecka, Schaufel, Schieler, Schlunder, Schmidt, Schneider, Schneider, Schröder, Schumacher, Sclafani, Seckel, Seunarine, Shefali, Silva, Snihur, Soedingrekso, Soldin, Song, Spiczak, Spiering, Stachurska, Stamatikos, Stanev, Stein, Stettner, Steuer, Stezelberger, Stokstad, Stößl, Strotjohann, Stürwald, Stuttard, Sullivan, Taboada, Tenholt, Ter-Antonyan, Terliuk, Tilav, Tollefson, Tomankova, Tönnis, Toscano, Tosi, Trettin, Tselengidou, Tung, Turcati, Turcotte, Turley, Ty, Unger, Elorrieta, Usner, Vandenbroucke, Driessche, van Eijk, van Eijndhoven, van Santen, Verpoest, Veske, Vraeghe, Walck, Wallace,
  Wallraff, Wandkowsky, Watson, Weaver, Weindl, Weiss, Weldert, Wendt, Werthebach, Whelan, Whitehorn, Wiebe, Wiebusch, Wille, Williams, Wills, Wolf, Wood, Wood, Woschnagg, Wrede, Xu, Xu, Xu, Yanez, Yodh, Yoshida, Yuan, \& Zöcklein}]{Aartsen_2020}
Aartsen, M.~G., Ackermann, M., Adams, J., {et~al.} 2020, The Astrophysical Journal Letters, 898, L10, \dodoi{10.3847/2041-8213/ab9d24}

\bibitem[{Aasi {et~al.}(2015)Aasi, Abbott, Abbott, Abbott, Abernathy, Ackley, Adams, Adams, \& Addesso}]{LIGO}
Aasi, J., Abbott, B.~P., Abbott, R., {et~al.} 2015, Classical and Quantum Gravity, 32, 074001, \dodoi{10.1088/0264-9381/32/7/074001}

\bibitem[{{Abac} {et~al.}(2024){Abac}, {Abbott}, {Abouelfettouh}, {Acernese}, {Ackley}, {Adhicary}, {et~al.}}]{2024ApJ...970L..34A}
{Abac}, A.~G., {Abbott}, R., {Abouelfettouh}, I., {et~al.} 2024, \apjl, 970, L34, \dodoi{10.3847/2041-8213/ad5beb}

\bibitem[{Abbasi {et~al.}(2023)Abbasi, Ackermann, Adams, Aggarwal, Aguilar, Ahlers, Ahrens, Alameddine, Alves, Amin, Andeen, Anderson, Anton, Argüelles, Asali, Ashida, Athanasiadou, Axani, Bai, V., Baricevic, Bartos, Barwick, Basu, Bay, Beatty, Becker, Tjus, Beise, Bellenghi, Benda, BenZvi, Berley, Bernardini, Besson, Binder, Bindig, Blaufuss, Blot, Bontempo, Book, Borowka, Böser, Botner, Böttcher, Bourbeau, Bradascio, Braun, Brinson, Bron, Brostean-Kaiser, Burley, Busse, Campana, Carnie-Bronca, Chen, Chen, Chirkin, Choi, Clark, Classen, Coleman, Collin, Connolly, Conrad, Coppin, Correa, Countryman, Cowen, Cross, Dappen, Dave, Clercq, DeLaunay, López, Dembinski, Deoskar, Desai, Desiati, de~Vries, de~Wasseige, DeYoung, Diaz, Díaz-Vélez, Dittmer, Dujmovic, DuVernois, Ehrhardt, Eller, Engel, Erpenbeck, Evans, Evenson, Fan, Fazely, Fedynitch, Feigl, Fiedlschuster, Fienberg, Finley, Fischer, Fox, Franckowiak, Friedman, Fritz, Fürst, Gaisser, Gallagher, Ganster, Garcia, Garrappa, Gerhardt, Ghadimi, Glaser,
  Glauch, Glüsenkamp, Goehlke, Gonzalez, Goswami, Grant, Grégoire, Griswold, Günther, Gutjahr, Haack, Hallgren, Halliday, Halve, Halzen, Hamdaoui, Minh, Hanson, Hardin, Harnisch, Hatch, Haungs, Helbing, Hellrung, Henningsen, Heuermann, Hickford, Hill, Hill, Hoffman, Hoshina, Hou, Huber, Hultqvist, Hünnefeld, Hussain, Hymon, In, Iovine, Ishihara, Jansson, Japaridze, Jeong, Jin, Jones, Kang, Kang, Kang, Kappes, Kappesser, Kardum, Karg, Karl, Karle, Katz, Kauer, Kelley, Kheirandish, Kin, Kiryluk, Klein, Kochocki, Koirala, Kolanoski, Kontrimas, Köpke, Kopper, Koskinen, Koundal, Kovacevich, Kowalski, Kozynets, Krupczak, Kun, Kurahashi, Lad, Gualda, Larson, Lauber, Lazar, Lee, Leonard, Leszczyńska, Lincetto, Liu, Liubarska, Lohfink, Love, Mariscal, Lu, Lucarelli, Ludwig, Luszczak, Lyu, Ma, Madsen, Mahn, Makino, Mancina, Sainte, Mariş, Márka, Márka, Marsee, Martinez-Soler, Maruyama, McElroy, McNally, Mead, Meagher, Mechbal, Medina, Meier, Meighen-Berger, Merckx, Micallef, Mockler, Montaruli, Moore, Morse,
  Moulai, Mukherjee, Naab, Nagai, Naumann, Necker, Neumann, Niederhausen, Nisa, Nowicki, Pollmann, Oehler, Oeyen, Olivas, Orsoe, Osborn, O’Sullivan, Pandya, Pankova, Park, Parker, Paudel, Paul, de~los Heros, Peters, Peterson, Philippen, Pieper, Pizzuto, Plum, Popovych, Porcelli, Rodriguez, Pries, Przybylski, Raab, Rack-Helleis, Rameez, Rawlins, Rechav, Rehman, Reichherzer, Renzi, Resconi, Reusch, Rhode, Richman, Riedel, Roberts, Robertson, Rodan, Roellinghoff, Rongen, Rott, Ruhe, Ruohan, Ryckbosch, Cantu, Safa, Saffer, Salazar-Gallegos, Sampathkumar, Herrera, Sandrock, Santander, Sarkar, Sarkar, Satalecka, Schaufel, Schieler, Schindler, Schlueter, Schmidt, Schneider, Schröder, Schumacher, Schwefer, Sclafani, Seckel, Seunarine, Sharma, Shefali, Shimizu, Silva, Oliveira, Skrzypek, Smithers, Snihur, Soedingrekso, Sogaard, Soldin, Spannfellner, Spiczak, Spiering, Stamatikos, Stanev, Stein, Stezelberger, Stürwald, Stuttard, Sullivan, Sullivan, Taboada, Ter-Antonyan, Thompson, Thwaites, Tilav, Tollefson,
  Tönnis, Toscano, Tosi, Trettin, Tung, Turcotte, Twagirayezu, Ty, Elorrieta, Upshaw, Valtonen-Mattila, Vandenbroucke, van Eijndhoven, Vannerom, van Santen, Vara, Veitch-Michaelis, Verpoest, Veske, Walck, Wang, Watson, Weaver, Weigel, Weindl, Weldert, Wendt, Werthebach, Weyrauch, Whitehorn, Wiebusch, Willey, Williams, Wolf, Wrede, Wulff, Xu, Yanez, Yildizci, Yoshida, Yu, Yuan, Zhang, Zhelnin, \& Collaboration}]{Abbasi_2023}
Abbasi, R., Ackermann, M., Adams, J., {et~al.} 2023, The Astrophysical Journal, 944, 80, \dodoi{10.3847/1538-4357/aca5fc}

\bibitem[{Abbott {et~al.}(2019)Abbott, Abbott, Abbott, Abraham, Acernese, Ackley, Adams, Adhikari, Adya, Affeldt, Agathos, Agatsuma, Aggarwal, Aguiar, Aiello, Ain, Ajith, Allen, Allocca, Aloy, Altin, Amato, Ananyeva, Anderson, Anderson, Angelova, Antier, Appert, Arai, Araya, Areeda, Arène, Arnaud, Arun, Ascenzi, Ashton, Aston, Astone, Aubin, Aufmuth, AultONeal, Austin, Avendano, Avila-Alvarez, Babak, Bacon, Badaracco, Bader, Bae, Baker, Baldaccini, Ballardin, Ballmer, Banagiri, Barayoga, Barclay, Barish, Barker, Barkett, Barnum, Barone, Barr, Barsotti, Barsuglia, Barta, Bartlett, Bartos, Bassiri, Basti, Bawaj, Bayley, Bazzan, Bécsy, Bejger, Belahcene, Bell, Beniwal, Berger, Bergmann, Bernuzzi, Bero, Berry, Bersanetti, Bertolini, Betzwieser, Bhandare, Bidler, Bilenko, Bilgili, Billingsley, Birch, Birney, Birnholtz, Biscans, Biscoveanu, Bisht, Bitossi, Bizouard, Blackburn, Blackman, Blair, Blair, Blair, Bloemen, Bode, Boer, Boetzel, Bogaert, Bondu, Bonilla, Bonnand, Booker, Boom, Booth, Bork, Boschi, Bose,
  Bossie, Bossilkov, Bosveld, Bouffanais, Bozzi, Bradaschia, Brady, Bramley, Branchesi, Brau, Briant, Briggs, Brighenti, Brillet, Brinkmann, Brisson, Brockill, Brooks, Brown, Brunett, Buikema, Bulik, Bulten, Buonanno, Buskulic, Bustamante~Rosell, Buy, Byer, Cabero, Cadonati, Cagnoli, Cahillane, Calderón~Bustillo, Callister, Calloni, Camp, Campbell, Canepa, Cannon, Cao, Cao, Capocasa, Carbognani, Caride, Carney, Carullo, Casanueva~Diaz, Casentini, Caudill, Cavaglià, Cavalier, Cavalieri, Cella, Cerdá-Durán, Cerretani, Cesarini, Chaibi, Chakravarti, Chamberlin, Chan, Chao, Charlton, Chase, Chassande-Mottin, Chatterjee, Chaturvedi, Chatziioannou, Cheeseboro, Chen, Chen, Chen, Cheng, Cheong, Chia, Chincarini, Chiummo, Cho, Cho, Cho, Christensen, Chu, Chua, Chung, Chung, Ciani, Ciobanu, Ciolfi, Cipriano, Cirone, Clara, Clark, Clearwater, Cleva, Cocchieri, Coccia, Cohadon, Cohen, Colgan, Colleoni, Collette, Collins, Cominsky, Constancio, Conti, Cooper, Corban, Corbitt, Cordero-Carrión, Corley, Cornish, Corsi,
  Cortese, Costa, Cotesta, Coughlin, Coughlin, Coulon, Countryman, Couvares, Covas, Cowan, Coward, Cowart, Coyne, Coyne, Creighton, Creighton, Cripe, Croquette, Crowder, Cullen, Cumming, Cunningham, Cuoco, Canton, Dálya, Danilishin, D’Antonio, Danzmann, Dasgupta, Da~Silva~Costa, Datrier, Dattilo, Dave, Davier, Davis, Daw, DeBra, Deenadayalan, Degallaix, De~Laurentis, Deléglise, Del~Pozzo, DeMarchi, Demos, Dent, De~Pietri, Derby, De~Rosa, De~Rossi, DeSalvo, de~Varona, Dhurandhar, Díaz, Dietrich, Di~Fiore, Di~Giovanni, Di~Girolamo, Di~Lieto, Ding, Di~Pace, Di~Palma, Di~Renzo, Dmitriev, Doctor, Donovan, Dooley, Doravari, Dorrington, Downes, Drago, Driggers, Du, Ducoin, Dupej, Dwyer, Easter, Edo, Edwards, Effler, Ehrens, Eichholz, Eikenberry, Eisenmann, Eisenstein, Essick, Estelles, Estevez, Etienne, Etzel, Evans, Evans, Fafone, Fair, Fairhurst, Fan, Farinon, Farr, Farr, Fauchon-Jones, Favata, Fays, Fazio, Fee, Feicht, Fejer, Feng, Fernandez-Galiana, Ferrante, Ferreira, Ferreira, Ferrini, Fidecaro, Fiori,
  Fiorucci, Fishbach, Fisher, Fishner, Fitz-Axen, Flaminio, Fletcher, Flynn, Fong, Font, Forsyth, Fournier, Frasca, Frasconi, Frei, Freise, Frey, Frey, Fritschel, Frolov, Fulda, Fyffe, Gabbard, Gadre, Gaebel, Gair, Gammaitoni, Ganija, Gaonkar, Garcia, García-Quirós, Garufi, Gateley, Gaudio, Gaur, Gayathri, Gemme, Genin, Gennai, George, George, Gergely, Germain, Ghonge, Ghosh, Ghosh, Ghosh, Giacomazzo, Giaime, Giardina, Giazotto, Gill, Giordano, Glover, Godwin, Goetz, Goetz, Goncharov, González, Gonzalez~Castro, Gopakumar, Gorodetsky, Gossan, Gosselin, Gouaty, Grado, Graef, Granata, Grant, Gras, Grassia, Gray, Gray, Greco, Green, Green, Gretarsson, Groot, Grote, Grunewald, Gruning, Guidi, Gulati, Guo, Gupta, Gupta, Gustafson, Gustafson, Haegel, Halim, Hall, Hall, Hamilton, Hammond, Haney, Hanke, Hanks, Hanna, Hannam, Hannuksela, Hanson, Hardwick, Haris, Harms, Harry, Harry, Haster, Haughian, Hayes, Healy, Heidmann, Heintze, Heitmann, Hello, Hemming, Hendry, Heng, Hennig, Heptonstall, Hernandez~Vivanco,
  Heurs, Hild, Hinderer, Hoak, Hochheim, Hofman, Holgado, Holland, Holt, Holz, Hopkins, Horst, Hough, Howell, Hoy, Hreibi, Huang, Huerta, Huet, Hughey, Hulko, Husa, Huttner, Huynh-Dinh, Idzkowski, Iess, Ingram, Inta, Intini, Irwin, Isa, Isac, Isi, Iyer, Izumi, Jacqmin, Jadhav, Jani, Janthalur, Jaranowski, Jenkins, Jiang, Johnson, Johnson-McDaniel, Jones, Jones, Jones, Jonker, Ju, Junker, Kalaghatgi, Kalogera, Kamai, Kandhasamy, Kang, Kanner, Kapadia, Karki, Karvinen, Kashyap, Kasprzack, Katsanevas, Katsavounidis, Katzman, Kaufer, Kawabe, Keerthana, Kéfélian, Keitel, Kennedy, Key, Khalili, Khan, Khan, Khan, Khan, Khazanov, Khursheed, Kijbunchoo, Kim, Kim, Kim, Kim, Kim, Kim, Kimball, King, King, Kinley-Hanlon, Kirchhoff, Kissel, Kleybolte, Klika, Klimenko, Knowles, Koch, Koehlenbeck, Koekoek, Koley, Kondrashov, Kontos, Koper, Korobko, Korth, Kowalska, Kozak, Kringel, Krishnendu, Królak, Kuehn, Kumar, Kumar, Kumar, Kumar, Kuo, Kutynia, Kwang, Lackey, Lai, Lam, Landry, Lane, Lang, Lange, Lantz, Lanza,
  Lartaux-Vollard, Lasky, Laxen, Lazzarini, Lazzaro, Leaci, Leavey, Lecoeuche, Lee, Lee, Lee, Lee, Lee, Lee, Lehmann, Lenon, Leroy, Letendre, Levin, Li, Li, Li, Li, Lin, Linde, Linker, Littenberg, Liu, Liu, Lo, Lockerbie, London, Longo, Lorenzini, Loriette, Lormand, Losurdo, Lough, Lousto, Lovelace, Lower, Lück, Lumaca, Lundgren, Lynch, Ma, Macas, Macfoy, MacInnis, Macleod, Macquet, Magaña-Sandoval, Magaña~Zertuche, Magee, Majorana, Maksimovic, Malik, Man, Mandic, Mangano, Mansell, Manske, Mantovani, Marchesoni, Marion, Márka, Márka, Markakis, Markosyan, Markowitz, Maros, Marquina, Marsat, Martelli, Martin, Martin, Martynov, Mason, Massera, Masserot, Massinger, Masso-Reid, Mastrogiovanni, Matas, Matichard, Matone, Mavalvala, Mazumder, McCann, McCarthy, McClelland, McCormick, McCuller, McGuire, McIver, McManus, McRae, McWilliams, Meacher, Meadors, Mehmet, Mehta, Meidam, Melatos, Mendell, Mercer, Mereni, Merilh, Merzougui, Meshkov, Messenger, Messick, Metzdorff, Meyers, Miao, Michel, Middleton, Mikhailov,
  Milano, Miller, Miller, Millhouse, Mills, Milovich-Goff, Minazzoli, Minenkov, Mishkin, Mishra, Mistry, Mitra, Mitrofanov, Mitselmakher, Mittleman, Mo, Moffa, Mogushi, Mohapatra, Montani, Moore, Moraru, Moreno, Morisaki, Mours, Mow-Lowry, Mukherjee, Mukherjee, Mukherjee, Mukund, Mullavey, Munch, Muñiz, Muratore, Murray, Nagar, Nardecchia, Naticchioni, Nayak, Neilson, Nelemans, Nelson, Nery, Neunzert, Ng, Ng, Nguyen, Nichols, Nielsen, Nissanke, Nitz, Nocera, North, Nuttall, Obergaulinger, Oberling, O’Brien, O’Dea, Ogin, Oh, Oh, Ohme, Ohta, Okada, Oliver, Oppermann, Oram, O’Reilly, Ormiston, Ortega, O’Shaughnessy, Ossokine, Ottaway, Overmier, Owen, Pace, Pagano, Page, Pai, Pai, Palamos, Palashov, Palomba, Pal-Singh, Pan, Pang, Pang, Pankow, Pannarale, Pant, Paoletti, Paoli, Papa, Parida, Parker, Pascucci, Pasqualetti, Passaquieti, Passuello, Patil, Patricelli, Pearlstone, Pedersen, Pedraza, Pedurand, Pele, Penn, Perego, Perez, Perreca, Pfeiffer, Phelps, Phukon, Piccinni, Pichot, Piergiovanni, Pillant,
  Pinard, Pirello, Pitkin, Poggiani, Pong, Ponrathnam, Popolizio, Porter, Powell, Prajapati, Prasad, Prasai, Prasanna, Pratten, Prestegard, Privitera, Prodi, Prokhorov, Puncken, Punturo, Puppo, Pürrer, Qi, Quetschke, Quinonez, Quintero, Quitzow-James, Raab, Radkins, Radulescu, Raffai, Raja, Rajan, Rajbhandari, Rakhmanov, Ramirez, Ramos-Buades, Rana, Rao, Rapagnani, Raymond, Razzano, Read, Regimbau, Rei, Reid, Reitze, Ren, Ricci, Richardson, Richardson, Ricker, Riemenschneider, Riles, Rizzo, Robertson, Robie, Robinet, Rocchi, Rolland, Rollins, Roma, Romanelli, Romano, Romel, Romie, Rose, Rosińska, Rosofsky, Ross, Rowan, Rüdiger, Ruggi, Rutins, Ryan, Sachdev, Sadecki, Sakellariadou, Salafia, Salconi, Saleem, Salemi, Samajdar, Sammut, Sanchez, Sanchez, Sanchis-Gual, Sandberg, Sanders, Santiago, Sarin, Sassolas, Sathyaprakash, Saulson, Sauter, Savage, Schale, Scheel, Scheuer, Schmidt, Schnabel, Schofield, Schönbeck, Schreiber, Schulte, Schutz, Schwalbe, Scott, Scott, Seidel, Sellers, Sengupta, Sennett,
  Sentenac, Sequino, Sergeev, Setyawati, Shaddock, Shaffer, Shahriar, Shaner, Shao, Sharma, Shawhan, Shen, Shink, Shoemaker, Shoemaker, ShyamSundar, Siellez, Sieniawska, Sigg, Silva, Singer, Singh, Singhal, Sintes, Sitmukhambetov, Skliris, Slagmolen, Slaven-Blair, Smith, Smith, Somala, Son, Sorazu, Sorrentino, Souradeep, Sowell, Spencer, Srivastava, Srivastava, Staats, Stachie, Standke, Steer, Steinke, Steinlechner, Steinlechner, Steinmeyer, Stevenson, Stocks, Stone, Stops, Strain, Stratta, Strigin, Strunk, Sturani, Stuver, Sudhir, Summerscales, Sun, Sunil, Suresh, Sutton, Swinkels, Szczepańczyk, Tacca, Tait, Talbot, Talukder, Tanner, Tápai, Taracchini, Tasson, Taylor, Thies, Thomas, Thomas, Thondapu, Thorne, Thrane, Tiwari, Tiwari, Tiwari, Toland, Tonelli, Tornasi, Torres-Forné, Torrie, Töyrä, Travasso, Traylor, Tringali, Trovato, Trozzo, Trudeau, Tsang, Tse, Tso, Tsukada, Tsuna, Tuyenbayev, Ueno, Ugolini, Unnikrishnan, Urban, Usman, Vahlbruch, Vajente, Valdes, van Bakel, van Beuzekom, van~den Brand,
  Van Den~Broeck, Vander-Hyde, van Heijningen, van~der Schaaf, van Veggel, Vardaro, Varma, Vass, Vasúth, Vecchio, Vedovato, Veitch, Veitch, Venkateswara, Venugopalan, Verkindt, Vetrano, Viceré, Viets, Vine, Vinet, Vitale, Vo, Vocca, Vorvick, Vyatchanin, Wade, Wade, Wade, Walet, Walker, Wallace, Walsh, Wang, Wang, Wang, Wang, Wang, Ward, Warden, Warner, Was, Watchi, Weaver, Wei, Weinert, Weinstein, Weiss, Wellmann, Wen, Wessel, Weßels, Westhouse, Wette, Whelan, White, Whiting, Whittle, Wilken, Williams, Williamson, Willis, Willke, Wimmer, Winkler, Wipf, Wittel, Woan, Woehler, Wofford, Worden, Wright, Wu, Wysocki, Xiao, Yamamoto, Yancey, Yang, Yap, Yazback, Yeeles, Yu, Yu, Yuen, Yvert, Zadrożny, Zanolin, Zappa, Zelenova, Zendri, Zevin, Zhang, Zhang, Zhang, Zhao, Zhou, Zhou, Zhu, Zimmerman, Zlochower, Zucker, \& Zweizig}]{Abbott_2019}
Abbott, B., Abbott, R., Abbott, T., {et~al.} 2019, Physical Review X, 9, \dodoi{10.1103/physrevx.9.031040}

\bibitem[{Abbott(2017{\natexlab{a}})}]{Abbott_2017}
Abbott, B.~a. 2017{\natexlab{a}}, Physical Review Letters, 119, \dodoi{10.1103/physrevlett.119.161101}

\bibitem[{{Abbott} {et~al.}(2017){Abbott}, {Abbott}, {Abbott}, {Acernese}, {Ackley}, {Adams}, {Adams}, {Addesso}, {Adhikari}, {Adya}, {Affeldt}, {Afrough}, {Agarwal}, {Agathos}, {Agatsuma}, {(INTEGRAL}, {et~al.}}]{grb_2017}
{Abbott}, B.~P., {Abbott}, R., {Abbott}, T.~D., {et~al.} 2017, APJL, 848, L13, \dodoi{10.3847/2041-8213/aa920c}

\bibitem[{Abbott(2017{\natexlab{b}})}]{Abbott_2017_2}
Abbott, B. P. e.~a. 2017{\natexlab{b}}, The Astrophysical Journal Letters, 848, L13, \dodoi{10.3847/2041-8213/aa920c}

\bibitem[{Abbott(2017{\natexlab{c}})}]{Abbott_2017_3}
---. 2017{\natexlab{c}}, The Astrophysical Journal Letters, 848, L12, \dodoi{10.3847/2041-8213/aa91c9}

\bibitem[{Abbott(2017{\natexlab{d}})}]{Abbott_2017_4}
---. 2017{\natexlab{d}}, The Astrophysical Journal Letters, 850, L39, \dodoi{10.3847/2041-8213/aa9478}

\bibitem[{Abbott {et~al.}(2020)Abbott, Abbott, Abraham, Acernese, Ackley, Adams, Adhikari, Adya, Affeldt, Agathos, Agatsuma, Aggarwal, Aguiar, Aich, Aiello, Ain, Ajith, Akcay, Allen, Allocca, Altin, Amato, Anand, Ananyeva, Anderson, Anderson, Angelova, Ansoldi, Antier, Appert, Arai, Araya, Areeda, Arène, Arnaud, Aronson, Arun, Asali, Ascenzi, Ashton, Aston, Astone, Aubin, Aufmuth, AultONeal, Austin, Avendano, Babak, Bacon, Badaracco, Bader, Bae, Baer, Baird, Baldaccini, Ballardin, Ballmer, Bals, Balsamo, Baltus, Banagiri, Bankar, Bankar, Barayoga, Barbieri, Barish, Barker, Barkett, Barneo, Barone, Barr, Barsotti, Barsuglia, Barta, Bartlett, Bartos, Bassiri, Basti, Bawaj, Bayley, Bazzan, Bécsy, Bejger, Belahcene, Bell, Beniwal, Benjamin, Benkel, Bentley, Bergamin, Berger, Bergmann, Bernuzzi, Berry, Bersanetti, Bertolini, Betzwieser, Bhandare, Bhandari, Bidler, Biggs, Bilenko, Billingsley, Birney, Birnholtz, Biscans, Bischi, Biscoveanu, Bisht, Bissenbayeva, Bitossi, Bizouard, Blackburn, Blackman, Blair,
  Blair, Blair, Bobba, Bode, Boer, Boetzel, Bogaert, Bondu, Bonilla, Bonnand, Booker, Boom, Bork, Boschi, Bose, Bossilkov, Bosveld, Bouffanais, Bozzi, Bradaschia, Brady, Bramley, Branchesi, Brau, Breschi, Briant, Briggs, Brighenti, Brillet, Brinkmann, Brito, Brockill, Brooks, Brooks, Brown, Brunett, Bruno, Bruntz, Buikema, Bulik, Bulten, Buonanno, Buskulic, Byer, Cabero, Cadonati, Cagnoli, Cahillane, Bustillo, Callaghan, Callister, Calloni, Camp, Canepa, Cannon, Cao, Cao, Carapella, Carbognani, Caride, Carney, Carullo, Diaz, Casentini, Castañeda, Caudill, Cavaglià, Cavalier, Cavalieri, Cella, Cerdá-Durán, Cesarini, Chaibi, Chakravarti, Chan, Chan, Chao, Charlton, Chase, Chassande-Mottin, Chatterjee, Chaturvedi, Chatziioannou, Chen, Chen, Chen, Cheng, Cheong, Chia, Chiadini, Chierici, Chincarini, Chiummo, Cho, Cho, Cho, Christensen, Chu, Chua, Chung, Chung, Ciani, Ciecielag, Cieślar, Ciobanu, Ciolfi, Cipriano, Cirone, Clara, Clark, Clearwater, Clesse, Cleva, Coccia, Cohadon, Cohen, Colleoni, Collette,
  Collins, Colpi, Constancio, Conti, Cooper, Corban, Corbitt, Cordero-Carrión, Corezzi, Corley, Cornish, Corre, Corsi, Cortese, Costa, Cotesta, Coughlin, Coughlin, Coulon, Countryman, Couvares, Covas, Coward, Cowart, Coyne, Coyne, Creighton, Creighton, Cripe, Croquette, Crowder, Cudell, Cullen, Cumming, Cummings, Cunningham, Cuoco, Curylo, Canton, Dálya, Dana, Daneshgaran-Bajastani, D’Angelo, Danilishin, D’Antonio, Danzmann, Darsow-Fromm, Dasgupta, Datrier, Dattilo, Dave, Davier, Davies, Davis, Daw, DeBra, Deenadayalan, Degallaix, Laurentis, Deléglise, Delfavero, Lillo, Pozzo, DeMarchi, D’Emilio, Demos, Dent, Pietri, Rosa, Rossi, DeSalvo, Varona, Dhurandhar, Díaz, Diaz-Ortiz, Dietrich, Fiore, Fronzo, Giorgio, Giovanni, Giovanni, Girolamo, Lieto, Ding, Pace, Palma, Renzo, Divakarla, Dmitriev, Doctor, Donovan, Dooley, Doravari, Dorrington, Downes, Drago, Driggers, Du, Ducoin, Dupej, Durante, D’Urso, Dwyer, Easter, Eddolls, Edelman, Edo, Edy, Effler, Ehrens, Eichholz, Eikenberry, Eisenmann,
  Eisenstein, Ejlli, Errico, Essick, Estelles, Estevez, Etienne, Etzel, Evans, Evans, Ewing, Fafone, Fairhurst, Fan, Farinon, Farr, Farr, Fauchon-Jones, Favata, Fays, Fazio, Feicht, Fejer, Feng, Fenyvesi, Ferguson, Fernandez-Galiana, Ferrante, Ferreira, Ferreira, Fidecaro, Fiori, Fiorucci, Fishbach, Fisher, Fittipaldi, Fitz-Axen, Fiumara, Flaminio, Floden, Flynn, Fong, Font, Forsyth, Fournier, Frasca, Frasconi, Frei, Freise, Frey, Frey, Fritschel, Frolov, Fronzè, Fulda, Fyffe, Gabbard, Gadre, Gaebel, Gair, Galaudage, Ganapathy, Ganguly, Gaonkar, García-Quirós, Garufi, Gateley, Gaudio, Gayathri, Gemme, Genin, Gennai, George, George, Gergely, Ghonge, Ghosh, Ghosh, Ghosh, Giacomazzo, Giaime, Giardina, Gibson, Gier, Gill, Glanzer, Gniesmer, Godwin, Goetz, Goetz, Gohlke, Goncharov, González, Gopakumar, Gossan, Gosselin, Gouaty, Grace, Grado, Granata, Grant, Gras, Grassia, Gray, Gray, Greco, Green, Green, Gretarsson, Griggs, Grignani, Grimaldi, Grimm, Grote, Grunewald, Gruning, Guidi, Guimaraes, Guixé, Gulati,
  Guo, Gupta, Gupta, Gupta, Gustafson, Gustafson, Haegel, Halim, Hall, Hamilton, Hammond, Haney, Hanke, Hanks, Hanna, Hannam, Hannuksela, Hansen, Hanson, Harder, Hardwick, Haris, Harms, Harry, Harry, Hasskew, Haster, Haughian, Hayes, Healy, Heidmann, Heintze, Heinze, Heitmann, Hellman, Hello, Hemming, Hendry, Heng, Hennes, Hennig, Heurs, Hild, Hinderer, Hoback, Hochheim, Hofgard, Hofman, Holgado, Holland, Holt, Holz, Hopkins, Horst, Hough, Howell, Hoy, Huang, Hübner, Huerta, Huet, Hughey, Hui, Husa, Huttner, Huxford, Huynh-Dinh, Idzkowski, Iess, Inchauspe, Ingram, Intini, Isac, Isi, Iyer, Jacqmin, Jadhav, Jadhav, James, Jani, Janthalur, Jaranowski, Jariwala, Jaume, Jenkins, Jiang, Johns, Johnson-McDaniel, Jones, Jones, Jones, Jones, Jones, Jonker, Ju, Junker, Kalaghatgi, Kalogera, Kamai, Kandhasamy, Kang, Kanner, Kapadia, Karki, Kashyap, Kasprzack, Kastaun, Katsanevas, Katsavounidis, Katzman, Kaufer, Kawabe, Kéfélian, Keitel, Keivani, Kennedy, Key, Khadka, Khalili, Khan, Khan, Khan, Khazanov, Khetan,
  Khursheed, Kijbunchoo, Kim, Kim, Kim, Kim, Kim, Kim, Kim, Kimball, King, Kinley-Hanlon, Kirchhoff, Kissel, Kleybolte, Klimenko, Knowles, Knyazev, Koch, Koehlenbeck, Koekoek, Koley, Kondrashov, Kontos, Koper, Korobko, Korth, Kovalam, Kozak, Kringel, Krishnendu, Królak, Krupinski, Kuehn, Kumar, Kumar, Kumar, Kumar, Kumar, Kuo, Kutynia, Lackey, Laghi, Lalande, Lam, Lamberts, Landry, Landry, Lane, Lang, Lange, Lantz, Lanza, Rosa, Lartaux-Vollard, Lasky, Laxen, Lazzarini, Lazzaro, Leaci, Leavey, Lecoeuche, Lee, Lee, Lee, Lee, Lee, Lehmann, Leroy, Letendre, Levin, Li, Li, li, Li, Li, Linde, Linker, Linley, Littenberg, Liu, Liu, Llorens-Monteagudo, Lo, Lockwood, London, Longo, Lorenzini, Loriette, Lormand, Losurdo, Lough, Lousto, Lovelace, Lück, Lumaca, Lundgren, Ma, Macas, Macfoy, MacInnis, Macleod, MacMillan, Macquet, Hernandez, Magaña-Sandoval, Magee, Majorana, Maksimovic, Malik, Man, Mandic, Mangano, Mansell, Manske, Mantovani, Mapelli, Marchesoni, Marion, Márka, Márka, Markakis, Markosyan, Markowitz,
  Maros, Marquina, Marsat, Martelli, Martin, Martin, Martinez, Martynov, Masalehdan, Mason, Massera, Masserot, Massinger, Masso-Reid, Mastrogiovanni, Matas, Matichard, Mavalvala, Maynard, McCann, McCarthy, McClelland, McCormick, McCuller, McGuire, McIsaac, McIver, McManus, McRae, McWilliams, Meacher, Meadors, Mehmet, Mehta, Villa, Melatos, Mendell, Mercer, Mereni, Merfeld, Merilh, Merritt, Merzougui, Meshkov, Messenger, Messick, Metzdorff, Meyers, Meylahn, Mhaske, Miani, Miao, Michaloliakos, Michel, Middleton, Milano, Miller, Millhouse, Mills, Milotti, Milovich-Goff, Minazzoli, Minenkov, Mishkin, Mishra, Mistry, Mitra, Mitrofanov, Mitselmakher, Mittleman, Mo, Mogushi, Mohapatra, Mohite, Molina-Ruiz, Mondin, Montani, Moore, Moraru, Morawski, Moreno, Morisaki, Mours, Mow-Lowry, Mozzon, Muciaccia, Mukherjee, Mukherjee, Mukherjee, Mukherjee, Mukund, Mullavey, Munch, Muñiz, Murray, Nagar, Nardecchia, Naticchioni, Nayak, Neil, Neilson, Nelemans, Nelson, Nery, Neunzert, Ng, Ng, Nguyen, Nguyen, Nichols, Nichols,
  Nissanke, Nocera, Noh, North, Nothard, Nuttall, Oberling, O’Brien, Oganesyan, Ogin, Oh, Oh, Ohme, Ohta, Okada, Oliver, Olivetto, Oppermann, Oram, O’Reilly, Ormiston, Ortega, O’Shaughnessy, Ossokine, Osthelder, Ottaway, Overmier, Owen, Pace, Pagano, Page, Pagliaroli, Pai, Pai, Palamos, Palashov, Palomba, Pan, Panda, Pang, Pankow, Pannarale, Pant, Paoletti, Paoli, Parida, Parker, Pascucci, Pasqualetti, Passaquieti, Passuello, Patricelli, Payne, Pearlstone, Pechsiri, Pedersen, Pedraza, Pele, Penn, Perego, Perez, Périgois, Perreca, Perriès, Petermann, Pfeiffer, Phelps, Phukon, Piccinni, Pichot, Piendibene, Piergiovanni, Pierro, Pillant, Pinard, Pinto, Piotrzkowski, Pirello, Pitkin, Plastino, Poggiani, Pong, Ponrathnam, Popolizio, Porter, Powell, Prajapati, Prasai, Prasanna, Pratten, Prestegard, Principe, Prodi, Prokhorov, Punturo, Puppo, Pürrer, Qi, Quetschke, Quinonez, Raab, Raaijmakers, Radkins, Radulesco, Raffai, Rafferty, Raja, Rajan, Rajbhandari, Rakhmanov, Ramirez, Ramos-Buades, Rana, Rao,
  Rapagnani, Raymond, Razzano, Read, Regimbau, Rei, Reid, Reitze, Rettegno, Ricci, Richardson, Richardson, Ricker, Riemenschneider, Riles, Rizzo, Robertson, Robinet, Rocchi, Rodriguez-Soto, Rolland, Rollins, Roma, Romanelli, Romano, Romel, Romero-Shaw, Romie, Rose, Rose, Rose, Rosińska, Rosofsky, Ross, Rowan, Rowlinson, Roy, Roy, Roy, Ruggi, Rutins, Ryan, Sachdev, Sadecki, Sakellariadou, Salafia, Salconi, Saleem, Salemi, Samajdar, Sanchez, Sanchez, Sanchis-Gual, Sanders, Santiago, Santos, Sarin, Sassolas, Sathyaprakash, Sauter, Savage, Savant, Sawant, Sayah, Schaetzl, Schale, Scheel, Scheuer, Schmidt, Schnabel, Schofield, Schönbeck, Schreiber, Schulte, Schutz, Schwarm, Schwartz, Scott, Scott, Seidel, Sellers, Sengupta, Sennett, Sentenac, Sequino, Sergeev, Setyawati, Shaddock, Shaffer, Shahriar, Sharma, Sharma, Shawhan, Shen, Shikauchi, Shink, Shoemaker, Shoemaker, Shukla, ShyamSundar, Siellez, Sieniawska, Sigg, Singer, Singh, Singh, Singha, Singhal, Sintes, Sipala, Skliris, Slagmolen, Slaven-Blair, Smetana,
  Smith, Smith, Somala, Son, Soni, Sorazu, Sordini, Sorrentino, Souradeep, Sowell, Spencer, Spera, Srivastava, Srivastava, Staats, Stachie, Standke, Steer, Steinhoff, Steinke, Steinlechner, Steinlechner, Steinmeyer, Stevenson, Stocks, Stops, Stover, Strain, Stratta, Strunk, Sturani, Stuver, Sudhagar, Sudhir, Summerscales, Sun, Sunil, Sur, Suresh, Sutton, Swinkels, Szczepańczyk, Tacca, Tait, Talbot, Tanasijczuk, Tanner, Tao, Tápai, Tapia, San~Martin, Tasson, Taylor, Tenorio, Terkowski, Thirugnanasambandam, Thomas, Thomas, Thompson, Thondapu, Thorne, Thrane, Tinsman, Saravanan, Tiwari, Tiwari, Tiwari, Toland, Tonelli, Tornasi, Torres-Forné, Torrie, e~Melo, Töyrä, Trail, Travasso, Traylor, Tringali, Tripathee, Trovato, Trudeau, Tsang, Tse, Tso, Tsukada, Tsuna, Tsutsui, Turconi, Ubhi, Ueno, Ugolini, Unnikrishnan, Urban, Usman, Utina, Vahlbruch, Vajente, Valdes, Valentini, Bakel, Beuzekom, Brand, Broeck, Vander-Hyde, Schaaf, Heijningen, Veggel, Vardaro, Varma, Vass, Vasúth, Vecchio, Vedovato, Veitch, Veitch,
  Venkateswara, Venugopalan, Verkindt, Veske, Vetrano, Viceré, Viets, Vinciguerra, Vine, Vinet, Vitale, Vivanco, Vo, Vocca, Vorvick, Vyatchanin, Wade, Wade, Wade, Walet, Walker, Wallace, Wallace, Walsh, Wang, Wang, Wang, Ward, Warden, Warner, Was, Watchi, Weaver, Wei, Weinert, Weinstein, Weiss, Wellmann, Wen, Weßels, Westhouse, Wette, Whelan, Whiting, Whittle, Wilken, Williams, Willis, Willke, Winkler, Wipf, Wittel, Woan, Woehler, Wofford, Wong, Wright, Wu, Wysocki, Xiao, Yamamoto, Yang, Yang, Yang, Yap, Yazback, Yeeles, Yu, Yu, Yuen, Zadrożny, Zadrożny, Zanolin, Zelenova, Zendri, Zevin, Zhang, Zhang, Zhang, Zhao, Zhao, Zhou, Zhou, Zhu, Zimmerman, Zucker, Zweizig, Collaboration, \& Collaboration}]{Abbott_2020_GW190814}
Abbott, R., Abbott, T.~D., Abraham, S., {et~al.} 2020, The Astrophysical Journal Letters, 896, L44, \dodoi{10.3847/2041-8213/ab960f}

\bibitem[{{Abbott} {et~al.}(2021){Abbott}, {Abbott}, {Abraham}, {Acernese}, {Ackley}, {Adams}, {Adams}, {Adhikari}, {Adya}, {Affeldt}, {Agarwal}, {Agathos}, {Agatsuma}, {Aggarwal}, {Aguiar}, {Aiello}, {Ain}, {Ajith}, {Akutsu}, {Aleman}, {Allen}, {Allocca}, {Altin}, {Amato}, {Anand}, {Ananyeva}, {Anderson}, {Anderson}, {Ando}, {Angelova}, {Ansoldi}, {Antelis}, {Antier}, {Appert}, {Arai}, {Arai}, {Arai}, {Araki}, {Araya}, {Araya}, {Areeda}, {Ar{\`e}ne}, {Aritomi}, {Arnaud}, {Aronson}, {Arun}, {Asada}, {Asali}, {Ashton}, {Aso}, {Aston}, {Astone}, {Aubin}, {Aufmuth}, {Aultoneal}, {Austin}, {Babak}, {Badaracco}, {Bader}, {Bae}, {Bae}, {Baer}, {Bagnasco}, {Bai}, {Baiotti}, {Baird}, {Bajpai}, {Ligo Scientific Collaboration}, {VIRGO Collaboration}, \& {KAGRA Collaboration}}]{2021ApJ...915L...5A}
{Abbott}, R., {Abbott}, T.~D., {Abraham}, S., {et~al.} 2021, \apjl, 915, L5, \dodoi{10.3847/2041-8213/ac082e}

\bibitem[{Abbott {et~al.}(2023)Abbott, Abbott, Acernese, Ackley, Adams, Adhikari, Adhikari, Adya, Affeldt, Agarwal, Agathos, Agatsuma, Aggarwal, Aguiar, Aiello, Ain, Ajith, Akcay, Akutsu, Albanesi, Allocca, Altin, Amato, Anand, Anand, Ananyeva, Anderson, Anderson, Ando, Andrade, Andres, Andri\ifmmode~\acute{c}\else \'{c}\fi{}, Angelova, Ansoldi, Antelis, Antier, Appert, Arai, Arai, Arai, Araki, Araya, Araya, Areeda, Ar\`ene, Aritomi, Arnaud, Arogeti, Aronson, Arun, Asada, Asali, Ashton, Aso, Assiduo, Aston, Astone, Aubin, Austin, Babak, Badaracco, Bader, Badger, Bae, Bae, Baer, Bagnasco, Bai, Baiotti, Baird, Bajpai, Ball, Ballardin, Ballmer, Balsamo, Baltus, Banagiri, Bankar, Barayoga, Barbieri, Barish, Barker, Barneo, Barone, Barr, Barsotti, Barsuglia, Barta, Bartlett, Barton, Bartos, Bassiri, Basti, Bawaj, Bayley, Baylor, Bazzan, B\'ecsy, Bedakihale, Bejger, Belahcene, Benedetto, Beniwal, Bennett, Bentley, BenYaala, Bergamin, Berger, Bernuzzi, Berry, Bersanetti, Bertolini, Betzwieser, Beveridge, Bhandare,
  Bhardwaj, Bhattacharjee, Bhaumik, Bilenko, Billingsley, Bini, Birney, Birnholtz, Biscans, Bischi, Biscoveanu, Bisht, Biswas, Bitossi, Bizouard, Blackburn, Blair, Blair, Blair, Bobba, Bode, Boer, Bogaert, Boldrini, Bonavena, Bondu, Bonilla, Bonnand, Booker, Boom, Bork, Boschi, Bose, Bose, Bossilkov, Boudart, Bouffanais, Bozzi, Bradaschia, Brady, Bramley, Branch, Branchesi, Brandt, Brau, Breschi, Briant, Briggs, Brillet, Brinkmann, Brockill, Brooks, Brooks, Brown, Brunett, Bruno, Bruntz, Bryant, Bulik, Bulten, Buonanno, Buscicchio, Buskulic, Buy, Byer, Davies, Cadonati, Cagnoli, Cahillane, Bustillo, Callaghan, Callister, Calloni, Cameron, Camp, Canepa, Canevarolo, Cannavacciuolo, Cannon, Cao, Cao, Capocasa, Capote, Carapella, Carbognani, Carlin, Carney, Carpinelli, Carrillo, Carullo, Carver, Diaz, Casentini, Castaldi, Caudill, Cavagli\`a, Cavalier, Cavalieri, Ceasar, Cella, Cerd\'a-Dur\'an, Cesarini, Chaibi, Chakravarti, Subrahmanya, Champion, Chan, Chan, Chan, Chan, Chan, Chandra, Chanial, Chao,
  Chapman-Bird, Charlton, Chase, Chassande-Mottin, Chatterjee, Chatterjee, Chatterjee, Chaturvedi, Chaty, Chatziioannou, Chen, Chen, Chen, Chen, Chen, Chen, Chen, Chen, Cheng, Cheong, Cheung, Chia, Chiadini, Chiang, Chiarini, Chierici, Chincarini, Chiofalo, Chiummo, Cho, Cho, Choudhary, Choudhary, Christensen, Chu, Chu, Chu, Chua, Chung, Ciani, Ciecielag, Cie\ifmmode~\acute{s}\else \'{s}\fi{}lar, Cifaldi, Ciobanu, Ciolfi, Cipriano, Cirone, Clara, Clark, Clark, Clarke, Clearwater, Clesse, Cleva, Coccia, Codazzo, Cohadon, Cohen, Cohen, Colleoni, Collette, Colombo, Colpi, Compton, Constancio, Conti, Cooper, Corban, Corbitt, Cordero-Carri\'on, Corezzi, Corley, Cornish, Corre, Corsi, Cortese, Costa, Cotesta, Coughlin, Coulon, Countryman, Cousins, Couvares, Coward, Cowart, Coyne, Coyne, Creighton, Creighton, Criswell, Croquette, Crowder, Cudell, Cullen, Cumming, Cummings, Cunningham, Cuoco, Cury\l{}o, Dabadie, Canton, Dall'Osso, D\'alya, Dana, DaneshgaranBajastani, D'Angelo, Danila, Danilishin, D'Antonio, Danzmann,
  Darsow-Fromm, Dasgupta, Datrier, Dattilo, Dave, Davier, Davis, Davis, Daw, de~Alarc\'on, Dean, DeBra, Deenadayalan, Degallaix, De~Laurentis, Del\'eglise, Del~Favero, De~Lillo, De~Lillo, Del~Pozzo, DeMarchi, De~Matteis, D'Emilio, Demos, Dent, Depasse, De~Pietri, De~Rosa, De~Rossi, DeSalvo, De~Simone, Dhurandhar, D\'{\i}az, Diaz-Ortiz, Didio, Dietrich, Di~Fiore, Di~Fronzo, Di~Giorgio, Di~Giovanni, Di~Giovanni, Di~Girolamo, Di~Lieto, Ding, Di~Pace, Di~Palma, Di~Renzo, Divakarla, Dmitriev, Doctor, D'Onofrio, Donovan, Dooley, Doravari, Dorrington, Drago, Driggers, Drori, Ducoin, Dupej, Durante, D'Urso, Duverne, Dwyer, Eassa, Easter, Ebersold, Eckhardt, Eddolls, Edelman, Edo, Edy, Effler, Eguchi, Eichholz, Eikenberry, Eisenmann, Eisenstein, Ejlli, Engelby, Enomoto, Errico, Essick, Estell\'es, Estevez, Etienne, Etzel, Evans, Evans, Ewing, Fafone, Fair, Fairhurst, Farah, Farinon, Farr, Farr, Farrow, Fauchon-Jones, Favaro, Favata, Fays, Fazio, Feicht, Fejer, Fenyvesi, Ferguson, Fernandez-Galiana, Ferrante, Ferreira,
  Fidecaro, Figura, Fiori, Fishbach, Fisher, Fittipaldi, Fiumara, Flaminio, Floden, Fong, Font, Fornal, Forsyth, Franke, Frasca, Frasconi, Frederick, Freed, Frei, Freise, Frey, Fritschel, Frolov, Fronz\'e, Fujii, Fujikawa, Fukunaga, Fukushima, Fulda, Fyffe, Gabbard, Gabella, Gadre, Gair, Gais, Galaudage, Gamba, Ganapathy, Ganguly, Gao, Gaonkar, Garaventa, Garc\'{\i}a, Garc\'{\i}a-N\'u\~nez, Garc\'{\i}a-Quir\'os, Garufi, Gateley, Gaudio, Gayathri, Ge, Gemme, Gennai, George, George, Gerberding, Gergely, Gewecke, Ghonge, Ghosh, Ghosh, Ghosh, Ghosh, Giacomazzo, Giacoppo, Giaime, Giardina, Gibson, Gier, Giesler, Giri, Gissi, Glanzer, Gleckl, Godwin, Goetz, Goetz, Gohlke, Golomb, Goncharov, Gonz\'alez, Gopakumar, Gosselin, Gouaty, Gould, Grace, Grado, Granata, Granata, Grant, Gras, Grassia, Gray, Gray, Greco, Green, Green, Gretarsson, Gretarsson, Griffith, Griffiths, Griggs, Grignani, Grimaldi, Grimm, Grote, Grunewald, Gruning, Guerra, Guidi, Guimaraes, Guix\'e, Gulati, Guo, Guo, Gupta, Gupta, Gupta, Gustafson,
  Gustafson, Guzman, Ha, Haegel, Hagiwara, Haino, Halim, Hall, Hamilton, Hammond, Han, Haney, Hanks, Hanna, Hannam, Hannuksela, Hansen, Hansen, Hanson, Harder, Hardwick, Haris, Harms, Harry, Harry, Hartwig, Hasegawa, Haskell, Hasskew, Haster, Hattori, Haughian, Hayakawa, Hayama, Hayes, Healy, Heidmann, Heidt, Heintze, Heinze, Heinzel, Heitmann, Hellman, Hello, Helmling-Cornell, Hemming, Hendry, Heng, Hennes, Hennig, Hennig, Hernandez, Hernandez~Vivanco, Heurs, Hild, Hill, Himemoto, Hines, Hiranuma, Hirata, Hirose, Hochheim, Hofman, Hohmann, Holcomb, Holland, Holley-Bockelmann, Hollows, Holmes, Holt, Holz, Hong, Hopkins, Hough, Hourihane, Howell, Hoy, Hoyland, Hreibi, Hsieh, Hsu, Huang, Huang, Huang, Huang, Huang, Huang, H\"ubner, Huddart, Hughey, Hui, Hui, Husa, Huttner, Huxford, Huynh-Dinh, Ide, Idzkowski, Iess, Ikenoue, Imam, Inayoshi, Ingram, Inoue, Ioka, Isi, Isleif, Ito, Itoh, Iyer, Izumi, JaberianHamedan, Jacqmin, Jadhav, Jadhav, James, Jan, Jani, Janquart, Janssens, Janthalur, Jaranowski, Jariwala,
  Jaume, Jenkins, Jenner, Jeon, Jeunon, Jia, Jin, Johns, Johnson-McDaniel, Jones, Jones, Jones, Jones, Jones, Jonker, Ju, Jung, Jung, Junker, Juste, Kaihotsu, Kajita, Kakizaki, Kalaghatgi, Kalogera, Kamai, Kamiizumi, Kanda, Kandhasamy, Kang, Kanner, Kao, Kapadia, Kapasi, Karat, Karathanasis, Karki, Kashyap, Kasprzack, Kastaun, Katsanevas, Katsavounidis, Katzman, Kaur, Kawabe, Kawaguchi, Kawai, Kawasaki, K\'ef\'elian, Keitel, Key, Khadka, Khalili, Khan, Khazanov, Khetan, Khursheed, Kijbunchoo, Kim, Kim, Kim, Kim, Kim, Kim, Kimball, Kimura, Kinley-Hanlon, Kirchhoff, Kissel, Kita, Kitazawa, Kleybolte, Klimenko, Knee, Knowles, Knyazev, Koch, Koekoek, Kojima, Kokeyama, Koley, Kolitsidou, Kolstein, Komori, Kondrashov, Kong, Kontos, Koper, Korobko, Kotake, Kovalam, Kozak, Kozakai, Kozu, Kringel, Krishnendu, Kr\'olak, Kuehn, Kuei, Kuijer, Kulkarni, Kumar, Kumar, Kumar, Kumar, Kume, Kuns, Kuo, Kuo, Kuromiya, Kuroyanagi, Kusayanagi, Kuwahara, Kwak, Lagabbe, Laghi, Lalande, Lam, Lamberts, Landry, Lane, Lang, Lange,
  Lantz, La~Rosa, Lartaux-Vollard, Lasky, Laxen, Lazzarini, Lazzaro, Leaci, Leavey, Lecoeuche, Lee, Lee, Lee, Lee, Lee, Lee, Lehmann, Lema\^{\i}tre, Leonardi, Leroy, Letendre, Levesque, Levin, Leviton, Leyde, Li, Li, Li, Li, Li, Li, Lin, Lin, Lin, Lin, Lin, Linde, Linker, Linley, Littenberg, Liu, Liu, Liu, Liu, Llamas, Llorens-Monteagudo, Lo, Lockwood, Loh, London, Longo, Lopez, Portilla, Lorenzini, Loriette, Lormand, Losurdo, Lott, Lough, Lousto, Lovelace, Lucaccioni, L\"uck, Lumaca, Lundgren, Luo, Lynam, Macas, MacInnis, Macleod, MacMillan, Macquet, Hernandez, Magazz\`u, Magee, Maggiore, Magnozzi, Mahesh, Majorana, Makarem, Maksimovic, Maliakal, Malik, Man, Mandic, Mangano, Mango, Mansell, Manske, Mantovani, Mapelli, Marchesoni, Marchio, Marion, Mark, M\'arka, M\'arka, Markakis, Markosyan, Markowitz, Maros, Marquina, Marsat, Martelli, Martin, Martin, Martinez, Martinez, Martinez, Martinovic, Martynov, Marx, Masalehdan, Mason, Massera, Masserot, Massinger, Masso-Reid, Mastrogiovanni, Matas, Mateu-Lucena,
  Matichard, Matiushechkina, Mavalvala, McCann, McCarthy, McClelland, McClincy, McCormick, McCuller, McGhee, McGuire, McIsaac, McIver, McRae, McWilliams, Meacher, Mehmet, Mehta, Meijer, Melatos, Melchor, Mendell, Menendez-Vazquez, Menoni, Mercer, Mereni, Merfeld, Merilh, Merritt, Merzougui, Meshkov, Messenger, Messick, Meyers, Meylahn, Mhaske, Miani, Miao, Michaloliakos, Michel, Michimura, Middleton, Milano, Miller, Miller, Miller, Millhouse, Mills, Milotti, Minazzoli, Minenkov, Mio, Mir, Miravet-Ten\'es, Mishra, Mishra, Mistry, Mitra, Mitrofanov, Mitselmakher, Mittleman, Miyakawa, Miyamoto, Miyazaki, Miyo, Miyoki, Mo, Modafferi, Moguel, Mogushi, Mohapatra, Mohite, Molina, Molina-Ruiz, Mondin, Montani, Moore, Moraru, Morawski, More, Moreno, Moreno, Mori, Morisaki, Moriwaki, Morr\'as, Mours, Mow-Lowry, Mozzon, Muciaccia, Mukherjee, Mukherjee, Mukherjee, Mukherjee, Mukherjee, Mukund, Mullavey, Munch, Mu\~niz, Murray, Musenich, Muusse, Nadji, Nagano, Nagano, Nagar, Nakamura, Nakano, Nakano, Nakashima, Nakayama,
  Napolano, Nardecchia, Narikawa, Naticchioni, Nayak, Nayak, Negishi, Neil, Neilson, Nelemans, Nelson, Nery, Neubauer, Neunzert, Ng, Ng, Nguyen, Nguyen, Nguyen, Quynh, Ni, Nichols, Nishizawa, Nissanke, Nitoglia, Nocera, Norman, North, Nozaki, Siles, Nuttall, Oberling, O'Brien, Obuchi, O'Dell, Oelker, Ogaki, Oganesyan, Oh, Oh, Oh, Ohashi, Ohishi, Ohkawa, Ohme, Ohta, Okada, Okutani, Okutomi, Olivetto, Oohara, Ooi, Oram, O'Reilly, Ormiston, Ormsby, Ortega, O'Shaughnessy, O'Shea, Oshino, Ossokine, Osthelder, Otabe, Ottaway, Overmier, Pace, Pagano, Page, Pagliaroli, Pai, Pai, Palamos, Palashov, Palomba, Pan, Pan, Panda, Pang, Pang, Pankow, Pannarale, Pant, Panther, Paoletti, Paoli, Paolone, Parisi, Park, Park, Parker, Pascucci, Pasqualetti, Passaquieti, Passuello, Patel, Pathak, Patricelli, Patron, Paul, Payne, Pedraza, Pegoraro, Pele, Arellano, Penn, Perego, Pereira, Pereira, Perez, P\'erigois, Perkins, Perreca, Perri\`es, Petermann, Petterson, Pfeiffer, Pham, Phukon, Piccinni, Pichot, Piendibene, Piergiovanni,
  Pierini, Pierro, Pillant, Pillas, Pilo, Pinard, Pinto, Pinto, Piotrzkowski, Piotrzkowski, Pirello, Pitkin, Placidi, Planas, Plastino, Pluchar, Poggiani, Polini, Pong, Ponrathnam, Popolizio, Porter, Poulton, Powell, Pracchia, Pradier, Prajapati, Prasai, Prasanna, Pratten, Principe, Prodi, Prokhorov, Prosposito, Prudenzi, Puecher, Punturo, Puosi, Puppo, P\"urrer, Qi, Quetschke, Quitzow-James, Qutob, Raab, Raaijmakers, Radkins, Radulesco, Raffai, Rail, Raja, Rajan, Ramirez, Ramirez, Ramos-Buades, Rana, Rapagnani, Rapol, Ray, Raymond, Raza, Razzano, Read, Rees, Regimbau, Rei, Reid, Reid, Reitze, Relton, Renzini, Rettegno, Reza, Rezac, Ricci, Richards, Richardson, Richardson, Riemenschneider, Riles, Rinaldi, Rink, Rizzo, Robertson, Robie, Robinet, Rocchi, Rodriguez, Rolland, Rollins, Romanelli, Romano, Romel, Romero-Rodr\'{\i}guez, Romero-Shaw, Romie, Ronchini, Rosa, Rose, Rosi\ifmmode~\acute{n}\else \'{n}\fi{}ska, Ross, Rowan, Rowlinson, Roy, Roy, Roy, Rozza, Ruggi, Ruiz-Rocha, Ryan, Sachdev, Sadecki, Sadiq,
  Sago, Saito, Saito, Sakai, Sakai, Sakellariadou, Sakuno, Salafia, Salconi, Saleem, Salemi, Samajdar, Sanchez, Sanchez, Sanchez, Sanchis-Gual, Sanders, Sanuy, Saravanan, Sarin, Sassolas, Satari, Sathyaprakash, Sato, Sato, Sauter, Savage, Sawada, Sawant, Sawant, Sayah, Schaetzl, Scheel, Scheuer, Schiworski, Schmidt, Schmidt, Schnabel, Schneewind, Schofield, Sch\"onbeck, Schulte, Schutz, Schwartz, Scott, Scott, Seglar-Arroyo, Sekiguchi, Sekiguchi, Sellers, Sengupta, Sentenac, Seo, Sequino, Sergeev, Setyawati, Shaffer, Shahriar, Shams, Shao, Sharma, Sharma, Shawhan, Shcheblanov, Shibagaki, Shikauchi, Shimizu, Shimoda, Shimode, Shinkai, Shishido, Shoda, Shoemaker, Shoemaker, ShyamSundar, Sieniawska, Sigg, Singer, Singh, Singh, Singha, Sintes, Sipala, Skliris, Slagmolen, Slaven-Blair, Smetana, Smith, Smith, Soldateschi, Somala, Somiya, Son, Soni, Soni, Sordini, Sorrentino, Sorrentino, Sotani, Soulard, Souradeep, Sowell, Spagnuolo, Spencer, Spera, Srinivasan, Srivastava, Srivastava, Staats, Stachie, Steer,
  Steinhoff, Steinlechner, Steinlechner, Stevenson, Stops, Stover, Strain, Strang, Stratta, Strunk, Sturani, Stuver, Sudhagar, Sudhir, Sugimoto, Suh, Sullivan, Sullivan, Summerscales, Sun, Sun, Sunil, Sur, Suresh, Sutton, Suzuki, Suzuki, Swinkels, Szczepa\ifmmode~\acute{n}\else \'{n}\fi{}czyk, Szewczyk, Tacca, Tagoshi, Tait, Takahashi, Takahashi, Takamori, Takano, Takeda, Takeda, Talbot, Talbot, Tanaka, Tanaka, Tanaka, Tanaka, Tanaka, Tanasijczuk, Tanioka, Tanner, Tao, Tao, Mart\'{\i}n, Taranto, Tasson, Telada, Tenorio, Terhune, Terkowski, Thirugnanasambandam, Thomas, Thomas, Thomas, Thompson, Thondapu, Thorne, Thrane, Tiwari, Tiwari, Tiwari, Toivonen, Toland, Tolley, Tomaru, Tomigami, Tomura, Tonelli, Torres-Forn\'e, Torrie, e~Melo, T\"oyr\"a, Trapananti, Travasso, Traylor, Trevor, Tringali, Tripathee, Troiano, Trovato, Trozzo, Trudeau, Tsai, Tsai, Tsang, Tsang, Tsao, Tse, Tso, Tsubono, Tsuchida, Tsukada, Tsuna, Tsutsui, Tsuzuki, Turbang, Turconi, Tuyenbayev, Ubhi, Uchikata, Uchiyama, Udall, Ueda, Uehara,
  Ueno, Ueshima, Unnikrishnan, Uraguchi, Urban, Ushiba, Utina, Vahlbruch, Vajente, Vajpeyi, Valdes, Valentini, Valsan, van Bakel, van Beuzekom, van~den Brand, Van Den~Broeck, Vander-Hyde, van~der Schaaf, van Heijningen, Vanosky, van Putten, van Remortel, Vardaro, Vargas, Varma, Vas\'uth, Vecchio, Vedovato, Veitch, Veitch, Venneberg, Venugopalan, Verkindt, Verma, Verma, Veske, Vetrano, Vicer\'e, Vidyant, Viets, Vijaykumar, Villa-Ortega, Vinet, Virtuoso, Vitale, Vo, Vocca, von Reis, von Wrangel, Vorvick, Vyatchanin, Wade, Wade, Wagner, Walet, Walker, Wallace, Wallace, Walsh, Wang, Wang, Wang, Ward, Warner, Was, Washimi, Washington, Watchi, Weaver, Webster, Weinert, Weinstein, Weiss, Weller, Weller, Wellmann, Wen, We\ss{}els, Wette, Whelan, White, Whiting, Whittle, Wilken, Williams, Williams, Williams, Williamson, Willis, Willke, Wilson, Winkler, Wipf, Wlodarczyk, Woan, Woehler, Wofford, Wong, Wu, Wu, Wu, Wu, Wysocki, Xiao, Xu, Yamada, Yamamoto, Yamamoto, Yamamoto, Yamamoto, Yamashita, Yamazaki, Yang, Yang,
  Yang, Yang, Yang, Yap, Yeeles, Yelikar, Ying, Yokogawa, Yokoyama, Yokozawa, Yoo, Yoshioka, Yu, Yu, Yuzurihara, Zadro\ifmmode~\dot{z}\else \.{z}\fi{}ny, Zanolin, Zeidler, Zelenova, Zendri, Zevin, Zhan, Zhang, Zhang, Zhang, Zhang, Zhang, Zhao, Zhao, Zhao, Zhao, Zheng, Zhou, Zhou, Zhu, Zhu, Zimmerman, Zlochower, Zucker, \& Zweizig}]{PhysRevX.13.041039}
Abbott, R., Abbott, T.~D., Acernese, F., {et~al.} 2023, Phys. Rev. X, 13, 041039, \dodoi{10.1103/PhysRevX.13.041039}

\bibitem[{Abbott {et~al.}(2024)Abbott, Abbott, Acernese, Ackley, Adams, Adhikari, Adhikari, Adya, Affeldt, Agarwal, Agathos, Agatsuma, Aggarwal, Aguiar, Aiello, Ain, Ajith, Albanesi, Allocca, Altin, Amato, Anand, Anand, Ananyeva, Anderson, Anderson, Andrade, Andres, Andri\ifmmode~\acute{c}\else \'{c}\fi{}, Angelova, Ansoldi, Antelis, Antier, Appert, Arai, Araya, Areeda, Ar\`ene, Arnaud, Aronson, Arun, Asali, Ashton, Assiduo, Aston, Astone, Aubin, Austin, Babak, Badaracco, Bader, Badger, Bae, Baer, Bagnasco, Bai, Baird, Ball, Ballardin, Ballmer, Balsamo, Baltus, Banagiri, Bankar, Barayoga, Barbieri, Barish, Barker, Barneo, Barone, Barr, Barsotti, Barsuglia, Barta, Bartlett, Barton, Bartos, Bassiri, Basti, Bawaj, Bayley, Baylor, Bazzan, B\'ecsy, Bedakihale, Bejger, Belahcene, Benedetto, Beniwal, Bennett, Bentley, BenYaala, Bergamin, Berger, Bernuzzi, Berry, Bersanetti, Bertolini, Betzwieser, Beveridge, Bhandare, Bhardwaj, Bhattacharjee, Bhaumik, Bilenko, Billingsley, Bini, Birney, Birnholtz, Biscans, Bischi,
  Biscoveanu, Bisht, Biswas, Bitossi, Bizouard, Blackburn, Blair, Blair, Blair, Bobba, Bode, Boer, Bogaert, Boldrini, Bonavena, Bondu, Bonilla, Bonnand, Booker, Boom, Bork, Boschi, Bose, Bose, Bossilkov, Boudart, Bouffanais, Bozzi, Bradaschia, Brady, Bramley, Branch, Branchesi, Brau, Breschi, Briant, Briggs, Brillet, Brinkmann, Brockill, Brooks, Brooks, Brown, Brunett, Bruno, Bruntz, Bryant, Bulik, Bulten, Buonanno, Buscicchio, Buskulic, Buy, Byer, Cadonati, Cagnoli, Cahillane, Bustillo, Callaghan, Callister, Calloni, Cameron, Camp, Canepa, Canevarolo, Cannavacciuolo, Cannon, Cao, Capote, Carapella, Carbognani, Carlin, Carney, Carpinelli, Carrillo, Carullo, Carver, Diaz, Casentini, Castaldi, Caudill, Cavagli\`a, Cavalier, Cavalieri, Ceasar, Cella, Cerd\'a-Dur\'an, Cesarini, Chaibi, Chakravarti, Subrahmanya, Champion, Chan, Chan, Chan, Chan, Chandra, Chanial, Chao, Charlton, Chase, Chassande-Mottin, Chatterjee, Chatterjee, Chatterjee, Chattopadhyay, Chaturvedi, Chaty, Chatziioannou, Chen, Chen, Chen, Chen,
  Chen, Cheng, Cheong, Cheung, Chia, Chiadini, Chiarini, Chierici, Chincarini, Chiofalo, Chiummo, Cho, Cho, Choudhary, Choudhary, Christensen, Chu, Chua, Chung, Ciani, Ciecielag, Cie\ifmmode~\acute{s}\else \'{s}\fi{}lar, Cifaldi, Ciobanu, Ciolfi, Cipriano, Cirone, Clara, Clark, Clark, Clarke, Clearwater, Clesse, Cleva, Coccia, Codazzo, Cohadon, Cohen, Cohen, Colleoni, Collette, Colombo, Colpi, Compton, Constancio, Conti, Cooper, Corban, Corbitt, Cordero-Carri\'on, Corezzi, Corley, Cornish, Corre, Corsi, Cortese, Costa, Cotesta, Coughlin, Coulon, Countryman, Cousins, Couvares, Coward, Cowart, Coyne, Coyne, Creighton, Creighton, Criswell, Croquette, Crowder, Cudell, Cullen, Cumming, Cummings, Cunningham, Cuoco, Cury\l{}o, Dabadie, Canton, Dall'Osso, D\'alya, Dana, DaneshgaranBajastani, D'Angelo, Danila, Danilishin, D'Antonio, Danzmann, Darsow-Fromm, Dasgupta, Datrier, Datta, Dattilo, Dave, Davier, Davies, Davis, Davis, Daw, Dean, DeBra, Deenadayalan, Degallaix, De~Laurentis, Del\'eglise, Del~Favero, De~Lillo,
  De~Lillo, Del~Pozzo, DeMarchi, De~Matteis, D'Emilio, Demos, Dent, Depasse, De~Pietri, De~Rosa, De~Rossi, DeSalvo, De~Simone, Dhurandhar, D\'{\i}az, Diaz-Ortiz, Didio, Dietrich, Di~Fiore, Di~Fronzo, Di~Giorgio, Di~Giovanni, Di~Giovanni, Di~Girolamo, Di~Lieto, Ding, Di~Pace, Di~Palma, Di~Renzo, Divakarla, Divyajyoti, Dmitriev, Doctor, D'Onofrio, Donovan, Dooley, Doravari, Dorrington, Drago, Driggers, Drori, Ducoin, Dupej, Durante, D'Urso, Duverne, Dwyer, Eassa, Easter, Ebersold, Eckhardt, Eddolls, Edelman, Edo, Edy, Effler, Eichholz, Eikenberry, Eisenmann, Eisenstein, Ejlli, Engelby, Errico, Essick, Estell\'es, Estevez, Etienne, Etzel, Evans, Evans, Ewing, Fafone, Fair, Fairhurst, Fanning, Farah, Farinon, Farr, Farr, Farrow, Fauchon-Jones, Favaro, Favata, Fays, Fazio, Feicht, Fejer, Fenyvesi, Ferguson, Fernandez-Galiana, Ferrante, Ferreira, Fidecaro, Figura, Fiori, Fishbach, Fisher, Fittipaldi, Fiumara, Flaminio, Floden, Fong, Font, Fornal, Forsyth, Franke, Frasca, Frasconi, Frederick, Freed, Frei, Freise,
  Frey, Fritschel, Frolov, Fronz\'e, Fulda, Fyffe, Gabbard, Gabella, Gadre, Gair, Gais, Galaudage, Gamba, Ganapathy, Ganguly, Gaonkar, Garaventa, Garc\'{\i}a, Garc\'{\i}a-N\'u\~nez, Garc\'{\i}a-Quir\'os, Garufi, Gateley, Gaudio, Gayathri, Gemme, Gennai, George, George, Gerberding, Gergely, Gewecke, Ghonge, Ghosh, Ghosh, Ghosh, Ghosh, Giacomazzo, Giacoppo, Giaime, Giardina, Gibson, Gier, Giesler, Giri, Gissi, Glanzer, Gleckl, Godwin, Goetz, Goetz, Gohlke, Goncharov, Gonz\'alez, Gopakumar, Gosselin, Gouaty, Gould, Grace, Grado, Granata, Granata, Grant, Gras, Grassia, Gray, Gray, Greco, Green, Green, Gretarsson, Gretarsson, Griffith, Griffiths, Griggs, Grignani, Grimaldi, Grimm, Grote, Grunewald, Gruning, Guerra, Guidi, Guimaraes, Guix\'e, Gulati, Guo, Guo, Gupta, Gupta, Gupta, Gustafson, Gustafson, Guzman, Haegel, Halim, Hall, Hamilton, Hammond, Haney, Hanks, Hanna, Hannam, Hannuksela, Hansen, Hansen, Hanson, Harder, Hardwick, Haris, Harms, Harry, Harry, Hartwig, Haskell, Hasskew, Haster, Haughian, Hayes,
  Healy, Heidmann, Heidt, Heintze, Heinze, Heinzel, Heitmann, Hellman, Hello, Helmling-Cornell, Hemming, Hendry, Heng, Hennes, Hennig, Hennig, Hernandez, Vivanco, Heurs, Hild, Hill, Hines, Hochheim, Hofman, Hohmann, Holcomb, Holland, Holley-Bockelmann, Hollows, Holmes, Holt, Holz, Hopkins, Hough, Hourihane, Howell, Hoy, Hoyland, Hreibi, Hsu, Huang, H\"ubner, Huddart, Hughey, Hui, Husa, Huttner, Huxford, Huynh-Dinh, Idzkowski, Iess, Ingram, Isi, Isleif, Iyer, JaberianHamedan, Jacqmin, Jadhav, Jadhav, James, Jan, Jani, Janquart, Janssens, Janthalur, Jaranowski, Jariwala, Jaume, Jenkins, Jenner, Jeunon, Jia, Johns, Johnson-McDaniel, Jones, Jones, Jones, Jones, Jones, Jonker, Ju, Junker, Juste, Kalaghatgi, Kalogera, Kamai, Kandhasamy, Kang, Kanner, Kao, Kapadia, Kapasi, Karat, Karathanasis, Karki, Kashyap, Kasprzack, Kastaun, Katsanevas, Katsavounidis, Katzman, Kaur, Kawabe, K\'ef\'elian, Keitel, Key, Khadka, Khalili, Khan, Khazanov, Khetan, Khursheed, Kijbunchoo, Kim, Kim, Kim, Kim, Kim, Kimball, Kinley-Hanlon,
  Kirchhoff, Kissel, Kleybolte, Klimenko, Knee, Knowles, Knyazev, Koch, Koekoek, Koley, Kolitsidou, Kolstein, Komori, Kondrashov, Kontos, Koper, Korobko, Kovalam, Kozak, Kringel, Krishnendu, Kr\'olak, Kuehn, Kuei, Kuijer, Kulkarni, Kumar, Kumar, Kumar, Kumar, Kuns, Kuwahara, Lagabbe, Laghi, Lalande, Lam, Lamberts, Landry, Lane, Lang, Lange, Lantz, La~Rosa, Lartaux-Vollard, Lasky, Laxen, Lazzarini, Lazzaro, Leaci, Leavey, Lecoeuche, Lee, Lee, Lee, Lee, Lehmann, Lema\^{\i}tre, Leroy, Letendre, Levesque, Levin, Leviton, Leyde, Li, Li, Li, Li, Li, Linde, Linker, Linley, Littenberg, Liu, Liu, Liu, Llamas, Llorens-Monteagudo, Lo, Lockwood, London, Longo, Lopez, Portilla, Lorenzini, Loriette, Lormand, Losurdo, Lott, Lough, Lousto, Lovelace, Lucaccioni, L\"uck, Lumaca, Lundgren, Lynam, Macas, MacInnis, Macleod, MacMillan, Macquet, Hernandez, Magazz\`u, Magee, Maggiore, Magnozzi, Mahesh, Majorana, Makarem, Maksimovic, Maliakal, Malik, Man, Mandic, Mangano, Mango, Mansell, Manske, Mantovani, Mapelli, Marchesoni,
  Marion, Mark, M\'arka, M\'arka, Markakis, Markosyan, Markowitz, Maros, Marquina, Marsat, Martelli, Martin, Martin, Martinez, Martinez, Martinez, Martinovic, Martynov, Marx, Masalehdan, Mason, Massera, Masserot, Massinger, Masso-Reid, Mastrogiovanni, Matas, Mateu-Lucena, Matichard, Matiushechkina, Mavalvala, McCann, McCarthy, McClelland, McClincy, McCormick, McCuller, McGhee, McGuire, McIsaac, McIver, McRae, McWilliams, Meacher, Mehmet, Mehta, Meijer, Melatos, Melchor, Mendell, Menendez-Vazquez, Menoni, Mercer, Mereni, Merfeld, Merilh, Merritt, Merzougui, Meshkov, Messenger, Messick, Meyers, Meylahn, Mhaske, Miani, Miao, Michaloliakos, Michel, Middleton, Milano, Miller, Miller, Miller, Millhouse, Mills, Milotti, Minazzoli, Minenkov, Mir, Miravet-Ten\'es, Mishra, Mishra, Mistry, Mitra, Mitrofanov, Mitselmakher, Mittleman, Mo, Moguel, Mogushi, Mohapatra, Mohite, Molina, Molina-Ruiz, Mondin, Montani, Moore, Moraru, Morawski, More, Moreno, Moreno, Morisaki, Mours, Mow-Lowry, Mozzon, Muciaccia, Mukherjee,
  Mukherjee, Mukherjee, Mukherjee, Mukherjee, Mukund, Mullavey, Munch, Mu\~niz, Murray, Musenich, Muusse, Nadji, Nagar, Napolano, Nardecchia, Naticchioni, Nayak, Nayak, Neil, Neilson, Nelemans, Nelson, Nery, Neubauer, Neunzert, Ng, Ng, Nguyen, Nguyen, Nguyen, Nichols, Nissanke, Nitoglia, Nocera, Norman, North, Nuttall, Oberling, O'Brien, O'Dell, Oelker, Oganesyan, Oh, Oh, Ohme, Ohta, Okada, Olivetto, Oram, O'Reilly, Ormiston, Ormsby, Ortega, O'Shaughnessy, O'Shea, Ossokine, Osthelder, Ottaway, Overmier, Pace, Pagano, Page, Pagliaroli, Pai, Pai, Palamos, Palashov, Palomba, Pan, Panda, Pang, Pankow, Pannarale, Pant, Panther, Paoletti, Paoli, Paolone, Park, Parker, Pascucci, Pasqualetti, Passaquieti, Passuello, Patel, Pathak, Patricelli, Patron, Patrone, Paul, Payne, Pedraza, Pegoraro, Pele, Penn, Perego, Pereira, Pereira, Perez, P\'erigois, Perkins, Perreca, Perri\`es, Petermann, Petterson, Pfeiffer, Pham, Phukon, Piccinni, Pichot, Piendibene, Piergiovanni, Pierini, Pierro, Pillant, Pillas, Pilo, Pinard, Pinto,
  Pinto, Piotrzkowski, Pirello, Pitkin, Placidi, Planas, Plastino, Pluchar, Poggiani, Polini, Pong, Ponrathnam, Popolizio, Porter, Poulton, Powell, Pracchia, Pradier, Prajapati, Prasai, Prasanna, Pratten, Principe, Prodi, Prokhorov, Prosposito, Prudenzi, Puecher, Punturo, Puosi, Puppo, P\"urrer, Qi, Quetschke, Quitzow-James, Raab, Raaijmakers, Radkins, Radulesco, Raffai, Rail, Raja, Rajan, Ramirez, Ramirez, Ramos-Buades, Rana, Rapagnani, Rapol, Ray, Raymond, Raza, Razzano, Read, Rees, Regimbau, Rei, Reid, Reid, Reitze, Relton, Renzini, Rettegno, Reza, Rezac, Ricci, Richards, Richardson, Richardson, Riemenschneider, Riles, Rinaldi, Rink, Rizzo, Robertson, Robie, Robinet, Rocchi, Rodriguez, Rolland, Rollins, Romanelli, Romano, Romel, Romero-Rodr\'{\i}guez, Romero-Shaw, Romie, Ronchini, Rosa, Rose, Rosell, Rosi\ifmmode~\acute{n}\else \'{n}\fi{}ska, Ross, Rowan, Rowlinson, Roy, Roy, Roy, Rozza, Ruggi, Ruiz-Rocha, Ryan, Sachdev, Sadecki, Sadiq, Sakellariadou, Salafia, Salconi, Saleem, Salemi, Samajdar, Sanchez,
  Sanchez, Sanchez, Sanchis-Gual, Sanders, Sanuy, Saravanan, Sarin, Sassolas, Satari, Sauter, Savage, Sawant, Sawant, Sayah, Schaetzl, Scheel, Scheuer, Schiworski, Schmidt, Schmidt, Schnabel, Schneewind, Schofield, Sch\"onbeck, Schulte, Schutz, Schwartz, Scott, Scott, Seglar-Arroyo, Sellers, Sengupta, Sentenac, Seo, Sequino, Sergeev, Setyawati, Shaffer, Shahriar, Shams, Sharma, Sharma, Shawhan, Shcheblanov, Shikauchi, Shoemaker, Shoemaker, ShyamSundar, Sieniawska, Sigg, Singer, Singh, Singh, Singha, Sintes, Sipala, Skliris, Slagmolen, Slaven-Blair, Smetana, Smith, Smith, Soldateschi, Somala, Son, Soni, Soni, Sordini, Sorrentino, Sorrentino, Soulard, Souradeep, Sowell, Spagnuolo, Spencer, Spera, Srinivasan, Srivastava, Srivastava, Staats, Stachie, Steer, Steinhoff, Steinlechner, Steinlechner, Stevenson, Stops, Stover, Strain, Strang, Stratta, Strunk, Sturani, Stuver, Sudhagar, Sudhir, Suh, Summerscales, Sun, Sun, Sunil, Sur, Suresh, Sutton, Swinkels, Szczepa\ifmmode~\acute{n}\else \'{n}\fi{}czyk, Szewczyk,
  Tacca, Tait, Talbot, Talbot, Tanasijczuk, Tanner, Tao, Tao, Mart\'{\i}n, Taranto, Tasson, Tenorio, Terhune, Terkowski, Thirugnanasambandam, Thomas, Thomas, Thomas, Thompson, Thondapu, Thorne, Thrane, Tiwari, Tiwari, Tiwari, Toivonen, Toland, Tolley, Tonelli, Torres-Forn\'e, Torrie, e~Melo, T\"oyr\"a, Trapananti, Travasso, Traylor, Trevor, Tringali, Tripathee, Troiano, Trovato, Trozzo, Trudeau, Tsai, Tsai, Tsang, Tse, Tso, Tsukada, Tsuna, Tsutsui, Turbang, Turconi, Ubhi, Udall, Ueno, Unnikrishnan, Urban, Utina, Vahlbruch, Vajente, Vajpeyi, Valdes, Valentini, Valsan, van Bakel, van Beuzekom, van~den Brand, Van Den~Broeck, Vander-Hyde, van~der Schaaf, van Heijningen, Vanosky, van Remortel, Vardaro, Vargas, Varma, Vas\'uth, Vecchio, Vedovato, Veitch, Veitch, Venneberg, Venugopalan, Verkindt, Verma, Verma, Veske, Vetrano, Vicer\'e, Vidyant, Viets, Vijaykumar, Villa-Ortega, Vinet, Virtuoso, Vitale, Vo, Vocca, von Reis, von Wrangel, Vorvick, Vyatchanin, Wade, Wade, Wagner, Walet, Walker, Wallace, Wallace, Walsh,
  Wang, Wang, Ward, Warner, Was, Washington, Watchi, Weaver, Webster, Weinert, Weinstein, Weiss, Weller, Weller, Wellmann, Wen, We\ss{}els, Wette, Whelan, White, Whiting, Whittle, Wilken, Williams, Williams, Williamson, Willis, Willke, Wilson, Winkler, Wipf, Wlodarczyk, Woan, Woehler, Wofford, Wong, Wu, Wysocki, Xiao, Yamamoto, Yang, Yang, Yang, Yang, Yap, Yeeles, Yelikar, Ying, Yoo, Yu, Yu, Zadro\ifmmode~\dot{z}\else \.{z}\fi{}ny, Zanolin, Zelenova, Zendri, Zevin, Zhang, Zhang, Zhang, Zhang, Zhao, Zhao, Zhao, Zhou, Zhou, Zhu, Zimmerman, Zlochower, Zucker, \& Zweizig}]{2024PhRvD.109b2001A}
---. 2024, Phys. Rev. D, 109, 022001, \dodoi{10.1103/PhysRevD.109.022001}

\bibitem[{Acernese {et~al.}(2015)Acernese, Agathos, Agatsuma, Aisa, Allemandou, Allocca, Amarni, Astone, Balestri, Ballardin, \& et~al.}]{Virgo}
Acernese, F., Agathos, M., Agatsuma, K., {et~al.} 2015, Classical and Quantum Gravity, 32, 024001, \dodoi{10.1088/0264-9381/32/2/024001}

\bibitem[{{Ackley} {et~al.}(2024){Ackley}, {Dyer}, {Lyman}, {O'Neill}, {Jimenez-Ibarra}, {F.}, {Kumar}, {Godson}, {Killestein}, {Gompertz}, {Kennedy}, {Belkin}, {Rana}, {Steeghs}, {Galloway}, {Dhillon}, {O'Brien}, {Ramsay}, {Noysena}, {Kotak}, {Breton}, {Palle}, {E.}, {Pollacco}, \& {GOTO Collaboration}}]{2024GCN.36257....1A}
{Ackley}, K., {Dyer}, M.~J., {Lyman}, J., {et~al.} 2024, GRB Coordinates Network, 36257, 1

\bibitem[{Ahumada {et~al.}(2024)Ahumada, Anand, Coughlin, Gupta, Kasliwal, Karambelkar, Stein, Waratkar, Swain, Jegou~du Laz, Anumarlapudi, Andreoni, Bulla, Srinivasaragavan, Toivonen, Wold, Bellm, Cenko, Kaplan, Sollerman, Bhalerao, Perley, Salgundi, Suresh, Hinds, Reusch, Necker, Cook, Pletskova, Singer, Banerjee, Barna, Copperwheat, Healy, Kiendrebeogo, Kumar, Kumar, Pezzella, Sagués-Carracedo, Sravan, Bloom, Chen, Graham, Helou, Laher, Mahabal, Purdum, Anupama, Barway, Basu, Raman, \& Roychowdhury}]{ahumada2024searchinggravitationalwaveoptical}
Ahumada, T., Anand, S., Coughlin, M.~W., {et~al.} 2024, Publications of the Astronomical Society of the Pacific, 136, 114201, \dodoi{10.1088/1538-3873/ad8265}

\bibitem[{{Ahumada} {et~al.}(2024){Ahumada}, {Stein}, {Swain}, {Salgundi}, {Suresh}, {Waratkar}, {Pathak}, {Bellm}, {Kasliwal}, {Wold}, {Anand}, {Karambelkar}, {Du Laz}, {Reusch}, {Andreoni}, {Bhalerao}, {Cenko}, {Coughlin}, {Kaplan}, {Necker}, {Perley}, {Sollerman}, {Ztf}, \& {Growth Collaborations}}]{2024GCN.36310....1A}
{Ahumada}, T., {Stein}, R., {Swain}, V., {et~al.} 2024, GRB Coordinates Network, 36310, 1

\bibitem[{{Alexander} {et~al.}(2017){Alexander}, {Berger}, {Fong}, {Williams}, {Guidorzi}, {Margutti}, {Metzger}, {Annis}, {Blanchard}, {Brout}, {Brown}, {Chen}, {Chornock}, {Cowperthwaite}, {Drout}, {Eftekhari}, {Frieman}, {Holz}, {Nicholl}, {Rest}, {Sako}, {Soares-Santos}, \& {Villar}}]{Alexander_2017}
{Alexander}, K.~D., {Berger}, E., {Fong}, W., {et~al.} 2017, APJL, 848, L21, \dodoi{10.3847/2041-8213/aa905d}

\bibitem[{Almualla {et~al.}(2021)Almualla, Anand, Coughlin, Dietrich, Guessoum, Sagués~Carracedo, Ahumada, Andreoni, Antier, Bellm, Bulla, \& Singer}]{10.1093/mnras/stab1090}
Almualla, M., Anand, S., Coughlin, M.~W., {et~al.} 2021, Monthly Notices of the Royal Astronomical Society, 504, 2822, \dodoi{10.1093/mnras/stab1090}

\bibitem[{{Anand} {et~al.}(2021){Anand}, {Coughlin}, {Kasliwal}, {Bulla}, {Ahumada}, {Sagu{\'e}s Carracedo}, {Almualla}, {Andreoni}, {Stein}, {Foucart}, {Singer}, {Sollerman}, {Bellm}, {Bolin}, {Caballero-Garc{\'\i}a}, {Castro-Tirado}, {Cenko}, {De}, {Dekany}, {Duev}, {Feeney}, {Fremling}, {Goldstein}, {Golkhou}, {Graham}, {Guessoum}, {Hankins}, {Hu}, {Kong}, {Kool}, {Kulkarni}, {Kumar}, {Laher}, {Masci}, {Mr{\'o}z}, {Nissanke}, {Porter}, {Reusch}, {Riddle}, {Rosnet}, {Rusholme}, {Serabyn}, {S{\'a}nchez-Ram{\'\i}rez}, {Rigault}, {Shupe}, {Smith}, {Soumagnac}, {Walters}, \& {Valeev}}]{2021NatAs...5...46A}
{Anand}, S., {Coughlin}, M.~W., {Kasliwal}, M.~M., {et~al.} 2021, Nature Astronomy, 5, 46, \dodoi{10.1038/s41550-020-1183-3}

\bibitem[{{Andrade} {et~al.}(2024){Andrade}, {Gokuldass}, {Klotz}, {Masek}, {Rajabov}, {Karpov}, {Prouza}, {Antier}, {Coughlin}, {Hello}, {Melo}, {Turpin}, {Pradier}, {Duverne}, \& {Guessoum}}]{2024GCN.36284....1A}
{Andrade}, C., {Gokuldass}, P., {Klotz}, A., {et~al.} 2024, GRB Coordinates Network, 36284, 1

\bibitem[{{Andrews}(2024)}]{2024Mirro...6....6A}
{Andrews}, J. 2024, The NOIRLab Mirror, 6, 6

\bibitem[{Antier {et~al.}(2020)Antier, Agayeva, Almualla, Awiphan, Baransky, Barynova, Beradze, Blažek, Boër, Burkhonov, Christensen, Coleiro, Corre, Coughlin, Crisp, Dietrich, Ducoin, Duverne, Marchal-Duval, Gendre, Gokuldass, Eggenstein, Eymar, Hello, Howell, Ismailov, Kann, Karpov, Klotz, Kochiashvili, Lachaud, Leroy, Lin, Li, Mašek, Mo, Menard, Morris, Noysena, Orange, Prouza, Rattanamala, Sadibekova, Saint-Gelais, Serrau, Simon, Stachie, Thöne, Tillayev, Turpin, Postigo, Vasylenko, Vidadi, Was, Wang, Zhang, Zhang, \& Zhang}]{Antier_2020}
Antier, S., Agayeva, S., Almualla, M., {et~al.} 2020, Monthly Notices of the Royal Astronomical Society, 497, 5518–5539, \dodoi{10.1093/mnras/staa1846}

\bibitem[{Ashton(2019)}]{Ashton_2019}
Ashton, G. e.~a. 2019, The Astrophysical Journal Supplement Series, 241, 27, \dodoi{10.3847/1538-4365/ab06fc}

\bibitem[{Aubin {et~al.}(2021)Aubin, Brighenti, Chierici, Estevez, Greco, Guidi, Juste, Marion, Mours, Nitoglia, Sauter, \& Sordini}]{Aubin2021-ja}
Aubin, F., Brighenti, F., Chierici, R., {et~al.} 2021, Class. Quantum Gravity, 38, 095004

\bibitem[{{Barbary} {et~al.}(2016){Barbary}, {Barclay}, {Biswas}, {Craig}, {Feindt}, {Friesen}, {Goldstein}, {Jha}, {Rodney}, {Sofiatti}, {Thomas}, \& {Wood-Vasey}}]{2016ascl.soft11017B}
{Barbary}, K., {Barclay}, T., {Biswas}, R., {et~al.} 2016, {SNCosmo: Python library for supernova cosmology}, Astrophysics Source Code Library, record ascl:1611.017

\bibitem[{Baumann {et~al.}(2024)Baumann, Marchand, Pineau, \& Boch}]{matthieu_baumann_2024_14205461}
Baumann, M., Marchand, M., Pineau, F.-X., \& Boch, T. 2024, cds-astro/mocpy: v0.17.1, v0.17.1,  Zenodo, \dodoi{10.5281/zenodo.14205461}

\bibitem[{{Becerra} {et~al.}(2024){Becerra}, {Watson}, {Dichiara}, {Troja}, {Butler}, {Lee}, {Kutyrev}, {Dimitrova}, {Garcia Cifuentes}, {Angulo Valdez}, {Pereyra}, \& {Lopez}}]{2024GCN.36256....1B}
{Becerra}, R.~L., {Watson}, A.~M., {Dichiara}, S., {et~al.} 2024, GRB Coordinates Network, 36256, 1

\bibitem[{{Bhattacharya} {et~al.}(2019){Bhattacharya}, {Kumar}, \& {Smoot}}]{2019MNRAS.486.5289B}
{Bhattacharya}, M., {Kumar}, P., \& {Smoot}, G. 2019, \mnras, 486, 5289, \dodoi{10.1093/mnras/stz1147}

\bibitem[{Biscoveanu {et~al.}(2019)Biscoveanu, Vitale, \& Haster}]{Biscoveanu_2019}
Biscoveanu, S., Vitale, S., \& Haster, C.-J. 2019, The Astrophysical Journal Letters, 884, L32, \dodoi{10.3847/2041-8213/ab479e}

\bibitem[{{Breeveld} {et~al.}(2023){Breeveld}, {Kuin}, {Oates}, {Brown}, {De Pasquale}, {Gronwall}, {Klingler}, {Marshall}, {Page}, {Siegel}, {Beardmore}, {Bernardini}, {Campana}, {Cenko}, {D'A{\`\i}}, {D'Avanzo}, {D'Elia}, {Dichiara}, {Evans}, {Eyles-Ferris}, {Hartmann}, {Kennea}, {Laha}, {Lien}, {Melandri}, {Nousek}, {O'Brien}, {Osborne}, {Page}, {Palmer}, {Ronchini}, {Sbarrato}, {Sbarufatti}, {Tagliaferri}, {Tohuvavohu}, {Troja}, \& {Swift Team}}]{2023GCN.33851....1B}
{Breeveld}, A.~A., {Kuin}, N.~P.~M., {Oates}, S.~R., {et~al.} 2023, GRB Coordinates Network, 33851, 1

\bibitem[{{Brethauer} {et~al.}(2024){Brethauer}, {Kasen}, {Margutti}, \& {Chornock}}]{Brethauer2024}
{Brethauer}, D., {Kasen}, D., {Margutti}, R., \& {Chornock}, R. 2024, \apj, 975, 213, \dodoi{10.3847/1538-4357/ad7d83}

\bibitem[{Brown {et~al.}(2013)Brown, Baliber, Bianco, Bowman, Burleson, Conway, Crellin, Depagne, De~Vera, Dilday, Dragomir, Dubberley, Eastman, Elphick, Falarski, Foale, Ford, Fulton, Garza, Gomez, Graham, Greene, Haldeman, Hawkins, Haworth, Haynes, Hidas, Hjelstrom, Howell, Hygelund, Lister, Lobdill, Martinez, Mullins, Norbury, Parrent, Paulson, Petry, Pickles, Posner, Rosing, Ross, Sand, Saunders, Shobbrook, Shporer, Street, Thomas, Tsapras, Tufts, Valenti, Vander~Horst, Walker, White, \& Willis}]{Brown_2013}
Brown, T.~M., Baliber, N., Bianco, F.~B., {et~al.} 2013, Publications of the Astronomical Society of the Pacific, 125, 1031–1055, \dodoi{10.1086/673168}

\bibitem[{Bulla(2019)}]{10.1093/mnras/stz2495}
Bulla, M. 2019, Monthly Notices of the Royal Astronomical Society, 489, 5037, \dodoi{10.1093/mnras/stz2495}

\bibitem[{{Bulla}(2023)}]{Bulla2023}
{Bulla}, M. 2023, \mnras, 520, 2558, \dodoi{10.1093/mnras/stad232}

\bibitem[{Burns(2020)}]{Burns_2020}
Burns, E. 2020, Living Reviews in Relativity, 23, \dodoi{10.1007/s41114-020-00028-7}

\bibitem[{{Cabrera} {et~al.}(2024){Cabrera}, {Palmese}, {Hu}, {O'Connor}, {Ford}, {McKernan}, {Andreoni}, {Ahumada}, {Amsellem}, {Busmann}, {Clark}, {Coughlin}, {Dadiani}, {Diaz}, {Graham}, {Gruen}, {Kunnumkai}, {Postiglione}, {Riffeser}, {Sommer}, \& {Valdes}}]{cabrera2024searchingelectromagneticemissionagn}
{Cabrera}, T., {Palmese}, A., {Hu}, L., {et~al.} 2024, \prd, 110, 123029, \dodoi{10.1103/PhysRevD.110.123029}

\bibitem[{Chabanat {et~al.}(1998)Chabanat, Bonche, Haensel, Meyer, \& Schaeffer}]{CHABANAT1998231}
Chabanat, E., Bonche, P., Haensel, P., Meyer, J., \& Schaeffer, R. 1998, Nuclear Physics A, 635, 231, \dodoi{https://doi.org/10.1016/S0375-9474(98)00180-8}

\bibitem[{{Chandra} {et~al.}(2024){Chandra}, {Gupta}, {Gamba}, {Kashyap}, {Chattopadhyay}, {Gonzalez}, {Bernuzzi}, \& {Sathyaprakash}}]{2024arXiv240503841C}
{Chandra}, K., {Gupta}, I., {Gamba}, R., {et~al.} 2024, arXiv e-prints, arXiv:2405.03841, \dodoi{10.48550/arXiv.2405.03841}

\bibitem[{Chatterjee {et~al.}(2020)Chatterjee, Ghosh, Brady, Kapadia, Miller, Nissanke, \& Pannarale}]{Chatterjee_2020}
Chatterjee, D., Ghosh, S., Brady, P.~R., {et~al.} 2020, The Astrophysical Journal, 896, 54, \dodoi{10.3847/1538-4357/ab8dbe}

\bibitem[{{Chaudhary} {et~al.}(2024){Chaudhary}, {Toivonen}, {Waratkar}, {Mo}, {Chatterjee}, {Antier}, {Brockill}, {Coughlin}, {Essick}, {Ghosh}, {Morisaki}, {Baral}, {Baylor}, {Adhikari}, {Brady}, {Cabourn Davies}, {Dal Canton}, {Cavaglia}, {Creighton}, {Choudhary}, {Chu}, {Clearwater}, {Davis}, {Dent}, {Drago}, {Ewing}, {Godwin}, {Guo}, {Hanna}, {Huxford}, {Harry}, {Katsavounidis}, {Kovalam}, {Li}, {Magee}, {Marx}, {Meacher}, {Messick}, {Morice-Atkinson}, {Pace}, {De Pietri}, {Piotrzkowski}, {Roy}, {Sachdev}, {Singer}, {Singh}, {Szczepanczyk}, {Tang}, {Trevor}, {Tsukada}, {Villa-Ortega}, {Wen}, \& {Wysocki}}]{2023arXiv230804545S}
{Chaudhary}, S.~S., {Toivonen}, A., {Waratkar}, G., {et~al.} 2024, Proceedings of the National Academy of Science, 121, e2316474121, \dodoi{10.1073/pnas.2316474121}

\bibitem[{Chu {et~al.}(2022)Chu, Kovalam, Wen, Slaven-Blair, Bosveld, Chen, Clearwater, Codoreanu, Du, Guo, Guo, Kim, Li, Oloworaran, Panther, Powell, Sengupta, Wette, \& Zhu}]{Chu2022-nw}
Chu, Q., Kovalam, M., Wen, L., {et~al.} 2022, Phys. Rev. D., 105

\bibitem[{{Cook} {et~al.}(2023{\natexlab{a}}){Cook}, {Ebert}, {Helou}, {Mazzarella}, {Schmitz}, \& {Singer}}]{2023GCN.33820....1C}
{Cook}, D.~O., {Ebert}, R., {Helou}, G., {et~al.} 2023{\natexlab{a}}, GRB Coordinates Network, 33820, 1

\bibitem[{{Cook} {et~al.}(2023{\natexlab{b}}){Cook}, {Ebert}, {Helou}, {Mazzarella}, {Schmitz}, {Singer}, \& {NASA/IPAC Extragalactic Database Team}}]{2023GCN.34092....1C}
---. 2023{\natexlab{b}}, GRB Coordinates Network, 34092, 1

\bibitem[{{Cook} {et~al.}(2024){Cook}, {Ebert}, {Helou}, {Mazzarella}, {\%Schmitz}, {Singer}, \& {NED team}}]{2024GCN.36243....1C}
---. 2024, GRB Coordinates Network, 36243, 1

\bibitem[{{Coughlin} {et~al.}(2020){Coughlin}, {Dietrich}, {Antier}, {Bulla}, {Foucart}, {Hotokezaka}, {Raaijmakers}, {Hinderer}, \& {Nissanke}}]{2020MNRAS.492..863C}
{Coughlin}, M.~W., {Dietrich}, T., {Antier}, S., {et~al.} 2020, \mnras, 492, 863, \dodoi{10.1093/mnras/stz3457}

\bibitem[{{Coughlin} {et~al.}(2023){Coughlin}, {Bloom}, {Nir}, {Antier}, {du Laz}, {van der Walt}, {Crellin-Quick}, {Culino}, {Duev}, {Goldstein}, {Healy}, {Karambelkar}, {Lilleboe}, {Shin}, {Singer}, {Ahumada}, {Anand}, {Bellm}, {Dekany}, {Graham}, {Kasliwal}, {Kostadinova}, {Kiendrebeogo}, {Kulkarni}, {Jenkins}, {LeBaron}, {Mahabal}, {Neill}, {Parazin}, {Peloton}, {Perley}, {Riddle}, {Rusholme}, {van Santen}, {Sollerman}, {Stein}, {Turpin}, {Wold}, {Amat}, {Bonnefon}, {Bonnefoy}, {Flament}, {Kerkow}, {Kishore}, {Jani}, {Mahanty}, {Liu}, {Llinares}, {Makarison}, {Olli{\'e}ric}, {Perez}, {Pont}, \& {Sharma}}]{2023ApJS..267...31C}
{Coughlin}, M.~W., {Bloom}, J.~S., {Nir}, G., {et~al.} 2023, \apjs, 267, 31, \dodoi{10.3847/1538-4365/acdee1}

\bibitem[{{Coulter} {et~al.}(2017){Coulter}, {Kilpatrick}, {Siebert}, {Foley}, {Shappee}, {Drout}, {Simon}, {Piro}, {Rest}, \& {One-Meter Two-Hemisphere (1M2H) Collaboration}}]{2017GCN.21529....1C}
{Coulter}, D.~A., {Kilpatrick}, C.~D., {Siebert}, M.~R., {et~al.} 2017, GRB Coordinates Network, 21529, 1

\bibitem[{{Coulter} {et~al.}(2023){Coulter}, {Kilpatrick}, {Rojas-Bravo}, {Foley}, {Kong}, {O'Connor}, {Piro}, {Rest}, \& {1M2H Collaboration}}]{2023GCN.33829....1C}
{Coulter}, D.~A., {Kilpatrick}, C.~D., {Rojas-Bravo}, C., {et~al.} 2023, GRB Coordinates Network, 33829, 1

\bibitem[{{Coulter} {et~al.}(2024){Coulter}, {Kilpatrick}, {Jones}, {Foley}, {Filippenko}, {Zheng}, {Swift}, {Rahman}, {Stacey}, {Piro}, {Rojas-Bravo}, {Anais Vilchez}, {Mu{\~n}oz-Elgueta}, {Arcavi}, {Dimitriadis}, {Siebert}, {Bloom}, {Bustamante-Rosell}, {Clever}, {Davis}, {Kutcka}, {Macias}, {McGill}, {Qui{\~n}onez}, {Ramirez-Ruiz}, {Siellez}, {Tinyanont}, {Cenko}, {Drout}, {Hausen}, {Jacobson-Gal{\'a}n}, {Howell}, {Kasen}, {McCully}, {Rest}, {Taggart}, \& {Valenti}}]{2024arXiv240415441C}
{Coulter}, D.~A., {Kilpatrick}, C.~D., {Jones}, D.~O., {et~al.} 2024, arXiv e-prints, arXiv:2404.15441, \dodoi{10.48550/arXiv.2404.15441}

\bibitem[{{Dal Canton} {et~al.}(2021){Dal Canton}, {Nitz}, {Gadre}, {Cabourn Davies}, {Villa-Ortega}, {Dent}, {Harry}, \& {Xiao}}]{Canton_2021}
{Dal Canton}, T., {Nitz}, A.~H., {Gadre}, B., {et~al.} 2021, The Astrophysical Journal, 923, 254, \dodoi{10.3847/1538-4357/ac2f9a}

\bibitem[{{Dalessi} \& {Fermi-GBM Team}(2023)}]{2023GCN.34095....1D}
{Dalessi}, S., \& {Fermi-GBM Team}. 2023, GRB Coordinates Network, 34095, 1

\bibitem[{{D{\'a}lya} {et~al.}(2022){D{\'a}lya}, {D{\'\i}az}, {Bouchet}, {Frei}, {Jasche}, {Lavaux}, {Macas}, {Mukherjee}, {P{\'a}lfi}, {de Souza}, {Wandelt}, {Bilicki}, \& {Raffai}}]{2022MNRAS.514.1403D}
{D{\'a}lya}, G., {D{\'\i}az}, R., {Bouchet}, F.~R., {et~al.} 2022, \mnras, 514, 1403, \dodoi{10.1093/mnras/stac1443}

\bibitem[{{D'Avanzo} {et~al.}(2018){D'Avanzo}, {Campana}, {Salafia}, {Ghirlanda}, {Ghisellini}, {Melandri}, {Bernardini}, {Branchesi}, {Chassande-Mottin}, {Covino}, {D'Elia}, {Nava}, {Salvaterra}, {Tagliaferri}, \& {Vergani}}]{D_Avanzo_2018}
{D'Avanzo}, P., {Campana}, S., {Salafia}, O.~S., {et~al.} 2018, Astronomy and Astrophysics, 613, L1, \dodoi{10.1051/0004-6361/201832664}

\bibitem[{{de Wet, S.} {et~al.}(2021){de Wet, S.}, {Groot, P. J.}, {Bloemen, S.}, {Le Poole, R.}, {Klein-Wolt, M.}, {Körding, E.}, {McBride, V.}, {Paterson, K.}, {Pieterse, D. L. A.}, {Vreeswijk, P. M.}, \& {Woudt, P.}}]{de_Wet_2021}
{de Wet, S.}, {Groot, P. J.}, {Bloemen, S.}, {et~al.} 2021, A\&A, 649, A72, \dodoi{10.1051/0004-6361/202040231}

\bibitem[{Dietrich \& Ujevic(2017)}]{Dietrich_2017}
Dietrich, T., \& Ujevic, M. 2017, Classical and Quantum Gravity, 34, 105014, \dodoi{10.1088/1361-6382/aa6bb0}

\bibitem[{{Dobie} {et~al.}(2023){Dobie}, {Gulati}, {Murphy}, \& {Jagwar Collaboration.}}]{2023GCN.33834....1D}
{Dobie}, D., {Gulati}, A., {Murphy}, T., \& {Jagwar Collaboration.} 2023, GRB Coordinates Network, 33834, 1

\bibitem[{Ducoin {et~al.}(2020)Ducoin, Corre, Leroy, \& Le Floch}]{10.1093/mnras/staa114}
Ducoin, J.-G., Corre, D., Leroy, N., \& Le Floch, E. 2020, Monthly Notices of the Royal Astronomical Society, 492, 4768, \dodoi{10.1093/mnras/staa114}

\bibitem[{{Ducoin} {et~al.}(2024){Ducoin}, {Antier}, {Klotz}, {Karpov}, {Andrade}, {Gokuldass}, {Masek}, {Rajabov}, {Coughlin}, {Duverne}, {Hello}, {Melo}, {Pradier}, {Turpin}, {Akl}, {Guillot}, {Abe}, {Agabi}, {Deloupy}, {Takey}, {Shokry}, {Elhosseiny}, {Molham}, {Tawfeek}, {Noysena}, {Tanasan}, {Freeberg}, {Jacquiery}, {Vignes}, {Eggenstein}, {Husseno-Desenonges}, {Gokuldass}, {Masek}, {Rajabov}, {Report}, \& {Grandma Collaboration.}}]{2024GCN.36299....1D}
{Ducoin}, J.~G., {Antier}, S., {Klotz}, A., {et~al.} 2024, GRB Coordinates Network, 36299, 1

\bibitem[{Dudi {et~al.}(2022)Dudi, Dietrich, Rashti, Brügmann, Steinhoff, \& Tichy}]{Dudi_2022}
Dudi, R., Dietrich, T., Rashti, A., {et~al.} 2022, Physical Review D, 105, \dodoi{10.1103/physrevd.105.064050}

\bibitem[{Dye {et~al.}(2017)Dye, Lawrence, Read, Fan, Kerr, Varricatt, Furnell, Edge, Irwin, Hambly, Lucas, Almaini, Chambers, Green, Hewett, Liu, McGreer, Best, Zhang, Sutorius, Froebrich, Magnier, Hasinger, Lederer, Bold, \& Tedds}]{Dye2017}
Dye, S., Lawrence, A., Read, M.~A., {et~al.} 2017, Monthly Notices of the Royal Astronomical Society, 473, 5113, \dodoi{10.1093/mnras/stx2652}

\bibitem[{{East} {et~al.}(2015){East}, {Paschalidis}, \& {Pretorius}}]{2015ApJ...807L...3E}
{East}, W.~E., {Paschalidis}, V., \& {Pretorius}, F. 2015, \apjl, 807, L3, \dodoi{10.1088/2041-8205/807/1/L3}

\bibitem[{{Evans} {et~al.}(2023){Evans}, {Kennea}, {Tohuvavohu}, {Beardmore}, {Bernardini}, {Breeveld}, {Campana}, {Cenko}, {D'Avanzo}, {Eyles-Ferris}, {Gronwall}, {Hartmann}, {Klingler}, {Kuin}, {Lien}, {Marshall}, {Melandri}, {Nousek}, {Oates}, {O'Brien}, {Osborne}, {Page}, {Page}, {Sbarufatti}, {Siegel}, {Tagliaferri}, {Troja}, \& {Swift Team}}]{2023GCN.33824....1E}
{Evans}, P.~A., {Kennea}, J.~A., {Tohuvavohu}, A., {et~al.} 2023, GRB Coordinates Network, 33824, 1

\bibitem[{{Evans} {et~al.}(2024){Evans}, {Page}, {Kennea}, {Tohuvavohu}, {Cenko}, {Eyles-Ferris}, {Beardmore}, {Bernardini}, {Breeveld}, {Campana}, {De Pasquale}, {Dichiara}, {D'Avanzo}, {D'A{\`\i}}, {D'Elia}, {Gronwall}, {Hartmann}, {Klingler}, {Kuin}, {Laha}, {Oates}, {Osborne}, {O'Brien}, {Page}, {Raman}, {Ronchini}, {Sbarrato}, {Sbarufatti}, {Siegel}, {Tagliaferri}, {Troja}, \& {Swift Team}}]{2024GCN.36278....1E}
{Evans}, P.~A., {Page}, K.~L., {Kennea}, J.~A., {et~al.} 2024, GRB Coordinates Network, 36278, 1

\bibitem[{{Ewing} {et~al.}(2024){Ewing}, {Huxford}, {Singh}, {Tsukada}, {Hanna}, {Huang}, {Joshi}, {Li}, {Magee}, {Messick}, {Pace}, {Ray}, {Sachdev}, {Sakon}, {Tapia}, {Adhicary}, {Baral}, {Baylor}, {Cannon}, {Caudill}, {Chaudhary}, {Coughlin}, {Cousins}, {Creighton}, {Essick}, {Fong}, {George}, {Godwin}, {Harada}, {Kennington}, {Kuwahara}, {Meacher}, {Morisaki}, {Mukherjee}, {Niu}, {Posnansky}, {Toivonen}, {Tsutsui}, {Ueno}, {Viets}, {Wade}, {Wade}, \& {Waratkar}}]{2024PhRvD.109d2008E}
{Ewing}, B., {Huxford}, R., {Singh}, D., {et~al.} 2024, \prd, 109, 042008, \dodoi{10.1103/PhysRevD.109.042008}

\bibitem[{Foucart(2012)}]{PhysRevD.86.124007}
Foucart, F. 2012, Phys. Rev. D, 86, 124007, \dodoi{10.1103/PhysRevD.86.124007}

\bibitem[{{Foucart} {et~al.}(2018){Foucart}, {Hinderer}, \& {Nissanke}}]{2018PhRvD..98h1501F}
{Foucart}, F., {Hinderer}, T., \& {Nissanke}, S. 2018, \prd, 98, 081501, \dodoi{10.1103/PhysRevD.98.081501}

\bibitem[{{Freire} \& {Wex}(2024)}]{Radiopulsars}
{Freire}, P., \& {Wex}, N. 2024, Living Reviews in Relativity

\bibitem[{{Frostig} {et~al.}(2022){Frostig}, {Biscoveanu}, {Mo}, {Karambelkar}, {Dal Canton}, {Chen}, {Kasliwal}, {Katsavounidis}, {Lourie}, {Simcoe}, \& {Vitale}}]{Frostig2022}
{Frostig}, D., {Biscoveanu}, S., {Mo}, G., {et~al.} 2022, \apj, 926, 152, \dodoi{10.3847/1538-4357/ac4508}

\bibitem[{Frostig {et~al.}(2024)Frostig, Burdge, De, Fur{\'e}sz, Haworth, Hinrichsen, Karambelkar, Kasliwal, Lourie, Malonis, Mo, Simcoe, Soto, \& Stein}]{WINTER2024}
Frostig, D., Burdge, K.~B., De, K., {et~al.} 2024, in Ground-based and Airborne Instrumentation for Astronomy X, ed. J.~J. Bryant, K.~Motohara, \& J.~R.~D. Vernet, Vol. 13096, International Society for Optics and Photonics (SPIE), 130963J, \dodoi{10.1117/12.3019165}

\bibitem[{{Fryer} {et~al.}(2024){Fryer}, {Hungerford}, {Wollaeger}, {Miller}, {De}, {Fontes}, {Korobkin}, {Kedia}, {Ristic}, \& {O'Shaughnessy}}]{Fryer2024}
{Fryer}, C.~L., {Hungerford}, A.~L., {Wollaeger}, R.~T., {et~al.} 2024, \apj, 961, 9, \dodoi{10.3847/1538-4357/ad1036}

\bibitem[{{Fulton} {et~al.}(2023){Fulton}, {Nicholl}, {Smith}, {Srivastav}, {Young}, {McCollum}, {Moore}, {Sim}, {Weston}, {Sheng}, {Shingles}, {Sommer}, {Aamer}, {Smartt}, {Stevance}, {Rhodes}, {Andersson}, {Denneau}, {Tonry}, {Weiland}, {Lawrence}, {Siverd}, {Erasmus}, {Koorts}, {Anderson}, {Jordan}, {Suc}, {Rest}, {Chen}, \& {Stubbs}}]{2023GCN.33830....1F}
{Fulton}, M.~D., {Nicholl}, M., {Smith}, K.~W., {et~al.} 2023, GRB Coordinates Network, 33830, 1

\bibitem[{Glendenning \& Moszkowski(1991)}]{PhysRevLett.67.2414}
Glendenning, N.~K., \& Moszkowski, S.~A. 1991, Phys. Rev. Lett., 67, 2414, \dodoi{10.1103/PhysRevLett.67.2414}

\bibitem[{{Goldstein} {et~al.}(2017){Goldstein}, {Veres}, {Burns}, {Briggs}, {Hamburg}, {Kocevski}, {Wilson-Hodge}, {Preece}, {Poolakkil}, {Roberts}, {Hui}, {Connaughton}, {Racusin}, {von Kienlin}, {Dal Canton}, {Christensen}, {Littenberg}, {Siellez}, {Blackburn}, {Broida}, {Bissaldi}, {Cleveland}, {Gibby}, {Giles}, {Kippen}, {McBreen}, {McEnery}, {Meegan}, {Paciesas}, \& {Stanbro}}]{Goldstein_2017}
{Goldstein}, A., {Veres}, P., {Burns}, E., {et~al.} 2017, APJL, 848, L14, \dodoi{10.3847/2041-8213/aa8f41}

\bibitem[{Gorski {et~al.}(2005)Gorski, Hivon, Banday, Wandelt, Hansen, Reinecke, \& Bartelmann}]{Gorski_2005}
Gorski, K.~M., Hivon, E., Banday, A.~J., {et~al.} 2005, The Astrophysical Journal, 622, 759–771, \dodoi{10.1086/427976}

\bibitem[{Gottlieb {et~al.}(2023)Gottlieb, Issa, Jacquemin-Ide, Liska, Foucart, Tchekhovskoy, Metzger, Quataert, Perna, Kasen, Duez, Kidder, Pfeiffer, \& Scheel}]{gottlieb2023largescaleevolutionsecondslongrelativistic}
Gottlieb, O., Issa, D., Jacquemin-Ide, J., {et~al.} 2023, Large-scale Evolution of Seconds-long Relativistic Jets from Black Hole-Neutron Star Mergers.
\newblock \doarXiv{2306.14947}

\bibitem[{{Groot} {et~al.}(2024){Groot}, {Bloemen}, {Pieterse}, {Tranin}, {Arcavi}, {Vreeswijk}, \& {Blackgem/Meerlicht Consortium}}]{2024GCN.36237....1G}
{Groot}, P.~J., {Bloemen}, S., {Pieterse}, D., {et~al.} 2024, GRB Coordinates Network, 36237, 1

\bibitem[{Groot {et~al.}(2024)Groot, Bloemen, Vreeswijk, van Roestel, Jonker, Nelemans, Klein-Wolt, Lepoole, Pieterse, Rodenhuis, Boland, Haverkorn, Aerts, Bakker, Balster, Bekema, Dijkstra, Dolron, Elswijk, van Elteren, Engels, Fokker, de~Haan, Hahn, ter Horst, Lesman, Kragt, Morren, Nillissen, Pessemier, Raskin, de~Rijke, Scheers, Schuil, Timmer, Antunes~Amaral, Arancibia-Rojas, Arcavi, Blagorodnova, Biswas, Breton, Dawson, Dayal, De~Wet, Duffy, Faris, Fausnaugh, Gal-Yam, Geier, Horesh, Johnston, Katusiime, Kelley, Kosakowski, Kupfer, Leloudas, Levan, Modiano, Mogawana, Munday, Paice, Patat, Pelisoli, Ramsay, Ranaivomanana, Ruiz-Carmona, Schaffenroth, Scaringi, Stoppa, Street, Tranin, Uzundag, Valenti, Veresvarska, Vuc̆ković, Wichern, Wijers, Wijnands, \& Zimmerman}]{Groot_2024}
Groot, P.~J., Bloemen, S., Vreeswijk, P.~M., {et~al.} 2024, Publications of the Astronomical Society of the Pacific, 136, 115003, \dodoi{10.1088/1538-3873/ad8b6a}

\bibitem[{Grove {et~al.}(2020)Grove, Cheung, Kerr, Mitchell, Phlips, Woolf, Wulf, Briggs, Wilson-Hodge, Kocevski, \& Perkins}]{grove2020glowbuglowcosthighsensitivitygammaray}
Grove, J.~E., Cheung, C.~C., Kerr, M., {et~al.} 2020, Glowbug, a Low-Cost, High-Sensitivity Gamma-Ray Burst Telescope.
\newblock \doarXiv{2009.11959}

\bibitem[{{Hallinan} {et~al.}(2017){Hallinan}, {Corsi}, {Mooley}, {Hotokezaka}, {Nakar}, {Kasliwal}, {Kaplan}, {Frail}, {Myers}, {Murphy}, {De}, {Dobie}, {Allison}, {Bannister}, {Bhalerao}, {Chandra}, {Clarke}, {Giacintucci}, {Ho}, {Horesh}, {Kassim}, {Kulkarni}, {Lenc}, {Lockman}, {Lynch}, {Nichols}, {Nissanke}, {Palliyaguru}, {Peters}, {Piran}, {Rana}, {Sadler}, \& {Singer}}]{2017Sci...358.1579H}
{Hallinan}, G., {Corsi}, A., {Mooley}, K.~P., {et~al.} 2017, Science, 358, 1579, \dodoi{10.1126/science.aap9855}

\bibitem[{{HAWC Collaboration}(2024)}]{2024GCN.36286....1H}
{HAWC Collaboration}. 2024, GRB Coordinates Network, 36286, 1

\bibitem[{Hayashi {et~al.}(2022)Hayashi, Fujibayashi, Kiuchi, Kyutoku, Sekiguchi, \& Shibata}]{Hayashi_2022}
Hayashi, K., Fujibayashi, S., Kiuchi, K., {et~al.} 2022, Physical Review D, 106, \dodoi{10.1103/physrevd.106.023008}

\bibitem[{Hayashi {et~al.}(2023)Hayashi, Kiuchi, Kyutoku, Sekiguchi, \& Shibata}]{Hayashi_2023}
Hayashi, K., Kiuchi, K., Kyutoku, K., Sekiguchi, Y., \& Shibata, M. 2023, Physical Review D, 107, \dodoi{10.1103/physrevd.107.123001}

\bibitem[{{Heinzel} {et~al.}(2021){Heinzel}, {Coughlin}, {Dietrich}, {Bulla}, {Antier}, {Christensen}, {Coulter}, {Foley}, {Issa}, \& {Khetan}}]{Heinzel2021}
{Heinzel}, J., {Coughlin}, M.~W., {Dietrich}, T., {et~al.} 2021, \mnras, 502, 3057, \dodoi{10.1093/mnras/stab221}

\bibitem[{{Hosseinzadeh} {et~al.}(2024{\natexlab{a}}){Hosseinzadeh}, {Paterson}, {Rastinejad}, {Shrestha}, {Daly}, {Lundquist}, {Sand}, {Fong}, {Bostroem}, {Hall}, {Wyatt}, {Gibbs}, {Christensen}, {Lindstrom}, {Nation}, {Chatelain}, \& {McCully}}]{2024ApJ...964...35H}
{Hosseinzadeh}, G., {Paterson}, K., {Rastinejad}, J.~C., {et~al.} 2024{\natexlab{a}}, \apj, 964, 35, \dodoi{10.3847/1538-4357/ad2170}

\bibitem[{{Hosseinzadeh} {et~al.}(2024{\natexlab{b}}){Hosseinzadeh}, {Shrestha}, {Kilpatrick}, {Rastinejad}, {Pearson}, {Andrews}, {Sand}, {Bostroem}, {Daly}, {Fong}, {Lundquist}, {Paterson}, {Wyatt}, {Rankin}, {Gibbs}, \& {Carson Fuls}}]{2024GCN.36266....1H}
{Hosseinzadeh}, G., {Shrestha}, M., {Kilpatrick}, C.~D., {et~al.} 2024{\natexlab{b}}, GRB Coordinates Network, 36266, 1

\bibitem[{{Hu} {et~al.}(2024){Hu}, {Cabrera}, {O'Connor}, {Andreoni}, {Palmese}, {Kunnumkai}, \& {Gw-Mmads Team}}]{2024GCN.36273....1H}
{Hu}, L., {Cabrera}, T., {O'Connor}, B., {et~al.} 2024, GRB Coordinates Network, 36273, 1

\bibitem[{{IceCube Collaboration}(2023{\natexlab{a}})}]{2023GCN.33814....1I}
{IceCube Collaboration}. 2023{\natexlab{a}}, GRB Coordinates Network, 33814, 1

\bibitem[{{IceCube Collaboration}(2023{\natexlab{b}})}]{2023GCN.33907....1I}
---. 2023{\natexlab{b}}, GRB Coordinates Network, 33907, 1

\bibitem[{{IceCube Collaboration}(2023{\natexlab{c}})}]{2023GCN.33980....1I}
---. 2023{\natexlab{c}}, GRB Coordinates Network, 33980, 1

\bibitem[{{IceCube Collaboration}(2024)}]{2024GCN.36410....1I}
---. 2024, GRB Coordinates Network, 36410, 1

\bibitem[{{Im} {et~al.}(2015){Im}, {Choi}, \& {Kim}}]{10.5303/JKAS.2015.48.4.207}
{Im}, M., {Choi}, C., \& {Kim}, K. 2015, Journal of Korean Astronomical Society, 48, 207, \dodoi{10.5303/JKAS.2015.48.4.207}

\bibitem[{Im {et~al.}(2020)Im, Paek, Kim, \& Lim}]{Im_Paek_Kim_Lim_2020}
Im, M., Paek, G. S.~H., Kim, J., \& Lim, G. 2020, Proceedings of the International Astronomical Union, 16, 207–210, \dodoi{10.1017/S1743921322000503}

\bibitem[{{Im} {et~al.}(2021){Im}, {Kim}, {Lee}, {Lee}, {Pak}, {Shim}, {Sung}, {Kang}, {Kim}, {Heo}, {Hinse}, {Ishiguro}, {Lim}, {Ly}, {Paek}, {Seo}, {Yoon}, {Woo}, {Ahn}, {Cho}, {Choi}, {Han}, {Hwang}, {Ji}, {Lee}, {Lee}, {Lee}, {Kim}, {Kim}, {Kim}, {Kim}, {Jeong}, {Park}, {Paek}, {Kim}, \& {Park}}]{2021JKAS...54...89I}
{Im}, M., {Kim}, Y., {Lee}, C.-U., {et~al.} 2021, Journal of Korean Astronomical Society, 54, 89, \dodoi{10.5303/JKAS.2021.54.3.89}

\bibitem[{{Jayaraman} {et~al.}(2023){Jayaraman}, {Fausnaugh}, {Mo}, {Katsavounidis}, {Vanderspek}, \& {Ricker}}]{2023GCN.33878....1J}
{Jayaraman}, R., {Fausnaugh}, M.~M., {Mo}, G., {et~al.} 2023, GRB Coordinates Network, 33878, 1

\bibitem[{{Jeong} {et~al.}(2024){Jeong}, {Im}, {Paek}, {Chang}, {Lee}, {Kim}, {Lee}, \& {Gecko Team}}]{2024GCN.36343....1J}
{Jeong}, M., {Im}, M., {Paek}, G. S.~H., {et~al.} 2024, GRB Coordinates Network, 36343, 1

\bibitem[{{Jhawar} {et~al.}(2024){Jhawar}, {Wouters}, {Pang}, {Bulla}, {Coughlin}, \& {Dietrich}}]{Jhawar2024}
{Jhawar}, S., {Wouters}, T., {Pang}, P. T.~H., {et~al.} 2024, arXiv e-prints, arXiv:2410.21978, \dodoi{10.48550/arXiv.2410.21978}

\bibitem[{Kapadia {et~al.}(2020)Kapadia, Caudill, Creighton, Farr, Mendell, Weinstein, Cannon, Fong, Godwin, Lo, Magee, Meacher, Messick, Mohite, Mukherjee, \& Sachdev}]{Kapadia_2020}
Kapadia, S.~J., Caudill, S., Creighton, J. D.~E., {et~al.} 2020, Classical and Quantum Gravity, 37, 045007, \dodoi{10.1088/1361-6382/ab5f2d}

\bibitem[{{Karambelkar}(2023)}]{2023TNSTR1278....1K}
{Karambelkar}, V. 2023, Transient Name Server Discovery Report, 2023-1278, 1

\bibitem[{{Karambelkar} {et~al.}(2024){Karambelkar}, {Mo}, {Stein}, {Lourie}, {Frostig}, {Ahumada}, {Simcoe}, \& {Kasliwal}}]{2024GCN.36248....1K}
{Karambelkar}, V., {Mo}, G., {Stein}, R., {et~al.} 2024, GRB Coordinates Network, 36248, 1

\bibitem[{Kasen {et~al.}(2013)Kasen, Badnell, \& Barnes}]{Kasen_2013}
Kasen, D., Badnell, N.~R., \& Barnes, J. 2013, The Astrophysical Journal, 774, 25, \dodoi{10.1088/0004-637x/774/1/25}

\bibitem[{{Kawaguchi} {et~al.}(2024){Kawaguchi}, {Domoto}, {Fujibayashi}, {Hayashi}, {Hamidani}, {Shibata}, {Tanaka}, \& {Wanajo}}]{2024arXiv240415027K}
{Kawaguchi}, K., {Domoto}, N., {Fujibayashi}, S., {et~al.} 2024, arXiv e-prints, arXiv:2404.15027, \dodoi{10.48550/arXiv.2404.15027}

\bibitem[{Kawaguchi {et~al.}(2020)Kawaguchi, Shibata, \& Tanaka}]{Kawaguchi_2020}
Kawaguchi, K., Shibata, M., \& Tanaka, M. 2020, The Astrophysical Journal, 889, 171, \dodoi{10.3847/1538-4357/ab61f6}

\bibitem[{{Kawai} {et~al.}(2023){Kawai}, {Negoro}, {Nakajima}, {Mihara}, {Sugita}, {Serino}, {Hiramatsu}, {Nishikawa}, {Kawakubo}, \& {MAXI Team}}]{2023GCN.34088....1K}
{Kawai}, N., {Negoro}, H., {Nakajima}, M., {et~al.} 2023, GRB Coordinates Network, 34088, 1

\bibitem[{{Kawai} {et~al.}(2024){Kawai}, {Negoro}, {Nakajima}, {Mihara}, {Sugita}, {Serino}, {Kawakubo}, {Hiramatsu}, {Nishikawa}, {Kondo}, \& {MAXI Team}}]{2024GCN.36238....1K}
---. 2024, GRB Coordinates Network, 36238, 1

\bibitem[{{Keinan} \& {Arcavi}(2024)}]{2024arXiv240517558K}
{Keinan}, I., \& {Arcavi}, I. 2024, arXiv e-prints, arXiv:2405.17558, \dodoi{10.48550/arXiv.2405.17558}

\bibitem[{{Kiendrebeogo} {et~al.}(2023){Kiendrebeogo}, {Farah}, {Foley}, {Gray}, {Kunert}, {Puecher}, {Toivonen}, {VandenBerg}, {Anand}, {Ahumada}, {Karambelkar}, {Coughlin}, {Dietrich}, {Kam}, {Pang}, {Singer}, \& {Sravan}}]{2023ApJ...958..158K}
{Kiendrebeogo}, R.~W., {Farah}, A.~M., {Foley}, E.~M., {et~al.} 2023, \apj, 958, 158, \dodoi{10.3847/1538-4357/acfcb1}

\bibitem[{Klimenko {et~al.}(2016)Klimenko, Vedovato, Drago, Salemi, Tiwari, Prodi, Lazzaro, Ackley, Tiwari, Da~Silva, \& Mitselmakher}]{Klimenko2016-zr}
Klimenko, S., Vedovato, G., Drago, M., {et~al.} 2016, Phys. Rev. D., 93

\bibitem[{{Konno} {et~al.}(2024){Konno}, {Garrappa}, {Ofek}, {Ben-Ami}, {Polishook}, \& {LAST Collaboration}}]{2024GCN.36264....1K}
{Konno}, R., {Garrappa}, S., {Ofek}, E.~O., {et~al.} 2024, GRB Coordinates Network, 36264, 1

\bibitem[{Kornilov {et~al.}(2011)Kornilov, Lipunov, Gorbovskoy, Belinski, Kuvshinov, Tyurina, Shatsky, Sankovich, Krylov, Balanutsa, Chazov, Kuznetsov, Zimnuhov, Senik, Tlatov, Parkhomenko, Dormidontov, Krushinsky, Zalozhnyh, Popov, Yazev, Budnev, Ivanov, Konstantinov, Gress, Chvalaev, Yurkov, Sergienko, \& Kudelina}]{Kornilov_2011}
Kornilov, V.~G., Lipunov, V.~M., Gorbovskoy, E.~S., {et~al.} 2011, Experimental Astronomy, 33, 173–196, \dodoi{10.1007/s10686-011-9280-z}

\bibitem[{{Kr{\"u}ger} \& {Foucart}(2020)}]{2020PhRvD.101j3002K}
{Kr{\"u}ger}, C.~J., \& {Foucart}, F. 2020, \prd, 101, 103002, \dodoi{10.1103/PhysRevD.101.103002}

\bibitem[{{Kumar} {et~al.}(2024){Kumar}, {Hiramatsu}, {Berger}, {Villar}, {Soto}, {Yadavalli}, {Hussaini}, {Gagliano}, \& {Ransome}}]{2024GCN.36263....1K}
{Kumar}, H., {Hiramatsu}, D., {Berger}, E., {et~al.} 2024, GRB Coordinates Network, 36263, 1

\bibitem[{{Kunnumkai} {et~al.}(2024{\natexlab{a}}){Kunnumkai}, {Palmese}, {Bulla}, {Dietrich}, {Farah}, \& {Pang}}]{2024arXiv240910651K}
{Kunnumkai}, K., {Palmese}, A., {Bulla}, M., {et~al.} 2024{\natexlab{a}}, arXiv e-prints, arXiv:2409.10651, \dodoi{10.48550/arXiv.2409.10651}

\bibitem[{{Kunnumkai} {et~al.}(2024{\natexlab{b}}){Kunnumkai}, {Palmese}, {Farah}, {Bulla}, {Dietrich}, {Pang}, {Anand}, {Andreoni}, {Cabrera}, \& {Connor}}]{2024arXiv241113673K}
{Kunnumkai}, K., {Palmese}, A., {Farah}, A.~M., {et~al.} 2024{\natexlab{b}}, arXiv e-prints, arXiv:2411.13673, \dodoi{10.48550/arXiv.2411.13673}

\bibitem[{Kyutoku {et~al.}(2015)Kyutoku, Ioka, Okawa, Shibata, \& Taniguchi}]{Kyutoku_2015}
Kyutoku, K., Ioka, K., Okawa, H., Shibata, M., \& Taniguchi, K. 2015, Physical Review D, 92, \dodoi{10.1103/physrevd.92.044028}

\bibitem[{{Kyutoku} {et~al.}(2015){Kyutoku}, {Ioka}, {Okawa}, {Shibata}, \& {Taniguchi}}]{2015PhRvD..92d4028K}
{Kyutoku}, K., {Ioka}, K., {Okawa}, H., {Shibata}, M., \& {Taniguchi}, K. 2015, \prd, 92, 044028, \dodoi{10.1103/PhysRevD.92.044028}

\bibitem[{Kyutoku {et~al.}(2013)Kyutoku, Ioka, \& Shibata}]{PhysRevD.88.041503}
Kyutoku, K., Ioka, K., \& Shibata, M. 2013, Phys. Rev. D, 88, 041503, \dodoi{10.1103/PhysRevD.88.041503}

\bibitem[{Lackey {et~al.}(2014)Lackey, Kyutoku, Shibata, Brady, \& Friedman}]{Lackey_2014}
Lackey, B.~D., Kyutoku, K., Shibata, M., Brady, P.~R., \& Friedman, J.~L. 2014, Physical Review D, 89, \dodoi{10.1103/physrevd.89.043009}

\bibitem[{{Lamoureux} {et~al.}(2023){Lamoureux}, {Ducoin}, {Tillayev}, {Hello}, {Inasaridze}, {Natsvlishvili}, {Kochiashvili}, {Beradze}, {Aivazyan}, {Burkhonov}, {Rajabov}, {Ehgamberdiev}, {Sadibekova}, {Hainic}, {R{\"u}nger}, {Melo}, {Almeida}, {Fraga}, {Corradi}, {Sasaki}, {Navarete}, {Wang}, {Zhu}, {Wang}, {Zeng}, {Iskandar}, {Antier}, {Gervasoni}, {Pormente}, {Coughlin}, {Karpov}, {Klotz}, {Pradier}, {Turpin}, {Leroy}, {Masek}, {Gurbanov}, \& {Grandma Collaboration}}]{2023GCN.34130....1L}
{Lamoureux}, M., {Ducoin}, J.~G., {Tillayev}, Y., {et~al.} 2023, GRB Coordinates Network, 34130, 1

\bibitem[{{Lattimer} \& {Schramm}(1974)}]{1974ApJ...192L.145L}
{Lattimer}, J.~M., \& {Schramm}, D.~N. 1974, \apjl, 192, L145, \dodoi{10.1086/181612}

\bibitem[{{Lattimer} \& {Schramm}(1976)}]{1976ApJ...210..549L}
---. 1976, \apj, 210, 549, \dodoi{10.1086/154860}

\bibitem[{Leggett {et~al.}(2020)Leggett, Cross, \& Hambly}]{10.1093/mnras/staa310}
Leggett, S.~K., Cross, N. J.~G., \& Hambly, N.~C. 2020, Monthly Notices of the Royal Astronomical Society, 493, 2568, \dodoi{10.1093/mnras/staa310}

\bibitem[{Li \& Paczyński(1998)}]{Li_1998}
Li, L.-X., \& Paczyński, B. 1998, The Astrophysical Journal, 507, L59–L62, \dodoi{10.1086/311680}

\bibitem[{Li {et~al.}(2020)Li, Han, Tang, Wang, Hu, Yuan, Fan, \& Wei}]{li2020gw190426152155mergerneutronstarblack}
Li, Y.-J., Han, M.-Z., Tang, S.-P., {et~al.} 2020, GW190426\_152155: a merger of neutron star-black hole or low mass binary black holes?
\newblock \doarXiv{2012.04978}

\bibitem[{{LIGO Scientific Collaboration} {et~al.}(2023{\natexlab{a}}){LIGO Scientific Collaboration}, {VIRGO Collaboration}, \& {Kagra Collaboration}}]{2023GCN.33813....1L}
{LIGO Scientific Collaboration}, {VIRGO Collaboration}, \& {Kagra Collaboration}. 2023{\natexlab{a}}, GRB Coordinates Network, 33813, 1

\bibitem[{{LIGO Scientific Collaboration} {et~al.}(2023{\natexlab{b}}){LIGO Scientific Collaboration}, {VIRGO Collaboration}, \& {Kagra Collaboration}}]{2024GCN.34086}
---. 2023{\natexlab{b}}, GRB Coordinates Network, 34086, 1

\bibitem[{{LIGO Scientific Collaboration} {et~al.}(2023{\natexlab{c}}){LIGO Scientific Collaboration}, {VIRGO Collaboration}, \& {Kagra Collaboration}}]{2023GCN.33884....1L}
---. 2023{\natexlab{c}}, GRB Coordinates Network, 33884, 1

\bibitem[{{LIGO Scientific Collaboration} {et~al.}(2023{\natexlab{d}}){LIGO Scientific Collaboration}, {VIRGO Collaboration}, \& {Kagra Collaboration}}]{2023GCN.34087....1L}
---. 2023{\natexlab{d}}, GRB Coordinates Network, 34087, 1

\bibitem[{{Ligo Scientific Collaboration} {et~al.}(2023){Ligo Scientific Collaboration}, {VIRGO Collaboration}, \& {Kagra Collaboration}}]{2023GCN.33816....1L}
{Ligo Scientific Collaboration}, {VIRGO Collaboration}, \& {Kagra Collaboration}. 2023, GRB Coordinates Network, 33816, 1

\bibitem[{{LIGO Scientific Collaboration} {et~al.}(2024{\natexlab{a}}){LIGO Scientific Collaboration}, {VIRGO Collaboration}, \& {Kagra Collaboration}}]{2024GCN.36236....1L}
{LIGO Scientific Collaboration}, {VIRGO Collaboration}, \& {Kagra Collaboration}. 2024{\natexlab{a}}, GRB Coordinates Network, 36236, 1

\bibitem[{{LIGO Scientific Collaboration} {et~al.}(2024{\natexlab{b}}){LIGO Scientific Collaboration}, {VIRGO Collaboration}, \& {Kagra Collaboration}}]{2024GCN.36240....1L}
---. 2024{\natexlab{b}}, GRB Coordinates Network, 36240, 1

\bibitem[{{LIGO Scientific Collaboration} {et~al.}(2024{\natexlab{c}}){LIGO Scientific Collaboration}, {VIRGO Collaboration}, \& {Kagra Collaboration}}]{2024GCN.36812....1L}
---. 2024{\natexlab{c}}, GRB Coordinates Network, 36812, 1

\bibitem[{{LIGO Scientific Collaboration} {et~al.}(2024{\natexlab{d}}){LIGO Scientific Collaboration}, {VIRGO Collaboration}, \& {Kagra Collaboration}}]{2024GCN.36669}
---. 2024{\natexlab{d}}, GRB Coordinates Network, 36669, 1

\bibitem[{{LIGO Scientific Collaboration} {et~al.}(2024{\natexlab{e}}){LIGO Scientific Collaboration}, {VIRGO Collaboration}, \& {Kagra Collaboration}}]{2024GCN.36704}
---. 2024{\natexlab{e}}, GRB Coordinates Network, 36704, 1

\bibitem[{{LIGO Scientific Collaboration} {et~al.}(2024{\natexlab{f}}){LIGO Scientific Collaboration}, {Virgo Collaboration}, \& {KAGRA Collaboration}}]{ligo_scientific_collaboration_2024_10845779}
{LIGO Scientific Collaboration}, {Virgo Collaboration}, \& {KAGRA Collaboration}. 2024{\natexlab{f}}, {Observation of Gravitational Waves from the Coalescence of a 2.5-4.5 Msun Compact Object and a Neutron Star --- Data Release},  Zenodo, \dodoi{10.5281/zenodo.10845779}

\bibitem[{{LIGO Scientific Collaboration} {et~al.}(2025){LIGO Scientific Collaboration}, {VIRGO Collaboration}, \& {Kagra Collaboration}}]{2025GCN.39175....1L}
{LIGO Scientific Collaboration}, {VIRGO Collaboration}, \& {Kagra Collaboration}. 2025, GRB Coordinates Network, 39175, 1

\bibitem[{{LIGO Scientific Collaboration} {et~al.}(2017){LIGO Scientific Collaboration}, {Virgo Collaboration}, {IceCube Collaboration}, {AstroSat Team}, {Insight-HXMT Collaboration}, {ANTARES Collaboration}, {Swift Collaboration}, {AGILE Team}, {DES Collaboration}, {GRAWITA Telescope}, {ATCA Telescope}, {AGILE Team}, {ASKAP Group}, {et~al.}}]{mma170817}
{LIGO Scientific Collaboration}, {Virgo Collaboration}, {IceCube Collaboration}, {et~al.} 2017, APJL, 848, L12, \dodoi{10.3847/2041-8213/aa91c9}

\bibitem[{{Lipunov} {et~al.}(2023{\natexlab{a}}){Lipunov}, {Kornilov}, {Gorbovskoy}, {Zhirkov}, {Tyurina}, {Balanutsa}, {Kuznetsov}, {Vlasenko}, {Antipov}, {Zimnukhov}, {Senik}, {Minkina}, {Chasovnikov}, {Topolev}, {Kuvshinov}, {Cheryasov}, {Kechin}, {Podesta}, {Lopez}, {Podesta}, {Francile}, {Rebolo}, {Serra}, {Buckley}, {Gres}, {Budnev}, {Carrasco}, {Valdes}, {Chavushyan}, {Patino Alvarez}, {Martinez}, {Corella}, {Rodriguez}, {Tlatov}, {Dormidontov}, {Gabovich}, \& {Yurkov}}]{2023GCN.33895....1L}
{Lipunov}, V., {Kornilov}, V., {Gorbovskoy}, E., {et~al.} 2023{\natexlab{a}}, GRB Coordinates Network, 33895, 1

\bibitem[{{Lipunov} {et~al.}(2023{\natexlab{b}}){Lipunov}, {Kornilov}, {Gorbovskoy}, {Zhirkov}, {Tyurina}, {Balanutsa}, {Kuznetsov}, {Vlasenko}, {Antipov}, {Zimnukhov}, {Senik}, {Minkina}, {Chasovnikov}, {Topolev}, {Kuvshinov}, {Cheryasov}, {Kechin}, {Podesta}, {Lopez}, {Podesta}, {Francile}, {Rebolo}, {Serra}, {Buckley}, {Gres}, {Budnev}, {Carrasco}, {Valdes}, {Chavushyan}, {Patino Alvarez}, {Martinez}, {Corella}, {Rodriguez}, {Tlatov}, {Dormidontov}, {Gabovich}, \& {Yurkov}}]{2023GCN.34093....1L}
---. 2023{\natexlab{b}}, GRB Coordinates Network, 34093, 1

\bibitem[{{Lipunov} {et~al.}(2024){Lipunov}, {Kornilov}, {Gorbovskoy}, {Zhirkov}, {Tyurina}, {Balanutsa}, {Kuznetsov}, {Vlasenko}, {Antipov}, {Zimnukhov}, {Senik}, {Minkina}, {Chasovnikov}, {Topolev}, {Kuvshinov}, {Cheryasov}, {Kechin}, {Tselik}, {Sosnovskij}, {Podesta}, {Lopez}, {Podesta}, {Francile}, {Rebolo}, {Serra}, {Buckley}, {Gress}, {Budnev}, {Ershova}, {Carrasco}, {Valdes}, {Chavushyan}, {Patino Alvarez}, {Martinez}, {Corella}, {Rodriguez}, {Tlatov}, {Dormidontov}, {Yurkov}, \& {Gabovich}}]{2024GCN.36234....1L}
---. 2024, GRB Coordinates Network, 36234, 1

\bibitem[{{Longo} {et~al.}(2023){Longo}, {Tavani}, {Verrecchia}, {Pittori}, {Cardillo}, {Casentini}, {Foffano}, {Piano}, {Ursi}, {Lucarelli}, {Baroncelli}, {Bulgarelli}, {Ciabattoni}, {di Piano}, {Fioretti}, {Panebianco}, {Parmiggiani}, {Pilia}, \& {Agile Team}}]{2023GCN.33894....1L}
{Longo}, F., {Tavani}, M., {Verrecchia}, F., {et~al.} 2023, GRB Coordinates Network, 33894, 1

\bibitem[{Lourie {et~al.}(2020)Lourie, Baker, Burruss, Egan, Furész, Frostig, Garcia-Zych, Ganciu, Haworth, Hinrichsen, Kasliwal, Karambelkar, Malonis, Simcoe, \& Zolkower}]{Lourie_2020}
Lourie, N.~P., Baker, J.~W., Burruss, R.~S., {et~al.} 2020, in Ground-based and Airborne Instrumentation for Astronomy VIII, ed. C.~J. Evans, J.~J. Bryant, \& K.~Motohara (SPIE), 55, \dodoi{10.1117/12.2561210}

\bibitem[{{Lovelace} {et~al.}(2013){Lovelace}, {Duez}, {Foucart}, {Kidder}, {Pfeiffer}, {Scheel}, \& {Szil{\'a}gyi}}]{2013CQGra..30m5004L}
{Lovelace}, G., {Duez}, M.~D., {Foucart}, F., {et~al.} 2013, Classical and Quantum Gravity, 30, 135004, \dodoi{10.1088/0264-9381/30/13/135004}

\bibitem[{{Lundquist} {et~al.}(2019){Lundquist}, {Paterson}, {Fong}, {Sand}, {Andrews}, {Shivaei}, {Daly}, {Valenti}, {Yang}, {Christensen}, {Gibbs}, {Shelly}, {Wyatt}, {Eskandari}, {Kuhn}, {Amaro}, {Arcavi}, {Behroozi}, {Butler}, {Chomiuk}, {Corsi}, {Drout}, {Egami}, {Fan}, {Foley}, {Frye}, {Gabor}, {Green}, {Grier}, {Guzman}, {Hamden}, {Howell}, {Jannuzi}, {Kelly}, {Milne}, {Moe}, {Nugent}, {Olszewski}, {Palazzi}, {Paschalidis}, {Psaltis}, {Reichart}, {Rest}, {Rossi}, {Schroeder}, {Smith}, {Smith}, {Spekkens}, {Strader}, {Stark}, {Trilling}, {Veillet}, {Wagner}, {Weiner}, {Wheeler}, {Williams}, \& {Zabludoff}}]{2019ApJ...881L..26L}
{Lundquist}, M.~J., {Paterson}, K., {Fong}, W., {et~al.} 2019, \apjl, 881, L26, \dodoi{10.3847/2041-8213/ab32f2}

\bibitem[{Lynch {et~al.}(2017)Lynch, Vitale, Essick, Katsavounidis, \& Robinet}]{Lynch2017-yk}
Lynch, R., Vitale, S., Essick, R., Katsavounidis, E., \& Robinet, F. 2017, Phys. Rev. D., 95

\bibitem[{Mardini {et~al.}(2023)Mardini, Frebel, Betre, Jacobson, Norris, \& Christlieb}]{mardini2023metalpoorstarsobservedmagellan}
Mardini, M.~K., Frebel, A., Betre, L., {et~al.} 2023, Metal-poor stars observed with the Magellan Telescope. IV. Neutron-capture element signatures in 27 main-sequence stars.
\newblock \doarXiv{2305.05363}

\bibitem[{Martineau {et~al.}(2024)Martineau, Foucart, Scheel, Duez, Kidder, \& Pfeiffer}]{Martineau:2024zur}
Martineau, T., Foucart, F., Scheel, M., {et~al.} 2024.
\newblock \doarXiv{2405.06819}

\bibitem[{Matas {et~al.}(2020)Matas, Dietrich, Buonanno, Hinderer, Pürrer, Foucart, Boyle, Duez, Kidder, Pfeiffer, \& Scheel}]{Matas_2020}
Matas, A., Dietrich, T., Buonanno, A., {et~al.} 2020, Physical Review D, 102, \dodoi{10.1103/physrevd.102.043023}

\bibitem[{{Meegan}(2009)}]{2009ApJ...702..791M}
{Meegan}, C. e.~a. 2009, \apj, 702, 791, \dodoi{10.1088/0004-637X/702/1/791}

\bibitem[{{M{\'e}sz{\'a}ros} {et~al.}(2019){M{\'e}sz{\'a}ros}, {Fox}, {Hanna}, \& {Murase}}]{2019NatRP...1..585M}
{M{\'e}sz{\'a}ros}, P., {Fox}, D.~B., {Hanna}, C., \& {Murase}, K. 2019, Nature Reviews Physics, 1, 585, \dodoi{10.1038/s42254-019-0101-z}

\bibitem[{Metzger(2019)}]{Metzger_2019}
Metzger, B.~D. 2019, Living Reviews in Relativity, 23, \dodoi{10.1007/s41114-019-0024-0}

\bibitem[{{Mo} {et~al.}(2023){Mo}, {Jayaraman}, {Fausnaugh}, {Katsavounidis}, {Ricker}, \& {Vanderspek}}]{2023ApJ...948L...3M}
{Mo}, G., {Jayaraman}, R., {Fausnaugh}, M., {et~al.} 2023, \apjl, 948, L3, \dodoi{10.3847/2041-8213/acca70}

\bibitem[{{Morokuma} {et~al.}(2024{\natexlab{a}}){Morokuma}, {Tominaga}, {Tanaka}, {Yoshida}, \& {Japanese Collaboration for Gravitational-Wave Electro-Magnetic Follow-up (J-GEM) Collaboration}}]{2024GCN.36265....1M}
{Morokuma}, T., {Tominaga}, N., {Tanaka}, I., {Yoshida}, M., \& {Japanese Collaboration for Gravitational-Wave Electro-Magnetic Follow-up (J-GEM) Collaboration}. 2024{\natexlab{a}}, GRB Coordinates Network, 36265, 1

\bibitem[{{Morokuma} {et~al.}(2024{\natexlab{b}}){Morokuma}, {Tominaga}, {Yanagisawa}, {Tanaka}, {Yoshida}, {Matsubayashi}, {Akitaya}, {Hamada}, {Suzuki}, {Higuchi}, {Sasada}, {Seki}, {Takahashi}, {Joshima}, {Hagio}, {Kubo}, {Honda}, {Takahashi}, {Ohshima}, {Kawabata}, {Nakaoka}, {Itoh}, {Kokubo}, {Hayatsu}, {Hanayama}, {Kanai}, {Oasa}, {Murata}, {Taguchi}, {Ohta}, {Kawabata}, {Maeda}, {Kusune}, {Niino}, {Sekiguchi}, {Tanaka}, {Utsumi}, \& {Japanese Collaboration for Gravitational-Wave Electro-Magnetic Follow-up Collaboration}}]{2024GCN.36302....1M}
{Morokuma}, T., {Tominaga}, N., {Yanagisawa}, K., {et~al.} 2024{\natexlab{b}}, GRB Coordinates Network, 36302, 1

\bibitem[{{Nicholl} \& {Andreoni}(2024)}]{2024arXiv241018274N}
{Nicholl}, M., \& {Andreoni}, I. 2024, arXiv e-prints, arXiv:2410.18274, \dodoi{10.48550/arXiv.2410.18274}

\bibitem[{{Oates} {et~al.}(2023){Oates}, {Breeveld}, {Brown}, {De Pasquale}, {Gronwall}, {Klingler}, {Kuin}, {Marshall}, {Page}, {Siegel}, {Beardmore}, {Bernardini}, {Campana}, {Cenko}, {\%D'Ai}, {D'Avanzo}, {D'Elia}, {Dicharia}, {Evans}, {\%Eyles-Ferris}, {Hartman}, {Kennea}, {Laha}, {Nousek}, {O'Brien}, {Osborne}, {Page}, {Palmer}, {Ronchini}, {Sbarrato}, {Sbarufatti}, {Tagliaferri}, {Tohuvavohu}, {\%Troja}, \& {Swift Team}}]{2023GCN.33832....1O}
{Oates}, S.~R., {Breeveld}, A.~A., {Brown}, P., {et~al.} 2023, GRB Coordinates Network, 33832, 1

\bibitem[{{Paek} {et~al.}(2023{\natexlab{a}}){Paek}, {Im}, {Jeong}, {Chang}, {Choi}, {Kim}, {Khalouei}, {Lee}, \& {Gecko Team}}]{2023GCN.33833....1P}
{Paek}, G. S.~H., {Im}, M., {Jeong}, M., {et~al.} 2023{\natexlab{a}}, GRB Coordinates Network, 33833, 1

\bibitem[{{Paek} {et~al.}(2023{\natexlab{b}}){Paek}, {Im}, {Seo}, {Choi}, {Moon}, {Yoon}, {Lee}, {Kim}, {Soojong Pak}, {Sung}, \& {Gecko Team}}]{2023GCN.34146....1P}
{Paek}, G. S.~H., {Im}, M., {Seo}, J., {et~al.} 2023{\natexlab{b}}, GRB Coordinates Network, 34146, 1

\bibitem[{Paek {et~al.}(2024)Paek, Im, Kim, Lim, Park, Choi, Kim, Barbieri, Salafia, Paek, Shin, Seo, Lee, Lee, Kim, \& Sung}]{Paek_2024}
Paek, G. S.~H., Im, M., Kim, J., {et~al.} 2024, The Astrophysical Journal, 960, 113, \dodoi{10.3847/1538-4357/ad0238}

\bibitem[{Paek {et~al.}(2025)Paek, Im, Jeong, Chang, Hur, Hong, Kim, Lee, Lee, Lee, Jung, Kim, Lee, Lee, \& Kim}]{paek2025geckofollowupobservationbinary}
Paek, G. S.~H., Im, M., Jeong, M., {et~al.} 2025, The Astrophysical Journal, 981, 38, \dodoi{10.3847/1538-4357/adaf99}

\bibitem[{{Palmese} {et~al.}(2024){Palmese}, {Hu}, {Cabrera}, {O'Connor}, {Kunnumkai}, {Andreoni}, \& {Gw-Mmads Team}}]{2024GCN.36245....1P}
{Palmese}, A., {Hu}, L., {Cabrera}, T., {et~al.} 2024, GRB Coordinates Network, 36245, 1

\bibitem[{Pannarale {et~al.}(2011)Pannarale, Tonita, \& Rezzolla}]{Pannarale_2011}
Pannarale, F., Tonita, A., \& Rezzolla, L. 2011, The Astrophysical Journal, 727, 95, \dodoi{10.1088/0004-637x/727/2/95}

\bibitem[{{Paterson}(2019)}]{2019IAUS..339..203P}
{Paterson}, K. 2019, in IAU Symposium, Vol. 339, Southern Horizons in Time-Domain Astronomy, ed. R.~E. {Griffin}, 203--203, \dodoi{10.1017/S1743921318002594}

\bibitem[{{Paterson} {et~al.}(2021){Paterson}, {Lundquist}, {Rastinejad}, {Fong}, {Sand}, {Andrews}, {Amaro}, {Eskandari}, {Wyatt}, {Daly}, {Bradley}, {Zhou-Wright}, {Valenti}, {Yang}, {Christensen}, {Gibbs}, {Shelly}, {Bilinski}, {Chomiuk}, {Corsi}, {Drout}, {Foley}, {Gabor}, {Garnavich}, {Grier}, {Hamden}, {Krantz}, {Olszewski}, {Paschalidis}, {Reichart}, {Rest}, {Smith}, {Strader}, {Trilling}, {Veillet}, {Wagner}, {Weiner}, \& {Zabludoff}}]{2021ApJ...912..128P}
{Paterson}, K., {Lundquist}, M.~J., {Rastinejad}, J.~C., {et~al.} 2021, \apj, 912, 128, \dodoi{10.3847/1538-4357/abeb71}

\bibitem[{{Pellegrino} {et~al.}(2024){Pellegrino}, {Arcavi}, {Howell}, {McCully}, {Andrews}, {Farah}, {Newsome}, {Padilla Gonzalez}, \& {Terreran}}]{2024GCN.36480....1P}
{Pellegrino}, C., {Arcavi}, I., {Howell}, D.~A., {et~al.} 2024, GRB Coordinates Network, 36480, 1

\bibitem[{Pian {et~al.}(2017)Pian, D’Avanzo, Benetti, Branchesi, Brocato, Campana, Cappellaro, Covino, D’Elia, Fynbo, Getman, Ghirlanda, Ghisellini, Grado, Greco, Hjorth, Kouveliotou, Levan, Limatola, Malesani, Mazzali, Melandri, Møller, Nicastro, Palazzi, Piranomonte, Rossi, Salafia, Selsing, Stratta, Tanaka, Tanvir, Tomasella, Watson, Yang, Amati, Antonelli, Ascenzi, Bernardini, Boër, Bufano, Bulgarelli, Capaccioli, Casella, Castro-Tirado, Chassande-Mottin, Ciolfi, Copperwheat, Dadina, De~Cesare, Di~Paola, Fan, Gendre, Giuffrida, Giunta, Hunt, Israel, Jin, Kasliwal, Klose, Lisi, Longo, Maiorano, Mapelli, Masetti, Nava, Patricelli, Perley, Pescalli, Piran, Possenti, Pulone, Razzano, Salvaterra, Schipani, Spera, Stamerra, Stella, Tagliaferri, Testa, Troja, Turatto, Vergani, \& Vergani}]{Pian_2017}
Pian, E., D’Avanzo, P., Benetti, S., {et~al.} 2017, Nature, 551, 67–70, \dodoi{10.1038/nature24298}

\bibitem[{{Piro} {et~al.}(2024{\natexlab{a}}){Piro}, {Simon}, {Polin}, {Coulter}, {Drout}, {Foley}, {Rojas-Bravo}, \& {Kilpatrick}}]{2024GCN.36267....1P}
{Piro}, A.~L., {Simon}, J.~D., {Polin}, A., {et~al.} 2024{\natexlab{a}}, GRB Coordinates Network, 36267, 1

\bibitem[{{Piro} {et~al.}(2024{\natexlab{b}}){Piro}, {Simon}, {Coulter}, {Drout}, {Foley}, {Rojas-Bravo}, {Kilpatrick}, {CGEM collaboration}, \& {1M2H collaboration}}]{2024GCN.36244....1P}
{Piro}, A.~L., {Simon}, J.~D., {Coulter}, D.~A., {et~al.} 2024{\natexlab{b}}, GRB Coordinates Network, 36244, 1

\bibitem[{Pratten {et~al.}(2021)Pratten, García-Quirós, Colleoni, Ramos-Buades, Estellés, Mateu-Lucena, Jaume, Haney, Keitel, Thompson, \& Husa}]{Pratten_2021}
Pratten, G., García-Quirós, C., Colleoni, M., {et~al.} 2021, Physical Review D, 103, \dodoi{10.1103/physrevd.103.104056}

\bibitem[{{Raaijmakers} {et~al.}(2021){Raaijmakers}, {Nissanke}, {Foucart}, {Kasliwal}, {Bulla}, {Fern{\'a}ndez}, {Henkel}, {Hinderer}, {Hotokezaka}, {Luko{\v{s}}i{\={u}}t{\.{e}}}, {Venumadhav}, {Antier}, {Coughlin}, {Dietrich}, \& {Edwards}}]{2021ApJ...922..269R}
{Raaijmakers}, G., {Nissanke}, S., {Foucart}, F., {et~al.} 2021, \apj, 922, 269, \dodoi{10.3847/1538-4357/ac222d}

\bibitem[{{Raman} {et~al.}(2023){Raman}, {Ronchini}, {Tohuvavohu}, {DeLaunay}, {Kennea}, \& {Parsotan}}]{2023GCN.34128....1R}
{Raman}, G., {Ronchini}, S., {Tohuvavohu}, A., {et~al.} 2023, GRB Coordinates Network, 34128, 1

\bibitem[{{Rastinejad} {et~al.}(2024){Rastinejad}, {Sand}, {Hosseinzadeh}, {Kilpatrick}, {Fong}, {Shrestha}, {Andrews}, {Bostroem}, {Daly}, {Lundquist}, {Paterson}, {Pearson}, \& {Wyatt}}]{2024GCN.36285....1R}
{Rastinejad}, J.~C., {Sand}, D.~J., {Hosseinzadeh}, G., {et~al.} 2024, GRB Coordinates Network, 36285, 1

\bibitem[{{Ricker} {et~al.}(2015){Ricker}, {Winn}, {Vanderspek}, {Latham}, {Bakos}, {Bean}, {Berta-Thompson}, {Brown}, {Buchhave}, {Butler}, {Butler}, {Chaplin}, {Charbonneau}, {Christensen-Dalsgaard}, {Clampin}, {Deming}, {Doty}, {De Lee}, {Dressing}, {Dunham}, {Endl}, {Fressin}, {Ge}, {Henning}, {Holman}, {Howard}, {Ida}, {Jenkins}, {Jernigan}, {Johnson}, {Kaltenegger}, {Kawai}, {Kjeldsen}, {Laughlin}, {Levine}, {Lin}, {Lissauer}, {MacQueen}, {Marcy}, {McCullough}, {Morton}, {Narita}, {Paegert}, {Palle}, {Pepe}, {Pepper}, {Quirrenbach}, {Rinehart}, {Sasselov}, {Sato}, {Seager}, {Sozzetti}, {Stassun}, {Sullivan}, {Szentgyorgyi}, {Torres}, {Udry}, \& {Villasenor}}]{2015JATIS...1a4003R}
{Ricker}, G.~R., {Winn}, J.~N., {Vanderspek}, R., {et~al.} 2015, Journal of Astronomical Telescopes, Instruments, and Systems, 1, 014003, \dodoi{10.1117/1.JATIS.1.1.014003}

\bibitem[{Roberts {et~al.}(2011)Roberts, Kasen, Lee, \& Ramirez-Ruiz}]{Roberts_2011}
Roberts, L.~F., Kasen, D., Lee, W.~H., \& Ramirez-Ruiz, E. 2011, The Astrophysical Journal, 736, L21, \dodoi{10.1088/2041-8205/736/1/l21}

\bibitem[{{Roberts} \& {Fermi-GBM Team}(2024)}]{2024GCN.36241....1R}
{Roberts}, O.~J., \& {Fermi-GBM Team}. 2024, GRB Coordinates Network, 36241, 1

\bibitem[{{Roming} {et~al.}(2005){Roming}, {Kennedy}, {Mason}, {Nousek}, {Ahr}, {Bingham}, {Broos}, {Carter}, {Hancock}, {Huckle}, {Hunsberger}, {Kawakami}, {Killough}, {Koch}, {McLelland}, {Smith}, {Smith}, {Soto}, {Boyd}, {Breeveld}, {Holland}, {Ivanushkina}, {Pryzby}, {Still}, \& {Stock}}]{2005SSRv..120...95R}
{Roming}, P. W.~A., {Kennedy}, T.~E., {Mason}, K.~O., {et~al.} 2005, \ssr, 120, 95, \dodoi{10.1007/s11214-005-5095-4}

\bibitem[{{Ronchini} {et~al.}(2024){Ronchini}, {Bala}, {Wood}, {Delaunay}, {Dichiara}, {Kennea}, {Parsotan}, {Raman}, {Tohuvavohu}, {Adhikari}, {Bhat}, {Biscoveanu}, {Bissaldi}, {Burns}, {Campana}, {Chandra}, {Cleveland}, {Dalessi}, {De Pasquale}, {Garc{\'\i}a-Bellido}, {Gasbarra}, {Giles}, {Gupta}, {Hartmann}, {Hristov}, {Hui}, {Kashyap}, {Kocevski}, {Mailyan}, {Malacaria}, {Nakano}, {Principe}, {Roberts}, {Sathyaprakash}, {Shao}, {Troja}, {Veres}, \& {Wilson-Hodge}}]{ronchini2024constrainingpossiblegammarayburst}
{Ronchini}, S., {Bala}, S., {Wood}, J., {et~al.} 2024, \apjl, 970, L20, \dodoi{10.3847/2041-8213/ad5d74}

\bibitem[{{Sarin} \& {Rosswog}(2024)}]{Sarin2024}
{Sarin}, N., \& {Rosswog}, S. 2024, \apjl, 973, L24, \dodoi{10.3847/2041-8213/ad739d}

\bibitem[{{Savchenko} {et~al.}(2023{\natexlab{a}}){Savchenko}, {Ferrigno}, {Rodi}, {Coleiro}, \& {Mereghetti}}]{2023GCN.33890....1S}
{Savchenko}, V., {Ferrigno}, C., {Rodi}, J., {Coleiro}, A., \& {Mereghetti}, S. 2023{\natexlab{a}}, GRB Coordinates Network, 33890, 1

\bibitem[{{Savchenko} {et~al.}(2023{\natexlab{b}}){Savchenko}, {Ferrigno}, {Rodi}, {Coleiro}, {Mereghetti}, \& {INTEGRAL Multi-MESSENGER Collaboration}}]{2023GCN.33815....1S}
{Savchenko}, V., {Ferrigno}, C., {Rodi}, J., {et~al.} 2023{\natexlab{b}}, GRB Coordinates Network, 33815, 1

\bibitem[{{Savchenko} {et~al.}(2024){Savchenko}, {Ferrigno}, {Rodi}, {Coleiro}, {Mereghetti}, \& {INTEGRAL Multi-MESSENGER Collaboration}}]{2024GCN.36247....1S}
---. 2024, GRB Coordinates Network, 36247, 1

\bibitem[{Savchenko {et~al.}(2017)Savchenko, Ferrigno, Kuulkers, Bazzano, Bozzo, Brandt, Chenevez, Courvoisier, Diehl, Domingo, Hanlon, Jourdain, von Kienlin, Laurent, Lebrun, Lutovinov, Martin-Carrillo, Mereghetti, Natalucci, Rodi, Roques, Sunyaev, \& Ubertini}]{Savchenko_2017}
Savchenko, V., Ferrigno, C., Kuulkers, E., {et~al.} 2017, The Astrophysical Journal Letters, 848, L15, \dodoi{10.3847/2041-8213/aa8f94}

\bibitem[{Schlafly \& Finkbeiner(2011)}]{Schlafly_2011}
Schlafly, E.~F., \& Finkbeiner, D.~P. 2011, The Astrophysical Journal, 737, 103, \dodoi{10.1088/0004-637X/737/2/103}

\bibitem[{{Serino} {et~al.}(2023){Serino}, {Sugita}, {Negoro}, {Kawai}, {Nakajima}, {Kobayashi}, {Tanaka}, {Soejima}, {Kudo}, {Mihara}, {Kawamuro}, {Yamada}, {Tamagawa}, {Matsuoka}, {Sakamoto}, {Hiramatsu}, {Nishikawa}, {Yoshida}, {Tsuboi}, {Urabe}, {Nawa}, {Nemoto}, {Shidatsu}, {Takahashi}, {Niwano}, {Sato}, {Higuchi}, {Yatsu}, {Nakahira}, {Ueno}, {Tomida}, {Ishikawa}, {Ogawa}, {Kurihara}, {Ueda}, {Setoguchi}, {Yoshitake}, {Nakatani}, {Yamauchi}, {Hagiwara}, {Umeki}, {Otsuki}, {Yamaoka}, {Kawakubo}, {Sugizaki}, {Iwakiri}, \& {MAXI Team}}]{2023GCN.33823....1S}
{Serino}, M., {Sugita}, S., {Negoro}, H., {et~al.} 2023, GRB Coordinates Network, 33823, 1

\bibitem[{Singer {et~al.}(2022)Singer, Parazin, Coughlin, Bloom, Crellin-Quick, Goldstein, \& van~der Walt}]{Singer_2022}
Singer, L.~P., Parazin, B., Coughlin, M.~W., {et~al.} 2022, The Astronomical Journal, 163, 209, \dodoi{10.3847/1538-3881/ac5ab8}

\bibitem[{Singer \& Price(2016)}]{Singer_2016}
Singer, L.~P., \& Price, L.~R. 2016, Physical Review D, 93, \dodoi{10.1103/physrevd.93.024013}

\bibitem[{{Smartt} {et~al.}(2017){Smartt}, {Chen}, {Jerkstrand}, {Coughlin}, {Kankare}, {Sim}, {Fraser}, {Inserra}, {Maguire}, {Chambers}, {Huber}, {Kr{\"u}hler}, {Leloudas}, {Magee}, {Shingles}, {Smith}, {Young}, {Tonry}, {Kotak}, {Gal-Yam}, {Lyman}, {Homan}, {Agliozzo}, {Anderson}, {Angus}, {Ashall}, {Barbarino}, {Bauer}, {Berton}, {Botticella}, {Bulla}, {Bulger}, {Cannizzaro}, {Cano}, {Cartier}, {Cikota}, {Clark}, {De Cia}, {Della Valle}, {Denneau}, {Dennefeld}, {Dessart}, {Dimitriadis}, {Elias-Rosa}, {Firth}, {Flewelling}, {Fl{\"o}rs}, {Franckowiak}, {Frohmaier}, {Galbany}, {Gonz{\'a}lez-Gait{\'a}n}, {Greiner}, {Gromadzki}, {Guelbenzu}, {Guti{\'e}rrez}, {Hamanowicz}, {Hanlon}, {Harmanen}, {Heintz}, {Heinze}, {Hernandez}, {Hodgkin}, {Hook}, {Izzo}, {James}, {Jonker}, {Kerzendorf}, {Klose}, {Kostrzewa-Rutkowska}, {Kowalski}, {Kromer}, {Kuncarayakti}, {Lawrence}, {Lowe}, {Magnier}, {Manulis}, {Martin-Carrillo}, {Mattila}, {McBrien}, {M{\"u}ller}, {Nordin}, {O'Neill}, {Onori}, {Palmerio}, {Pastorello},
  {Patat}, {Pignata}, {Podsiadlowski}, {Pumo}, {Prentice}, {Rau}, {Razza}, {Rest}, {Reynolds}, {Roy}, {Ruiter}, {Rybicki}, {Salmon}, {Schady}, {Schultz}, {Schweyer}, {Seitenzahl}, {Smith}, {Sollerman}, {Stalder}, {Stubbs}, {Sullivan}, {Szegedi}, {Taddia}, {Taubenberger}, {Terreran}, {van Soelen}, {Vos}, {Wainscoat}, {Walton}, {Waters}, {Weiland}, {Willman}, {Wiseman}, {Wright}, {Wyrzykowski}, \& {Yaron}}]{2017Natur.551...75S}
{Smartt}, S.~J., {Chen}, T.~W., {Jerkstrand}, A., {et~al.} 2017, \nat, 551, 75, \dodoi{10.1038/nature24303}

\bibitem[{{Smartt} {et~al.}(2024){Smartt}, {Nicholl}, {Srivastav}, {Huber}, {Chambers}, {Smith}, {Young}, {Fulton}, {Tonry}, {Stubbs}, {Denneau}, {Cooper}, {Aamer}, {Anderson}, {Andersson}, {Bulger}, {Chen}, {Clark}, {de Boer}, {Gao}, {Gillanders}, {Lawrence}, {Lin}, {Lowe}, {Magnier}, {Minguez}, {Moore}, {Rest}, {Shingles}, {Siverd}, {Smith}, {Stalder}, {Stevance}, {Wainscoat}, \& {Williams}}]{2024MNRAS.528.2299S}
{Smartt}, S.~J., {Nicholl}, M., {Srivastav}, S., {et~al.} 2024, \mnras, 528, 2299, \dodoi{10.1093/mnras/stae100}

\bibitem[{{Smith} {et~al.}(2020){Smith}, {Smartt}, {Young}, {Tonry}, {Denneau}, {Flewelling}, {Heinze}, {Weiland}, {Stalder}, {Rest}, {Stubbs}, {Anderson}, {Chen}, {Clark}, {Do}, {F{\"o}rster}, {Fulton}, {Gillanders}, {McBrien}, {O'Neill}, {Srivastav}, \& {Wright}}]{2020PASP..132h5002S}
{Smith}, K.~W., {Smartt}, S.~J., {Young}, D.~R., {et~al.} 2020, \pasp, 132, 085002, \dodoi{10.1088/1538-3873/ab936e}

\bibitem[{{Smith} {et~al.}(2024{\natexlab{a}}){Smith}, {Huber}, {Srivastav}, {Smartt}, {Chambers}, {Young}, {Nicholl}, {Fulton}, {Aamer}, {Angus}, {McCollum}, {Moore}, {Sim}, {Weston}, {Sheng}, {Ramsden}, {Shingles}, {Sommer}, {Gillanders}, {Stevance}, {Rhodes}, {Andersson}, {Schultz}, {de Boer}, {Herman}, {Fairlamb}, {Gao}, {Lin}, {Lowe}, {Magnier}, {Minguez}, {Smith}, {Wainscoat}, {Chen}, {Rest}, \& {Stubbs}}]{2024GCN.36258....1S}
{Smith}, K.~W., {Huber}, M., {Srivastav}, S., {et~al.} 2024{\natexlab{a}}, GRB Coordinates Network, 36258, 1

\bibitem[{{Smith} {et~al.}(2024{\natexlab{b}}){Smith}, {Huber}, {Srivastav}, {Smartt}, {Chambers}, {Young}, {Nicholl}, {Fulton}, {Aamer}, {Angus}, {McCollum}, {Moore}, {Sim}, {Weston}, {Sheng}, {Ramsden}, {Shingles}, {Sommer}, {Gillanders}, {Stevance}, {Rhodes}, {Andersson}, {Schultz}, {de Boer}, {Herman}, {Fairlamb}, {Gao}, {Lin}, {Lowe}, {Magnier}, {Minguez}, {Smith}, {Wainscoat}, {Chen}, {Rest}, \& {Stubbs}}]{2024GCN.36303....1S}
---. 2024{\natexlab{b}}, GRB Coordinates Network, 36303, 1

\bibitem[{{Srivastav} {et~al.}(2024){Srivastav}, {Smartt}, {Smith}, {Young}, {Nicholl}, {Fulton}, {Moore}, {Aamer}, {Angus}, {McCollum}, {Sim}, {Weston}, {Sheng}, {Shingles}, {Sommer}, {Gillanders}, {Stevance}, {Rhodes}, {Andersson}, {Denneau}, {Tonry}, {Weiland}, {Lawrence}, {Siverd}, {Erasmus}, {Koorts}, {Anderson}, {Jordan}, {Suc}, {Rest}, {Chen}, \& {Stubbs}}]{2024GCN.36251....1S}
{Srivastav}, S., {Smartt}, S.~J., {Smith}, K.~W., {et~al.} 2024, GRB Coordinates Network, 36251, 1

\bibitem[{Steeghs {et~al.}(2022)Steeghs, Galloway, Ackley, Dyer, Lyman, Ulaczyk, Cutter, Mong, Dhillon, O’Brien, Ramsay, Poshyachinda, Kotak, Nuttall, Pallé, Breton, Pollacco, Thrane, Aukkaravittayapun, Awiphan, Burhanudin, Chote, Chrimes, Daw, Duffy, Eyles-Ferris, Gompertz, Heikkilä, Irawati, Kennedy, Killestein, Kuncarayakti, Levan, Littlefair, Makrygianni, Marsh, Mata-Sanchez, Mattila, Maund, McCormac, Mkrtichian, Mullaney, Noysena, Patel, Rol, Sawangwit, Stanway, Starling, Strøm, Tooke, West, White, \& Wiersema}]{Steeghs_2022}
Steeghs, D., Galloway, D.~K., Ackley, K., {et~al.} 2022, Monthly Notices of the Royal Astronomical Society, 511, 2405–2422, \dodoi{10.1093/mnras/stac013}

\bibitem[{{Sugita} {et~al.}(2023{\natexlab{a}}){Sugita}, {Yoshida}, {Sakamoto}, {Kawakubo}, {Yamaoka}, {Nakahira}, {Asaoka}, {Torii}, {Akaike}, {Kobayashi}, {Shimizu}, {Tamura}, {Cannady}, {Cherry}, {Ricciarini}, {Marrocchesi}, \& {Calet Collaboration}}]{2023GCN.33897....1S}
{Sugita}, S., {Yoshida}, A., {Sakamoto}, T., {et~al.} 2023{\natexlab{a}}, GRB Coordinates Network, 33897, 1

\bibitem[{{Sugita} {et~al.}(2023{\natexlab{b}}){Sugita}, {Serino}, {Negoro}, {Kawai}, {Nakajima}, {Kobayashi}, {Tanaka}, {Soejima}, {Kudo}, {Mihara}, {Kawamuro}, {Yamada}, {Tamagawa}, {Matsuoka}, {Sakamoto}, {Hiramatsu}, {Nishikawa}, {Yoshida}, {Tsuboi}, {Urabe}, {Nawa}, {Nemoto}, {Shidatsu}, {Takahashi}, {Niwano}, {Sato}, {Higuchi}, {Yatsu}, {Nakahira}, {Ueno}, {Tomida}, {Ishikawa}, {Ogawa}, {Kurihara}, {Ueda}, {Setoguchi}, {Yoshitake}, {Nakatani}, {Yamauchi}, {Hagiwara}, {Umeki}, {Otsuki}, {Yamaoka}, {Kawakubo}, {Sugizaki}, {Iwakiri}, \& {MAXI Team}}]{2023GCN.33893....1S}
{Sugita}, S., {Serino}, M., {Negoro}, H., {et~al.} 2023{\natexlab{b}}, GRB Coordinates Network, 33893, 1

\bibitem[{{Sun} {et~al.}(2024){Sun}, {Li}, {Chen}, {Guan}, {Li}, {Jia}, {Zhao}, {Zhang}, {Ge}, {Cui}, {Feng}, {Li}, {Liu}, {Lu}, {Song}, {Wang}, {Xu}, {Han}, {Zhang}, {Zhao}, {Jin}, {Hu}, {Zhang}, {Cheng}, {Liu}, {Liu}, {Ling}, {Zhang}, {Huang}, {Pan}, {Xu}, {Yuan}, {Zhang}, {Fu}, {Zhang}, {Burwitz}, {Friedrich}, {Meidinger}, {Nandra}, {Rau}, {Kuulkers}, {Santovincenzo}, {O'Brien}, {Cordier}, \& {Einstein Probe Team}}]{2024GCN.36313....1S}
{Sun}, H., {Li}, D.~Y., {Chen}, Y., {et~al.} 2024, GRB Coordinates Network, 36313, 1

\bibitem[{{Tak} {et~al.}(2023){Tak}, {Uhm}, \& {Gillanders}}]{Tak2023}
{Tak}, D., {Uhm}, Z.~L., \& {Gillanders}, J.~H. 2023, \apj, 958, 121, \dodoi{10.3847/1538-4357/ad06b0}

\bibitem[{{Tak} {et~al.}(2024){Tak}, {Uhm}, \& {Gillanders}}]{Tak2024}
---. 2024, \apj, 967, 54, \dodoi{10.3847/1538-4357/ad3af4}

\bibitem[{{Tan} {et~al.}(2024){Tan}, {Cai}, {Xiong}, {Wang}, {Xue}, {Zheng}, {Guo}, {Zhang}, {Li}, {Li}, {Yi}, {Wang}, {Li}, \& {Gecam Team}}]{2024GCN.36300....1T}
{Tan}, W.-J., {Cai}, C., {Xiong}, S.-L., {et~al.} 2024, GRB Coordinates Network, 36300, 1

\bibitem[{Thompson {et~al.}(2020)Thompson, Fauchon-Jones, Khan, Nitoglia, Pannarale, Dietrich, \& Hannam}]{Thompson_2020}
Thompson, J.~E., Fauchon-Jones, E., Khan, S., {et~al.} 2020, Physical Review D, 101, \dodoi{10.1103/physrevd.101.124059}

\bibitem[{Thompson {et~al.}(2024)Thompson, Hamilton, London, Ghosh, Kolitsidou, Hoy, \& Hannam}]{Thompson_2024}
Thompson, J.~E., Hamilton, E., London, L., {et~al.} 2024, Physical Review D, 109, \dodoi{10.1103/physrevd.109.063012}

\bibitem[{{Tonry} {et~al.}(2018{\natexlab{a}}){Tonry}, {Denneau}, {Heinze}, {Stalder}, {Smith}, {Smartt}, {Stubbs}, {Weiland}, \& {Rest}}]{2018PASP..130f4505T}
{Tonry}, J.~L., {Denneau}, L., {Heinze}, A.~N., {et~al.} 2018{\natexlab{a}}, \pasp, 130, 064505, \dodoi{10.1088/1538-3873/aabadf}

\bibitem[{{Tonry} {et~al.}(2018{\natexlab{b}}){Tonry}, {Denneau}, {Flewelling}, {Heinze}, {Onken}, {Smartt}, {Stalder}, {Weiland}, \& {Wolf}}]{2018ApJ...867..105T}
{Tonry}, J.~L., {Denneau}, L., {Flewelling}, H., {et~al.} 2018{\natexlab{b}}, \apj, 867, 105, \dodoi{10.3847/1538-4357/aae386}

\bibitem[{{Troja} {et~al.}(2017){Troja}, {Piro}, {van Eerten}, {Wollaeger}, {Im}, {Fox}, {Butler}, {Cenko}, {et~al.}}]{Troja_2017}
{Troja}, E., {Piro}, L., {van Eerten}, H., {et~al.} 2017, \nat, 551, 71, \dodoi{10.1038/nature24290}

\bibitem[{Tsukada {et~al.}(2023)Tsukada, Joshi, Adhicary, George, Guimaraes, Hanna, Magee, Zimmerman, Baral, Baylor, Cannon, Caudill, Cousins, Creighton, Ewing, Fong, Godwin, Harada, Huang, Huxford, Kennington, Kuwahara, Li, Meacher, Messick, Morisaki, Mukherjee, Niu, Pace, Posnansky, Ray, Sachdev, Sakon, Singh, Tapia, Tsutsui, Ueno, Viets, Wade, \& Wade}]{PhysRevD.108.043004}
Tsukada, L., Joshi, P., Adhicary, S., {et~al.} 2023, Phys. Rev. D, 108, 043004, \dodoi{10.1103/PhysRevD.108.043004}

\bibitem[{{Verrecchia} {et~al.}(2023){Verrecchia}, {Tavani}, {Ursi}, {Pittori}, {Cardillo}, {Casentini}, {Foffano}, {Piano}, {Lucarelli}, {Baroncelli}, {Bulgarelli}, {Ciabattoni}, {di Piano}, {Fioretti}, {Panebianco}, {Parmiggiani}, {Pilia}, {Longo}, \& {Agile Team}}]{2023GCN.33826....1V}
{Verrecchia}, F., {Tavani}, M., {Ursi}, A., {et~al.} 2023, GRB Coordinates Network, 33826, 1

\bibitem[{Villa-Ortega {et~al.}(2022)Villa-Ortega, Dent, \& Barroso}]{Villa-Ortega:2022qdo}
Villa-Ortega, V., Dent, T., \& Barroso, A.~C. 2022, Mon. Not. Roy. Astron. Soc., 515, 5718, \dodoi{10.1093/mnras/stac2120}

\bibitem[{{Wang} {et~al.}(2024){Wang}, {Yang}, {Li}, \& {GRID Collaboration}}]{2024GCN.36495....1W}
{Wang}, C., {Yang}, Z., {Li}, L., \& {GRID Collaboration}. 2024, GRB Coordinates Network, 36495, 1

\bibitem[{{Waratkar} {et~al.}(2023{\natexlab{a}}){Waratkar}, {Bhalerao}, {Bhattacharya}, {Rao}, {Vadawale}, \& {AstroSat CZTI Collaboration}}]{2023GCN.33896....1W}
{Waratkar}, G., {Bhalerao}, V., {Bhattacharya}, D., {et~al.} 2023{\natexlab{a}}, GRB Coordinates Network, 33896, 1

\bibitem[{{Waratkar} {et~al.}(2023{\natexlab{b}}){Waratkar}, {Bhalerao}, {Dixit}, {Ahmad}, {Bhattacharya}, {Rao}, {Vadawale}, \& {AstroSat CZTI Collaboration}}]{2023GCN.33819....1W}
{Waratkar}, G., {Bhalerao}, V., {Dixit}, M., {et~al.} 2023{\natexlab{b}}, GRB Coordinates Network, 33819, 1

\bibitem[{{Waratkar} {et~al.}(2024){Waratkar}, {Bhalerao}, {Joshi}, {Tembhurnikar}, {Bhattacharya}, {Rao}, {Vadawale}, \& {AstroSat CZTI Collaboration}}]{2024GCN.36242....1W}
{Waratkar}, G., {Bhalerao}, V., {Joshi}, J., {et~al.} 2024, GRB Coordinates Network, 36242, 1

\bibitem[{Wiggins \& Lai(2000)}]{Wiggins_2000}
Wiggins, P., \& Lai, D. 2000, The Astrophysical Journal, 532, 530–539, \dodoi{10.1086/308565}

\bibitem[{Wiggins \& Lai(1999)}]{Wiggins1999TidalIB}
Wiggins, P.~A., \& Lai, D. 1999, The Astrophysical Journal, 532, 530 .
\newblock \url{https://api.semanticscholar.org/CorpusID:119409764}

\bibitem[{Wyatt {et~al.}(2020)Wyatt, Tohuvavohu, Arcavi, Lundquist, Howell, \& Sand}]{Wyatt_2020}
Wyatt, S.~D., Tohuvavohu, A., Arcavi, I., {et~al.} 2020, The Astrophysical Journal, 894, 127, \dodoi{10.3847/1538-4357/ab855e}

\bibitem[{{Wyatt} {et~al.}(2020){Wyatt}, {Tohuvavohu}, {Arcavi}, {Lundquist}, {Howell}, \& {Sand}}]{2020ApJ...894..127W}
{Wyatt}, S.~D., {Tohuvavohu}, A., {Arcavi}, I., {et~al.} 2020, \apj, 894, 127, \dodoi{10.3847/1538-4357/ab855e}

\bibitem[{Yama {et~al.}(2023)Yama, Suzuki, Miyazaki, Rakich, Yamawaki, Kirikawa, Kondo, Hirao, Koshimoto, \& Sumi}]{Yama_2023}
Yama, H., Suzuki, D., Miyazaki, S., {et~al.} 2023, Journal of Astronomical Instrumentation, 12, \dodoi{10.1142/s2251171723500046}

\bibitem[{{Yoshida} {et~al.}(2023){Yoshida}, {Sakamoto}, {Sugita}, {Kawakubo}, {Yamaoka}, {Nakahira}, {Asaoka}, {Torii}, {Akaike}, {Kobayashi}, {Shimizu}, {Tamura}, {Cannady}, {Cherry}, {Ricciarini}, {Marrocchesi}, \& {Calet Collaboration}}]{2023GCN.33872....1Y}
{Yoshida}, A., {Sakamoto}, T., {Sugita}, S., {et~al.} 2023, GRB Coordinates Network, 33872, 1

\bibitem[{{Zhang} {et~al.}(2024){Zhang}, {Wang}, {Yang}, {Pan}, {Xu}, {Jin}, {Li}, {Sun}, {Liu}, {Ling}, {Zhang}, {Chen}, {Cheng}, {Cui}, {Fan}, {Hu}, {Hu}, {Huang}, {Jin}, {Lian}, {Liu}, {Liu}, {Lv}, {Mao}, {Pan}, {Wang}, {Wen}, {Wu}, {Xu}, {Yuan}, {Zhang}, {Zhang}, {Zhang}, {Zhao}, {Shui}, {Zhang}, {Kuulkers}, {Santovincenzo}, {O'Brien}, {Nandra}, {Rau}, {Cordier}, \& {Einstein Probe Team}}]{2024GCN.36277....1Z}
{Zhang}, W.~J., {Wang}, Y.~L., {Yang}, H.~N., {et~al.} 2024, GRB Coordinates Network, 36277, 1

\bibitem[{{Zhu} {et~al.}(2021){Zhu}, {Wu}, {Yang}, {Zhang}, {Yu}, {Gao}, {Cao}, \& {Liu}}]{2021ApJ...921..156Z}
{Zhu}, J.-P., {Wu}, S., {Yang}, Y.-P., {et~al.} 2021, \apj, 921, 156, \dodoi{10.3847/1538-4357/ac19a7}

\end{thebibliography}
\bibliographystyle{aasjournal}

\appendix
\label{Appendixref}

\section{NSBH gravitational wave alert candidates}
\label{Events}

Below, are detailed NSBH gravitational wave alerts identified by the LIGO-Virgo-Kagra collaboration tagged as "NSBH candidates", which motivated follow-up with electromagnetic instruments. We describe here their alert properties and the associated follow-up made by the whole community with a combination of public and refined analysis coverage.

%\subsection{Coverage of publicly reported observations (no refined analyses)}

In Table~\ref{tab:coverage-public}, is presented the coverage \% of the most up-to-date sky localization area and order of magnitude of the upper limit for S230518h, GW230529, S230627c and S240422ed using additional public reports on each event (not used in the main part of the article). We only include here coverages totalling 2\% or more from 0 to 6 days post T$_0$ and per filter.

\begin{table*}
\hspace*{-3cm}
\centering
\begin{tabular}{|l|ll|ll|ll|l|}
\hline
Filter & \multicolumn{2}{c|}{ 0 - 1 day} &  \multicolumn{2}{c|}{1 - 2 day} & \multicolumn{2}{c|}{2 - 6 day} & Instruments \\
 & \% c.r & upper &  \% c.r & upper &  \% c.r & upper & \\
\hline
\multicolumn{8}{|c|}{\textbf{S230518h}}\\
\hline
clear & 6\% & 17  & - & - & - & - & MASTER \\
$u$-band  & 2 \% & 19  & 33 \% & 18.5 &  11 \% & 18.5  & Swift/UVOT - MeerLICHT \\
$i$-band  & $<$1\% & -  &  33 \% & 19 &  11 \% & 19  & MeerLICHT - SWOPE \\
$q$-band   & - & -  & 37 \% & 20 &  15 \% & 20  & MeerLICHT \\
\hline
%\multicolumn{9}{|c|}{\textbf{GW230529}}\\
%\hline
%& Clear & 1\% & 21  & $<1$ \% & 20 &  1\% & 21 & MASTER \\
%\hline
%\multicolumn{9}{|c|}{\textbf{S230627c}}\\
%\hline
%& Clear & $<$ 1 \% & 16  & - & - & - & - & MASTER \\
%\hline
\multicolumn{8}{|c|}{\textbf{S240422ed}}\\
\hline
%& clear & \fixme{25\%} & \fixme{19.3}  & - & - & - & - & MASTER - \fixme{CSS/SAGUARO - GRANDMA}  \\
%& $B$-band & $<$1\%  & 19.5 & - & -  & - & -  & GRANDMA & yes \\
$u$-band & 4\%  & 19 &$<$1\% & - & - & -  & Swift/UVOT \\
$q$-band & 69\%  & 19.5 & 80\% & 19.5 & 82\% & 20 & MeerLICHT, BlackGEM \\
$i$-band & 5\%  & 20.5 & 3\% & 22.0 & $<$1\% & - & Las Cumbres, Magellan \\
%& $I$-band & 2\%  & 16.8 & - & -  & - & -  & GRANDMA & yes \\
$J$-band & 10\%  & 21.5 & 14\% & 21.5 & - & -  & PRIME \\
\hline
\end{tabular}
\caption{Coverage \% of the most up-to-date sky localization area and order of magnitude of the upper limit for S230518h and S240422ed (only when totalling 2\% or more from 0 to 6 days post T$_0$ and per filter) using publicly reported information using \textquote{tiling and galaxy-targeting} observations. These observations are not included in the KN analysis. GW230529, S230627c results are not presented here as these observations cover less than 2\% from 0 to 6 days post T$_0$ and per filter. We do not include serendipitous coverage from the follow-up of gravitational-wave candidates as they represent quasi-null coverage. Finally, PS1, DDOTI and LAST observations are not reported while observations have been made. Coverages with less than 2\% of sky localization area are taken into the kilonova upper limit computation if available.}
\label{tab:coverage-public}
\end{table*}

\subsection{S230518h GW Alert and Observational Campaign}

Please find below information related to the observational campaign of S230518h.

\subsubsection{GW Alert}

On 2023 May 18 at 12:59:08.167 UTC ($T_0$), the International Gravitational-wave Network (IGWN) identified a gravitational-wave (GW) candidate event (GPS time: 1368449966.167). The GW source was detected by both LIGO Hanford Observatory (H1), LIGO Livingston Observatory (L1)%, and Virgo Observatory (V1) contributed to the localization
. It was identified by several online searches: GstLAL \citep{2024PhRvD.109d2008E}, PyCBC Live \citep{Canton_2021}, MBTAOnline \citep{Aubin2021-ja} and  analysis pipelines. S230518h was publicly circulated to the astronomical community \footnote{\url{https://gracedb.ligo.org/superevents/S230518h}}, less than $\sim$~50 seconds post detection and with a measured initial False Alarm Rate of %$3.2~\times~10^{-10}$ Hz, or about 
$\sim$one in~98 years. The initial classification given by the PyCBC data analysis pipeline demonstrated a high probability of an NSBH candidate ($\sim$86\%), with a likelihood of being non-astrophysical ($\sim$10\%), a BBH ($\sim$4\%), or a BNS merger ($<$1\%). The low latency alert system classified the GW alert candidate with an extremely high potential to have a neutron star involved in the binary system (HasNS $>$99\%) with a very low probability of having light associated with the creation of an accretion disk post-merger, using several equation of states of neutron stars (HasRemnant$<$1\%). Hence it possessed a low probability of matter being present outside the final compact object, using the masses and spins inferred from the signal. In addition, there was minimal support ($<$1\%) that one or both components were between 3 and 5 solar masses (HasMassGap) \citep{2023arXiv230804545S}. The sky localization size was initially about 1002 deg$^2$ (90\% confidence), with reference to the Bayestar skymap, on 2023-05-18 at 13:26:12 UTC. The a posteriori luminosity distance was estimated to be 276 $\pm$ 79\,Mpc, marginalized over the whole sky. Further analysis of the LIGO Hanford Observatory (H1) and LIGO Livingston Observatory (L1) data around the time of the compact binary merger was conducted, with parameter estimation done using Bilby \citep{2023GCN.33816....1L}. Approximately 8.4 days later, on 2023-05-26 at 22:09:11, the 90\% credible region covered an area of 460 deg$^2$ and is located with a posterior distance luminosity of 204 $\pm$ 57\,Mpc \citep{2023GCN.33884....1L}. %Results of observations and counterpart candidates were shared across the GCN %infrastructure, TNS, and TreasureMap, which facilitated effective reporting with minimal delays. 

\subsubsection{Prompt searches}
\paragraph{Neutrino} Ice-Cube neutrino observations did not find any track-like muon neutrino events in IceCube data \citep{Aartsen_2020} \citep{Abbasi_2023}, in a time range of 1000 seconds around the alert event time, and within the \textit{preliminary} sky localization area \citep{2023GCN.33814....1I}. Assuming an E$^{-2}$ spectrum (E$^2$ dN/dE), upper limits were given from 0.03 to 1.1 GeV cm$^{-2}$. Additional searches for track-like muon neutrino events were conducted in a time range of -0.1 day, +14 days from the alert event time. A p-value of 0.73 was reported, consistent with no significant excess of track events, and with an upper limit from  0.03 to 1.2 GeV cm$^{-2}$ \citep{2023GCN.33907....1I}.

\paragraph{HEN, Gamma-rays, and X-rays}
INTEGRAL/SPI-ACS did not find gamma-ray signal within $\pm$ 300~s around the GW trigger time, and provided a 3-sigma
upper limit on the 75-2000 keV fluence of 3.8$\times$10$^{-7}$ erg/cm$^2$, in a region that covers 50\% of the bayestar sky localization \citep{2023GCN.33815....1S}. Konus-Wind (KW) also observed the whole sky localization area of S230518h and found no significant ($>$5$\sigma$) excess over the background in 2 hours before and after $T_0$ and provided an upper limit of 7.6 $\times$ 10$^{-7}$ erg/cm$^2$ for a 3~s integration period in 20 - 1500 keV \citep{2021ApJ...921..156Z}. Moreover, no CALET Gamma-ray Burst Monitor (CGBM) onboard trigger occurred around $T_0$ and prior or after 60s, and very limited overlap with the sky localization area \citep{2023GCN.33872....1Y}. Glowbug observatory \citep{grove2020glowbuglowcosthighsensitivitygammaray} observed less than~$<$~1 \% of the sky localization area $\pm$ 30~s around $T_0$, and reported an upper limit of 10$^{-6}$ erg/cm$^2$/s in the 50-2000~keV. In addition, AGILE/MCAL found no significant event candidates within a time interval covering $\pm$ 2 sec from $T_0$ and obtained a 2 sigma upper limits, in the 0.4-1 MeV energy range, of the order of 10$^{-8}$ erg/cm$^2$ for a 300~s integration time \citep{2023GCN.33826....1V}. Similarly,  AstroSat CZTI reported an upper limit of gamma-ray signal within 1000s post-GW trigger, in 8.25$\times10^{-6}$ ergs/cm$^2$ in 20 - 200 keV using a 10~s time window search. In complement, Fermi-GBM did not detect any gamma-ray signal within a 30~s time window, and provide an upper limit at 3-sigma at 10$^{-7}$ erg/s/cm$^2$ in the 1 keV - 10 MeV \citep{2023GCN.33819....1W}.

%agile

\paragraph{Optical} TESS imaged 25\% of the sky localization area promptly to 11h post $T_0$ every 200s and reported an upper limit of 16 mag in the 600-1000 nm wavelength range \citep{2023GCN.33878....1J}.

\subsubsection{Tiling}

\paragraph{X-rays} MAXI/GSC imaged 70 \% of the bayestar sky localization from 0.4h to 1.3h post-GW trigger in 2-20 keV and did not find any significant X-ray detection brighter than 20 mCrab \citep{2023GCN.33823....1S}. Similarly, Swift/XRT also carried out 60 observations for a total of 6.0 deg$^2$ coverage on the sky, e.g 2\% of updated GW sky localization and reported 8 X-ray sources candidates \citep{2023GCN.33824....1E}. Among them, two are not referenced in X-ray catalogs, and four are cataloged, such very unlikely to be associated with the GW trigger, due to prior detections before the GW signal.

\paragraph{Optical} 
Tiling observations were conducted by MASTER-Net (around 6\% of the Bilby skymap covered in clear), GECKO (37\% of the Bilby skymap in the $R$ band by KMTNet and additional 11\% in $r$-band by RASA36), Meerlicht (37\%, 33\%, 33\% of the Bilby skymap covered in $q,u,i$-bands, respectively), Swift/UVOT (2\% of the Bilby skymap),
ATLAS (69\% and 71 \% of the Bilby skymap covered in the $o$ and $c$-band), Swope (1\% of the bilby skymap covered in the $i$-band). The total coverage of the Bilby S230518h skymap was 25\% in TESS filter, 37\% in $q-$band, 34\% in $u-$band, 37\% in $R-$band, 69\% in $o-$band, 71\% in $c-$band, 33\% in $i-$band and 6\% with no filter.

In space, Swift/UVOT observed 2\% of the bilby skymap from 0.1 to 0.4 day post $T_0$ in $u$-band. On the ground, Master trigger follow-up observations 3h post-GW trigger time, and cover 6\% of the GW sky localization of Bilby in 24 hours at sensitivity around 17 mag\footnote{see \href{https://master.sai.msu.ru/site/master2/ligo_1.php?id=11865}{here} for Master coverage}. About one hour later, the GECKO telescopes observed from 4.5 hrs up to 1.2 days after $T_0$ ((37\% of the Bilby skymap in the $R$ band by KMTNet and additional 11\% in $r$-band by RASA36, the latter not being reported here), with an upper limit of 22 mag in $R$ band (\citealt{2023GCN.33833....1P, paek2025geckofollowupobservationbinary}, ApJ, in press): they did not report any GW candidate counterpart. A few hours later, ATLAS observed 44\% of the bilby sky localization (from 9.4 to 10.1~h after $T_0$) with an upper limit in $o$-band around 17.5 to 18.5 mag. ATLAS team did not report any significant new transient in their field \citep{2023GCN.33830....1F}. At the same time, the Swope 1-m telescope triggered 10~h after T$0$ and covered in 1h about 1~\% of the sky localization area in $i$-band, with a depth of the order of 21.3. Swope reported three candidate counterparts: SSS23a (i-mag:20.9+/-0.2), SSS23b (i-mag:21.1+/-0.2), SSS23c (i-mag:20.1+/-0.2) \citep{2023GCN.33829....1C}. Finally, Meerlicht observed from 1.15 to 2.2 days post $T_0$ in $q,u,i$-band reported in \citet{2020ApJ...894..127W}.

\subsubsection{Galaxy-targeting}

%\citet{2023GCN.33820....1C} publicly reported a list of promising host galaxy candidates seven hours after the merger. GECKO conducted galaxy targeting strategies using \textit{GeclDigestor} (Paek 2023a).

GECKO's Lee Sang Gak Telescope \citep{10.5303/JKAS.2015.48.4.207} performed galaxy-target observations of 12 host galaxy candidates to depths of 19.0 to 20.8 in $r$-band and publicly reported the candidates seven hours after the merger \citep{2023GCN.33820....1C}, but no transients were found around these galaxies \citep{paek2025geckofollowupobservationbinary}.

\subsubsection{Candidates}

\paragraph{Swift X-ray candidates} The Australia Telescope Compact Array (ATCA) telescope conducted radio observations on the X-ray source S230518hX8 on 2023-05-20 with 2x2048 MHz bands centered on 5.5 and 9 GHz and found a radio source consistent with the candidate with flux densities of $\sim$350 uJy and $\sim$500 uJy at 5.5 GHz and 9 GHz. This source was also detected by the Rapid ASKAP Continuum Survey in observations on 2019-08-03, suggesting it is unrelated to S230518h \citep{2023GCN.33834....1D}.

\paragraph{Swope candidates}  TESS observed  11 h after $T_0$ the candidates and obtained an upper limit of 16 on average in 600 cm - 1000 nm \citep{2023GCN.33878....1J}. Similarly, Swift/UVOT observations of the candidates SSS23a, SSS23b, and SSS23c began 1.5 days after $T_0$\citep{2023GCN.33832....1O}, and led to the non-detection in the UVOT photometric system and provide upper limit the $m2$ filter of 20.2, 20.1, and 19.6 mag for SSS23a, SSS23b, and SSS23c respectively. 

\paragraph{MeerLicht candidates} Swift/UVOT imaged AT2023ixg and AT2023iyb, 73 hours after $T_0$, did not find any bright flux in the $UV$ band and report 3-$\sigma$ upper limits in $w2>$20.4 mag for AT2023ixg and $w2>$20.7 mag for AT2023iyb \citep{2023GCN.33851....1B}. Similarly, TESS images of the candidate, did not find any optical significant excess at 18 mag in average or brighter \citep{2023GCN.33878....1J}. Finally, ATCA reported an upper limit (3-$\sigma$) ranging from 90-120 uJy at 5.5 GHz and 50-90 uJy at 9 GHz for Swope candidates \citep{2023GCN.33834....1D}.

%\subsubsection{Inputs for our analysis}

%We report in Table~\ref{followupcoverage}, the coverage of the most-update sky localization area at different epochs and in different optical bands. We select in this Table for our results in section~\ref{kilonovaresults}. We will compare the clear filter to $R$. We do not have synthetic kilonovae lightcurve for $o,q$, and so we will ignore it for our analysis. All counterpart candidates' observations are not taken into account. Note that we need sometimes, to compute ourselves the coverage of the tiles and pointings, and we base this computation on public information in GCN that does not provide us with all the complete information (field of view, upper limit per tile, etc). 

\subsection{GW230529 (initially named S230529ay) GW Alert and Observational Campaign}
\label{GW230529}

Please find below information related to the observational campaign of GW230529.

\subsubsection{GW Alert}

On 2023 May 29 at 18:15:00.7 UTC the international Gravitational-wave network (IGWN) identified a gravitational-wave candidate (gps time: 1369419318.746) \citep{2024ApJ...970L..34A}. The GW source was detected by multiple pipelines with a false alarm rate of less than one per thousand years and circulated to the public 15 seconds post $T_0$.
 The last localization had a 90\% credible area of 24 100deg$^2$ due to only being observed by a single detector (L1) and the posteriori luminosity distance of 201$^{+102}_{-96}$ Mpc. GW230529 is the source of interest because the estimated primary mass is about
3.6 solar mass (sub-solar BH?), the secondary mass is between 1.2 and 2.0 solar mass (NS regime). The nature of this event (NSBH, or BNS) is extensively discussed in \citet{2024ApJ...970L..34A}.

\subsubsection{Prompt searches}

\paragraph{Neutrino followup} IceCube looked for track-like muon neutrino events with a sky localization consistent with GW sky localization area between -0.1 days and +14 days from the alert event time. IceCube reported a p-value of 0.38, consistent with no significant excess of track events, and with an upper limit from  0.03 to 1.2 GeV cm$^{-2}$ \citep{2023GCN.33980....1I}.

\paragraph{HEN, Gamma-rays, and X-rays} No CALET Gamma-ray Burst Monitor (CGBM) onboard trigger in the 10 - 100 GeV occurred around $T_0$ and prior or after 60s in the overwrap region with the LVK high probability localization region : the upper limit is 1.2$\times$10$^{-5}$ erg/cm$^2/s$ in the 20 c.r region \citep{2023GCN.33897....1S}.
INTEGRAL/SPI-ACS did not find gamma-ray signal within $\pm$ 300~s around the GW trigger time, and provided a 3-sigma upper limit on the 75-2000 keV fluence of 3.4$\times$10$^{-7}$ erg/cm$^2$, in a region that covers 50\% of the bayestar sky localization \citep{2023GCN.33890....1S}.  Moreover, 
In addition, AGILE/MCAL found no significant event candidates within a time interval covering $\pm$ 15 sec from $T_0$, and within 60 \% of the 90 \%  LVK skymap and obtained a 2 sigma upper limits, in the 0.4-1 MeV energy range, of the order of 10$^{-6}$ erg/cm$^2$ for a 300~s integration time \citep{2023GCN.33894....1L}. Similarly, AstroSat CZTI reported an upper limit of gamma-ray signal from T$_0$+1001 to T$_0$+5139 seconds, in e-06 ergs/cm$^2$ in 20 - 200 keV using 10~s time window search over the full sky \citep{2023GCN.33896....1W}. In complement, Fermi-GBM did not detect any gamma-ray signal with a 30~s time window, and provide an upper limit at 3-sigma at 10$^{-7}$ erg/s/cm$^2$ in the 1 keV - 10 MeV \citep{ronchini2024constrainingpossiblegammarayburst} in 100 \% of the sky localization. Finally, the Swift Burst Alert Telescope (BAT) observed 97.4\% of the localization probability at the time of the merger and found no evidence of a signal was found during this search \citep{ronchini2024constrainingpossiblegammarayburst}. In the joint search BAT+GBM, \citet{ronchini2024constrainingpossiblegammarayburst}  excludes with 90 \% of confidence, the presence of a
a top–hat jet structure and on-axis with an isotropic luminosity above 10E48 erg.s-1, in 1 keV–10 MeV.
 
\subsubsection{Tiling}
\paragraph{X-rays} MAXI observed 90 min (up to 1.4h) after T$_0$ 90 \% of the 90 \% c.r localization of the bilby skypmap and did not detect any significant X-ray counterpart at the level of 20 mCrab in the 2 - 20 keV \citep{2023GCN.33893....1S}.

\paragraph{Optical followup} Due to the large size of the sky localization, the event was poorly followed-up in the optical band. The total coverage of the Bilby GW230529 skymap was 16\% in $g-$band, 12\% in $r-$band, 24\% in $o-$band, 5\% in $i-$band; $L$-band 14\%, and $<$1\% with no filter. The SAGUARO program performed unfiltered serendipitous observations of 1.4\% of the Bilby skymap using the 1.5m Catalina Sky Survey (CSS) telescope, from 9.5h to 2.6 days after $T_0$, median limiting magnitudes around 20 \citep{2024GCN.36266....1H} in $G$-band. Followed by MASTER-Net that conducted observations between 9.84h to 6 days after $T_0$, with an upper limit up from 17 to 20 mag without filter and covered less than 1\% of the skymap during the first 24h \citep{2023GCN.33895....1L}. ATLAS covered 24\% of the bilby skymap in $o$-band considering observations from 0 to 6 days. GOTO covered 14\% of the bilby skymap in $L$-band from 0.18 days to 6 days.
Finally, ZTF obtained images in the $g$, $r$, $i$, bands about 10 hours up to 6 days after the event time, covering 16 \%, 12\% and 5\% in $g$, $r$,$i$  of sky localization area, with a median limiting magnitudes were $g$,$r$ = 21 and;$i$=20. One candidate was found, AT2023jtt \citep{2023TNSTR1278....1K}, inside the 95\% error region, with an upper limit up to 20.49±0.23 in $r$-band. Five other optical candidates were investigated by the ZTF but rejected \citep{ahumada2024searchinggravitationalwaveoptical}.

\subsubsection{Galaxy Targeting}

No public follow-up using galaxy targeting was published for this event.

\subsubsection{Candidates}

\paragraph{ZTF23aamnpce/2023jtt} -- ZTF23aamnpce/AT2023jtt did not received additional observation (unless PS1, \href{https://www.wis-tns.org/object/2023jtt}{see TNS}). Since, it is 0.5” from a galaxy with a photometric redshift of 0.22+/-0.06, suggesting it is probably not associated with the GW230529.

%\subsubsection{Inputs for our analysis}

%We report in Table~\ref{followupcoverage}, the coverage of the most-update sky localization area at different epochs and in different optical bands. We select in this Table for our results in section~\ref{kilonovaresults}. We will compare our synthetic KN light curves to clear, $g$, $r$, and $i$.

%In the previous observing runs, the LIGO-Virgo-KAGRA (LVK) collaboration detected several NSBH events, each contributing to the growing understanding of these phenomena. For instance, GW190814 was detected during O3 and had a secondary component with a mass of about 2.6 solar masses, straddling the boundary between the heaviest neutron stars and the lightest black holes \citep{2020ApJ...896L..44A,2024PhRvD.109b2001A}. The mass of GW230529's primary component falls within a similar mass range adding to the observations of low-mass black holes or heavy neutron stars. Another NSBH event from O3, which had a more massive black hole component and a neutron star around 1.5 solar masses is GW200115 \citep{2021ApJ...915L...5A}. Compared to GW230529, GW200115 had a more typical mass distribution for NSBH systems, highlighting the unique nature of GW230529's mass configuration. As the GW230529 was localized to a very broad sky area, it was challenging to pinpoint the exact location for electromagnetic follow-up observations. In contrast, other significant NSBH events from O1, O2, and O3 had better sky localization due to detections by multiple instruments. 

\subsection{S230627c}

Please find below information related to the observational campaign of S230627c.

\subsubsection{GW Alert}

On 2023 June 27 at 01:53:37.819 UTC, the International Gravitational-wave Network (IGWN) identified a gravitational-wave (GW) candidate event (gps time:1371866035.819) \citep{2024GCN.34086}. The GW source was detected by both LIGO Hanford Observatory and LIGO Livingston Observatory. It was identified by several online searches: CWB \citep{Klimenko2016-zr}, MBTA \citep{Aubin2021-ja}, GstLAL \citep{2024PhRvD.109d2008E}, oLIB \citep{Lynch2017-yk}, PyCBC Live \citep{Canton_2021}, and SPIIR \citep{Chu2022-nw} analysis pipelines. S230627c was publicly circulated to the astronomical community\footnote{\url{https://gracedb.ligo.org/superevents/S230627c}}, less than 50 seconds post-detection and with a measured initial False Alarm rate of %$3.168 \times 10^{-10}$ Hz or 
one per $\sim$100 years. The event was first classified as BBH ($>$99\%) and NSBH around 0.7\% by the GstLAL data analysis pipeline, whereas PyCBC Live demonstrated a probability of a Neutron Star-Black Hole candidate of (49\%), Black Hole-Black Hole candidate of (48\%) with a relatively small likelihood of being non-astrophysical (3\%), or a Binary Neutron Star Merger ($<$1\%) \citep{PhysRevD.108.043004,Kapadia_2020}. The low latency alert system classified the GW alert candidate as having relatively low potential to have a neutron star involved in the binary system (HasNS $<$1\%) as well as light associated with the creation of an accretion disk post-merger, using several equation of states of neutron stars (HasRemnant $<$1\%) \citep{Chatterjee_2020,2023arXiv230804545S}. Hence it possessed a low probability of matter being present outside the final compact object, using the masses and spins inferred from the signal. In addition, there was some support (14\%) that one or both components were between 3 and 5 solar masses (HasMassGap) \citep{2023arXiv230804545S}.  The refined 90\% credible region is well fit by an elongated ellipse with an area of 82 deg$^2$ and is located (if astrophysical) with a posterior distance luminosity of 291 $\pm$ 64\,Mpc. %Besides, the HasMassGap probability was updated to be 14\% on 2023-06-27T04:25:34, while the HasNS and HasRemnannt probabilities remained at $<$1\%.

\subsubsection{Prompt searches}

\paragraph{HEN, Gamma-rays, and X-rays}
Fermi-GBM did not detect any gamma-ray signal with a 30~s time window, and provided an upper limit at 3-sigma at 1.10$^{-7}$ erg/s/cm$^2$ in the 1 keV - 10 MeV, and in a 1s time window \citep{2023GCN.34095....1D}. Moreover, Swift/BAT was observing 99.2\% of the GW localization probability using the Bilby skymap at merger time, with the entirety of the GW 90\% credible region contained inside the coded FoV. After a search for emission on 8 timescales from 0.128s to 16.384s in the interval [-20,+20] seconds around the merger time, no evidence for a signal was found with a 5-sigma upper limit at 3.6$^{-8}$ erg/s/cm$^2$ in the 15 - 350 keV, and in a 1s time window\citep{2023GCN.34128....1R}. 

\paragraph{X-ray followup} X-ray observations (2-20 keV) were conducted by MAXI/GSC, which covered 64\% of the 90\% credible region of the Bayestar sky map from $T_0$+0 to $T_0$+1.5h, and which did not find any detection at 1-$\sigma$ averaged upper limit with a level of 140 mCrab at 2-20 keV \citep{2023GCN.34088....1K}.

\subsubsection{Galaxy-targeting}

\citet{2023GCN.34092....1C} reported, seven hours post-detection, a list of promising host galaxy candidates for the event. The GRANDMA network conducted a search of at least 45/1858 compatibles galaxies within the Bilby skymap, focusing on galaxies within the distance range 0.69–0.84 from $T_0$ \citep{2023GCN.34130....1L}. This effort involved galaxy-targeted observations of the LVK event using multiple telescopes, including the Abastumani-T70, UBAI-AZT-22, UBAI-NT60, UBAI-ST60, OST-CDK, NOWT, and OPD-60cm. The target galaxies were selected from the MANGROVE catalog \citep{10.1093/mnras/staa114}, based on compatibility within a 3$\sigma$ distance range consistent with the GW event. Despite this targeted approach, no significant candidates were detected.

The GRANDMA galaxy-targeted observations overlaid on the S230627c skymap, covered an area of less than 4\% of the Bilby localization region. In addition, the LOAO and KHAO telescopes from the GECKO network \citep{2023GCN.34146....1P} observed the seven highest-ranking host galaxy candidates identified from the GLADE+ catalog \citep{2022MNRAS.514.1403D}, beginning 0.11 days post-$T_0$.
%Figure~\ref{fig:GRANDMA-followup-S230627c} shows the galaxy-targeted observations overlaid on the S230627c skymap, with the total coverage area being less than 4\% of the Bilby localization region. In addition, the LOAO and KHAO telescopes from the GECKO network \citep{2023GCN.34146....1P} observed the seven highest-ranking host galaxy candidates identified from the GLADE+ catalog \citep{2022MNRAS.514.1403D}, beginning 0.11 days post-$T_0$

\subsubsection{Tiling}

\paragraph{Optical followup} %ztf candidates
The total coverage of the S230627c Bilby skymap from 0 to 6 days was $<$1\% in clear (median 15.7 mag), 88\% in $g-$band (median 21.0 mag), 88\% in $r-$band (median 21.0 mag), 8\% in $R-$band (median 19.7 mag), 28.4\% in $o-$band (median 18.4 mag) and 93\% in $L-band$ (median 19 mag). Among the earliest observations of the GW localization region were those by the Zwicky Transient Facility (ZTF), utilizing the Palomar 48-inch telescope equipped with the 47 square-degree ZTF camera. ZTF acquired images in the $g$- and $r$-bands within the Bilby skymap, starting approximately 2.2 hours after the LVK trigger. This coverage spanned 88.4\% of the probability enclosed in the localization region, based on the Bilby skymap\footnote{This coverage slightly differs from the reported 72\% of the 90\% credible region \citep{ahumada2024searchinggravitationalwaveoptical}.}. ZTF initially reported four candidates (details to follow) and, in an offline analysis, identified a complete list of nine candidates, none of which showed significant evidence of interest \citep{ahumada2024searchinggravitationalwaveoptical}. Further, optical follow-up was conducted by GECKO with the Chungbuk National University Observatory (CBNU) optical telescope \citep{2021JKAS...54...89I}, where one tile, representing the highest probability region, was observed on 2023-06-27T12:21:49 (0.44 days post-$T_0$) \citep{2023GCN.34146....1P}, reaching a median depth of 20.7 mag in the Johnson-$R$ filter. Additionally, MASTER reported observations in clear band with a limiting magnitude of 15, achieving less than 1\% of the sky localization area, starting observations 0.18 days post-$T_0$ \citep{2023GCN.34093....1L}. GOTO covered 93\% of the bilby skymap in $L$-band post T0  to 6 days. Finally, ATLAS commenced observations one day post-$T_0$, extending to six days post-$T_0$, covering 28.5\% of the Bilby sky localization area.

%The ZTF alert stream was queried using tools such as Kowalski, Fritz, emgwcave, AMPEL, and ZTFReST. The ZTF selection criteria involved identifying candidates with a minimum of two detections separated by at least 15 minutes to mitigate against moving objects. To refine the findings, the candidates were cross-matched with the Minor Planet Center database to identify known asteroids. Stellar sources were excluded and machine learning algorithms were employed. Additionally, it was ensured that no spatially coincident ZTF alerts were issued before the detection time of the LVK trigger. Forced photometry was conducted on the ZTF and ATLAS images, requiring no prior detections before the LVK trigger to validate their candidate selection process.

\subsubsection{Candidates}
Four sources met the ZTF criteria and were located within the 95\% error region, as detailed in Table~\ref{{ZTFref27c}} of the online analysis. These sources were subsequently observed by the community. A second epoch of ZTF observations revealed a flat evolution, effectively ruling them out as viable candidates for counterparts.

%GRANDMA conducted specific follow-up on the ZTF transient candidates ZTF23aaptudb/AT2023lxs and ZTF23aaptusa/AT2023lxx using the NOWT telescope \citep{2023GCN.34130....1L}. Additionally, forced photometry by ATLAS at the positions of these ZTF candidates yielded multiple 3-$\sigma$ detections at the location of AT2023lxu more than six days before the LVK trigger time. This early detection strongly suggests that AT2023lxu is a faint supernova (SN), ruling it out as a potential optical counterpart to S230627c. Furthermore, two 3-$\sigma$ detections of AT2023lxt were recorded before the LVK trigger, also excluding it as a viable candidate \citep{2023GCN.34096....1F}. Finally, the AS-32 telescope at the Abastumani Observatory observed the ZTF23aaptsuy/AT2023lxu transient on 2023-06-27 (UT) at 17:45:15 and again on 2023-07-01 (UT) at 17:56:01. These observations confirmed a flat light curve, supporting the exclusion of ZTF23aaptsuy/AT2023lxu as an optical counterpart to S230627c \citep{2023GCN.34137....1P}. Lastly, the SAO telescope observed ZTF23aaptsuy/AT2023lxu, achieving a maximum 5-$\sigma$ depth of 19.7 in the R band \citep{2023GCN.34146....1P}. 

%atlas
\begin{table}
\centering
\begin{tabular}{|l|l|l|l|l|l|l|>{\raggedright\arraybackslash}p{6cm}|}
\hline
ZTF Name &  IAU Name &  RA (deg) &  DEC (deg) & Filter & Mag & Mag Err  & Description \\
\hline
\hline
ZTF23aaptsuy & AT2023lxu & 160.20196 & +41.96817 & r' & 20.20 & 0.08 &  2.7" away from \texttt{WISEA J104048.69+415805.3} with a spectroscopic z=0.092961 (luminosity distance of 440\,Mpc), which is at the edge of the 3-$\sigma$ boundary of the LVK line of sight distance estimate.  \\
\hline
ZTF23aapttaw & AT2023lxt & 164.68981 & +60.95459 & r' & 21.11 & 0.20 &  0.236'' away from an LS source with a photo-z =\(0.254\pm0.12\), outside of the LVK volume  \\
\hline
ZTF23aaptudb & AT2023lxs & 166.55661 & +78.55964 & r' & 20.86 & 0.16 &  0.07'' away from a galaxy that has a Legacy Survey DR8 (LS; Duncan, 2022) photo-z =\(0.118\pm0.07\), suggesting lack of association with the LVK trigger. \\
\hline
ZTF23aaptusa & AT2023lxx & 162.04457 & +71.84141 & g' & 20.89 & 0.19 &  0.75'' away from a galaxy with an LS photo-z =\(0.175\pm0.044\), suggesting lack of association with the LVK trigger. The centroid position showed a slight dispersion in the three detections. \\
\hline
\end{tabular}
\label{ZTFref27c}
\caption{ZTF candidates to S230627c.}
\end{table}
.
%abao

%\begin{figure}
%    \centering
%    \includegraphics[width=\textwidth]{Figures/S230627c_galaxies_zoomed.png}
%    \caption{GRANDMA galaxy targeted followup of S230627c on top of the GW skymap}
%    \label{fig:GRANDMA-followup-S230627c}
%\end{figure}

\subsection{Low-significance candidate S240422ed}

Please find below information related to the observational campaign of S240422ed.

%\sa{To be added GRANDMA also participated to the follow-up reaching a maximum depth of g,r$>$19.9 (see Fig. ~\ref{fig:S240422ed_GRANDMA_follow-up_zommed} a representation of the follow-up done by the collaboration). It shows the tiling and galaxy targeting strategy done by GRANDMA to cover \ma{11.5}\% of S240422ed skymap.}

%\subsection{Event}
\subsubsection{GW Alert}

On April 22, 2024, at 21:35:13.417 UTC, the International Gravitational-wave Network (IGWN) identified a gravitational-wave candidate event (gps time: 1397856931.42) \citep{2024GCN.36236....1L}. The GW source was detected by both LIGO Hanford Observatory (H1), LIGO Livingston Observatory (L1), and Virgo Observatory (V1), which contributed to the localization. It was identified by several online searches: GstLAL \citep{2024PhRvD.109d2008E} and PyCBC Live \citep{Canton_2021} analysis pipelines. MBTA Online and CWB were online but did not find triggers above the public threshold\footnote{ https://emfollow.docs.ligo.org/userguide/}. S240422ed was first publicly distributed to the astronomical community \footnote{\url{https://gracedb.ligo.org/superevents/S240422ed}}, less than $\sim$ 50 seconds post detection and with a measured initial False Alarm rate of FAR = one per $10^5$ years. The initial classification given by GstLAL demonstrated a high probability of an NSBH candidate ($>$~99\%)  \citep{PhysRevD.108.043004,Kapadia_2020}, and with a great potential to have light associated with the creation of an accretion disk post-merger, using several equation of states of neutron stars (HasRemnant $>$~99\%) \citep{Chatterjee_2020,2023arXiv230804545S}.  The most updated sky localization area of the 90 \% credible region was 259 deg$^2$ and its distance of 188 $\pm$ 43 Mpc. However, LVK also reported noise transients in both LIGO Livingston and LIGO Hanford detectors data within 10 seconds of the event time, which may impact the measurement of the FAR. Two months later, S240422ed was classified as low significant (below the public threshold) with a revised False Alarm Rate by one in thirty-five days and with a refined classification of being non-Astrophysical (93\%) \citep{2024GCN.36812....1L}. If Astrophysical, the classification tends to be more a BNS merger than an NSBH. Further investigation with offline analysis will help to understand the nature of this event.

\subsubsection{Prompt searches}

\paragraph{Neutrinos} Ice-Cube neutrino observations did not find any track-like muon neutrino events in IceCube data, in a time range of -0.1 day, +14 days from the alert event time. A p-value of 0.41 was reported, consistent with no significant excess of track events, and with an upper limit from 0.014 to 0.57 GeV $cm^{-2}$ \citep{2024GCN.36410....1I}.

\paragraph{HEN, Gamma-rays, and X-rays}
Fermi-GBM observed 100\% of the bilby localization probability at the event time, however, there was no onboard trigger +/- 30 s around $T_0$, with a 3-sigma flux upper limits over 10-1000 keV of 10$^{-7}$ erg/s/cm$^2$ \citep{2024GCN.36241....1R}. AstroSat-CZTI also did not find any evidence of hard X-ray transients in the 20–200 keV energy range during a 100-second window around the trigger time and provided upper limits in e-07 erg/cm$^2$ using 10 s time window search \citep{2024GCN.36242....1W}. Similarly, INTEGRAL/SPI-ACS did not find a gamma-ray signal within $\pm$ 300 s around the GW trigger time and estimated a 3-sigma upper limit on the 75-2000 keV fluence of 1.9e-07 erg/cm$^2$ within the 50\% probability containment region of the source localization \citep{2024GCN.36247....1S}. GECAM-C monitored the full localization region and no candidates were found to 3-sigma upper limits of the GRB energy flux in 10 keV-2000 keV, in of 10$^{-7}$ erg/s/cm$^2$ \citep{2024GCN.36300....1T}. GRID searched for GRB candidates in a 50-sec window around the trigger time and did not find any GRB transients, to a 3 sigma flux upper limits over 10-1000 keV, weighted by GW localization probability in 10$^{-7}$ erg/s/cm$^2$ \citep{2024GCN.36495....1W}.

\subsubsection{Tiling}

\paragraph{Gamma-rays} HAWC collaboration searched the 95\% probability containment area from $T_0$-5dt to $T_0$+10dt, and no significant gamma-ray detection above the background was observed \citep{2024GCN.36286....1H}. 

\paragraph{X-rays}
MAXI/GSC covered 100\% of the 90\% credible region of the bayestar skymap from ($T_0$+0.9h to $T_0$+1.0 h) and did not find any significant X-ray detection in the one-orbit scan observation with a typical 1-sigma upper limit of 20 mCrab at 2-20 keV \citep{2024GCN.36238....1K}. Swift-XRT observed the LVC error region from 10-90 ks after the trigger, covering 10.1 deg$^2$ of the sky (e.g 6\% of the sky credible region of the bilby skymap), and identified 3 rank 2 candidate counterpart sources \citep{2024GCN.36278....1E}. These sources were followed up greatly by the community. EP-WXT covered ~90 deg$^2$ of the 90\% credible region of the updated LVK sky localization 830 s before the trigger time and no significant X-ray counterpart was detected with flux upper limits in 0.5-4 keV at the 90\% confidence level \citep{2024GCN.36277....1Z}. EP-FXT reported then the detection of a possible X-ray counterpart candidate, EP240426a,with an absorption-corrected flux in 0.5-10 keV of ~9.2 x 10 $^{-13}$ erg/s/cm$^2$ \citep{2024GCN.36313....1S}. 

\paragraph{Optical} The total coverage of the S240422ed Bilby skymap from 0 to 6 days was $<$1\% in $B-$band (median 20.5 mag), 84\%, in $g-$band (median 19.5 mag), 88\% in $q-$band (median 19.7 mag), 92\% in $r-$band (median 23.1 mag), 81\% in $R-$band (median 16.7 mag), 99\% in $o-$band (median 18.7 mag), 23\% in $i-$band (median 22 mag), 4\% in $I-$band (median 16.8 mag), 83\% in $z-$band (median 22.6 mag), 29\% in $J-$band (median 16.5 mag), 97\% in $L$-band (median mag 19.5), 19\% in $G$-band (19.5 median mag) and $<$1\% with no filter. %In total, 17 surveys covered the 90 \% credible region promptly to 6 days. 4/5 covered at least 25\% of the bilby skymap. 

In details\footnote{Note that all percentages of the credible sky localization area have been calculated from this original work.}, MASTER-Net conducted observations between 2 min to 1 day after $T_0$, with an upper limit around 17 mag without filter \citep{2024GCN.36234....1L}. MeerLICHT observed promptly post $T_0$ and BlackGEM started observations about 1h and half post $T_0$: they covered respectively 84\% of the bilby skymap in $q$-band from 0 to 6 days in 19.5 upper limit and 85 \% from 0 to 6 days in 19.8  \citep{2024GCN.36237....1G}. GOTO observed directly after $T_0$ up to 6 days in $L$-band with an upper limit of 19.5, and cover 97\% of the bilby skymap \citep{2024GCN.36257....1A}. ATLAS observed 99.5\% of the bilby skymap in $o$-band from 0.8h to 6 days \citep{2024GCN.36251....1S}. DECam observed around 84.3\% of the bilby skymap in $r$-band with an median upper limit of 23.1 mag, 83.2\% of the bilby skymap in $z$-band with an upper limit of 22.6 mag from 2h post $T_0$ to 6 days \citep{2024GCN.36245....1P} and reported various candidates \citep{2024GCN.36273....1H}. DDOTI observed around 42\% of the Bayestar skymap in $w$-band with an median upper limit of 19 mag \citep{2024GCN.36256....1B}. GRANDMA conducted tiled observations of the 11.7\% credible area utilizing the FRAM-Auger telescope, from 2.5h post $T_0$ in 2 hours with no filter and with an upper limit of 16.7 \citep{2024GCN.36284....1A}. WINTER imaged 15.7\% of the Bilby sky area in $J$-band (in 16.5 mag upper limit with 5 sigma), started 5.5h post $T_0$ for 2.5h \citep{2024GCN.36248....1K}. KMTNet of GECKO observed 3h post $T_0$ to 6 days maximum and covers 78\% of the sky localization area in $R$-band with a magnitude of 21.5 \citep{2024GCN.36343....1J}. ZTF obtained images in the $g$, $r$, $i$, bands about 6 hours up to 6 days after the event time, covering 84 \% , 83\% and 17 \% in $g,r,i$-bands  of sky localization area, with a median limiting magnitudes were $g,r,i$ $\simeq$ 20 (this work, \citep{2024GCN.36310....1A}). Magellan took observations in $i$-band from few hours to 6 days with a median of 22 mag as upper limit and reported one candidate \citep{2024GCN.36244....1P,2024GCN.36267....1P}. PS1 took data 8h post $T_0$ and during 2h and the night after, and covered a sky region totalling 66\% of the sky localization area in $i,z$-bands with respective depth of 20 mag in $i$ and 19.6 in $z$ the first night and of 20.5 mag in $i$ and 20.5 in $z$ the first night \citep{2024GCN.36258....1S,2024GCN.36303....1S}. LAST collaboration reported a total coverage of the sky localization area of 70 \% without filter with an upper limit of 18.5 \citep{2024GCN.36264....1K}. SAGUARO/CSS observed about 5 hours post $T_0$ 19.2\% of the sky localization area with an upper limit of 19.5 (Gaia G) \citep{2024GCN.36266....1H}. Finally PRIME started observation on 2023-04-23 17:01 with an upper limit of 21.5 in $J$-band and totalizing 21\% credible region of the bilby skymap. 

%\subsection{GRANDMA Follow-up of S240422ed}
%\subsubsection{Follow-up Overview}

%\begin{figure}
%    \centering
%    \includegraphics[width=\columnwidth]{Figures/S240422ed_GRANDMA_follow-up.pdf}
%    \caption{\textcolor{red}{TBD}}
%    \label{fig:S240422ed_GRANDMA_follow-up}
%\end{figure}

%\input{grandma_observations_all_counterparts.tex}
%\input{full_observations_description.txt}
%In response to the gravitational wave event S240422ed, the GRANDMA collaboration (Antier et al. 2020a) undertook a series of coordinated observational campaigns aimed at identifying and detecting optical counterparts, thanks to its operational platform SkyPortal (Coughlin et al. 2023) and associated partners. These efforts included tiling observations with the FRAM-Auger telescope and targeted galaxy observations with multiple telescopes across the globe. Additionally, GRANDMA followed up on specific counterpart candidates identified by Swift-XRT and other surveys. Despite comprehensive coverage and detailed analysis, no significant candidates were detected. The following section details the observations and results from each component of the campaign.

\subsubsection{Tiling}

%G, which features a 30cm ODK + CCD MII G4-16000 configuration and offers a field of view of 60' x 60', situated at the Pierre Auger Observatory in Malargue, Argentina. The FRAM-Auger observations took place from April 23, 2024, starting at 00:02:54 UTC, approximately 2.46 hours following the gravitational wave trigger time, and continued until 05:16:49 UTC on the same day. Covering 11.5\% of the localization region based on the Bilby skymap, two consecutive 2-minute-long exposures in the Johnson R filter were acquired at 132 individual pointings. After low-latency analysis procedures were initiated, comparing the acquired data against the PS1 catalog, no significant candidate was identified within the observed region down to the 16.7 magnitude in the R-band, which corresponds to the 5 sigma limit in the AB system. Thus, these observations did not yield any significant candidates indicative of an optical counterpart to S240422ed .(c/ analysis system). The data obtained from these observations, detailing the results and analysis, was publically reported in (c/ GCN 36284)

\subsubsection{Galaxy Targeting}

\citet{2023GCN.33820....1C} reported, 15 min post-detection, a list of promising about 4000 host galaxy candidates for the event (in the first skymap) and around 1500 host galaxy in the LVK S240422ed-4-Update sky localization \citep{2024GCN.36243....1C}. GRANDMA conducted a search of at least 45/2245 galaxies compatibles with the Bilby skymap \citep{2023GCN.34130....1L} using ASTEP, LesMakes-T60, KAO, TRT in Thailand, and amateurs telescopes from Kilonova-catcher (KNC-BBO,KNC-T30) \citep{2024GCN.36299....1D}. \citet{2024GCN.36263....1K} also searched for counterpart located in galaxies coincident with the 3D localization and found two optical counterpart candidates but finally rejected. Subaru/MOIRCS conducted near-infrared imaging observations of 105 galaxies contained in the bayestar localization area in $Y$ and $Ks$ \citep{2024GCN.36265....1M,2024GCN.36302....1M}. Finally, Las Cumbres Observatory targeted potential host galaxies  in a total of 17 deg$^2$ \citep{2024GCN.36480....1P}.

\subsubsection{Candidate Follow-up}
Overall, among the several candidate counterparts detected, approximately 46 were identified preliminarily. The optical counterparts, detected by DECam (28 candidates), GECKO (10 candidates), and Mallegan (3 candidates) instruments, characterized by faint magnitudes ranging from ~19 to 21 (AB), were accompanied by limited photometric data. 

X-ray follow-up investigations were conducted by, MAXI/GSC, Swift/XRT, EP-WXT, and EP-FXT. A rigorous vetting process was conducted involving numerous observations to estimate redshifts and analyze archival surveys for photometric activity preceding the merger event. For instance, SAGUARO performed a detailed vetting analysis of DECam candidates \citep{2024GCN.36285....1R}, and several teams vetted S240422ed's candidate counterparts in the optical band including WINTER, DECaPSs2, NOT, GRANDMA, GECKO, BLACKGEM, MeerLICHT, ENGRAVE/VLT, MOSFIRE, ANU 2.3m, 2.5m PRL Telescope, GOTO, BOOTES-4/MET, Swope, DDOTI, DBSP, GMOS-S, P200, and analysis of archival data in radio. Despite these extensive analyses no confirmed counterpart was identified or conclusively linked to S240422ed. Table 6 outlines all the candidate counterparts to S240422ed.

In Table~\ref{tab:Xandradio} below, are reported all transients: the text is extracted from all GCN reports related S240422ed.
\begin{center}

\scalefont{0.5}
%\begin{longtable}{|p{1.5cm}|p{0.5cm}|c|p{4.5cm}|p{2.5cm}|}
%\begin{longtable}{|p{2cm}|p{0.6cm}|c|p{6.6cm}|p{3.5cm}|}
\begin{longtable}{|p{2cm}|p{0.75cm}|p{1.8cm}|>{\raggedright\arraybackslash}p{13cm}|}
\hline
Candidate & GCNs & Discovery Date & Findings and Comments \\
\hline
\hline
\multicolumn{4}{|c|}{Optical-band} \\
\hline

AT 2024hdr & 3 & 2024-04-22 23:42:30.196 & \texttt{ATLAS} forced photometry detections extending over 200 days before the GW event examined. Host galaxy z=0.0416 based on weak emission line features. Considering the characteristics of nuclear transients and ruling out their association with the GW event: the candidate is not a GW counterpart. Not GW Counterpart  \\
\hline

AT 2024hdo & 5 & 2024-04-22 23:48:39.928 & Associated (Pcc = 0.002) with host galaxy \texttt{WISEJ080327.75-260039.2} from \texttt{GLADE} at z=\(0.09\pm0.02\) (D=$\sim$404 Mpc). Inconsistent host galaxy photometric redshift with distance inferred from the GW event. 
The spectrum revealed strong emission lines at z=0.0658 and broad P-Cygni H-alpha emission consistent with Type II SN at the same z. Unlikely GW association \\
\hline

AT 2024hdq & 3 & 2024-04-22 23:50:15.764 & Nuclear transient. \texttt{ZTF} detections indicated periodic behavior since 2022. Not GW Counterpart \\
\hline

AT 2024hfj & 1  & 2024-04-22 23:51:46.851
&  One candidate host is situated within 1 arcmin: \texttt{WISEA J075010.62-261059.0}. Source was not fast fading ($>$0.2 mag/day) and did not exhibit significant color evolution.  Likely unrelated to the GW event \\
\hline

AT 2024hdp & 3 &  2024-04-22 23:53:19.570 & Associated (Pcc = 0.001) with the host galaxy \texttt{WISEJ080210.31-271529.7} from \texttt{GLADE} at z = \(0.09 \pm 0.02 \) (D=$\sim$411 Mpc). \texttt{ATLAS} forced photometry detections were recorded $\sim$3-18 days before the GW event. Unlikely GW Association \\
\hline

AT 2024hdw & 1 & 2024-04-23 00:09:22.464 & Two candidate hosts situated to the north within 1 arcmin: \texttt{WISEA+J080141.03-292637.1} and \texttt{WISEA+J080142.38-292621.8}. Source was not fast fading ($>$0.2 mag/day) and did not exhibit significant color evolution. Likely unrelated to the GW event \\
\hline

AT 2024hdk & 2 & 2024-04-23 00:52:49.715 & Three marginal ($\sim$ $\sigma$) \texttt{ATLAS} forced photometry detections observed $\sim$3-5 days before the GW event, which ruled the candidate unrelated to the GW event alongside color and spectroscopic information. Unrelated to \texttt{S240422ed}. \\
\hline

AT 2024hel & 1  & 2024-04-23 00:59:08.448 &  Candidate host is identified at a small offset: \texttt{WISEA J083612.37-164424.5}. Source was not fast fading ($>$0.2 mag/day) and did not exhibit significant color evolution.  Likely unrelated to the GW event. \\
\hline

AT 2024hfr & 3 & 2024-04-23 01:06:53.381 & Source located within 0.3 arcsec from the object WISEA J084103.91-183532.4. Galaxy 2MASS photometric z=0.049. No indication of a fast (superior at 0.3 mag/day) rise/fade in its light curve based on preliminary photometry. Likely not associated with GW event. \\
\hline

AT 2024hek & 1  & 2024-04-23 01:23:00.960 & Candidate host is identified at a small offset, known as \texttt{WISEA J082713.16-201301.5}. Source was not fast fading ($>$0.2 mag/day) and did not exhibit significant color evolution. Likely unrelated to the GW event.  \\
\hline

AT 2024hdn & 5 & 2024-04-23 01:23:01.223 & Color ((r-J)corr $\sim$\(0.15 \pm 0.1 \ \rm mag \) and five \texttt{ATLAS} forced photometry detections 4-24 days before the GW event, indicated an unrelated status to the GW event. The 2D spectrum showed two emission lines at $\sim$10161 and $\sim$10679 \AA, observed to originate from a star formation knot or background galaxy, north of the candidate host. Not GW Counterpart. \\
\hline

AT 2024hdm & 4 & 2024-04-23 01:23:01.223 & Color ((r-J)corr$\sim$\(0.15\pm 0.1 \ \rm mag\)) and marginal (<5$\sigma$) \texttt{ATLAS} forced photometry detections recorded \~{}2 months - 4 days before the GW event indicate an unrelated status to the GW.
The candidate was found to be close to a galaxy with photometric z=$\sim$0.048 (suggests abs. mag.(Mr)= -16 at the time of discovery. Candidate detected with an apparent mag.(mJ)= \(20.4 \pm 0.1 \). Observations suggest consistency of transient with SN event. Disfavored as GW counterpart.  \\
\hline

AT 2024hfs & 3 & 2024-04-23 01:23:01.223 & No indication of a fast ($>$ 0.3 mag/day) rise/fade in its light curve based on preliminary photometry. Likely unrelated to the GW event.  \\
\hline

AT 2024hit & 1 & 2024-04-23 01:32:17.953 & Transient was not recovered after subtracting \texttt{VISTA} reference images. Spectrum obtained with \texttt{DBSP} between 3800-9000\AA matches that of an S0 galaxy at z=\(0.060 \pm 0.003\),(z is $\sim$3-$\sigma$) away from the GW mean distance). Possible spurious detections were observed in \texttt{DECam} images at the location of \texttt{AT2024hit}. Hence, it is likely \texttt{AT2024hit} is unrelated to \texttt{S240422ed}. Not GW Counterpart. \\
\hline

AT 2024hfo & 2 & 2024-04-23 03:02:10.736 & \texttt{ATLAS} forced photometry pre-detections observed with (SNR$\sim$10-100). No \texttt{ZTF} forced photometry pre-detections in the last 12 months. Lightcurve appears flat and source is offset from its host galaxy. No available redshift information. Disfavored as GW counterpart. \\
\hline

AT 2024hdl & 6 & 2024-04-23 03:26:35.635 & Marginal ($\sim$4-$\sigma$) \texttt{ATLAS} forced photometry detection recorded $\sim$4 days before the GW event. Multiple \texttt{ATLAS} forced photometry pre-detections found at the source location (SNR $\sim$4-120). No \texttt{ZTF} forced photometry pre-detections in the last 12 months. The source is estimated to have an abs. mag.=$\sim$-17.3 mag in the \texttt{DECam} z-band using \texttt{NED} photometric z=0.0552 (D=255 Mpc). Host galaxy z=0.0416.
A weak source was visible near \texttt{AT2024hdl}, with a host z=0.0659 based on AGN features. The light curve appears flat. 
%Detections before the GW event ruled the candidate unrelated to the GW event alongside color and spectroscopic information. 
 Disfavored as GW counterpart. Ruled out by the \texttt{ZTF} based on \texttt{ATLAS} force photometry. \\
\hline

AT 2024hiw & 1 & 2024-04-24 00:10:30.996 & Source is positioned 2.3 arcsec away from a galaxy.  \\
\hline

AT 2024hiu & 1 & 2024-04-24 01:01:31.429 & Source located 3.8 arcsec away from a galaxy. Based on \texttt{Gaia DR3} data and visual inspection: the source is presumed to be either a star superimposed on top of the galaxy or located very close to it. Check GCN 36351 \\
\hline

AT 2024hfq & 2  & 2024-04-24 00:22:27.689 &  Source located within 0.4 arcsec of the galaxy \texttt{WISEA J082508.82-244345.9}. Does not show a fast ($>$0.3 mag/day) rise or fade in its light curve from preliminary photometry. Likely not associated with GW event. \\
\hline

AT 2024hga & 2 & 2024-04-24 05:51:11.232 & Multiple \texttt{ATLAS} forced photometry pre-detections identified at source location (SNR$\sim$5-60). No \texttt{ZTF} forced photometry pre-detections in the last 12 months. The source appears to be located in the nucleus, with a flat light curve. No photometric z information is available. Color index (i-z)=$\sim$0, indicating little variation between these wavelengths.  Disfavored as GW counterpart. \\
\hline
 
AT 2024hgb & 2 & 2024-04-24 05:54:56.736 & Multiple \texttt{ATLAS} forced photometry pre-detections identified at source location (SNR$\sim$4-50). No \texttt{ZTF} forced photometry pre-detections in the last six months. The source is estimated to have an abs. mag.=-19.8 in z-band using a \texttt{NED} photometric z=0.099940 (D=475 Mpc). Disfavored as GW counterpart.  \\
\hline

AT 2024hft & 2 & 2024-04-24 05:59:38.400 & Multiple \texttt{ATLAS} forced Photometry pre-detections noted (SNR$\sim$4-9). Abs. mag.=$\sim$ -18.9 mag in z-band using \texttt{NED} photometric z=0.091 (D=430 Mpc). No \texttt{ZTF} forced photometry pre-detections in the last 12 months.
Source located within 0.4 arcsec of the galaxy \texttt{WISEA J082508.82-244345.9}. Disfavored as GW counterpart. \\
\hline

AT 2024hfu & 3 & 2024-04-24 06:03:25.632 & Source is offset from its host galaxy, with a \texttt{NED} photometric z=0.092748. Detected as \texttt{ZTF24aakvqha}, with approximated abs. mag.=-18 in the r-band, using the same \texttt{NED} photometric z. Lack of evolution observed in the r-band. The host galaxy z=0.126. Disfavored as GW counterpart. \\
\hline

AT 2024hgc & 2  & 2024-04-24 06:06:15.840 & \texttt{ATLAS} forced photometry pre-detections observed (SNR$\sim$10-50). No \texttt{ZTF} forced photometry pre-detections in the last 12 months.
The source is located on the host galaxy, but no photometric z information is available. The light curve appears flat. Disfavored as GW counterpart. \\
\hline

AT 2024hfx & 5 & 2024-04-24 06:28:18.624 & Presence of a single clear archival detection at the position of \texttt{AT2024hfx} in \texttt{ZTF} on April 19, 2021, labeled as 'bogus' in the \texttt{ZTF} Stamp Classifier, coincided with the position of \texttt{AT2024hfx}. Multiple forced photometry detections were made by \texttt{ATLAS} and \texttt{ZTF}. \texttt{PS1 STRM} classified \texttt{AT2024hfx} as a QSO.
Consistent with a CV/dwarf nova event considering its rising lightcurve, faint host, and i-z color close to 0. Disfavored as GW counterpart. \\
\hline

AT 2024hgl & 2 & 2024-04-24 17:32:29.000 & Pre-detections confirmed in \texttt{ATLAS} forced photometry (SNR$\sim$4-19), and the latest detections on Jan. 7, 2024, and Mar. 18, 2024. Pre-detection observed in \texttt{ZTF} on Nov. 10, 2023. Inspection of the public
ATLAS archive shows no recent brightening or flaring at the source
location.
Despite pre-detections, no source was detected at the candidate position with 5-\(\sigma\) forced photometry by \texttt{GOTO-S}. Not GW counterpart. \\
\hline

AT 2024hbf & 1 & 2024-04-23 00:12:28.224 &  Multiple uncatalogued galaxies present within 1 arcmin, visible in \texttt{DECaPS} imaging. Galaxy \texttt{2MFGC 06268} is situated at a 2-arcmin offset with a spectroscopic z=0.059677, consistent within $\sim$3-$\sigma$) with the GW distance. Source was not fast fading ($>$0.2 mag/day) and didn't exhibit significant color evolution. Likely unrelated to the GW event. \\
\hline

GECKO24a & 1 & - &  Comparable brightness and decay rate with AT2017gfo-like kilonova during early epochs. In the vicinity of a galaxy association with $\sim$1.7 arcsec separation. Not excluded yet. \\
\hline

GECKO24b & 1  & - &  
Ruled out as a kilonova candidate due to brightness and/or decay rate. Unlikely GW association. \\
\hline
GECKO24c & 1  & - & Identified as a moving object. Unlikely GW association. \\
\hline

GECKO24d & 1 & - &  Identified as a moving object. Unlikely GW association. \\
\hline

GECKO24e & 1  & - &  Ruled out as a kilonova candidate due to brightness and/or decay rate. Unlikely GW association. \\
\hline

GECKO24f & 1  & - &  Ruled out as a kilonova candidate due to brightness and/or decay rate. Unlikely GW association. \\
\hline

GECKO24g & 1 & - &  Its comparable brightness and decay rate with \texttt{AT2017gfo}-like kilonova during the early epoch and vicinity of a galaxy makes it unlikely to be associated with the GW. Ruled out as kilonovae candidates due to brightness and/or decay rate. \\
\hline

GECKO24h & 1 & - &  Its comparable brightness and decay rate with \texttt{AT2017gfo}-like kilonova during the early epoch and vicinity of a galaxy makes it unlikely to be associated with the GW.  Excluded as a candidate by the historical \texttt{ATLAS} activity.  \\
\hline

GECKO24i (2024hea) & 1  & 2024-04-23 00:12:51.552 &  Its comparable brightness and decay rate with \texttt{AT2017gfo}-like kilonova during early epoch and vicinity of a galaxy makes it unlikely to be associated with the GW. \\
\hline

GECKO24j (2024heb) & 1  & 2024-04-23 00:12:51.552 & Its comparable brightness and decay rate with \texttt{AT2017gfo}-like kilonova during the early epoch and vicinity of a galaxy makes it unlikely to be associated with the GW. \\

\hline
Mag24a  & 6 & - & Estimate brightness in the i-band was estimated as 23.3 mag at 20240423.98 UT. It is offset 2.1” west and 8.6” north from its apparent host galaxy, \texttt{WISEA J075605.75-225400.0}, which has a photometric redshift of z = 0.049 (D=213 Mpc)(i.e., within the current Bilby measured volume).  At this distance with a Milky Way extinction of Ai = 0.748 mag and no host-galaxy extinction, the absolute magnitude is Mi = -14.1 mag.  At this distance, \texttt{Mag24a} is offset by 9.2 kpc from the assumed host galaxy in projection. Not GW Counterpart. \\
\hline
\hline
\multicolumn{4}{|c|}{X-ray bands}\\
\hline
EP240426a & 14 & 2024-04-26 07:18:25 & Preliminary forced photometry suggested possible faint detection. The location of the transient is close to the nucleus ($\sim$0.36 arcsec). Variability due to possible AGN activity could not be excluded. No transient was found within an uncertainty of 10 arcsec. 
Observation on Apr. 26 revealed a residual source close to the nucleus of galaxy \texttt{2MASX J08072584-2927344}, with a mag.(z)= \(21.3 \pm 0.2\). Together with archival K-band NIR data and radio data, this suggested the source is due to AGN activity. Not GW Counterpart. \\
\hline

S240422edX190 & 7  & 2024-04-24 05:52:25 & No transient source found within XRT error circle in any of the fields. XRT detected an object that was determined not to be a real X-ray object. Source’s PSF: not consistent with that expected for the reported count rate. Not GW Counterpart. \\
%check gcn 36311/36301
\hline

S240422edX61 & 10  & 2024-04-24 05:46:49 & Initially detected with a count rate of \(8.3 \pm 3.0 \)x\(10^-2\)ct \(s^-1\). Undetected in latest \texttt{Swift-XRT} observation, with a 3-$\sigma$) upper limit of 1.02 x\(10^-2\) ct \(s^-1\). Within a radius of 20 arcsec, no new source was detected at the 5-$\sigma$) depth of 16.6 mag in J-band infrared observations by \texttt{WINTER} and \texttt{NOT} telescopes. Not GW counterpart. \\
\hline

S240422edX101 & 8  & 2024-04-24 05:49:29 & No transient source or significant X-ray counterpart found within the XRT error circle. Not GW Counterpart. \\
%check gcn 36311/36301
\hline

S240422edX255 & 1  & - & Observation of \texttt{EP240426a} 28.8 ks after its discovery by \texttt{EP-FXT} revealed this source's detection located 6.2" from the localization reported by \texttt{EP-FXT}. The source did not surpass the \texttt{RASS} 3-$\sigma$) upper limit, indicating consistency with background noise levels. \\
\hline
\hline
\multicolumn{4}{|c|}{Infra-red Counterparts} \\
\hline
JGEM24a & 5  & 2024-04-23 & Discovered by NIR Y-band observations targeted for galaxy \texttt{GL080850-243120}. The optical spectrum of the potential host galaxy of \texttt{J-GEM24a} revealed spectral lines (Ca II H and K and Na I D) identified at z=0.055. This z was inconsistent with the estimated GW distance, \(170 \pm 42\) Mpc at the 2 $\sigma$ level. Source was not detected. Not GW Counterpart. \\

\hline
\hline
\caption{Summary of all optical and X-ray candidates found during S240422ed follow-up.}
\label{tab:Xandradio}
\end{longtable}

\end{center}

\section{Synthetic light curves of kilonovae}
\label{appendix-filter}
The Bulla 2019 - Anand 2021 model computes the synthetic kilonova light curves for 27 different bands. We compare these observations with the synthetic light curves to the observations by selecting the corresponding filters (see Table~\ref{tab:various-filter}).

\begin{table}[h]
    \centering
    \begin{tabular}{|c|c|}
    \hline
       Filter of observations  &  Filter of KN model\\
       \hline
       \hline
       $u$,$g$,$r$,$i$,$z$ & $u$,$g$,$r$,$i$,$z$ (sdss) \\
       \hline
       J & swope2::J \\
       \hline
       Johnson R & bessellr \\
       \hline
       Gaia G & gaia::g \\
       \hline
       TESS filter & tess \\
       \hline
       C/open & ps1:open \\
       \hline
       $L$ (GOTO) & gotol \\
       \hline
       $c$ (ATLAS filter) & atlasc \\
       \hline
       $o$  (ATLAS filter) & atlaso \\
       \hline
       XRT & - \\
       \hline
       q & - \\
       \hline
    \end{tabular}
    \caption{Filters of fields and corresponding ones in which the light curves from POSSIS are computed. These filters and their associated transmissions function can be found in \url{https://sncosmo.readthedocs.io/en/stable/bandpass-list.html} \citep{2016ascl.soft11017B}.}
    \label{tab:various-filter}
\end{table}

\section{Optimisation of the detection of the kilonova: additional filters}
\label{optimisation-appendix}
Below, please find Fig.~\ref{fig:opti-CcLG-filters} and~\ref{fig:opti-tess-filters}, observational strategy figures similar to Fig.~\ref{fig:observation-peak-time} and Fig~\ref{fig:opti-other-filters} but with additional filters. We show observations of S230518h in $tess$ filter separately in Fig.~\ref{fig:opti-tess-filters} as TESS follow-up is not included for GW230529, S230627c and S240422ed analysis but only for S230518h. We have decided to present this result as TESS covered almost 100\% of the peak time distribution.

\begin{figure*}
    \centering
    \includegraphics[width=\textwidth]{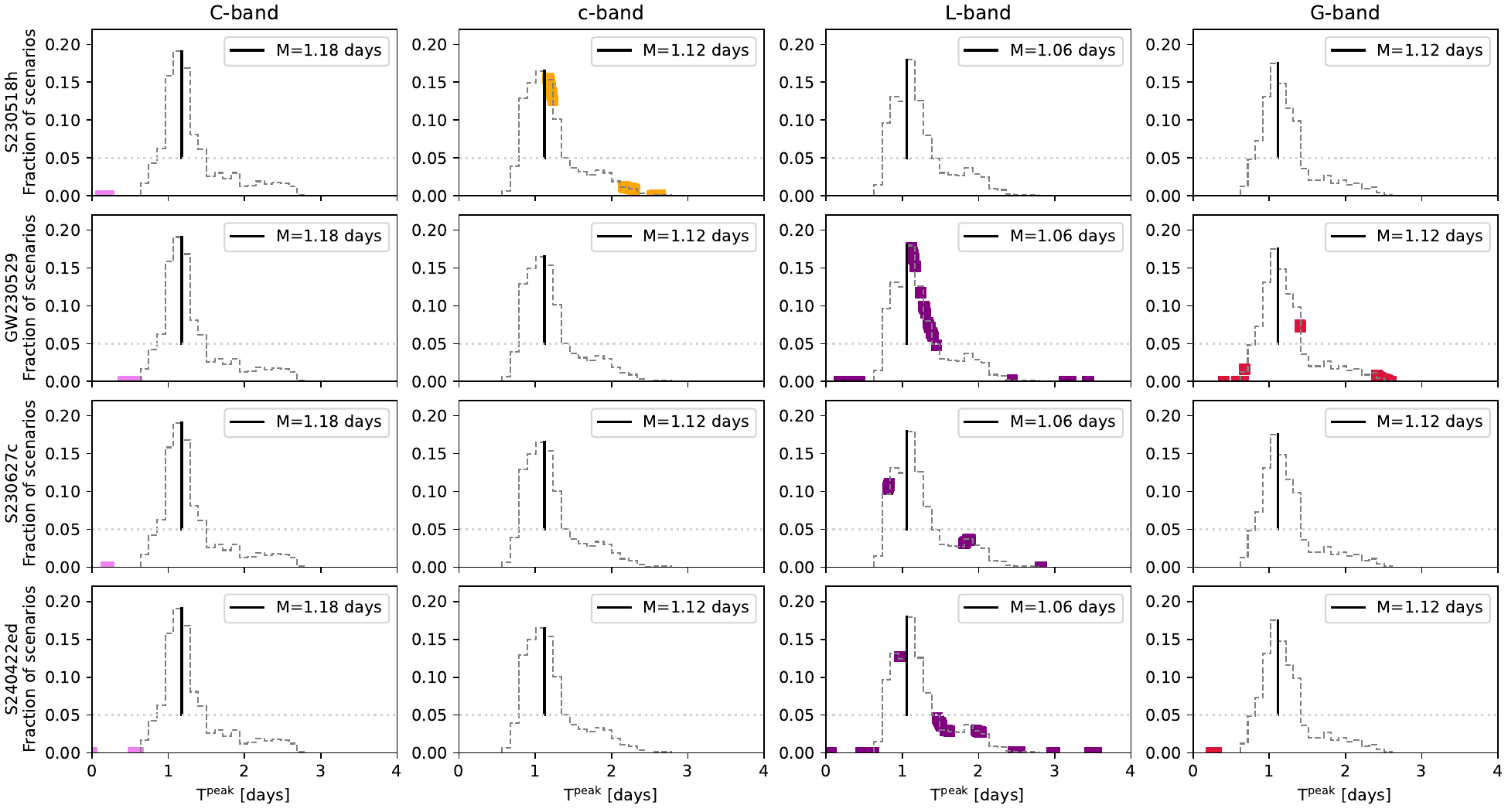}
    \caption{Comparison between the peak time luminosity of our kilonova population in $C$ (left column), $c$ (second column), $L$ (third column), and $G$ (fourth column)-bands and the time of optical observations for S230518h (first row), S230529ay (second row), S230627c (third row), S240422ed (fourth row). The dashed gray line represents the distribution of peak time considering all $m_{dyn}$-$m_{wind}$-$\theta$ scenarios. The solid black line represents the median of the peak time distribution considering only bins containing more than 5\% of the distribution. Observations of the community are shown in color squares. All observations in $c$-band are done by ATLAS.}
    \label{fig:opti-CcLG-filters}
\end{figure*}

\begin{figure*}
    \centering
    \includegraphics[width=0.3\textwidth]{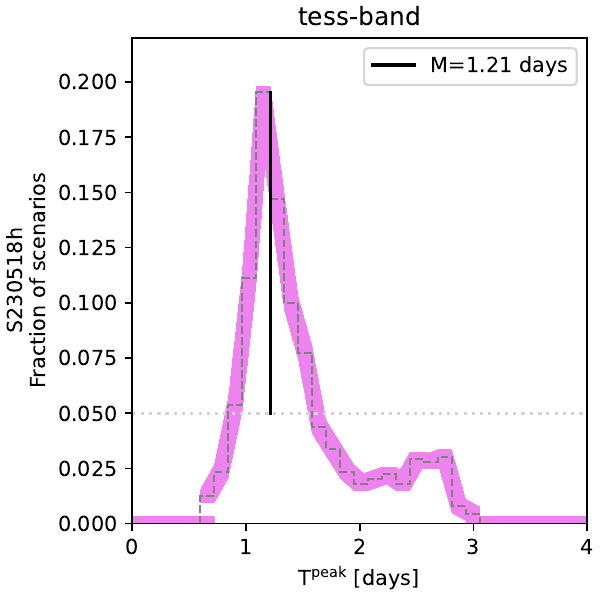}
    \caption{Comparison between the peak time luminosity of our kilonova population in $tess$ filter and the time of optical observations for S230518h. The dashed gray line represents the distribution of peak time considering all $m_{dyn}$-$m_{wind}$-$\theta$ scenarios. The solid black line represents the median of the peak time distribution considering only bins containing more than 5\% of the distribution. Observations of TESS are shown in color squares.}
    \label{fig:opti-tess-filters}
\end{figure*}

\section{Ejecta masses}
\label{appendix-ejecta-mass}
Below, please find additional ejecta masses plots using $SLy$ and $H4$ but with a fixed $\xi$ proportion of unbound material from the disk.

\begin{figure*}
    \centering
        \includegraphics[width=\textwidth]{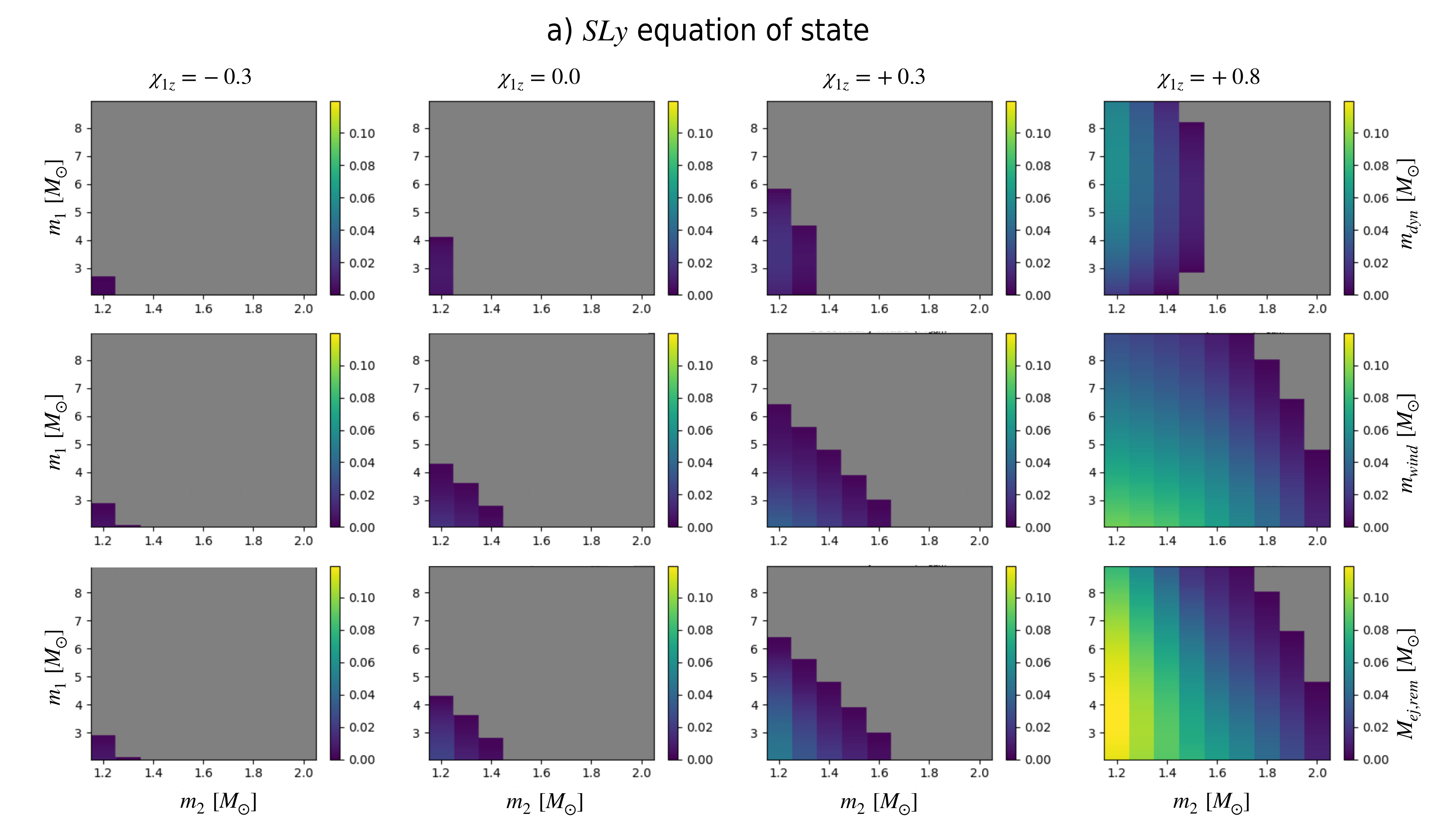}
        \hspace*{+0.2cm}
        \includegraphics[width=\textwidth]{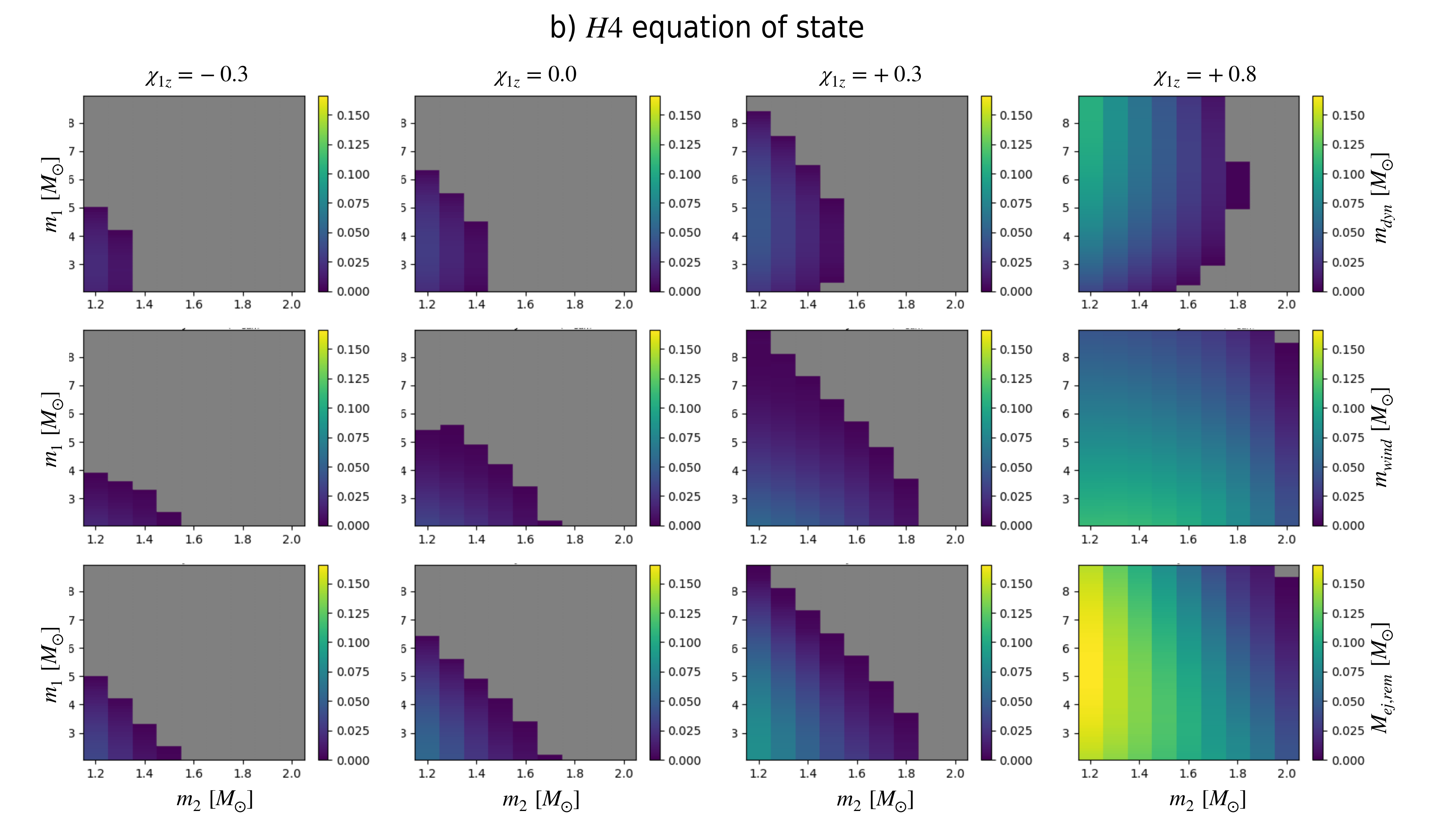}
    \caption{Ejecta masses (top, Dynamical middle Wind and bottom Total given a certain spin component of the black hole aligned with the orbital angular momentum. We consider no spin for the NS. $\xi$ the proportion of unbound material from the disk fixed to 0.3. a) Using $SLy$ equation of state of matter (for more compact NS) b) Using $H4$ equation of state of matter (for less compact NS).  }
    \label{fig:Map_ejecta_SLy_xi0-3}
\end{figure*}

\begin{figure*}
    \centering
        \includegraphics[width=\textwidth]{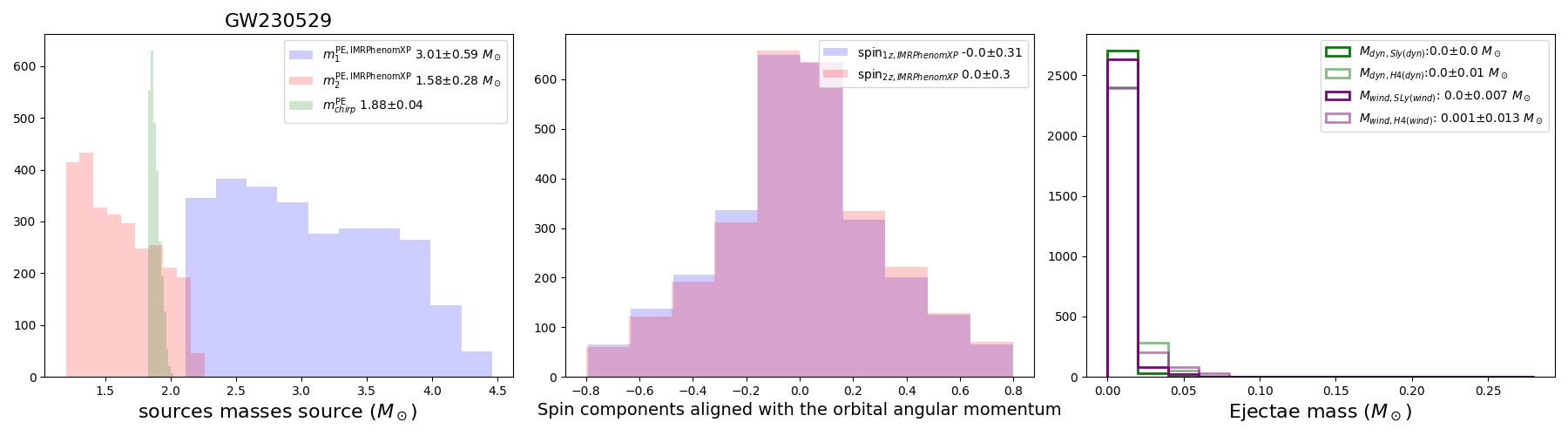}
    \caption{Parameter estimation of GW230529 from IMRPhenomXP waveform model: Masses distributions (left), spins components aligned with the orbital momentum (middle), resulting computation of ejecta mass (right).
    %Ejecta masses (Dynamical and Wind ejecta mass using $SLy$ and $H4$ equation of state of matter. We consider no spin for the NS. The fraction $\xi$ of the disk that is eventually unbound is calculated as a function of the mass ratio, as described in the main text. We show results a) Masses distributions from  IMRPhenomXP waveform model b) spins components aligned with the orbital momentum c) Computation of ejecta mass  
    }
    \label{fig:GW230629ay_PE}
\end{figure*}

Figure~\ref{fig:GW230629ay_PE} represents the posteriors distributions and computation of the Dynamical and Wind ejecta mass from posterior samples of masses and spins components aligned with the orbital momentum of GW230529 \citep{2024ApJ...970L..34A}.

\section{Coverage of GW skymaps}
\label{coverage-later-time}
\begin{figure*}
    \centering
    \includegraphics[width=0.45\linewidth]{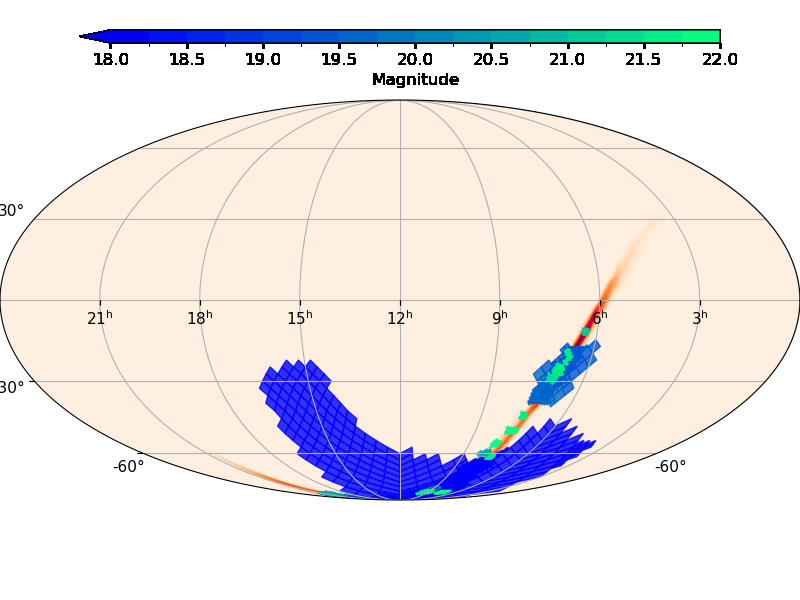}
    \includegraphics[width=0.45\linewidth]{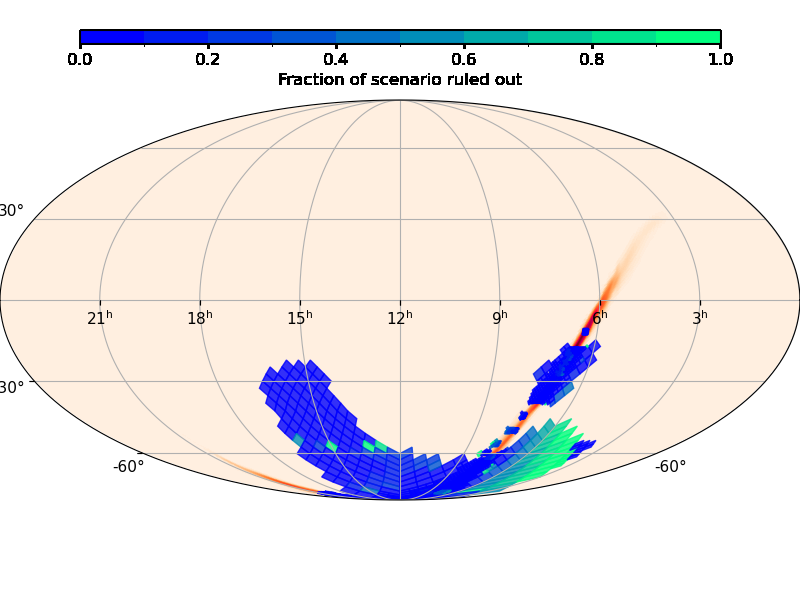}
    \includegraphics[width=0.45\linewidth]{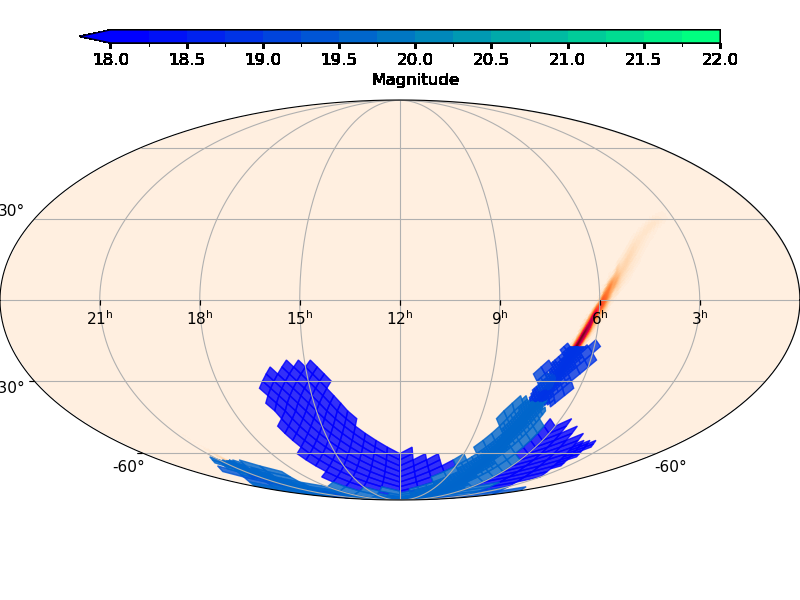}
    \includegraphics[width=0.45\linewidth]{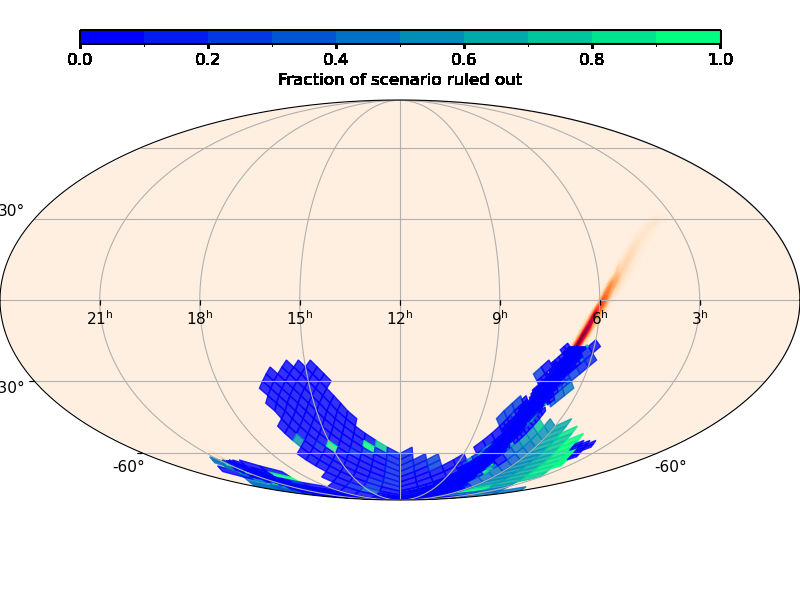}
    \caption{S230518h skymap with pixels colored by the observation' deepest magnitude covering them (left) and by the fraction of scenarios incompatible with observations covering them (right). The initial GW skymap is shown in reddish color. Observations are taken between 1 and 2 days (top) and between 2 and 6 days (bottom).}
    \label{fig:appendix-S230518h-skymaps}
\end{figure*}

\begin{figure*}
    \centering
    \includegraphics[width=0.45\linewidth]{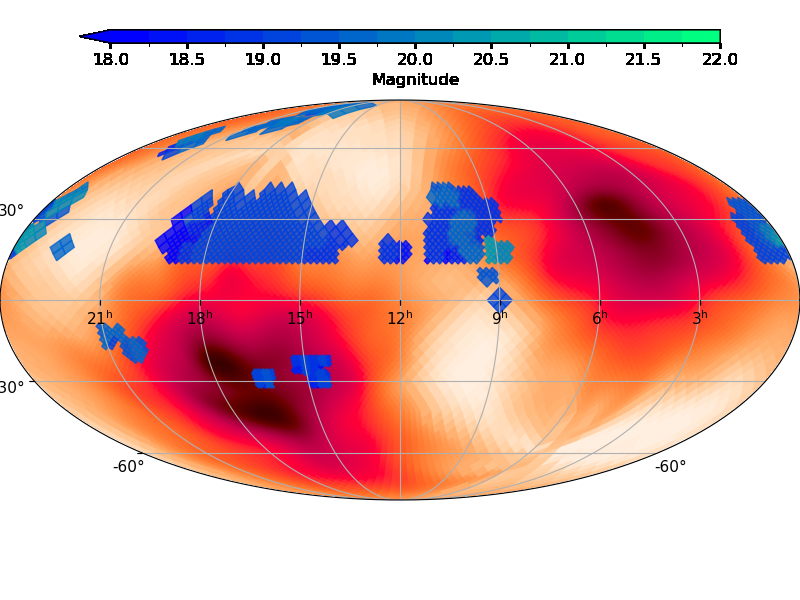}
    \includegraphics[width=0.45\linewidth]{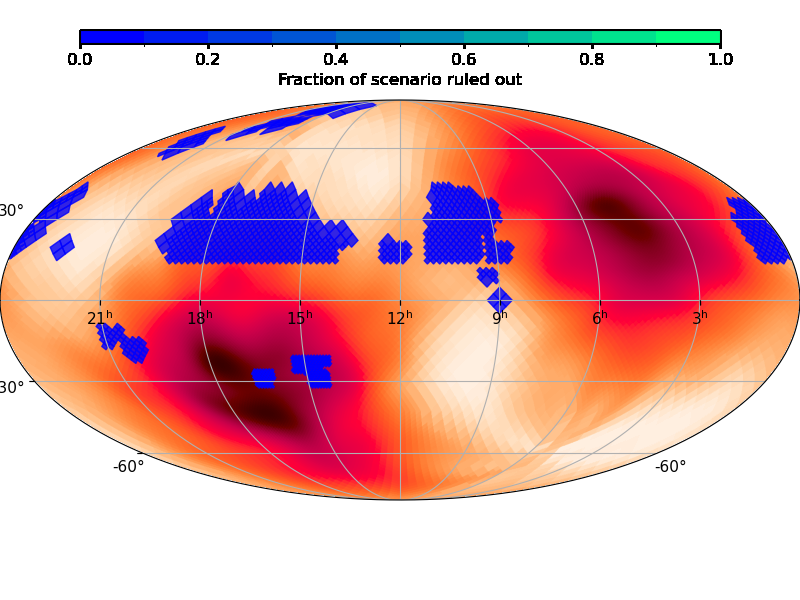}
    \includegraphics[width=0.45\linewidth]{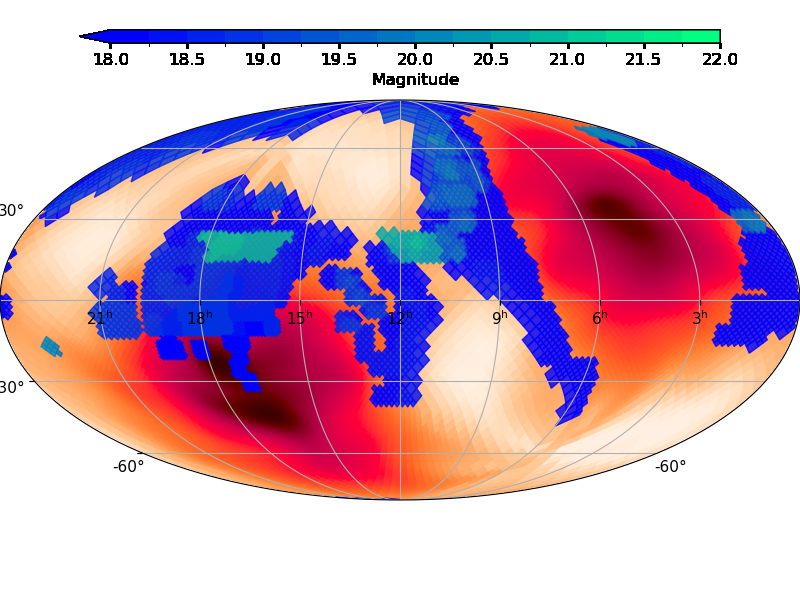}
    \includegraphics[width=0.45\linewidth]{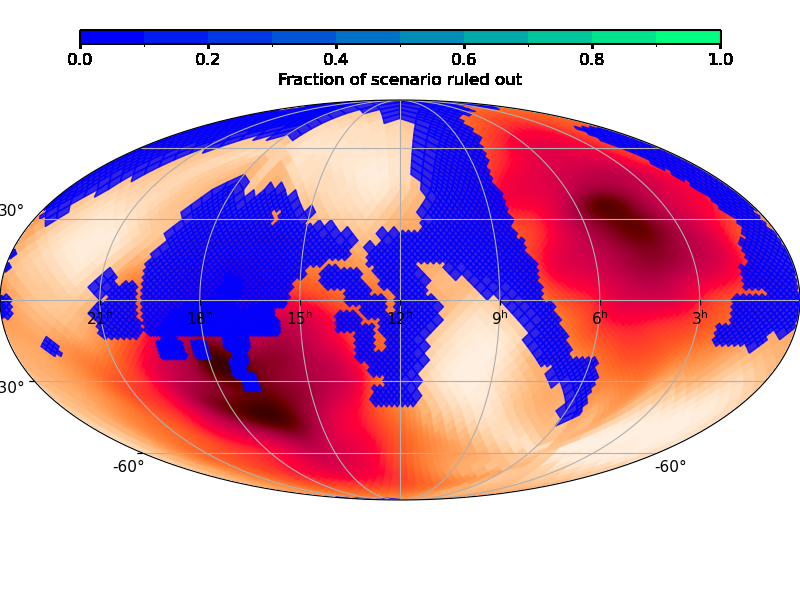}
    \caption{GW230529 skymap with pixels colored by the observation' deepest magnitude covering them (left) and by the fraction of scenarios incompatible with observations covering them (right). The initial GW skymap is shown in reddish color. Observations are taken between 1 and 2 days (top) and between 2 and 6 days (bottom).}
    \label{fig:appendix-GW230529ay-skymaps}
\end{figure*}

\begin{figure*}
    \centering
    \includegraphics[width=0.45\linewidth]{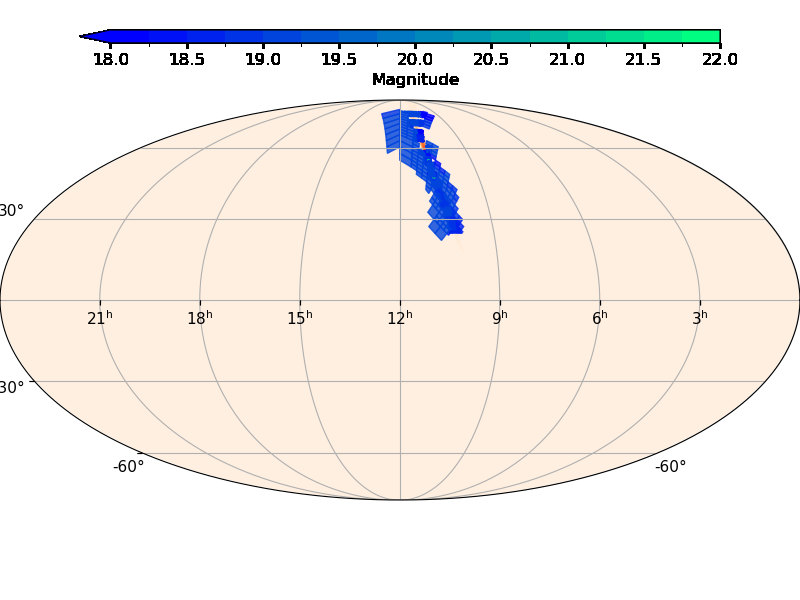}
    \includegraphics[width=0.45\linewidth]{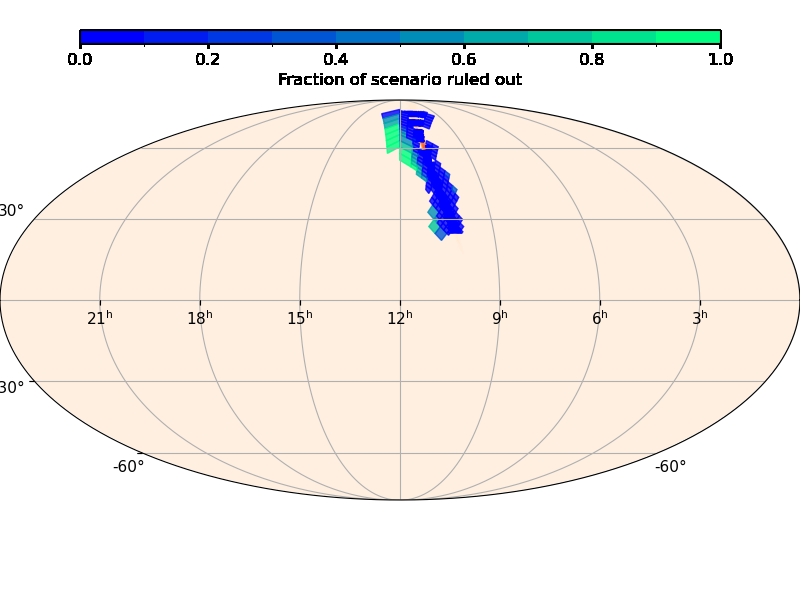}
    \includegraphics[width=0.45\linewidth]{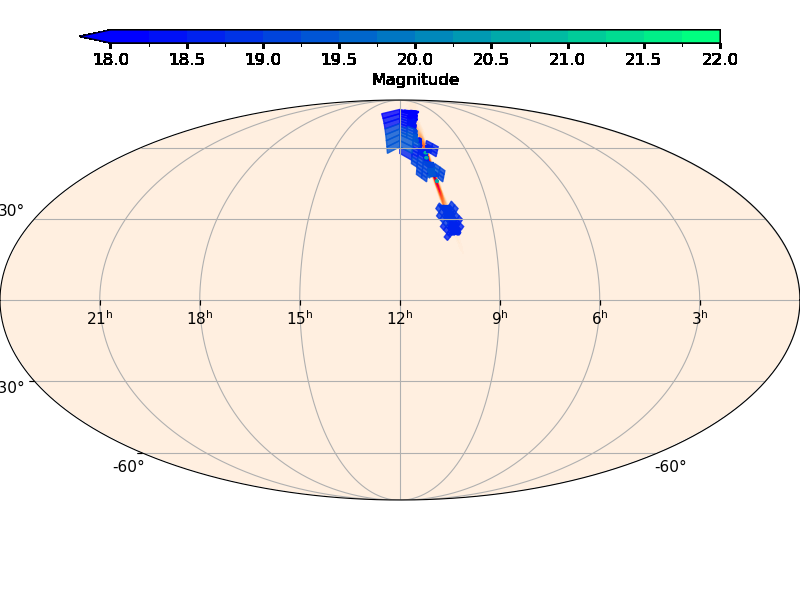}
    \includegraphics[width=0.45\linewidth]{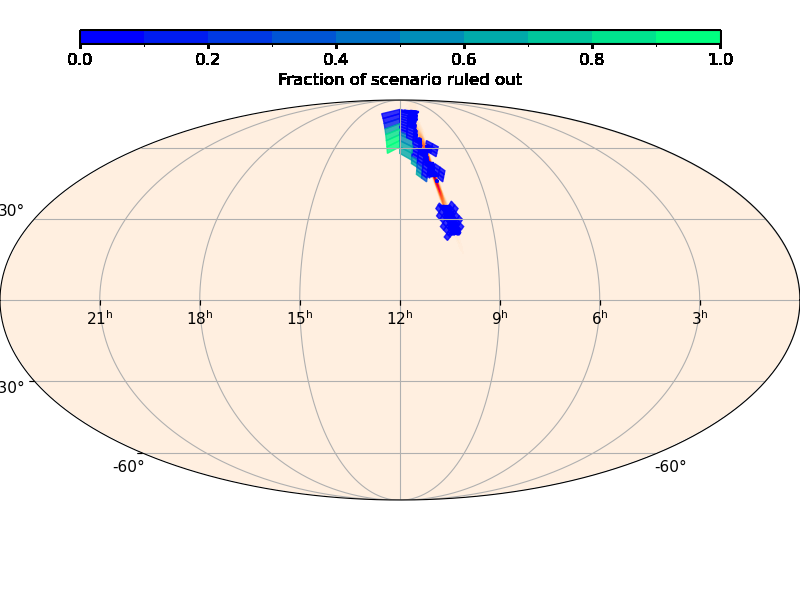}
    \caption{S230627c skymap with pixels colored by the observation's deepest magnitude covering them (left) and by the fraction of scenarios incompatible with observations covering them (right). The initial GW skymap is shown in reddish color. Observations are taken between 1 and 2 days (top) and between 2 and 6 days (bottom).}
    \label{fig:appendix-S230627c-skymaps}
\end{figure*}

\begin{figure*}
    \centering
    \includegraphics[width=0.45\linewidth]{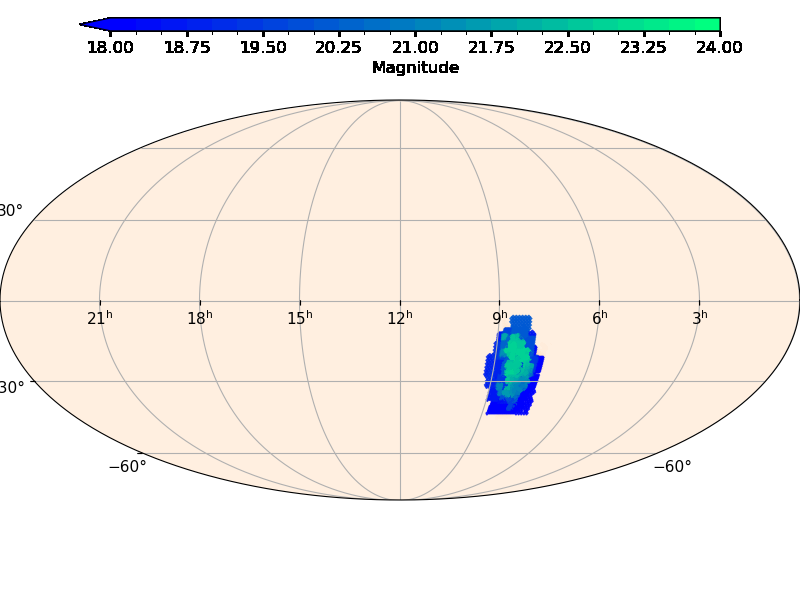}
    \includegraphics[width=0.45\linewidth]{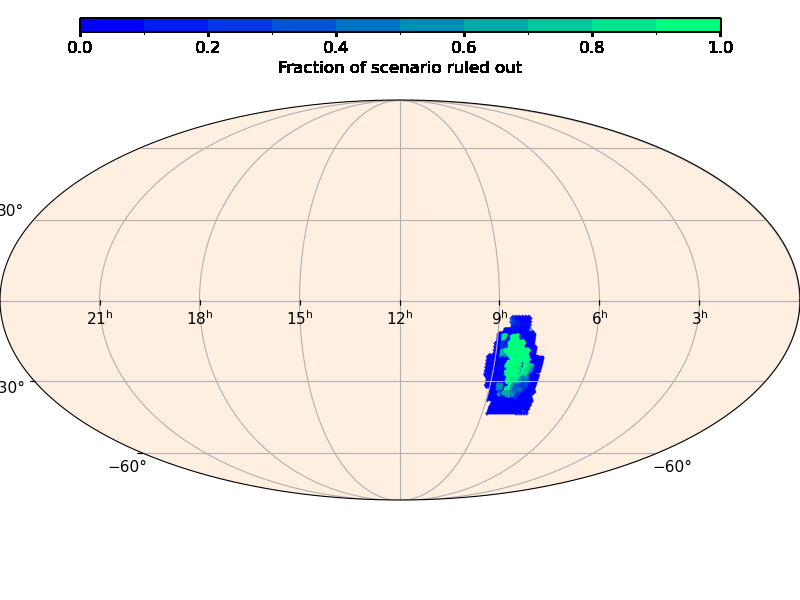}
    \includegraphics[width=0.45\linewidth]{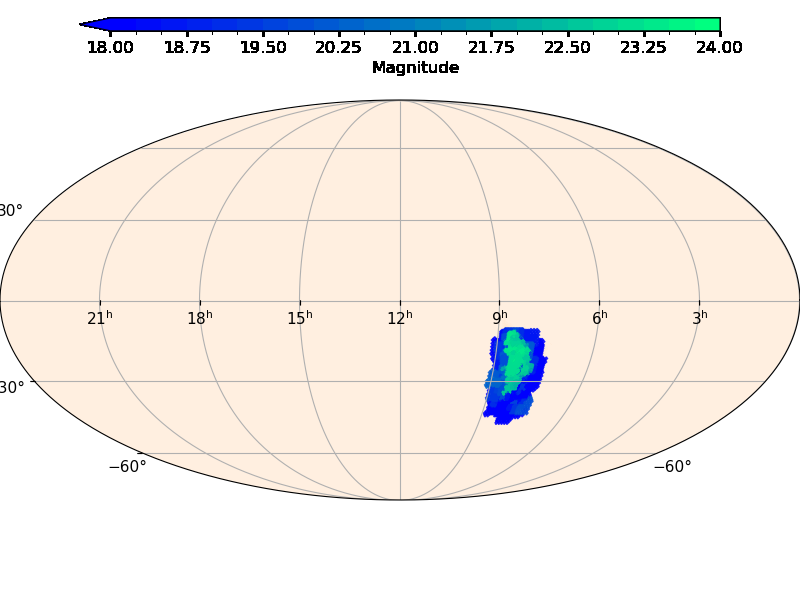}
    \includegraphics[width=0.45\linewidth]{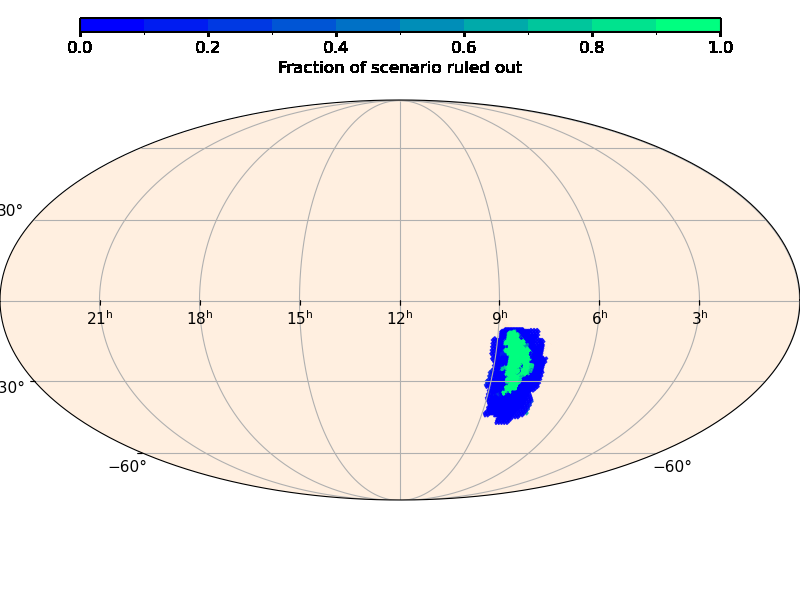}
    \caption{S240422ed skymap with pixels colored by the observation's deepest magnitude covering them (left) and by the fraction of scenarios incompatible with observations covering them (right). Observations are taken between 1 and 2 days (top) and between 2 and 6 days (bottom) and fully cover S240422ed skymap.}
    \label{fig:appendix-S240422ed-skymaps}
\end{figure*}

Below, please find the GW skymaps with pixels colored by the observation's deepest magnitude covering them and by the fraction of scenario incompatibles with observations for the time taken between 1 and 2 days and 2 and 6 days. Fig.~\ref{fig:appendix-S230518h-skymaps} corresponds to S230518h, Fig.~\ref{fig:appendix-GW230529ay-skymaps} to GW230529, Fig.~\ref{fig:appendix-S230627c-skymaps} to S230627c and Fig.~\ref{fig:appendix-S240422ed-skymaps} to the low-significance candidate S240422ed.
%\subsection{S230518h}

\end{document}